\documentclass[12pt, twoside, openright,a4paper]{book}
\usepackage{amsmath,amsfonts,bm}
\usepackage{graphicx}
\newcommand{\be}{\begin{equation}}
\newcommand{\bs}{\begin{split}}
\newcommand{\phd}{\phantom{\dagger}}

\def\vet{\mathbf}
\def\uar{\uparrow}
\def\dar{\downarrow}

\def\a{\alpha}
\def\b{\beta}
\def\g{\gamma}
\def\G{\Gamma}
\def\d{\delta}
\def\D{\Delta}
\def\e{\epsilon}
\def\eps{\varepsilon}

\def\t{\tau}
\def\r{\rho}
\def\w{\omega}
\def\W{\Omega}
\def\l{\lambda}

\def\s{\sigma}

\usepackage{fancyhdr}
\title{\bf Dynamics of Shuttle Devices}

\author{ Andrea Donarini}

\begin{document}

%\maketitle

\begin{titlepage}

$\,$ \vspace{5cm}
\begin{center}
\LARGE{ Dynamics of Shuttle Devices}
\end{center}
\vspace{.5cm}
\begin{center}
\Large{Andrea Donarini}
\end{center}
%\vspace{.5cm}
\begin{center}
Ph.D. Thesis\\
{\it Technical University of Denmark}
\end{center}
\vspace{1.5cm}
\begin{center}
\end{center}

\newpage
\thispagestyle{empty}
\phantom{phd}
\newpage
\thispagestyle{empty}

\noindent {\huge \bf Foreword}

\vspace{2cm}

This thesis is submitted in candidacy for the Ph.D. degree within
the \emph{Physics Program} at the Technical University of Denmark.
The thesis describes part of the work that I have carried out
under supervision of Antti-Pekka Jauho from the Department of
Micro and Nanotechnology.

I'm grateful to my supervisor for the very stimulating atmosphere
that he has been able to create and maintain within his group and
also for his gentle but firm guidance. I want to thank all the
members of the Theoretical Nanotechnology Group at the MIC
Department and in particular Dr. Tom\'a\v{s} Novotn\'y and
Christian Flindt for the close and intense collaboration that has
generated a large part of the results presented in this thesis.

I would like to thank Prof.~Timo Eirola for having introduced me
to the subject of the iterative numerical methods and for his
precious help in the solution of some of the numerical problems I
encounter in my research. I want also to thank Dr. Tobias Brandes
and Neill Lambert for the nice physics discussions I could have
with them during the period I spent collaborating with them in
Manchester. I want to thank Christian Flindt also for the
enthusiasm and the very positive attitude that he has been always
able to spread within the ``Shuttle Group''. He has also been an
indefatigable reader of the proofs and deserves many thanks for
that. Finally I also want to thank my family for their far but
constant support and all the Italian friends in Copenhagen for
being very patient and encouraging me during the writing period.

\begin{flushright}
\vspace{2cm}
{Lyngby, September 3, 2004}\\
\vspace{.5cm}
{Andrea Donarini}
\end{flushright}

\newpage
\thispagestyle{empty} \phantom{phd}
\newpage
\thispagestyle{empty}

\noindent {\huge \bf Preface}
 \vspace{2cm}

Much interest has been drawn in recent years to the concept and
realization of Nanoelectromechanical systems (NEMS). NEMS are
nanoscale devices that combine mechanical and electrical dynamics
in a strong interplay. The shuttle devices are a particular kind
of Nanoelectromechanical systems.  The characteristic component
that gives the name to these devices is an oscillating quantum dot
of nanometer size that transfers electrons one-by-one between a
source and a drain lead. The device represents the nano-scale
analog of an electromechanical bell in which a metallic ball
placed between the plates of a capacitor starts to oscillate when
a high voltage is applied to the plates.

This thesis contains the description and analysis of the dynamics
of two realizations of quantum shuttle devices. We describe the
dynamics using the Generalized Master Equation approach: a
well-suited method to treat this kind of open quantum systems. We
also classify the operating modes in three different regimes: the
tunneling, the shuttling and the coexistence regime. The
characterization of these regimes is given in terms of three
investigation tools: Wigner distribution functions, current and
current-noise. The essential dynamics of these regimes is captured
by three simplified models whose derivation from the full
description is possible due to the time scale separation of the
particular regime. We also obtain from these simplified models a
more intuitive picture of the variety of different dynamics
exhibited by the shuttle devices.

\begin{flushright}
\vspace{2cm}
{Lyngby, October 10, 2004}\\
\vspace{.5cm}
{Andrea Donarini}
\end{flushright}
\end{titlepage}

\newpage
\thispagestyle{empty}
\phantom{phd}
\newpage
\thispagestyle{empty}

\tableofcontents

\clearpage{\pagestyle{empty}\cleardoublepage}

%%%%%%%%%%%%%%%%%%%%%%%%%%%%%%%%%%%%%%%%%%%%%%%%%%%%%%%
%               INTRODUCTION
%%%%%%%%%%%%%%%%%%%%%%%%%%%%%%%%%%%%%%%%%%%%%%%%%%%%%%%
\chapter{Introduction}

In this chapter we give a short introduction to the world of
nanoelectromechanical systems. We then focus our attention on a
particular kind of device called electron shuttle. We sketch the
basic operating regime and give an overview of the different
theoretical models that have been proposed to describe the
dynamics of such devices. We report on the two main realizations of
shuttle devices and close the chapter with an outline of the
contents of this thesis.

\section{NEMS}

Much interest has been drawn in recent years on the concepts and
realization of Nanoelectromechanical systems (NEMS). NEMS are
nanoscale devices that combine mechanical and electrical dynamics
in a strong interplay. This property makes them interesting both
from a technological and fundamental point of view. They are
extremely sensitive mass and position detectors. Due to their very
high mechanical frequency one can even think of using them as the
basis for new form of mechanical computers. From the point of view
of fundamental research they represent extremely good tools to
probe directly the basic quantum mechanical laws. They could
represent the first man-made structures on which the mechanical
zero point fluctuation can be detected. They also rise the
question on the limiting dimension for persistence of mechanical
coherence. In general one of the fascinating aspects of these
objects is their mesoscopic character: they share with the
macroscopic world the large number of atoms of which they are made
(typically of millions of atoms) but on the other hand their
behaviour is (or should be) already significantly determined by
quantum mechanics.

\section{A new transport regime}

The shuttle devices are a particular kind of NEMS.  The
characteristic component that gives the name to these devices is
an oscillating object of nanometer size that transfers electrons
one-by-one between a source and a drain lead. The device
represents the nano-scale analog of an electromechanical bell in
which a metallic ball placed between the plates of a capacitor
starts to oscillate when a high voltage is applied to the plates.
The oscillations are sustained by the external bias that pumps
energy into the mechanical system: when the ball is in contact
with the negatively biased plate it gets charged and the
electrostatic field drives it towards the other capacitor plate
where the ball releases the electrons and returns back due to the
oscillator restoring forces\footnote{Due to the large amount of
electrons in this macroscopic realization the ball gets positively
charged at the second plate by loosing some extra electrons and
the restoring force contains also an electrostatic component. The
system is perfectly symmetric under commutation of charge sign.}
and the cycle starts again.

In the first proposal \cite{gor-prl-98} of a shuttle device  the
movable carrier is a metallic grain confined into a harmonic
potential by elastically deformable organic molecular links
attached to the leads. The transfer of charge is governed by
tunneling events, the tunneling amplitude being modulated by the
position of the oscillating grain. The exponential dependence of
the tunneling amplitude of the grain position leads to an
alternating opening and closing of the left and right tunneling
channels that resembles the charging and discharging dynamics of
the macroscopic analog.

Different models for shuttle devices have been proposed in the
literature since this first seminal work by Gorelik {\it et al.}\
\cite{gor-prl-98}. The mechanical degree of freedom has been
treated classically (using harmonic
\cite{isa-phb-98,wei-epl-99,nis-prb-01,pis-prb-04} or more
realistic potentials \cite{nor-prb-04}) and quantum mechanically
\cite{arm-prb-02,nov-prl-03,fed-prl-04}. Armour and MacKinnon
proposed a model with the oscillating grain flanked by two static
quantum dots \cite{arm-prb-02,nara,modena,christian}. More
generally the shuttling mechanism has been applied to Cooper pair
transport \cite{rom-prb-03,rom-preprint-04} and pumping of
superconducting phase \cite{gor-nat-01} or magnetic polarization
\cite{gor-preprint-04}.

The essential feature of the nano-scale realization is the
quantity transferred per cycle (a charge up to $10^{10}$ electrons
for a macroscopic bell) that is scaled  down to 1 quantum unit
(electron, spin, Cooper pair in the different realizations). We
can already guess the basic properties of the shuttle transport:

\begin{enumerate}

\item
Charge-position correlation: the shuttling dot loads the charge on
one side and transfers it on the other side, it releases it and
returns back to the starting point;

\item
Matching between electronic and mechanical characteristic times
(non-adiabaticity);

\item
Quantized current determined by the mechanical frequency;

\item
Low current fluctuations: the  stochasticity of the tunneling
event is suppressed due to  an interplay between mechanical and
electrical properties. The tunneling event is only probable
 at some particular short time periods fixed by the mechanical
dynamics (i.e. when the oscillating dot is close to a specific
lead).

\end{enumerate}
\section{Experimental implementations}

An experimental realization of the shuttle device has been
produced by Erbe {\it et al.}~\cite{erb-prl-01}. The structure
consists of a cantilever with a quantum island at the top placed
between source and  drain leads. Two lateral gates can set the
cantilever into motion via a capacitive coupling. An ac voltage
applied at these gates makes the cantilever vibrating and brings
the tip alternatively closer to the source or drain lead and thus
allows the shuttling of electrons.

%%%%%%%%%%%%%%%%%%%%%%%%%%%%%%%%%%
% FIGURE
%%%%%%%%%%%%%%%%%%%%%%%%%%%%%%%%%%
\begin{figure}
 \begin{center}
  \includegraphics[angle=0,width=.7\textwidth]{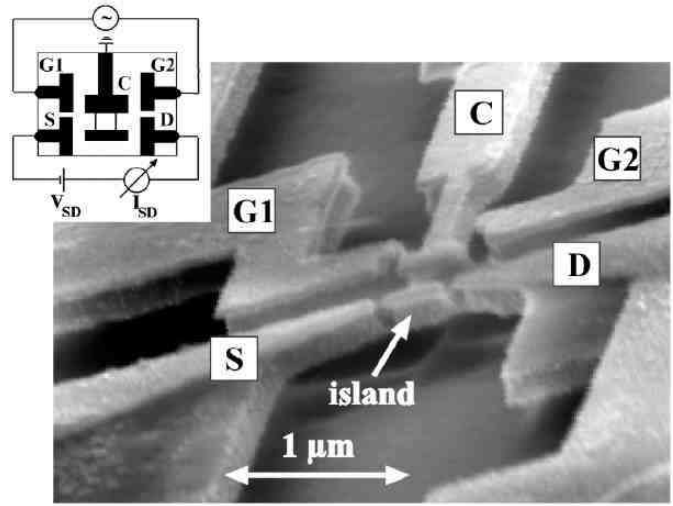}
    \caption{\small \textit{Electron micrograph of the nano-mechanical resonator.
    The cantilever (C) can be set into motion by applying an ac-voltage between the two gates G1
    and G2, and by applying a bias across the metallic tip of the cantilever (the island) electrons
    are shuttled from the source (S) to the drain (D). The picture is taken from \cite{erb-prl-01}.}
    \label{fig:erbe}}
 \end{center}
\end{figure}
%%%%%%%%%%%%%%%%%%%%%%%%%%%%%%%%%%
The device (shown in Fig. \ref{fig:erbe}) is built out of
silicon-on-insulator (SOI) materials (using Au for the metallic
parts) using etch mask techniques and optical and electron beam
lithography. The cantilever is $~1\mu m$ long and the
corresponding resonant frequency is of the order of $100 {\,\rm
MHz}$. The source electrode and the cantilever are at an average
distance of approximately $0.1 \,\mu m$. Shuttling experiments
have been performed by Erbe {\it et al.}~at different
temperatures. For experiments at $4.2 \,{\rm K}$ and $12 \,{\rm
K}$ they measured a pronounced peak in the current through the
cantilever for a driving frequency of approximately $120 \,{\rm
MHz}$ corresponding to the natural frequency of the first mode of
the oscillator. The peak in the current corresponds to a rate of
shuttle success of about one electron per 9 mechanical cycles. The
Erbe experiment is very close to the original proposal by Gorelik.
The only difference is in the external driving of the mechanical
oscillations. In the original proposal the bias was time
independent and the driving induced by the electrostatic force on
the charged oscillating island. The initially stochastic tunneling
events would eventually cause the shuttling instability and drive
the system into a self-sustained mechanical oscillation combined
with periodic charging and discharging events.

%%%%%%%%%%%%%%%%%%%%%%%%%%%%%%%%%%
% FIGURE
%%%%%%%%%%%%%%%%%%%%%%%%%%%%%%%%%%
\begin{figure}
 \begin{center}
  \includegraphics[angle=0,width=.7\textwidth]{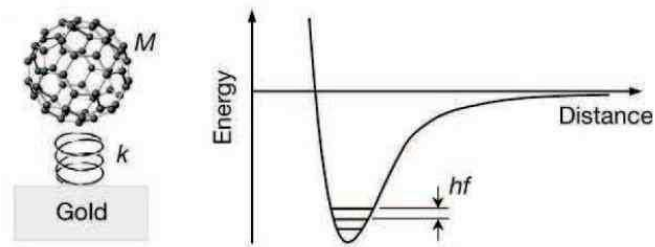}
    \caption{\small \textit{The C$_{60}$ experiment by Park {\it et al.} The C$_{60}$ molecule (of
    mass M) can be considered as being attached to a spring with spring constant $k$
    and corresponding quantized excitation energy $hf$ of the order of 5 meV.
    The figure is taken from \cite{par-nat-00}.}
    \label{fig:park}}
 \end{center}
\end{figure}
%%%%%%%%%%%%%%%%%%%%%%%%%%%%%%%%%%

Another experiment often mentioned  in the context of quantum
shuttles is the C$_{60}$-experiment by Park {\it et al.}\ \cite{par-nat-00} In this
experiment a C$_{60}$ molecule is deposited in the gap between two
gold electrodes. The gap, produced with break junction technique,
has a width of 1 nm. The molecule (of diameter 0.7 nm) is bound to
the electrodes due to van der Waals interaction. Around the
equilibrium position the potential can be approximated by a
harmonic potential and the molecule can be considered as attached
by springs. We show a schematic representation of this idea in
Fig.~\ref{fig:park}. In the experiment Park {\it et al.}~sweep the
voltage across the junction and register sudden increases in the
current. The steps are separated by 5 meV. Since the lowest
internal excitation energy of the C$_{60}$ molecule is $35$ meV
one concludes that the slower center of mass motion could be
involved in the process. This hypothesis is confirmed by the fact
that the separation between the steps is independent of the charge
on the C$_{60}$ molecule and the theoretical estimate for the
energy corresponding to the center of mass oscillation in the van
der Waals potential is exactly 5 meV. The IV-curve measured in
these experiments can be interpreted in terms of shuttling
\cite{fed-epl-02}, but also alternative explanations have been
promoted \cite{boe-epl-01, mcc-prb-03, bra-prb-03}. Whether we are
in presence of coherent or incoherent shuttling transport and to
what extent the shuttling mechanism is involved in this set-up
cannot be completely clarified by current measurement. The
current-current correlation (also called current noise) was
proposed by Fedorets {\it et al.}\ \cite{fed-epl-02} for a better
understanding of the underlying dynamics of the current jumps. For
this reason the calculation of noise in shuttle devices has been
performed by many different groups. We proposed a completely
quantum formulation \cite{nov-prl-04} that is also explained in
detail in this thesis.

\section{This thesis}

This thesis contains the description and analysis of the dynamics
of two realizations of quantum shuttle devices. The models we
consider describe both the mechanical and electrical degrees of
freedom quantum mechanically. For the single dot quantum shuttle
we extended an existing classical model proposed by Gorelik {\it et al.}\ \cite{gor-prl-98}.
For the triple dot case we adopted the model invented by Armour
and MacKinnon \cite{arm-prb-02}. In the following we outline the contents of the
thesis:

In Chapter 2 we introduce the two models called Single Dot Quantum
Shuttle and Triple Dot Quantum Shuttle, the first being the
quantum extension also for the mechanical degree of freedom of the
model originally proposed by Gorelik {\it et al.}\
\cite{gor-prl-98} while the second is the model invented by
Armour and MacKinnon \cite{arm-prb-02}. Also in this model the
oscillator is treated quantum mechanically and the central moving
dot is flanked by two static dots.

We dedicate Chapter 3 to the derivation of the Generalized Master Equation (GME) that describes
the shuttle device dynamics. Due to the different coupling
strengths we treat the mechanical and electrical baths with two
different approaches. The Gurvitz approach for the electrical and
the standard Born-Markov approximation for the mechanical bath.
The derivation \`a la Gurvitz represents a large part of this
chapter. In order to facilitate the understanding of this
non-standard method and appreciate our extension to the shuttle
device we have given a long introduction in which we analyze in
great detail simpler models with increasing physical complexity.
This shows on one hand the essence of the derivation in simpler
cases and also underlines the potentiality of this approach. An
important aspect of this method is also to be a natural prelude to
full counting statistics since it naturally produces a GME that
counts the number of electron which tunneled through the device at
a certain time. We close the chapter with the description of the
numerical iterative method that we adopted for the calculation of
the stationary solution of the GME.

In Chapter 4  we introduce the concept of Wigner distribution
function and derive the corresponding Klein-Kramers equation for the
SDQS starting from the GME that we obtained in the previous
chapter. The Wigner function description is motivated by the
effort to keep the complete quantum treatment we achieved with the
GME without losing as much as possible the intuitive classical
picture and with the possibility to handle the quantum-classical
correspondence.

Chapter 5 is dedicated to the definition and application of the
three investigation tools we have chosen to analyze the properties
of shuttle devices: the charge resolved phase-space distribution,
the current and the current-noise. The phase space analysis
reveals the shuttling transition and the charge position
correlation typical of this operating regime. It also gives a very
clean way to appreciate ``geometrically'' the quantum to classical
transition of the shuttling behaviour for different device
realizations. The second investigation tool that we consider is
the current. From the current calculation we obtain also in the
quantum treatment the quantized value of one electron per cycle
found in the semiclassical treatments of similar devices. We then
present current-noise calculations based on the MacDonald formula.
The derivation is strongly dependent on the derivation \`a la
Gurvitz of the $n$-resolved GME. The low noise quasi-deterministic
behaviour of the shuttling transport is clear from the extremely
low Fano factors found for this regime. In general we are able
with all the three investigation tools to identify three operating
regimes of shuttle devices: the tunneling, shuttling and
coexistence regimes.

Chapter 6 is dedicated to a qualitative description of the these
regimes and also to the identification of the relation between
different times and length scales that define the three regimes in
terms of the model parameters.

In Chapter 7 we consider separately the tunneling, shuttling and
coexistence regime. The specific separation of time scales allows
us to identify the relevant variables and describe each regime by
a specific simplified model. Models for the tunneling, shuttling
and coexistence regime are analyzed in this chapter. We also give
a comparison with the full description in terms of Wigner
distributions, current and current-noise to prove that the models,
at least in the limits set by the chosen investigation tools,
capture the relevant dynamics.

A summary of the arguments treated in this thesis opens Chapter 8.
We conclude with a list of some of the open questions that could
encourage a continuation of the present work.

\clearpage{\pagestyle{empty}\cleardoublepage}

%%%%%%%%%%%%%%%%%%%%%%%%%%%%%%%%%%%%%%%%%%%%%%%%%%%%%%%%%%%
%                      MODEL
%%%%%%%%%%%%%%%%%%%%%%%%%%%%%%%%%%%%%%%%%%%%%%%%%%%%%%%%%%%
%
\chapter{The models}
We describe in this chapter two models of quantum shuttles: the Single-Dot Quantum Shuttle and the Triple-Dot Quantum Shuttle.
Due to the nanometer size of these devices we decide to treat quantum mechanically not only the electrical but also the mechanical dynamics. This approach was suggested by the work of Armour and MacKinnon \cite{arm-prb-02} for the triple dot device and implemented for the first time by us in the single dot device.
\section{Single-Dot Quantum Shuttle}
\begin{figure}
 \begin{center}
  \includegraphics[angle=0,width=.7\textwidth]{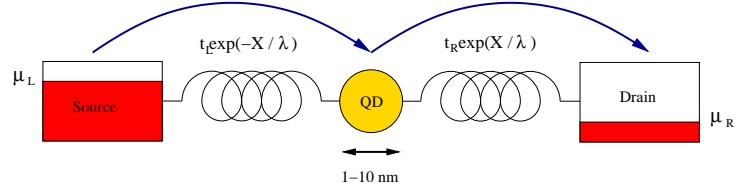}
  \caption{\small  \textit{Schematic representation of the Single-dot Shuttle: electrons tunnel
        from the left lead at chemical potential ($\mu_L$) to the quantum
        dot and eventually to the right lead at lower chemical potential $\mu_R$.
        The position dependent tunneling amplitudes are indicated.
        $X$ is the displacement from the equilibrium position.
        The springs represent the harmonic potential in which the central dot
        can move.}}
  \label{fig:SdS_model}
 \end{center}
\end{figure}

The Single-Dot Quantum Shuttle (SDQS) consists of a movable
quantum dot (QD) suspended between source and drain leads. One can
imagine the dot attached to the tip of a cantilever or connected
to the leads by some soft legands or embedded into an elastic
matrix. In the model the center of mass of the nanoparticle is
confined to a potential that, at least for small diplacements from
its equilibrium position, can be considered harmonic. We give a
schematic visualization of the device in figure
\ref{fig:SdS_model}.

%%%%%%%%%%%%%%%%%%%%%%%%%%%%%%%%%%%%%%%%%%%%%%%%%%%%%%%%%%%%%%%%%%%%
% Figure
%%%%%%%%%%%%%%%%%%%%%%%%%%%%%%%%%%%%%%%%%%%%%%%%%%%%%%%%%%%%%%%%%%%%
%%%%%%%%%%%%%%%%%%%%%%%%%%%%%%%%%%%%%%%%%%%%%%%%%%%%%%%%%%%%%%%%%%%%

Due to its small diameter, the QD has a very small capacitance and
thus a charging energy that exceeds (in the most recent
realizations almost at room temperature \cite{sch-apl-04}) the
thermal energy $k_B T$\footnote{A quick estimate of the charging
energy can be obtained for an isolated 2D metallic disk: $e^2/C =
e^2/(8 \e_r \e_0 R)$ where R is the disk radius and $\e_r =13$ in
GaAs. For a dot of radius 10 nm this yields $e^2/C = 20{\rm meV} =
k_B230 {\rm K}$ }. For this reason we assume that only one excess electron can occupy the device (Coulomb blockade) and we describe the electronic
state of the central dot as a two-level system (empty/charged).
Electrons can tunnel between leads and dot with tunneling
amplitudes which are exponentially dependent on the position of
the central island. This is due to the exponentially
decreasing/increasing overlapping of the electronic wave
functions.

The Hamiltonian of the model reads:

\be
    H =H_{\rm sys}+H_{\rm leads}+H_{\rm bath}
      +H_{\rm tun}+H_{\rm int}
\label{eq:SdS-Ham0}
\end{equation}
where
\be
 \bs
  &H_{\rm sys} =\frac{\hat{p}^2}{2 m}
               +\frac{1}{2}m \w^2 \hat{x}^2
               +(\eps_1- e\mathcal{E} \hat{x})c_1^{\dag}c_1^{\phd}\\
 &H_{\rm leads} = \sum_{k}(\eps_{l_k}
               c^{\dagger}_{l_k}c^{\phd}_{l_k}
               +\eps_{r_k}
               c^{\dagger}_{r_k}c^{\phd}_{r_k})\\
 &H_{\rm tun} = \sum_{k}[T_{l}(\hat{x}) c^{\dagger}_{l_k}c_1^{\phd} +
                         T_{r}(\hat{x}) c^{\dagger}_{r_k}c_1^{\phd}] + h.c.\\
 &H_{\rm bath} + H_{\rm int }= {\rm generic \; heat \; bath}
 \end{split}
\label{eq:SdS-Ham}
\end{equation}

Using the language of quantum optics we call the
movable grain alone \emph{the system}. This is then coupled
to two electric baths (the leads) and a generic heat bath. The system
is described by a single electronic level of energy $\eps_1$ and a harmonic
oscillator of mass $m$ and frequency $\w$. When the dot is charged the electrostatic force
($e \mathcal{E} $) acts on the grain and gives the \emph{electrical
influence} on the mechanical dynamics. The electric field $\mathcal{E}$ is
generated by the voltage drop between left and right lead. In our
model, though, it is kept as an external parameter, also in view
of the fact that we will always assume the potential drop to be
much larger than any other energy scale of the system (with the only exception of the charging energy of the dot). The
operator form $\hat{x}, \hat{p}$ for the mechanical variables is
due to the quantum treatment of the harmonic oscillator. In terms
of creation and annihilation operators for oscillator excitations
we would write:

\be \bs
 &\hat{x} = \sqrt{\frac{\hbar}{2 m \w}}(d^{\dagger}+d)\\
 &\hat{p} = i \sqrt{\frac{\hbar m \w}{2}}(d^{\dagger}-d)\\
 &\frac{\hat{p}^2}{2 m} + \frac{1}{2}m \w^2 \hat{x}^2 = \hbar \w
 \left(d^{\dagger}d + \frac{1}{2} \right)
\end{split}
\end{equation}

The leads are Fermi seas kept at two different chemical potentials
($\mu_L$ and $\mu_R$) by the external applied voltage ($\D V =
(\mu_L - \mu_R)/e$ ). The oscillator is immersed into a
dissipative environment that we model as a collection of bosons
and is coupled to that by a weak bilinear interaction:
\be \bs
 H_{\rm bath} &= \sum_{\vet{q}} \hbar \w_{\vet{q}}{d_{\vet{q}}}^{\dagger} d_{\vet{q}}\\
 H_{\rm int}  &= \sum_{\vet{q}} \hbar g({d_{\vet{q}}} + {d_{\vet{q}}}^{\dagger}) (d + d^{\dagger})
\end{split}
\end{equation}
where the bosons have been labelled by their wave number
$\vet{q}$. The damping rate is given by:
\be
 \gamma(\w) = 2 \pi g^2 D(\w)
\end{equation}
where $D(\w)$ is the density of states for the bosonic bath at the frequency of the system oscillator. A bath that generates a frequency independent $\gamma$ is called Ohmic.

The coupling to the electric baths is introduced by the tunneling
Hamiltonian $H_{\rm tun}$. The tunneling amplitudes $T_{l}(\hat{x})$ and
$T_{r}(\hat{x})$ depend exponentially on the position operator
$\hat{x}$ and represent the \emph{mechanical feedback} on the
electrical dynamics:

\be
 T_{l,r}(\hat{x})=t_{l,r}\exp(\mp\hat{x}/\lambda)
\end{equation}
where $\lambda$ is the tunneling length. The tunneling
rates from and to the leads ($\bar{\Gamma}_{L,R}$) can be expressed
in terms of the amplitudes:

\be
  \bar{\Gamma}_{L,R}=\langle \G_{L,R}(\hat{x})\rangle =\left \langle \frac{2 \pi}{\hbar}D_{L,R}\exp\left(\mp \frac{2 \hat{x}}{\lambda}\right)
 |t_{l,r}|^2 \right\rangle
\end{equation}
where $D_{L,R}$ are the densities of states of the left and right
lead respectively and the average is taken with respect to the quantum
state of the oscillator.

The model presents three relevant time scales: the period of the
oscillator $2\pi/\w$, the inverse of the damping rate $1/\gamma$
and the average injection/ejection time $1/\bar{\Gamma}_{L,R}$. It
is possible also to identify three important length scales: the
zero point uncertainty $\D x_z =\sqrt{\frac{\hbar}{2m\w}}$, the
tunneling length $\lambda$ and the displaced oscillator
equilibrium position $d=\frac{e\mathcal{E}}{m \w^2}$. Different relations
between time and length scales distinguish different operating
regimes of the SDQS.
\section{Triple-Dot Quantum Shuttle }
The Triple-Dot Quantum Shuttle (TDQS) was proposed by Armour and
MacKinnon \cite{arm-prb-02}. The system consists of an array of
three QD's: a movable dot, that we assume confined to a harmonic
potential, flanked by two static ones. Relying on low temperature
and on the low capacitance of the system with respect to the
leads, we again assume strong Coulomb blockade: only one electron
at a time can occupy the three-dot device. The Hamiltonian for the
model reads:

\be
    H =H_{\rm sys}+ H_{\rm leads} + H_{\rm bath} + H_{\rm tun} + H_{\rm int}
\end{equation}

\begin{figure}
 \begin{center}
  \includegraphics[angle=0,width=.7\textwidth]{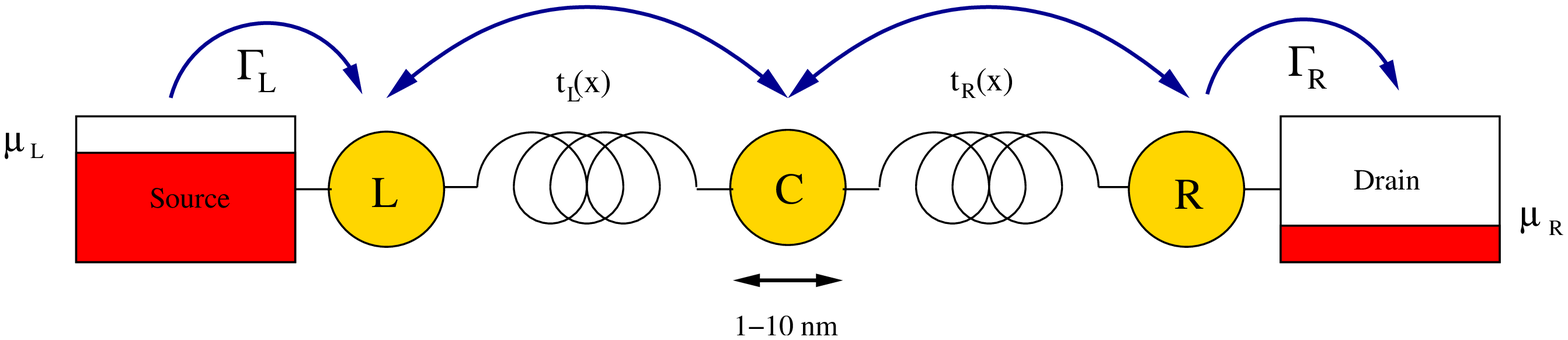}
  \label{fig:TdS_model}
  \caption{\small \textit{Schematic representation of the Triple-dot Shuttle:
  the leads and the three-dot array are represented. The arrows
  mimic the electrical dynamics. Single and double arrows indicate
  that the tunneling from and to the lead is always in a given
  direction and incoherent while the internal dynamics of the
  system is subject to coherent oscillations. The mechanical
  motion of the central dot confined to a harmonic potential
  is represented by the springs.}}
 \end{center}
\end{figure}

Only the system and tunneling part of the Hamiltonian differ from
the one dot case:
\be
 \bs
  H_{\rm sys} =&
   \e_0 |0\rangle\!\langle 0|
   + \frac{\D V}{2} |L\rangle \! \langle L|
   - \frac{\D V}{2 x_0}\hat{x}  |C\rangle\!\langle C|
   - \frac{\D V}{2} |R\rangle \! \langle R|
   + \hbar \w \left(d^{\dag}d + \frac{1}{2}\right)\\
   &+t_R(\hat{x})(|C \rangle\!\langle R| + |R \rangle\!\langle C|)
    +t_L(\hat{x})(|C \rangle\!\langle L| + |L \rangle\!\langle C|)\\
   H_{\rm tun} =&
   \sum_{k}T_{l}( c^{\dagger}_{l_k}|0\rangle\!\langle L| +
                  c^{\phd}_{l_k}   |L\rangle\!\langle 0|)
          +T_{r}( c^{\dagger}_{r_k}|0\rangle\!\langle R| +
                  c^{\phd}_{r_k}   |R\rangle\!\langle 0|)
 \end{split}
 \label{eq:TdS-Ham}
\end{equation}
where $H_{\rm osc}$ is the harmonic-oscillator Hamiltonian,
$|\a\rangle, \a=0,L,C,R$ are the vectors that span the electronic
part of the system Hilbert space . The tunable injection and
ejection energies (the energy levels of the outer dots, that we
can assume fixed by external gates) simulate a
controlled bias through the device ($\D V$) and the position
dependent tunneling amplitudes are now between elements
\emph{within} the system. These amplitude are assumed to be
exponentially dependent on the position of the central dot
$t_L(\hat{x})= -V_0 e^{-(x_0 + \hat{x})/\lambda}$ and
$t_R(\hat{x})= -V_0 e^{-(x_0 - \hat{x})/\lambda}$. Tunneling from
the leads is allowed only to the nearest dot and the corresponding
tunneling amplitude is independent of the position of the
oscillator. The ``device bias'' $\D V$ also gives rise to an
electrostatic force on the central dot, when charged. A schematic
representation of the Triple-dot Shuttle is given in figure
\ref{fig:TdS_model}.

For reasons that will become clearer in the following, we assume
that all the energy levels of the system (except the Coulomb
charging energy that ensures the strong Coulomb blockade regime)
lie well inside the bias window. In practice we will take the
limit $\mu_L \to \infty$ and $\mu_R \to -\infty$. This is
reflected in the directional flow of electrons from the source and
to the drain.

%%%%%%%%%%%%%%%%%%%%%%%%%%%%%%%%%%%%%%%%%%%%%%%%%%%%%%%%%%%%%%%%%
% Figure
%%%%%%%%%%%%%%%%%%%%%%%%%%%%%%%%%%%%%%%%%%%%%%%%%%%%%%%%%%%%%%%%%
%%%%%%%%%%%%%%%%%%%%%%%%%%%%%%%%%%%%%%%%%%%%%%%%%%%%%%%%%%%%%%%%%%%
\clearpage{\pagestyle{empty}\cleardoublepage}

%%%%%%%%%%%%%%%%%%%%%%%%%%%%%%%%%%%%%%%%%%%%%%%%%%%%%%%%%%%
%                      GME
%%%%%%%%%%%%%%%%%%%%%%%%%%%%%%%%%%%%%%%%%%%%%%%%%%%%%%%%%%%

\chapter{Generalized Master Equation}\label{sec:GME}

The state of a physical system is determined by the measurement of
a certain number of observables. Repeated measurements of a given
observable always return the same expectation value when the
system is in an eigenstate for that particular observable. The
uncertainty principle ensures us that, for quantum systems, there
are incompatible observables that can not be measured at the same
time with indefinite precision.

Given a generic quantum system $\mathcal{S}$ and a complete set of
compatible observables $\mathcal{A}_i$ \cite{destri}, an
eigenstate of the system for all observables is defined by the set
of the corresponding expectation values, i.e. the quantum numbers
$a_i$. Each of the possible sets of expectation values is
associated with an eigenvector in the Hilbert space of the system.
More precisely the Hilbert space of the system is spanned by the
eigenvectors of a complete set of compatible observables.

A pure state of the system is represented by a radius (class of
equivalence of normalized
vectors with arbitrary phase) of this
Hilbert space. We call $|\psi\rangle$
a representative vector
of a radius. Observables are associated to Hermitian operators
on the Hilbert space of the system. The dynamics of the quantum
system is governed by the Hamiltonian operator, i.e. the operator
associated to the observable energy. Given an initial vector, the
Schr\"{o}dinger equation prescribes the evolution of this vector at
all times:

\be
 i \hbar \frac{d}{dt} |\psi(t)\rangle = \mathcal{H} |\psi(t)\rangle
 \label{eq:Schroedinger}
\end{equation}
with the initial condition $|\psi(0)\rangle = |\psi_0\rangle$. An
equivalent formulation of the dynamics can be given in terms of
projector operators $|\psi \rangle \! \langle \psi|$. A projector
is independent from the arbitrary phase of the vector $|\psi
\rangle$, it is then equivalent to a radius of the Hilbert space
and represents a \emph{pure state} of the system. Using the
Leibnitz theorem for derivatives and the Schr\"odinger equation we
derive the equation of Liouville-von Neumann:

\be
 \frac{d \rho}{dt} = -\frac{i}{\hbar}[\mathcal{H},\rho]
 \label{eq:von-Neumann}
\end{equation}
where $\rho \equiv |\psi \rangle \! \langle \psi|$, $[A,B] \equiv
AB - BA$ is the commutator of the operators $A$ and $B$. The
operator $\rho$ is usually called \emph{density operator}. For
each basis of the Hilbert space all the operators have a matrix
representation. The matrix that corresponds to the density
operator is called \emph{density matrix}. Each vector of the basis
of the Hilbert space corresponds to a particular eigenstate of the
system defined by a set of quantum numbers. The diagonal elements
of the density matrix are called \emph{populations}. Each
population represents the probability that the system in the pure
state $|\psi\rangle\!\langle \psi| $ is in the eigenstate defined
by the corresponding set of quantum numbers. The trace of the
density matrix is one and supports this probabilistic
interpretation. The off-diagonal terms of the density matrix are
the \emph{coherencies} of the system.  They reflect the linear
structure of the Hilbert space. A linear combination of
eigenvectors gives rise to a pure state with non-zero coherencies.

Not all density operators correspond to pure states.
A convex linear combination of pure states $|\psi_n \rangle
\! \langle \psi_n |, n=1,...,N$ is called \emph{statistical
mixture}:
\be
 \rho = \sum_{n=1}^{N} P_n |\psi_n \rangle \! \langle \psi_n |
\end{equation}
where $ P_n \in [0,1),\sum_n P_n = 1$. This is an incoherent
superposition of pure states. Also statistical mixtures obey the
Liuoville-von Neumann equation of motion (\ref{eq:von-Neumann}).

The \emph{master equation} is an equation of motion for the
populations. It is a coarse grained\footnote{In the sense that it
describes the effective dynamics on a time scale long compared to
the typical times of the fastest processes in the physical
system.} equation that neglects coherencies. It was derived the
first time by Pauli under the assumption that coherencies have
random phases in time due to fast molecular dynamics. It reads:

\be
 \frac{dP_n(t)}{dt} = \sum_m[\Gamma_{nm}P_m(t)-\Gamma_{mn}P_n(t)]
\end{equation}
where $P_n$ is the population\footnote{Since a density matrix
without coherencies is a statistical mixture of eigenstates we
have adopted the notation $P_n \equiv \rho_{nn}$} of the
eigenstate $n$ and $\Gamma_{nm}$ is the rate of probability flow
from eigenstate $n$ to $m$ \cite{huang}.

\section{Coherent dynamics of small open systems}
The master equation is usually derived for models in which
a``small'' system with few degrees of freedom is in interaction
with a ``large'' bath with effectively an infinite number of
degrees of freedom. The Liouville von-Neumann equation of motion
for the total density matrix is very complicated to solve and
actually contains too much information since it also takes into
account coherencies of the bath. It is useful to average it over
bath variables and obtain an equation of motion for the density
matrix of the system (the \emph{reduced density matrix}). With no
further simplification this equation is called {\bf Generalized
Master Equation} (GME) since it involves not only the populations
but also the coherencies of the small subsystem. The derivation of
the GME from the equation of Liouville-von Neumann is far from
trivial and also non-universal: it involves a series of
approximations justified by the physical properties of the model
at hand. Despite the apparent similarities, the two equations are
deeply different: the equation of Liouville-von Neumann describes
the reversible dynamics of a closed system; the GME, instead,
describes the irreversible dynamics of an open system that
continuously exchange energy with the bath\footnote{How can
irreversibility be derived from reversibility? The solution of
this dilemma lies in time scales: system$+$bath recurrence time is
``infinite'' on the time scale of the system. The GME holds on the
time scales of the system.}.

Shuttle devices are small systems coupled to different baths
(leads, thermal bath) but they maintain a high degree of
correlation between electrical and mechanical degrees of
freedom captured by the coherencies of the reduced density matrix.
The GME seems to be a good candidate for the description of
their dynamics.

In the next two sections we will derive two GMEs using two
different approaches. They are both necessary for the description
of the shuttling devices since they correspond to the different
coupling of the system to the mechanical and electrical baths.

\section{Quantum optical derivation}\label{sec:QOptical}

The harmonic oscillator weakly coupled to a bosonic bath  is a
typical problem analyzed in quantum optics. This model well
describes in shuttling devices the interaction of the mechanical
degree of freedom of the NEMS with its environment. Following
section $5.1$ of the book ``Quantum Noise'' by C. W. Gardiner and
P. Zoller \cite{gardiner} we start considering a small system S
coupled to a large bath B described by the generic Hamiltonian:

\be
 H = H_{\rm S} + H_{\rm B} + H_{\rm I}
\end{equation}
where $H_{\rm S}$ and $H_{\rm B}$ respectively describe the
dynamics of the decoupled system and bath  and $H_{\rm I}$
represents the interaction between the two that we assume weak.
The density operator $\rho$ (the state of the system$+$bath)
satisfies, in the Schr\"odinger picture, the equation of
Liouville-von Neumann:

\be
 \dot{\rho} = -\frac{i}{\hbar}[H_{\rm S} + H_{\rm B} + H_{\rm I},\rho]
 \label{eq:Liouvillespecial}
\end{equation}
The state of the system is described by the reduced density matrix $\s$:

\be
 \s = {\rm Tr}_{\rm B}\{\rho\}
 \label{eq:defreduced}
\end{equation}
where ${\rm Tr}_{\rm B}$ indicates the partial trace over the bath
degrees of freedom. Our task is to derive from
(\ref{eq:Liouvillespecial}) an equation of motion for the reduced
density matrix $\s$.

\subsection{Interaction picture}
We start by going to the interaction picture and we use as
non-interacting Hamiltonian $H_{\rm S}+H_{\rm B}$. We indicate all
the operators in the interaction picture with a tilde. The total
density operator in the interaction picture reads:

\be
\tilde{\rho}(t) = \exp\left[\frac{i}{\hbar}(H_{\rm S}+H_{\rm B})t\right] \rho(t)
                  \exp\left[-\frac{i}{\hbar}(H_{\rm S}+H_{\rm B})t\right]
\label{eq:definteraction}
\end{equation}
and obeys the equation of motion:

\be
 \dot{\tilde{\rho}}(t) = -\frac{i}{\hbar}[\tilde{H}_{\rm I}(t),\tilde{\rho}(t)]
 \label{eq:EOMinteraction}
\end{equation}
where

\be
 \tilde{H}_{\rm I}(t) = \exp\left[\frac{i}{\hbar}(H_{\rm S}+H_{\rm B})t\right] H_{\rm I}
                        \exp\left[-\frac{i}{\hbar}(H_{\rm S}+H_{\rm B})t\right]
\end{equation}
From (\ref{eq:defreduced}) and (\ref{eq:definteraction}) it follows that

\be
 \s(t) = {\rm Tr}_{\rm B}\left\{\exp\left[-\frac{i}{\hbar}(H_{\rm S}+H_{\rm B})t\right] \tilde{\rho}(t)
                          \exp\left[\frac{i}{\hbar}(H_{\rm S}+H_{\rm B})t\right] \right\}
\end{equation}
The exponentials of $H_{\rm B}$ can be cancelled using the cyclic
property of the trace since $H_{\rm B}$ depends only on bath
variables. We get

\be
 \s(t) = \exp\left[-\frac{i}{\hbar}H_{\rm S}t\right] \tilde{\s}(t)
         \exp\left[\frac{i}{\hbar}H_{\rm S}t\right]
\end{equation}
where

\be
 \tilde{\s}(t) \equiv {\rm Tr}_B\{\tilde{\rho}(t)\}
\end{equation}
In other terms the interaction picture for the reduced density
matrix is effectively obtained only from the non-interacting
Hamiltonian for the system $H_{\rm S}$.

\subsection{Initial conditions}
We assume that the system and the bath are initially independent,
the initial total density operator $\rho$ is then factorized into
the tensor product:

\be
 \rho(0) = \s(0) \otimes \rho_{\rm B}
\end{equation}
For definiteness we assume the bath to be in thermal equilibrium:
\be
 \rho_{\rm B} = \frac{e^{-\beta H_{\rm B}}}{{\rm Tr_B}e^{-\b H_{\rm B}}}
\end{equation}
where $\beta = 1/k_B T$ is the inverse temperature.

\subsection{Reformulation of the equation of motion}
It is most important for the derivation of the GME to recast the
original equation of motion in the interaction picture
(\ref{eq:EOMinteraction}) into an integro-differential form. The
integral from $0$ to $t$ of (\ref{eq:EOMinteraction}) reads

\be
 \tilde{\rho}(t) = \tilde{\rho}(0) - \frac{i}{\hbar} \int_0^t dt'
                  [\tilde{H}_{\rm I}(t'),\tilde{\rho}(t')]
\end{equation}
and inserted back into (\ref{eq:EOMinteraction}) itself gives
\be
 \dot{\tilde{\rho}}(t) =
  -\frac{i}{\hbar}[\tilde{H}_{\rm I}(t),\tilde{\rho}(0)]
  -\frac{1}{\hbar^2}\int_0^t dt'
  \tilde{H}_{\rm I}(t),[\tilde{H}_{\rm I}(t'),\tilde{\rho}(t')]]
 \label{eq:integrodiff}
\end{equation}

\subsection{Average over the bath variables}
We take the partial trace over the bath variables on both sides of
(\ref{eq:integrodiff}). From the definition of the reduced density
operator (\ref{eq:defreduced}) we obtain, for the LHS,
$\dot{\tilde{\s}}$. We assume that the first term in the RHS
vanishes, namely

\be
 {\rm Tr_B}[\tilde{H}_{\rm I},\tilde{\rho}(0)] = 0
\end{equation}
where $\tilde{\rho}(0) = \rho(0) = \s(0) \otimes \rho_{\rm B}$.
This means that the interaction Hamiltonian has a bath component
with zero average. It is not difficult to fulfill this condition
in general by a redefinition of the system and interaction
Hamiltonian that subtracts the average of the bath component from
the latter.

\subsection{Weak coupling}
We assume that $H_{\rm I}$ is only a small perturbation of $H_{\rm
S}$ and $H_{\rm B}$. This condition allows a factorization at all
times of the total density operator into its system and bath
components. The density operator of the bath is also taken as
constant in time:

\be
 \tilde{\rho}(t) \approx \tilde{\s}(t) \otimes \rho_{\rm B}
\end{equation}

The factorization assumption can be weakened. We introduce for
this purpose the notion of correlation function. Given a physical
system in the state described by the stationary density operator
$\rho$ and two operators $\tilde{O_{\rm 1}}$ and $\tilde{O_{\rm
2}}$ the correlation function between $\tilde{O_{\rm 1}}$ and
$\tilde{O_{\rm 2}}$ in this order and at times $t$ and $t'$ is:

\be
 C_{O_1O_2}(t,t') \equiv {\rm Tr}\{\rho
 \tilde{O_{\rm 1}}(t)
 \tilde{O_{\rm 2}}(t')\}
\end{equation}
where the trace is taken over the the Hilbert space of the system
and the operators are in Heisenberg picture. Returning to our
system-bath model, we assume that the interaction Hamiltonian is a
sum of operators in the form $F_{\rm S}A_{\rm B}$. The minimal
requirement for the weak coupling approximation is that the
correlation functions of the bath are not influenced by the state
of the system. In formulas:

\be
 {\rm Tr_B}\{[\tilde{A_{\rm B}}(t),
 [\tilde{A_{\rm B}}(t'),\tilde{\rho}(t')]]\}
 \approx  \tilde{\s}(t') \otimes {\rm Tr_B}\{[\tilde{A_{\rm B}}(t),
 [\tilde{A_{\rm B}}(t'),\rho_{\rm B}]]\}
 \label{eq:weakcouplingapprox}
\end{equation}

\subsection{Markov approximation}
The integro-differential equation for $\tilde{\sigma}$ obtained in
the weak coupling approximation is non-local in time. The state of the
system at time $t$ depends on the history of the model starting
from the initial time $t=0$. This is the meaning of the integral
on the RHS of the equation\footnote{Note that causality is
preserved since the state of the system at times $t'> t$ does
\emph{not} enter the integral.}
\be
 \dot{\tilde{\s}} = -\frac{1}{\hbar^2}\int_0^t dt'
 {\rm Tr_B}\{[\tilde{H_{\rm I}}(t),
 [\tilde{H_{\rm I}}(t'),\tilde{\s}(t')\otimes \rho_{\rm B}]] \}.
\end{equation}
obtained from (\ref{eq:weakcouplingapprox}) by tracing over bath
variables. Due to the different sizes and the weak coupling the
effects on the bath of the interaction with the system are
negligible. The bath is a classical macroscopic object in thermal
equilibrium. Its stationary state is a thermal state: an
incoherent statistical mixture of energy  eigenstates. The
coherencies in the bath introduced by the interaction with the
system decay on a time scale called the correlation time. This is
precisely the decaying time of the correlation functions of the
bath. If the correlation time of the bath is much shorter than the
typical time scale for the system dynamics\footnote{By system
dynamics we mean in this case the time evolution of the reduced
density operator in the interaction picture $\tilde{\s}(t)$. In
this sense it is the weak coupling to keep the system dynamics
slow.}, then we can make in (\ref{eq:weakcouplingapprox}) the
replacement: \be \tilde{\sigma}(t') \to \tilde{\sigma}(t)
\end{equation}
and obtain in this way a differential equation for $\tilde{\s}$.
Finally, if we are interested in the dynamics of the system for
times much longer than the bath correlation time, the lower
integration limit in (\ref{eq:weakcouplingapprox}) can be moved to
$-\infty$ since the initial bath correlation are irrelevant. With
this set of approximations the knowledge of the state of the
system at some time $t_0$ is enough to determine the state at all
times $t>t_0$. This property is called Markov property. In the weak
coupling limit and assuming short correlation times in the heat
bath we have derived the following GME for the reduced density
operator in the interaction picture:

\be
 \dot{\tilde{\s}} = -\frac{1}{\hbar^2} \int_{0}^{\infty} d\t
 {\rm Tr_B}\{[\tilde{H}_{\rm I}(t),[\tilde{H}_{\rm I}(t-\t),
 \tilde{\s}(t)\otimes \rho_{\rm B}]]\}
 \label{eq:QOpGME}
\end{equation}

To proceed further the precise knowledge of the model Hamiltonian
is required. In section \ref{subsec:driving_and_damping} we will
specialize this derivation of the GME to the description of the
dissipative environment of the shuttling devices.

\section{Derivation ``\`{a} la Gurvitz''}\label{sec:Gurvitz}
The tunneling coupling of the shuttling devices to their
electrical baths (the leads) is \emph{not} weak. It sets, on the
contrary, the time scale of the electrical dynamics that in the
shuttling regime is comparable with the period of the mechanical
oscillations in the system.

In the SDQS the tunneling amplitudes depend exponentially on the
displacement of the central dot from the equilibrium position. The
oscillations of the QD modify correspondingly the tunneling rates.
In the shuttling regime the following non-adiabatic condition is
fulfilled:
\be
 \bar{\Gamma}_{\rm L,R} = \frac{\w}{2 \pi}
\end{equation}
where the average can be interpreted as a classical average over
the stable limit cycle trajectory or quantum mechanically as an
expectation value in the stationary state. In both cases Coulomb
blockade must be taken into account for a correct evaluation of
the average rate\footnote{We will discuss the details in section
\ref{sec:the_three_regimes}.}.

In the TDQS the coupling to the leads is constant and represents a
tunable parameter of the model. Also for this device the cleanest
shuttling regime is achieved for a rather high coupling (and
associated tunneling rates comparable with the mechanical
frequency).

In 1996 S. Gurvitz and Ya. S. Prager proposed a microscopic
derivation of the GME\footnote{In the article they use the
expression ``rate equations''. Nevertheless the equations derived
fully involve coherencies.} for quantum transport with an
\emph{arbitrary} coupling to the leads \cite{gur-prb-96}.
Following their article we give now in detail the derivation of
the rate equations for transport of non-interacting spin-less
particles through a static single dot connected to
leads\footnote{It must be noticed that the problem of calculation
of current through arrays of static quantum dots has been analyzed
in more or less equivalent approach also by other authors. We cite
as example for similarity of results  Wegewijs and Nazarov
\cite{weg-prb-99}.}. Even if in this oversimplified case the
result is intuitive and could be guessed just using common sense,
the generic features of the derivation will appear. We extend the
result to particles with spin in strong Coulomb blockade and
eventually we conclude the section showing that also coherent
transport can be treated in this formalism and we derive the GME
for a static double dot device.

Let us consider a quantum dot connected to two electronic leads.
We neglect (at the moment) the spin degree of freedom and Coulomb
interaction between the electrons. The energy levels in the
macroscopic leads are very dense while they are discrete in the
microscopic QD. We assume that only one level in the dot
participates in the dynamics of the model\footnote{While the non--
interacting approximation for the leads can be understood in the
frame of Landau theory of quasi-particles, no interaction combined
with single level approximation for the QD is an
oversimplification with no physical explanation that \emph{must}
(and will) be relaxed.}.

\subsection{Many-body basis expansion}

The Hamiltonian for the model reads

\be
\bs
 H =&\sum_l \eps_l c_l^{\dagger}c_l^{\phd} + \eps_1 c_1^{\dagger}c_1^{\phd}
   + \sum_r \eps_r c_r^{\dagger}c_r^{\phd} \\
   &+\sum_l \W_l (c_l^{\dagger}c_1^{\phd} + c_1^{\dagger}c_l^{\phd})
   + \sum_r \W_r (c_r^{\dagger}c_1^{\phd} + c_1^{\dagger}c_r^{\phd})
\end{split}
\label{eq:HGurvitz1}
\end{equation}
where $c_l^{\dagger}, c_1^{\dagger}, c_r^{\dagger}$ create  a
particle in the left lead (energy level $\eps_l$), in the dot and
in the right lead (energy level $\eps_r$) respectively. $\eps_1$
is the energy level of the dot and $\W_{l\,(r)}$ the tunneling
amplitudes between the dot and the left (right) lead. For
temperatures much smaller than the Fermi energy of the leads we
can approximate their Fermi distributions by step functions. The
chemical potential of the left (right) lead is assumed much higher
(lower) than the dot energy level.

We identify the empty state $|0\rangle$ for the model  with the
condition of empty dot and leads filled up to their Fermi
energies. Then we gradually move electrons from the emitter to the
dot and finally to the collector and associate a new vector to
each state of this ``decaying chain''. We construct in this way an
infinite many-body basis that defines the Hilbert space for the
model\footnote{Some of the possible states of the device are
excluded from this Hilbert space: states with electrons excited
above the Fermi level of the left lead and/or holes below the
Fermi energy of the right lead. They are neglected because they
would anyway be hardly populated due to the fast relaxation of the
leads to their thermal states and the very low probability of
electron tunneling for energies so far from the resonant level of
the dot.}. The first few elements of the basis read:

\be
 |0\rangle \qquad
 c_1^{\dagger}   c_{l_1}^{\phd}|0\rangle \qquad
 c_{r_1}^{\dagger}c_{l_1}^{\phd}|0\rangle \qquad
 c_1^{\dagger}c_{r_1}^{\dagger}c_{l_1}^{\phd}c_{l_2}^{\phd}|0\rangle \qquad
 c_{r_2}^{\dagger}c_{r_1}^{\dagger}c_{l_1}^{\phd}c_{l_2}^{\phd}|0\rangle \, \ldots
\label{eq:basis}
\end{equation}
where we choose by convention to move all the creation  operators
to the left and the annihilation operators to the right. We also
assume with the two groups that $\eps_{l_i} < \eps_{l_{i+1}}$ and
$\eps_{r_i} < \eps_{r_{i+1}}$ to avoid double counting.
%
%%%%%%%%%%%%%%%%%%%%%%%%%%%%%%%%%%%%%%%%%%%%%%%%%%%%%%%%%%%%%%%%%%%
% Figure
%%%%%%%%%%%%%%%%%%%%%%%%%%%%%%%%%%%%%%%%%%%%%%%%%%%%%%%%%%%%%%%%%%%
\begin{figure}
 \begin{center}
  \includegraphics[angle=0,width=\textwidth]{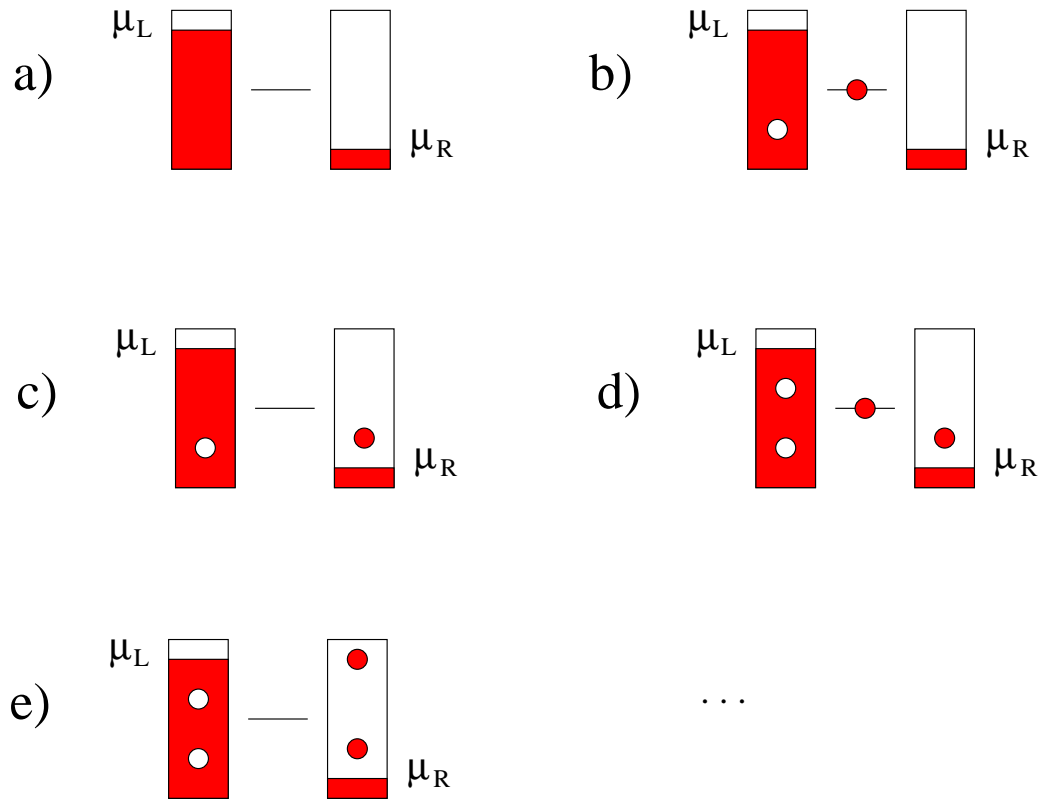}
  \label{fig:basis}
  \caption{\small \textit{Schematic representation of the first
  elements of the many-body basis for the single dot model. Electrons
  are progressively taken from the emitter and moved to the QD and finally
  to the collector. The vectors represented are in the order
${\rm a)} = |0\rangle$,
${\rm b)} = c_1^{\dagger}c_{l_1}|0\rangle$,
${\rm c)} = c_{r_1}^{\dagger}c_{l_1}|0\rangle$,
${\rm d)} = c_1^{\dagger}c_{r_1}^{\dagger}c_{l_1}c_{l_2}|0\rangle$,
${\rm e)} = c_{r_2}^{\dagger}c_{r_1}^{\dagger}c_{l_1}c_{l_2}|0\rangle$.}}
 \end{center}
\end{figure}
%%%%%%%%%%%%%%%%%%%%%%%%%%%%%%%%%%%%%%%%%%%%%%%%%%%%%%%%%%%%%%%%%%%%
The state of the model is described by the many-particle vector
$|\Psi(t)\rangle$ which we can expand over the basis:

\be
\bs
 |\Psi(t)\rangle =&\Big[ b_0(t) +
                   \sum_{l_1} b_{1l_1}(t) c_1^{\dagger}c_{l_1}^{\phd}
                   +\sum_{l_1,r_1}b_{l_1r_1}(t) c_{r_1}^{\dagger}c_{l_1}^{\phd} \\
                  &+ \!\sum_{l_1<l_2,r_1}b_{1l_1l_2r_1}(t)
                     c_1^{\dagger}c_{r_1}^{\dagger}c_{l_1}^{\phd}c_{l_2}^{\phd}
                   + \! \sum_{l_1<l_2,r_1<r_2}b_{l_1l_2r_1r_2}(t)
                     c_{r_2}^{\dagger}c_{r_1}^{\dagger}c_{l_1}^{\phd}c_{l_2}^{\phd}
                   +\ldots \Big]|0\rangle
\end{split}
\label{eq:expansion}
\end{equation}

\subsection{Recursive equation of motion for the coefficients}
The vector $|\Psi(t)\rangle$ obeys the Schr\"odinger equation
$i\hbar |\dot{\Psi}(t) \rangle = H |\Psi(t)\rangle$ and we impose
the initial condition $|\Psi(0)\rangle = |0\rangle$. In terms of
the coefficients of the expansion (\ref{eq:expansion}) we obtain
an infinite set of differential equations with the initial
condition $b_0(0)= 1$ and all the other coefficients equal to $0$
at time $t=0$.

Due to the quadratic form of the Hamiltonian, the infinite set of
differential equations for the coefficients $b$'s presents a
recursive structure: each coefficient is linked in its equation of
motion only to the previous and the next in the ``decaying
chain''(\ref{eq:expansion}). Since we are interested in keeping track
of the state of the dot we condense the full system of equations
into two equations for the generic coefficients
$b_{\{l_j\}_{j=1}^n\{r_j\}_{j=1}^n}(t)$ and
$b_{1\{l_j\}_{j=1}^{n+1}\{r_j\}_{j=1}^n}(t)$:
\be
\bs
 i\dot{b}_{\{l_j\}_{j=1}^n\{r_j\}_{j=1}^n} =&
        \left[\sum_{k=1}^n(\eps_{r_k}-\eps_{l_k})\right]
        b_{\{l_j\}_{j=1}^n\{r_j\}_{j=1}^n}\\
        &\!\!\!\!+\sum_{l_{n+1}}\W_{l_{n+1}}
        b_{1\{l_j\}_{j=1}^{n+1}\{r_j\}_{j=1}^n}
        +\sum_{k=1}^n(-1)^{k-n}\W_{r_k}
        b_{1\{l_j\}_{j=1}^n\{r_j\}_{j=1}^n \backslash \{r_k\}}\\
 i\dot{b}_{1\{l_j\}_{j=1}^{n+1}\{r_j\}_{j=1}^n} =&
      \left[\eps_1+\sum_{k=1}^n (\eps_{r_k}-\eps_{l_k})
       -\eps_{l_{n+1}}\right]
       b_{1\{l_j\}_{j=1}^{n+1}\{r_j\}_{j=1}^n}\\
      &\!\!\!\!+\sum_{r_{n+1}}\W_{r_{n+1}}
       b_{\{l_j\}_{j=1}^{n+1}\{r_j\}_{j=1}^{n+1}}
       +\sum_{k=1}^{n+1}(-1)^{k-n-1}\W_{l_k}
       b_{\{l_j\}_{j=1}^{n+1}\backslash \{l_k\}\{r_j\}_{j=1}^n}\\
\end{split}
\label{eq:diffsystem}
\end{equation}
where $\hbar=1$ and the sums over labels (e.g. $\sum_{l_{n+1}}$)
are continuous sums over all the possible energy levels of the
leads. We also used the shortened notations:

\be
\bs
 b_{\{l_j\}_{j=1}^n\{r_j\}_{j=1}^n} &\equiv
 b_{l_1l_2l_3\ldots l_nr_1r_2r_3\ldots r_n}(t)\\
 b_{1\{l_j\}_{j=1}^{n+1}\backslash \{l_k\}\{r_j\}_{j=1}^n} &\equiv
 b_{1l_1l_2l_3\ldots l_{k-1}l_{k+1}\ldots l_{n+1}r_1r_2r_3\ldots r_n}(t)
 \end{split}
\end{equation}

In order to  proceed in the derivation of the rate equations it is
most convenient to make the Laplace transform of the system of
differential equations (\ref{eq:diffsystem}). We obtain the system
of algebraic  equations:

\be
\bs
   &\left[E + \sum_{k=1}^n (\eps_{l_k}-\eps_{r_k})\right]
    \tilde{b}_{\{l_j\}_{j=1}^n\{r_j\}_{j=1}^n}(E)\\
   &-\sum_{l_{n+1}}\W_{l_{n+1}}
    \tilde{b}_{1\{l_j\}_{j=1}^{n+1}\{r_j\}_{j=1}^n}(E)
    -\sum_{k=1}^n(-1)^{k-n}\W_{r_k}
    \tilde{b}_{1\{l_j\}_{j=1}^{n}\{r_j\}_{j=1}^n\backslash \{r_k\}}(E)= i\d_{n0}\\
   &\left[E-\eps_1+\sum_{k=1}^n(\eps_{l_k}-\eps_{r_k})
    +\eps_{l_{n+1}}\right]
    \tilde{b}_{1\{l_j\}_{j=1}^{n+1}\{r_j\}_{j=1}^n}(E)\\
   &-\sum_{r_{n+1}}\W_{r_{n+1}}
    \tilde{b}_{\{l_j\}_{j=1}^{n+1}\{r_j\}_{j=1}^{n+1}}(E)
    -\sum_{k=1}^{n+1}(-1)^{k-n-1}\W_{l_k}
    \tilde{b}_{\{l_j\}_{j=1}^{n+1}\backslash \{l_k\}\{r_j\}_{j=1}^n}(E)=0\\
\end{split}
\label{eq:algebsystem}
\end{equation}
where the Laplace transformed coefficients are indicated with a
tilde and are functions of the variable $E$ (that we assume to
have an imaginary part to ensure convergence of the Laplace
integral). At this level the left-right asymmetry reveals itself
in the number of ``decay channels''. Each state of the chain
(\ref{eq:basis}) is coupled to an infinite number of right states
and only a finite number of left states. Since the couplings are
equivalent this results in a statistically definite direction of
motion for the electrons.

\subsection{Injection and ejection rates}
The continuous sums in (\ref{eq:algebsystem}) can  be simplified
using the recursive structure of the equation of motion. We
isolate in the second equation of (\ref{eq:algebsystem}) the coefficient
$\tilde{b}_{1\{l_j\}_{j=1}^{n+1}\{r_j\}_{j=1}^n}(E)$ and insert
the result into the first equation of (\ref{eq:algebsystem}). The
continuous sum results into two terms:

\be
 \sum_{l_{n+1}}\sum_{r_{n+1}} \frac{\W_{l_{n+1}}\W_{r_{n+1}}}
 {E -\eps_1+\sum_{k=1}^n (\eps_{l_k}-\eps_{r_k}) +\eps_{l_{n+1}}}
 \tilde{b}_{\{l_j\}_{j=1}^{n+1}\{r_j\}_{j=1}^{n+1}}
 \label{eq:vanish}
\end{equation}
and
\be
 \sum_{l_{n+1}}\sum_{k=1}^{n+1} (-1)^{k-n-1} \frac{\W_{l_{n+1}}\W_{l_k}}
 {E -\eps_1+\sum_{i=1}^n (\eps_{l_i}-\eps_{r_i}) +\eps_{l_{n+1}}}
 \tilde{b}_{\{l_j\}_{j=1}^{n+1}\backslash \{l_k\}\{r_j\}_{j=1}^n}
 \label{eq:rate1}
\end{equation}

Since the energy levels in the leads are dense we can substitute

\be
 \sum_{l_{n+1}} \to
 \int_{-\infty}^{+\infty}d\eps_{l_{n+1}}D_L(\eps_{l_{n+1}})
\end{equation}
where $D(\eps_{l_{n+1}})$ is the density of states in the left
lead calculated at energy $\eps_{l_{n+1}}$ and we have extended
the integration limits to infinity in the wide band and high bias
approximation. We can evaluate now the sum over $l_{n+1}$ in
(\ref{eq:vanish}) and (\ref{eq:rate1}) using residues method.
Since all the poles are in the same half plane we can neglect all
terms which are asymptotically $o(|\eps_{l_{n+1}}|^{-1})$ for
$|\eps_{l_{n+1}}| \to \infty $. It is clear from the algebraic
system (\ref{eq:algebsystem}) that a coefficient $\tilde{b}$ that
has among its indices $\eps_{l_{n+1}}$ behaves asymptotically at
least like $|\eps_{l_{n+1}}|^{-1}$. For this reason
(\ref{eq:vanish}) vanishes and only one term is left from
(\ref{eq:rate1}):

\be
 \tilde{b}_{\{l_j\}_{j=1}^{n}\{r_j\}_{j=1}^n}
 \int_{-\infty}^{+\infty}d\eps_{l+1} D_L(\eps_{l_{n+1}}) \frac{\W_{l_{n+1}}^2}
 {E -\eps_1+\sum_{i=1}^n (\eps_{l_i}-\eps_{r_i}) +\eps_{l_{n+1}}}
  \label{eq:rate2}
\end{equation}
For the evaluation of the integral is enough that the tunneling
amplitude $\W_{l_{n+1}}$ and the density of states
$D_L(\eps_{l_{n+1}})$ are analytical and non-zero where  the
denominator vanishes. We assume that they are constant to avoid
$n$-dependence in the tunneling rates and perform  the integral in
(\ref{eq:rate2}). The second equation (\ref{eq:algebsystem}) can
be treated analogously and the system reads:

\be
\bs
&\left[E + \sum_{k=1}^n (\eps_{l_k}-\eps_{r_k})
     +i\frac{\G_L}{2}\right]
     \tilde{b}_{\{l_j\}_{j=1}^n\{r_j\}_{j=1}^n}(E)\\
    &\phantom{aaaaa}-\sum_{k=1}^n(-1)^{k-n}\W_{r_k}
     \tilde{b}_{1\{l_j\}_{j=1}^{n}\{r_j\}_{j=1}^n\backslash \{r_k\}}(E)= i\d_{n0}\\
&\left[E -\eps_1+\sum_{k=1}^n (\eps_{l_k}-\eps_{r_k}) +\eps_{l_{n+1}}
     +i\frac{\G_R}{2}\right]
     \tilde{b}_{1\{l_j\}_{j=1}^{n+1}\{r_j\}_{j=1}^n}(E)\\
    &\phantom{aaaaa}-\sum_{k=1}^{n+1}(-1)^{k-n-1}\W_{l_k}
     \tilde{b}_{\{l_j\}_{j=1}^{n+1}\backslash \{l_k\}\{r_j\}_{j=1}^n}(E)=0\\
\end{split}
\label{eq:ratesystem}
\end{equation}
where we have introduced the \emph{injection} and \emph{ejection
rates} $\G_L$ and $\G_R$

\be
\bs
 \G_L &\equiv 2\pi D_L\W_L^2\\
 \G_R &\equiv 2\pi D_R\W_R^2
\end{split}
\end{equation}
whith the energy independent tunneling amplitudes ($\W_L$ and
$\W_R$) and density of states ($D_L$ and $D_R$)\footnote{We assume
real tunneling amplitudes  as it is also implied by the form of
the Hamiltonian (\ref{eq:HGurvitz1}). The most general case of
complex amplitude would result anyway in the tunneling rates
$\G_{L,R} \equiv 2\pi D_{L,R}|\W_{L,R}|^2$.}.

\subsection{The reduced density matrix}\label{sec:ReducedDensityMatrix}
The reduced density operator is defined as the trace over the bath
variables of the total density operator:

\be
 \s(t) = {\rm Tr}_{B}\{ |\Psi(t) \rangle \! \langle \Psi(t) | \}
\end{equation}

The matrix elements of the reduced density operators are explicitly

\be
 \s_{ij}(t) = \sum_{\{B\}} \langle i_S,B|\Psi(t)\rangle\!\langle\Psi(t)|B,j_S\rangle
\end{equation}
where $|i_S,B \rangle, i = 0,1$ is the vector that corresponds to
the empty or charged dot (the system) and a particular
configuration $B$ of the leads (the baths). We assume that the
bath state $B$ does not contain coherent superpositions of states
with different number of particles. This implies the vanishing of
coherencies in the reduced density matrix. It is useful to
organize the sum over the bath configurations according to the
number of extra electrons (holes) collected into the right (left)
lead.

\be
 \s_{ii}(t) = \sum_{n=0}^{\infty}
 \sum_{\{B_n\}}\langle i_S,B_n|\Psi(t)\rangle\!\langle\Psi(t)|B_n,i_S\rangle
 = \sum_{n=0}^{\infty}\s_{ii}^{(n)}(t)
 \label{eq:genRDM}
\end{equation}
where $B_n$ is a configuration of the baths with $n$ extra
electrons in the collector and we have introduced the
\emph{n-resolved} density matrix $\s^{(n)}$. Using the expansion
of the vector $|\Psi(t)\rangle$ in the many-body basis
(\ref{eq:basis}) we can express the $n$-resolved density matrix in
terms of the coefficients $b$. For the two non-vanishing elements:

\be
 \bs
  \s_{00}^{(n)} &= \sum_{\{B_n\}}|\langle 0_S,B_n|\Psi(t)\rangle|^2
  = \sum_{\{l_k\}\{r_k\}}|b_{\{l_k\}_{k=1}^n\{r_k\}_{k=1}^n}(t)|^2\\
  \s_{11}^{(n)} &= \sum_{\{B_n\}}|\langle 1_S,B_n|\Psi(t)\rangle|^2
  = \sum_{\{l_k\}\{r_k\}}|b_{1\{l_k\}_{k=1}^{n+1}\{r_k\}_{k=1}^n}(t)|^2
 \end{split}
 \label{eq:nRDM}
\end{equation}
where the sums are calculated over all the possible configurations
of indistinguishable particles (e.g. $\sum_{\{l_k\}} \equiv
\sum_{l_1<l_2<l_3<\ldots<l_n}$). The time-dependent matrix
elements of the reduced density matrix are connected to the
coefficients $\tilde{b}(E)$ by the inverse of the Laplace
transform:

\be
\bs
\s^{(n)}_{00}(t) &=
 \sum_{\{l_k\}\{r_k\}}\int\frac{dE dE'}{4 \pi^2}
 \tilde{b}_{\{l_k\}_{k=1}^n\{r_k\}_{k=1}^n}(E)
 \tilde{b}^*_{\{l_k\}_{k=1}^n\{r_k\}_{k=1}^n}(E')
 e^{i(E-E')t}\\
\s^{(n)}_{11}(t) &=
  \sum_{\{l_k\}\{r_k\}}\int\frac{dE dE'}{4 \pi^2}
  \tilde{b}_{1\{l_k\}_{k=1}^{n+1}\{r_k\}_{k=1}^n}(E)
  \tilde{b}^*_{1\{l_k\}_{k=1}^{n+1}\{r_k\}_{k=1}^n}(E')
  e^{i(E-E')t}\\
\end{split}
\label{eq:invLaplace}
\end{equation}
Apart from being a natural step in the derivation of the GME in
the Gurvitz approach, the $n$-resolved density matrix contains the
additional information on the number of electrons collected in the
resevoir at time $t$. This information is very useful to the
calculation of the current noise in the SDQS where the quantum
regression theorem can not be applied due to the form of the
current operator that involves both system and bath operators.

\subsection{Generalized Master Equation}\label{subsec:GME}
The equation of motion for the reduced density matrix is obtained
combining (\ref{eq:invLaplace}) and (\ref{eq:ratesystem}). First
we derive an equation of motion for the $n$-resolved reduced density
matrix $\s^{(n)}$. The case of the empty dot population with $n=0$
is special due to the particular choice of the initial condition
and we treat it separately. The starting point is the first of the
equations (\ref{eq:ratesystem}) specialized for $n=0$, namely:

\be
 \left( E + i \frac{\G_L}{2}\right)\tilde{b}_0(E) = i
\end{equation}
We taking the inverse Laplace transform and obtain

\be
 \dot{b}_0(t) = -\frac{\G_L}{2}b_0(t)
\end{equation}

The definition of $\s_{00}^{(0)}$ and the Leibnitz theorem for
derivatives lead to the conclusion:

\be
 \dot{\s}_{00}^{(0)} = \dot{b}_0b_0^* + b_0\dot{b}_0^*
                     = -\G_Lb_0b_0^* = -\G_L\s_{00}^{(0)}
 \label{eq:nGME0}
\end{equation}

This argument is applied also in the case  with $n \ne 0$ but the
structure of the equation is more complex and in general a final
continuous sum must be evaluated. We take the first equation in
(\ref{eq:ratesystem}) and multiply it by
$\tilde{b}^*_{\{l_j\}_{j=1}^n\{r_j\}_{j=1}^n}(E')$$e^{-i(E-E')t}$.
Then we subtract side by side the complex conjugate of the first
equation of (\ref{eq:ratesystem}) evaluated in $E'$ and multiplied
by $\tilde{b}_{\{l_j\}_{j=1}^n\{r_j\}_{j=1}^n}(E)$$e^{-i(E-E')t}$.
Finally we integrate in $dE$ and $dE'$ and sum over the bath
configurations with $n$ electrons in the collector. We repeat the
procedure also for the second equation in (\ref{eq:ratesystem})
and obtain:

\be
\bs
 &\sum_{\{l_j\}\{r_j\}}\!\int\frac{dE dE'}{4 \pi^2}\Big[(E-E'+i\G_L)
    \tilde{b}_{\{l_j\}_{j=1}^n\{r_j\}_{j=1}^n}(E)
  \tilde{b}^*_{\{l_j\}_{j=1}^n\{r_j\}_{j=1}^n}\!(E')e^{-i(E-E')t}\\
 &-2 \Im \Big(
 \sum_{k=1}^n(-1)^{k-n}\W_{r_k}
  \tilde{b}_{1\{l_j\}_{j=1}^n\{r_j\}_{j=1}^n\backslash \{r_k\}}(E)
 \tilde{b}^*_{\{l_j\}_{j=1}^n\{r_j\}_{j=1}^n}(E')e^{-i(E-E')t}
 \Big) \Big]=0
\end{split}
\label{eq:nGME11}
\end{equation}
for the first equation and similarly
\be
\bs
 &\sum_{\{l_j\}\{r_j\}}\!\int\frac{dE dE'}{4 \pi^2}\Big[(E-E'+i\G_R)
    \tilde{b}_{1\{l_j\}_{j=1}^{n+1}\{r_j\}_{j=1}^n}(E)
  \tilde{b}^*_{1\{l_j\}_{j=1}^{n+1}\{r_j\}_{j=1}^n}\!(E')e^{-i(E-E')t}\\
 &-2  \Im \Big(
 \sum_{k=1}^{n+1}(-1)^{k-n-1}\W_{l_k}
  \tilde{b}_{\{l_j\}_{j=1}^{n+1}\backslash \{l_k\}\{r_j\}_{j=1}^n}(E)
 \tilde{b}^*_{1\{l_j\}_{j=1}^{n+1}\{r_j\}_{j=1}^n}(E')e^{-i(E-E')t}
 \Big) \Big]=0\\
\end{split}
\label{eq:nGME12}
\end{equation}
for the second. $\Im$ indicates the imaginary part. In the
definition of the $n$-resolved reduced density matrix the two
coefficients $b$ correspond to the same bath configuration. The
finite sums in equations (\ref{eq:nGME11}) and (\ref{eq:nGME12})
still have coefficients with different bath configuration. Using
properties of the Laplace transform, the definition of $\s^{(n)}$
and the relations

\be
\bs
 \tilde{b}^*_{\{l_j\}_{j=1}^n\{r_j\}_{j=1}^n}(E') &=
 \frac{\sum_{k=1}^n (-1)^{k-n}\W_{r_k}
       \tilde{b}^*_{1\{l_j\}_{j=1}^n\{r_j\}_{j=1}^n\backslash \{r_k\}}(E')}
       {E'+\sum_{k=1}^n(\eps_{l_k} -\eps_{r_k}) -i\frac{\G_L}{2}}\\
 \tilde{b}^*_{1\{l_j\}_{j=1}^{n+1}\{r_j\}_{j=1}^n}(E') &=
 \frac{\sum_{k=1}^{n+1} (-1)^{k-n-1}\W_{l_k}
       \tilde{b}^*_{\{l_j\}_{j=1}^{n+1}\backslash \{l_k\}\{r_j\}_{j=1}^n}(E')}
       {E'-\eps_1+\sum_{k=1}^n(\eps_{l_k} -\eps_{r_k}) + \eps_{l_{n+1}} -i\frac{\G_R}{2}}
\end{split}
\end{equation}
obtained from (\ref{eq:ratesystem}), we transform
(\ref{eq:nGME11}) and (\ref{eq:nGME12}) into:

\be
\bs
 i(\dot{\s}_{00}^{(n)} + \G_L \s_{00}^{(n)}) =
 %%%%%%%%%
 \phantom{\Big[ 2 \Im\Big( \sum_{k,k'=1}^n
 \frac{(-1)^{k+k'}\W_{r_k}\W_{r_{k'}}}
 {E' + \sum_{k=1}^n(\eps_{l_k}-\eps_{r_k}) -i\frac{\G_L}{2}}}&\\
 %%%%%%%%%
 \sum_{\{l_j\}\{r_j\}} \int\frac{dE dE'}{4 \pi^2}
  \Big[ 2 \Im\Big( \sum_{k,k'=1}^n
 \frac{(-1)^{k+k'} \W_{r_k}\W_{r_{k'}}}
 {E' + \sum_{k=1}^n(\eps_{l_k}-\eps_{r_k}) -i\frac{\G_L}{2}}&\\
 \tilde{b}_{1\{l_j\}_{j=1}^n\{r_j\}_{j=1}^n\backslash \{r_k\}}(E)
 \tilde{b}^*_{1\{l_j\}_{j=1}^n\{r_j\}_{j=1}^n\backslash \{r_{k'}\}}(E')
         e^{-i(E-E')t}\Big)\Big]&
\end{split}
\end{equation}

and

\be
\bs
 i(\dot{\s}_{11}^{(n)} + \G_R \s_{11}^{(n)}) =
 %%%%%%%
 \phantom{\Big[ 2 \Im\Big( \sum_{k,k'=1}^{n+1}
 \frac{(-1)^{k+k'}\W_{l_k}\W_{l_{k'}}}
 {E' -\eps_1 + \sum_{k=1}^n(\eps_{l_k}-\eps_{r_k} + \eps_{l_{n+1}})
  -i\frac{\G_L}{2}}}&\\
 %%%%%%%
 \sum_{\{l_j\}\{r_j\}} \int\frac{dE dE'}{4 \pi^2}
  \Big[ 2 \Im\Big( \sum_{k,k'=1}^{n+1}
 \frac{(-1)^{k+k'}\W_{l_k}\W_{l_{k'}}}
 {E' -\eps_1 + \sum_{k=1}^n(\eps_{l_k}-\eps_{r_k} + \eps_{l_{n+1}})
  -i\frac{\G_L}{2}}&\\
 \tilde{b}_{\{l_j\}_{j=1}^{n+1}\backslash \{l_k\}\{r_j\}_{j=1}^n}(E)
 \tilde{b}^*_{\{l_j\}_{j=1}^{n+1}\backslash\{l_{k'}\}\{r_j\}_{j=1}^n}(E')
         e^{-i(E-E')t}\Big)\Big]&
\end{split}
\end{equation}

It is crucial at this point that $k = k'$. If $k \ne k'$ we can
eliminate the variables $r_{k}$ ($l_{k}$) from $\tilde{b}$ and
$r_{k'}$ ($l_{k'}$) from $\tilde{b}^*$, perform the integral over
one of the now common ``missing'' variables\footnote{Missing in
the sense that they have been eliminated from the coefficients
subcripts.} and obtain zero. We are left with the case $k = k'$.
We transform the sum over the ``missing'' variable $r_k$ ($l_k$)
into an integral in the corresponding energy. The discrete sum in
the index $k$ takes care of the integration limits and sets them
to infinity. The integral can be performed using residues methods
to get:

\be
\bs
 i[\dot{\s}_{00}^{(n)}(t) + \G_L \s_{00}^{(n)}(t)] &
 = i\G_R \sum_{\{l_j\}\{r_j\}}
 |b_{1\{l_j\}_{j=1}^n\{r_j\}_{j=1}^{n-1}}(t)|^2\\
 i[\dot{\s}_{11}^{(n)}(t) + \G_R \s_{11}^{(n)}(t)] &
 = i\G_L \sum_{\{l_j\}\{r_j\}}
 |b_{\{l_j\}_{j=1}^n\{r_j\}_{j=1}^n}(t)|^2\\
\end{split}
\end{equation}

Finally we use the representation of the $n$-resolved reduced
density matrix (\ref{eq:nRDM}) and obtain the master equation:

\be
\bs
 \dot{\s}_{00}^{(n)} &= -\G_L \s_{00}^{(n)} + \G_R \s_{11}^{(n-1)}\\
 \dot{\s}_{11}^{(n)} &= -\G_R \s_{11}^{(n)} + \G_L \s_{00}^{(n)}
\end{split}
\label{eq:nGMEsimple}
\end{equation}
where we assume that $\s_{11}^{(-1)}\equiv 0$ to include into the
same compact form also the equation (\ref{eq:nGME0}) for $n=0$.
From this set of equations it is possible to determine the current
in the left and right leads. The current in the right lead is the
time derivative of the total number of electrons collected in the
right lead at time $t$:
\be
 I_R(t) = \dot{N}_R(t) = \sum_{n=0}^{\infty}
 n [\dot{\s}_{00}^{(n)}(t) + \dot{\s}_{11}^{(n)}(t)]
\end{equation}
Inserting (\ref{eq:nGMEsimple}) we obtain the intuitive result:

\be
 I_R(t) = \sum_{n=0}^{\infty}
 n \G_R [\s_{11}^{(n-1)}(t) - \s_{11}^{(n)}(t)] =
 \G_R \sum_{n=0}^{\infty} \s_{11}^{(n)}(t) = \G_R \s_{11}(t)
 \label{eq:e-current}
\end{equation}

For the calculation of the left lead current we have to start with
the analog of (\ref{eq:nGMEsimple}) but this time resolved for the
number of holes $h$ accumulated in the emitter:

\be
\bs
 \dot{\s}_{00}^{(h)} &= -\G_L \s_{00}^{(h)} + \G_R \s_{11}^{(h)}\\
 \dot{\s}_{11}^{(h)} &= -\G_R \s_{11}^{(h)} + \G_L \s_{00}^{(h-1)}
\end{split}
\label{eq:hGMEsimple}
\end{equation}
The left lead current reads:

\be
 I_L(t) = \dot{N}_L(t) = \G_L \s_{00}(t).
 \label{eq:h-current}
\end{equation}

The average over the bath degrees of freedom is completed by
summing (\ref{eq:nGMEsimple}) or (\ref{eq:hGMEsimple}) over all
the possible number of electrons (holes) collected in the right
(left) lead.

\be
\bs
 \dot{\s}_{00}&=-\G_L \s_{00} + \G_R \s_{11}\\
 \dot{\s}_{11}&=-\G_R \s_{11} + \G_L \s_{00}
\end{split}
\label{eq:GMEsimple}
\end{equation}

\vspace{1cm}

The system (\ref{eq:GMEsimple}) is a set of rate equations for a
two-state model. The empty and charged states are connected by
charging and discharging rates ($\G_L$ and $\G_R$ respectively)
and the variations in the populations of the two states is given
by a balance of incoming and outgoing currents. The stationary
solution of (\ref{eq:GMEsimple}) is achieved for $I_L =
I_R$\footnote{It is easy to verify that the condition of balanced
current is equivalent to the stationary condition $\dot{\s}_{00} =
\dot{\s}_{11} = 0$ }. This condition and the general sum rule
$\s_{00} + \s_{11} = 1$ give the stationary populations:
\be
\bs
 \s_{00}^{st} &= \frac{\G_R}{\G_L + \G_R}\\
 \s_{11}^{st} &= \frac{\G_L}{\G_L + \G_R}
\end{split}
\end{equation}
and the stationary current:
\be
 I^{st} = \frac{\G_L \G_R}{\G_L + \G_R}.
\end{equation}

\subsection{Spin and strong Coulomb blockade}
The rate equations (\ref{eq:GMEsimple}) are an intuitive result
that can be written simply using common sense. Nevertheless the
effort spent for their microscopic derivation is justified by the
possible generalizations that will lead us to the GME for shuttle
devices. First we want to relax the assumption of spin-less
non-interacting particles. The spin of the electrons can be very
easily taken into account if we assume strong Coulomb repulsion in
the dot. Due to a charging energy much larger than any other
energy in the model we assume that only one electron at a time can
occupy the dot. The Hamiltonian reads:

\be
\bs
 H =&\sum_{l,s} \eps_l c_{ls}^{\dagger}c_{ls}^{\phd} + \eps_1 c_{1s}^{\dagger}c_{1s}^{\phd}
   + \sum_{r,s}\eps_r c_{rs}^{\dagger}c_{rs}^{\phd} \\
   &+\sum_{l,s} \W_l (c_{ls}^{\dagger}c_{1s}^{\phd} + c_{1s}^{\dagger}c_{ls}^{\phd})
   + \sum_{r,s} \W_r (c_{rs}^{\dagger}c_{1s}^{\phd} + c_{1s}^{\dagger}c_{rs}^{\phd})\\
   &+ U c_{1s}^{\dagger}c_{1s}^{\phd}c_{1-s}^{\dagger}c_{1-s}^{\phd}
\end{split}
\label{eq:HGurvitz2}
\end{equation}
%
%%%%%%%%%%%%%%%%%%%%%%%%%%%%%%%%%%%%%%%%%%%%%%%%%%%%%%%%%%%%%%%%%%%
% Figure
%%%%%%%%%%%%%%%%%%%%%%%%%%%%%%%%%%%%%%%%%%%%%%%%%%%%%%%%%%%%%%%%%%%
\begin{figure}
 \begin{center}
  \includegraphics[angle=-90,width=.7\textwidth]{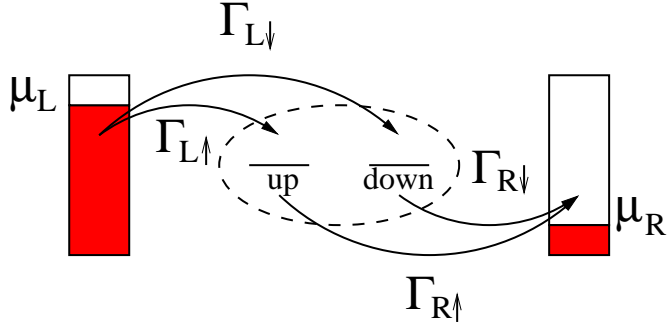}
  \caption{\small  \textit{Schematic representation of the dynamics of the
  single dot device with spin. The two spin species can tunnel
  to and from the two degenerate spin levels of the quantum dot with rates
  $\G_{L\uar},\G_{L\dar}$ and $\G_{R\uar},\G_{R\dar}$, respectively and without
  influencing each other. Due to Coulomb
  blockade only one electron at a time can occupy the dot.}}
 \end{center}
\end{figure}
%%%%%%%%%%%%%%%%%%%%%%%%%%%%%%%%%%%%%%%%%%%%%%%%%%%%%%%%%%%%%%%%%%%
%%%%%%%%%%%%%%%%%%%%%%%%%%%%%%%%%%%%%%%%%%%%%%%%%%%%%%%%%%%%%%%%%%%
% Figure
%%%%%%%%%%%%%%%%%%%%%%%%%%%%%%%%%%%%%%%%%%%%%%%%%%%%%%%%%%%%%%%%%%%
\begin{figure}
 \begin{center}
  \includegraphics[angle=-90,width=.7\textwidth]{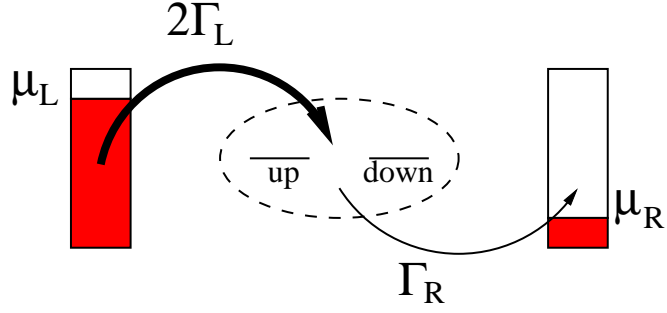}
  \caption{\small  \textit{If we neglect the information of the spin specie the
  system dynamics can be reduced to the one of an effective spinless system with asymmetric
  injection and ejection rates.}}
 \end{center}
\end{figure}
%%%%%%%%%%%%%%%%%%%%%%%%%%%%%%%%%%%%%%%%%%%%%%%%%%%%%%%%%%%%%%%%%%%
which is an extension of the Hamiltonian (\ref{eq:HGurvitz1})
where $s=\pm 1/2$ is the spin degree of freedom and $U$ the
charging energy of the double occupied dot. We take into account
the interaction in the definition of the Hilbert space by
discarding all the many-body states with double occupied dot. The
effective Hamiltonian that we consider is then quadratic, and the
Schr\"odinger equation projected onto the many-body basis gives rise
to a recursive set of equations similar to (\ref{eq:diffsystem}).
We have in this case three general equations corresponding to the
three different states of the quantum dot:

\be
\bs
 i\dot{b}_{\{\uar l_j\}_{j=1}^{n_{\uar}}
           \{\dar l_j\}_{j=1}^{n_{\dar}}}&_{
           \{\uar r_j\}_{j=1}^{n_{\uar}}
           \{\dar r_j\}_{j=1}^{n_{\dar}}} =\\
        &\left[\sum_{k=1}^{n_{\uar}+n_{\dar}}(\eps_{r_k}-\eps_{l_k})\right]
        b_{\{\uar l_j\}_{j=1}^{n_{\uar}}
           \{\dar l_j\}_{j=1}^{n_{\dar}}
           \{\uar r_j\}_{j=1}^{n_{\uar}}
           \{\dar r_j\}_{j=1}^{n_{\dar}}}\\
        &+\sum_{l_{n_{\uar}+1}}\W_{l_{n_{\uar}+1}}
        b_{\uar\{\uar l_j\}_{j=1}^{n_{\uar}+1}
               \{\dar l_j\}_{j=1}^{n_{\dar}}
               \{\uar r_j\}_{j=1}^{n_{\uar}}
               \{\dar r_j\}_{j=1}^{n_{\dar}}}\\
        &+\sum_{l_{n_{\dar}+1}}\W_{l_{n_{\dar}+1}}
        b_{\dar\{\uar l_j\}_{j=1}^{n_{\uar}}
               \{\dar l_j\}_{j=1}^{n_{\dar}+1}
               \{\uar r_j\}_{j=1}^{n_{\uar}}
               \{\dar r_j\}_{j=1}^{n_{\dar}}}\\
        &+\sum_{k=1}^{n_{\uar}}(-1)^{k-n_{\uar}}\W_{r_k}
        b_{\uar\{\uar l_j\}_{j=1}^{n_{\uar}}
               \{\dar l_j\}_{j=1}^{n_{\dar}}
               \{\uar r_j\}_{j=1}^{n_{\uar}}\backslash \{\uar r_k\}
               \{\dar r_j\}_{j=1}^{n_{\dar}}}\\
        &+\sum_{k=1}^{n_{\dar}}(-1)^{k-n_{\dar}}\W_{r_k}
        b_{\dar\{\uar l_j\}_{j=1}^{n_{\uar}}
               \{\dar l_j\}_{j=1}^{n_{\dar}}
               \{\uar r_j\}_{j=1}^{n_{\uar}}
               \{\dar r_j\}_{j=1}^{n_{\dar}}\backslash \{\dar r_k\}}\\
\end{split}
\label{eq:diffsystem21}
\end{equation}
for the empty dot coefficient,

\be
\bs
 i\dot{b}_{\uar\{\uar l_j\}_{j=1}^{n_{\uar}+1}
               \{\dar l_j\}_{j=1}^{n_{\dar}}}&_{
               \{\uar r_j\}_{j=1}^{n_{\uar}}
               \{\dar r_j\}_{j=1}^{n_{\dar}}} =\\
        &\left[\eps_1 + \sum_{k=1}^{n_{\uar}+n_{\dar}}
        (\eps_{r_k}-\eps_{l_k})-\eps_{l_{n_{\uar}+1}}\right]
            b_{\{\uar l_j\}_{j=1}^{n_{\uar}+1}
               \{\dar l_j\}_{j=1}^{n_{\dar}}
               \{\uar r_j\}_{j=1}^{n_{\uar}}
               \{\dar r_j\}_{j=1}^{n_{\dar}}}\\
        &+\sum_{r_{n_{\uar}+1}}\W_{r_{n_{\uar}+1}}
            b_{\{\uar l_j\}_{j=1}^{n_{\uar}+1}
               \{\dar l_j\}_{j=1}^{n_{\dar}}
               \{\uar r_j\}_{j=1}^{n_{\uar}+1}
               \{\dar r_j\}_{j=1}^{n_{\dar}}}\\
        &+\sum_{k=1}^{n_{\uar}+1}(-1)^{k-n_{\uar}-1}\W_{l_k}
            b_{\{\uar l_j\}_{j=1}^{n_{\uar}+1}\backslash \{\uar l_k\}
               \{\dar l_j\}_{j=1}^{n_{\dar}}
               \{\uar r_j\}_{j=1}^{n_{\uar}}
               \{\dar r_j\}_{j=1}^{n_{\dar}}}\\
\end{split}
\label{eq:diffsystem22}
\end{equation}
for the spin-up and finally

\be
\bs
 i\dot{b}_{\dar\{\uar l_j\}_{j=1}^{n_{\uar}}
               \{\dar l_j\}_{j=1}^{n_{\dar}+1}}&_{
               \{\uar r_j\}_{j=1}^{n_{\uar}}
               \{\dar r_j\}_{j=1}^{n_{\dar}}} =\\
       &\left[\eps_1 + \sum_{k=1}^{n_{\uar}+n_{\dar}}
       (\eps_{r_k}-\eps_{l_k})-\eps_{l_{n_{\dar}+1}}\right]
            b_{\{\uar l_j\}_{j=1}^{n_{\uar}}
               \{\dar l_j\}_{j=1}^{n_{\dar}+1}
               \{\uar r_j\}_{j=1}^{n_{\uar}}
               \{\dar r_j\}_{j=1}^{n_{\dar}}}\\
        &+\sum_{r_{n_{\uar}+1}}\W_{r_{n_{\dar}+1}}
            b_{\{\uar l_j\}_{j=1}^{n_{\uar}+1}
               \{\dar l_j\}_{j=1}^{n_{\dar}}
               \{\uar r_j\}_{j=1}^{n_{\uar}}
               \{\dar r_j\}_{j=1}^{n_{\dar}+1}}\\
        &+\sum_{k=1}^{n_{\dar}+1}(-1)^{k-n_{\dar}-1}\W_{l_k}
            b_{\{\uar l_j\}_{j=1}^{n_{\uar}}
               \{\dar l_j\}_{j=1}^{n_{\dar}+1}\backslash \{\dar l_k\}
               \{\uar r_j\}_{j=1}^{n_{\uar}}
               \{\dar r_j\}_{j=1}^{n_{\dar}}}
\end{split}
\label{eq:diffsystem23}
\end{equation}
for the spin-down coefficient. In the last three differential
equations (\ref{eq:diffsystem21}), (\ref{eq:diffsystem22}) and
(\ref{eq:diffsystem23}) we have extended the notation used in
equation (\ref{eq:diffsystem}) to take into account the spin
degree of freedom. Despite the heavy but complete notation that
keeps track of the four baths (two leads with two spin species per
lead) and the state of the dot, the same kind of arguments that we
used for the spin-less case bring us to the set of rate equations:
\be
\bs
\dot{\s}_{00}^{(n)} &= -(\G_{L\uar}+\G_{L\dar})\s_{00}^{(n)} +
                         \G_{R\uar}\s_{\uar\uar}^{(n-1)} +
                         \G_{R\dar}\s_{\dar\dar}^{(n-1)}\\
\dot{\s}_{\uar\uar}^{(n)} &= -\G_{R\uar}\s_{\uar\uar}^{(n)} +
                         \G_{L\uar}\s_{00}^{(n)}\\
\dot{\s}_{\dar\dar}^{(n)} &= -\G_{R\dar}\s_{\dar\dar}^{(n)} +
                         \G_{L\dar}\s_{00}^{(n)}
\end{split}
\label{eq:nGMEspin}
\end{equation}
where $\s_{\uar\uar}^{(n)}$ ($\s_{\dar\dar}^{(n)}$) is the
population of spin up (down) in the dot with $n = n_{\uar} +
n_{\dar}$ electrons in the collector and we have introduced the
spin-dependent injection and ejection rates:

\be
\bs
 \G_{L,R\uar} &= 2 \pi D_{L,R\uar}\W_{L,R}^2\\
 \G_{L,R\dar} &= 2 \pi D_{L,R\dar}\W_{L,R}^2
\end{split}
\end{equation}
The sum over the number of electrons in the collector gives:

\be
\bs
\dot{\s}_{00} &= -(\G_{L\uar}+\G_{L\dar})\s_{00} +
                         \G_{R\uar}\s_{\uar\uar} +
                         \G_{R\dar}\s_{\dar\dar}\\
\dot{\s}_{\uar\uar} &= -\G_{R\uar}\s_{\uar\uar}+
                        \G_{L\uar}\s_{00}\\
\dot{\s}_{\dar\dar} &= -\G_{R\dar}\s_{\dar\dar} +
                         \G_{L\dar}\s_{00}\\
\end{split}
\label{eq:GMEspin}
\end{equation}

The coherencies between different spin species in the QD (e.g.
$\s_{\uar\dar}$) vanish in this model because different spin
states of the dot correspond to different bath states and the only
way to have coherent superpositions in the dot would be to
maintain the same in the leads which instead are assumed (as
macroscopic objects) incoherent\footnote{Non-trivial spin
coherencies can be achieved for example by introducing a spin
dynamics in the dot. In that case different spin states on the dot
could correspond to the same bath state.}. If we are not
interested in the spin information on the dot we can introduce the
population for the charged dot $\s_{11}^{(n)} \equiv
\s_{\uar\uar}^{(n)} + \s_{\dar\dar}^{(n)}$. Assuming also
non-polarized leads (i.e. $D_{L,R\uar}=D_{L,R\dar}$) and,
consequently, tunneling rates independent of different spin
species the system of rate equations (\ref{eq:GMEspin}) becomes:
\be
\bs
 \dot{\s}_{00} &= -2\G_L\s_{00} + \G_R\s_{11}\\
 \dot{\s}_{11} &=  -\G_R\s_{11} +2\G_L\s_{00}
\end{split}
\label{eq:GMEspin2}
\end{equation}
where $\G_L = \G_{L\uar} = \G_{L\dar}$ and $\G_R = \G_{R\uar} =
\G_{R\dar}$. Comparing these rate equations with the ones
derived for the spin-less non-interacting model
(\ref{eq:GMEsimple}) we note that the only remaining signature of
the spin degree of freedom is in the injection rate. In the case
of identical leads the injection rate doubles the ejection rate.
This behaviour can be interpreted in terms of tunneling channels:
both spin species can tunnel in when the dot is empty, but once
the dot is charged with an electron of specific spin only that
species can tunnel out. At this level the spin degree of freedom
is just renormalizing the injection rate. Since this argument can
be repeated for any model in strong Coulomb blockade we will
restrict the derivation of the GME for shuttling devices to
spin-less non-interacting particles.

\subsection{Coherencies and double-dot model}
A simple example of a device that exhibits coherent transport is
represented by an array of two quantum dots located between a
source and a drain lead. We assume that the device is working in
strong Coulomb blockade (i.e. only one electron at a time can occupy
the device, either in the left or in the right dot). Electrons can
tunnel in the device from the emitter only to the left dot while
tunneling off is allowed only from the right dot. This condition
can be achieved due to the fact that the tunneling coupling to the
leads decreases exponentially with the distance and can be
neglected for the far lead. Also the two dots are in tunneling
contact. Since the transport must happen via tunneling between the
discrete levels of the dots we expect coherencies to play a role.
The Hamiltonian for the model reads:

\be
\bs
 H =& \eps_L |L\rangle\!\langle L|
     + \W_0( |L\rangle\!\langle R| + |R\rangle\!\langle L| )
     + \eps_R |R\rangle\!\langle R|\\
    &+ \sum_l \eps_l c_l^{\dagger}c_l^{\phd}
     + \sum_r \eps_r c_r^{\dagger}c_r^{\phd} \\
    &+ \sum_l \W_l
    (|L\rangle\!\langle 0|c_l^{\phd} + |0\rangle\!\langle L|c_l^{\dag})
     + \sum_r \W_r
    (|R\rangle\!\langle 0|c_r^{\phd} + |0\rangle\!\langle R|c_r^{\dag})
\end{split}
\label{eq:HGurvitzdouble}
\end{equation}
%
%%%%%%%%%%%%%%%%%%%%%%%%%%%%%%%%%%%%%%%%%%%%%%%%%%%%%%%%%%%%%%%%%%%
% Figure
%%%%%%%%%%%%%%%%%%%%%%%%%%%%%%%%%%%%%%%%%%%%%%%%%%%%%%%%%%%%%%%%%%%
\begin{figure}
 \begin{center}
  \includegraphics[angle=0,width=.8\textwidth]{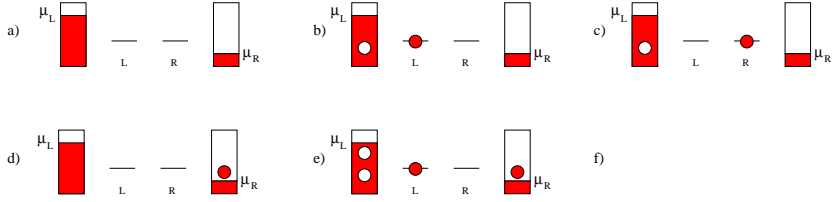}
  \caption{\small  \textit{Schematic representation of the first many-body basis elements
  for the double dot system.}}
 \end{center}
\end{figure}
%%%%%%%%%%%%%%%%%%%%%%%%%%%%%%%%%%%%%%%%%%%%%%%%%%%%%%%%%%%%%%%%%%%
We identify in the first line the system Hamiltonian with the
single energy levels of the left and right dot ($\eps_L$ and
$\eps_R$) and the tunneling amplitude $\W_0$. The second and third
lines describe respectively the electronic baths (the leads) and
their coupling to the device. The system can be found, as in the
spin model, in three different states: empty or occupied with an
electron either in the left or right dot. We associate to each of
these states the vectors $|0\rangle$, $|L\rangle$ and $|R\rangle$.
The Schr\"odinger equation projected onto the many-body basis for
the system+baths Hilbert space can be then represented by the
following three recursive differential equations for the expansion
coefficients:

\be
\bs
 i\dot{b}_{\{l_j\}_{j=1}^n\{r_j\}_{j=1}^n} =&
        \left[\sum_{k=1}^n(\eps_{r_k}-\eps_{l_k})\right]
        b_{\{l_j\}_{j=1}^n\{r_j\}_{j=1}^n}\\
        &\!\!\!+\sum_{l_{n+1}}\W_{l_{n+1}}
        b_{L\{l_j\}_{j=1}^{n+1}\{r_j\}_{j=1}^n}
        \!\!+\sum_{k=1}^n(-1)^{k-n}\W_{r_k}
        b_{R\{l_j\}_{j=1}^n\{r_j\}_{j=1}^n\backslash \{r_k\}}\\
 i\dot{b}_{L\{l_j\}_{j=1}^{n+1}\{r_j\}_{j=1}^n} =&
      \left[\eps_L+\sum_{k=1}^n (\eps_{r_k}-\eps_{l_k})
       -\eps_{l_{n+1}}\right]
       b_{L\{l_j\}_{j=1}^{n+1}\{r_j\}_{j=1}^n}\\
      &\!\!\!+\W_0
       b_{R\{l_j\}_{j=1}^{n+1}\{r_j\}_{j=1}^n}
       +\sum_{k=1}^{n+1}(-1)^{k-n-1}\W_{l_k}
       b_{\{l_j\}_{j=1}^{n+1}\backslash \{l_k\}\{r_j\}_{j=1}^n}\\
i\dot{b}_{R\{l_j\}_{j=1}^{n+1}\{r_j\}_{j=1}^n} =&
      \left[\eps_R+\sum_{k=1}^n (\eps_{r_k}-\eps_{l_k})
       -\eps_{l_{n+1}}\right]
       b_{R\{l_j\}_{j=1}^{n+1}\{r_j\}_{j=1}^n}\\
      &\!\!\!+\W_0
       b_{L\{l_j\}_{j=1}^{n+1}\{r_j\}_{j=1}^n}
       +\sum_{r_{n+1}}\W_{r_{n+1}}
       b_{\{l_j\}_{j=1}^{n+1}\{r_j\}_{j=1}^{n+1}}
\end{split}
\label{eq:diffsystemdouble}
\end{equation}

\noindent The Laplace  transform can be taken and the continuous
sums in the first and third equation be performed to give the set
of algebraic equations:
\be
\bs
&\left[E +\sum_{k=1}^n(\eps_{r_k}-\eps_{l_k}) + i\frac{\G_L}{2}\right]
        \tilde{b}_{\{l_j\}_{j=1}^n\{r_j\}_{j=1}^n}(E)\\
        &\phantom{aaaaa}-\sum_{k=1}^n(-1)^{k-n}\W_{r_k}
        \tilde{b}_{R\{l_j\}_{j=1}^n\{r_j\}_{j=1}^n\backslash \{r_k\}}(E)= i\d_{n0}\\
&\left[E + \eps_L+\sum_{k=1}^n (\eps_{r_k}-\eps_{l_k})
       -\eps_{l_{n+1}}\right]
       \tilde{b}_{L\{l_j\}_{j=1}^{n+1}\{r_j\}_{j=1}^n}(E)\\
       &\phantom{aaaaa}-\W_0
       \tilde{b}_{R\{l_j\}_{j=1}^{n+1}\{r_j\}_{j=1}^n}(E)
       -\sum_{k=1}^{n+1}(-1)^{k-n-1}\W_{l_k}
       \tilde{b}_{\{l_j\}_{j=1}^{n+1}\backslash \{l_k\}\{r_j\}_{j=1}^n}(E)=0\\
&\left[E + \eps_R+\sum_{k=1}^n (\eps_{r_k}-\eps_{l_k})
       -\eps_{l_{n+1}} + i\frac{\G_R}{2}\right]
       \tilde{b}_{R\{l_j\}_{j=1}^{n+1}\{r_j\}_{j=1}^n}(E)\\
       &\phantom{aaaaa}-\W_0
       \tilde{b}_{L\{l_j\}_{j=1}^{n+1}\{r_j\}_{j=1}^n}(E)=0
\end{split}
\label{eq:ratesystemdouble}
\end{equation}

In the double dot model some coherencies of the reduced density
matrix do not vanish since they correspond to different
``internal'' states of the dot and can be combined with the same
state of the baths. For example:
\be
\bs
 \s_{LR}^{(n)}(t) &=
 \sum_{\{B_n\}} \langle L, B_n | \Psi(t) \rangle \!
 \langle \Psi(t) | R, B_n \rangle\\
  &= \sum_{\{l_k\}\{r_k\}}
 b_{L\{l_k\}_{k=1}^{n+1}\{r_k\}_{k=1}^n}(t)
 b_{R\{l_k\}_{k=1}^{n+1}\{r_k\}_{k=1}^n}^*(t)
\end{split}
\end{equation}

The next step is an $n$-resolved GME for the reduced density
matrix $\s$. The equations for the populations are derived
following the procedure we explained for the single dot model:
\be
\bs
 \dot{\s}_{00}^{(n)} &= -\G_L \s_{00}^{(n)} + \G_R \s_{RR}^{(n-1)}\\
 \dot{\s}_{LL}^{(n)} &= -i\W_0\left(\s_{RL}^{(n)} -\s_{LR}^{(n)}\right)
               + \G_L \s_{00}^{n}\\
 \dot{\s}_{RR}^{(n)} &= -\G_R \s_{RR}^{(n)}
                        -i\W_0\left(\s_{LR}^{(n)} -\s_{RL}^{(n)}\right)
\end{split}
\label{eq:populationdouble}
\end{equation}
The coherencies  play an active role in the transport through the
quantum dots: the second and third equations in
(\ref{eq:populationdouble}) show that the left (right) dot can be
discharged (charged) only via  coherent transport. We concentrate
now on the equation for the coherence $\s_{LR}^{(n)}$. We take the
second equation in (\ref{eq:ratesystemdouble}) and multiply it by
$\tilde{b}_{R\{l_j\}_{j=1}^{n+1}\{r_j\}_{j=1}^n}^*(E')
e^{-i(E-E')t}$, then we subtract side by side the complex
conjugate of the third equation in (\ref{eq:ratesystemdouble})
evaluated in $E'$ and multiplied by
$\tilde{b}_{L\{l_j\}_{j=1}^{n+1}\{r_j\}_{j=1}^n}(E)
e^{-i(E-E')t}$. Finally we integrate over $dE$ and $dE'$ and sum
over the baths configurations with $n$ electrons in the collector.
Using the properties of the Laplace transform and the
representation of the reduced density matrix in terms of the
coefficients $b$ of the many-body expansion we obtain:
\be
\bs
 &i\dot{\s}_{LR}^{(n)}(t)
  + \Big(\eps_L -\eps_R +i\frac{\G_R}{2}\Big)\s_{LR}^{(n)}(t)
  -\W_0\Big[\s_{RR}^{(n)}(t) -\s_{LL}^{(n)}(t)\Big]\\
  &-\sum_{\{l_j\}\{r_j\}}\int\frac{dE dE'}{4 \pi^2}\Big[\sum_{k=1}^{n+1}
  (-1)^{k-n-1}\W_{l_k}
  \tilde{b}_{\{l_j\}_{j=1}^{n+1}\backslash \{l_k\}\{r_j\}_{j=1}^n}(E)\\
  &\tilde{b}_{R\{l_j\}_{j=1}^{n+1}\{r_j\}_{j=1}^{n}}^*(E')
  e^{-i(E-E')t}\Big] = 0
\end{split}
\label{eq:coherence}
\end{equation}
The last term in the LHS of equation (\ref{eq:coherence}) vanishes
since the integrand behaves asymptotically as
$o(|\eps_{l_j,r_j}|^{-1})$ in the limit $|\eps_{l_j,r_j}| \to
\infty$  for all the integrated variables $\eps_{l_j,r_j}$. The
average over the bath degrees of freedom is completed by the sum
over the number of electrons in the collector $n$ that leads to
the GME:
\be
\bs
 \dot{\s}_{00} &= -\G_L \s_{00}+\G_R \s_{RR}\\
 \dot{\s}_{LL} &= -i\W_0(\s_{RL}-\s_{LR}) + \G_L \s_{00}\\
 \dot{\s}_{RR} &= -\G_R \s_{RR}-i\W_0(\s_{LR}-\s_{RL})\\
 \dot{\s}_{LR} &= -i(\eps_L -\eps_R)\s_{LR}
                  -\textstyle{\frac{1}{2}}\G_R\s_{LR}
                  -i\W_0(\s_{RR}-\s_{LL})
\end{split}
\label{eq:GMEdouble}
\end{equation}
where we have omitted the equation for ${\s}_{RL}$ since $\s$ is a
hermitian operator on the Hilbert space of the system and $\s_{RL}
= \s_{LR}^*$. We can better visualize the coherent and incoherent
contribution to the GME using a matrix representation:

\be
 \dot{\s} = -i[H_{sys},\s] + \Xi[\s]
\end{equation}
where $\s$ is the density matrix

\be
\s =\left(
\begin{array}{ccc}
\s_{00} &\s_{0L} &\s_{0R} \\
\s_{L0} &\s_{LL} &\s_{LR} \\
\s_{R0} &\s_{RL} &\s_{RR} \\
\end{array}
\right),
\end{equation}
$H_{sys}$ is the Hamiltonian for the system extended to the empty
state for the system

\be
H_{sys} = \left(
\begin{array}{ccc}
0    & 0     & 0 \\
0    &\eps_L &\W_0 \\
0    &\W_0   &\eps_R \\
\end{array}
\right)
\end{equation}
and $\Xi[\bullet]$ is a linear super-operator that transform
operators on the Hilbert space of the system into operators on the
same space and acts on the density matrix:

\be
\Xi[\s] =\left(
\begin{array}{ccc}
-\G_L \s_{00} +\G_R \s_{RR}   & 0  & 0 \\
0  & \G_L \s_{00}  & -\frac{1}{2}\G_R\s_{LR} \\
0    &-\frac{1}{2}\G_R\s_{RL}   & -\G_R\s_{RR} \\
\end{array}
\right)
\end{equation}

\section{GME for shuttle devices}
\label{subsec:driving_and_damping}

The shuttle devices are in contact with two different kinds of
bath: two electrical baths (the leads) and a mechanical bath. We
assume that the electrical and mechanical baths act independently
on the device. This assumption splits the GME into two additive
components, one for each kind of bath. Due to the different
coupling strengths we derive the electrical component extending
the method proposed by Gurvitz and extensively presented in
Section \ref{sec:Gurvitz} while for the mechanical component we
adopt the weak coupling quantum optical derivation presented in
Section \ref{sec:QOptical}.

\subsection{Single Dot Quantum Shuttle}\label{sec:GMESDQS}

We start recalling the Hamiltonian for the SDQS:

\be
    H =H_{\rm sys}+H_{\rm leads}+H_{\rm bath}
      +H_{\rm tun}+H_{\rm int}
\end{equation}
where
\be
 \bs
  &H_{\rm sys} =\frac{\hat{p}^2}{2 m}
               +\frac{1}{2}m \w^2 \hat{x}^2
               +(\eps_1- e\mathcal{E} \hat{x})c_1^{\dag}c_1^{\phd}\\
 &H_{\rm leads} = \sum_{k}(\eps_{l_k}
               c^{\dagger}_{l_k}c^{\phd}_{l_k}
               +\eps_{r_k}
               c^{\dagger}_{r_k}c^{\phd}_{r_k})\\
 &H_{\rm tun} = \sum_{k}[T_{l}(\hat{x}) c^{\dagger}_{l_k}c_1^{\phd} +
                         T_{r}(\hat{x}) c^{\dagger}_{r_k}c_1^{\phd}] + h.c.\\
 &H_{\rm bath} + H_{\rm int }= {\rm generic \; heat \; bath}
 \end{split}
\end{equation}

Also the mechanical degree of freedom is treated quantum
mechanically. For example the position operator can be expressed
in the form:
\be
\hat{x} = \sqrt{\frac{\hbar}{2 m \w}}(d^{\dag}+d)
\end{equation}
where $d^{\dag}$ and $d$
are respectively the creation and annihilation operators for
the harmonic oscillator.
We neglect for a moment the mechanical bath and its coupling to
the system and start the Gurvitz analysis of the model dynamics.
The many-body basis introduced in (\ref{eq:basis}) must be
extended to take into account also the phononic excitations of the
system. For definiteness we choose the eigenvectors of the
oscillator Hamiltonian as a basis for the mechanical part. We
display the basis elements in the following table:

\begin{equation}
\begin{tabular}{|l|l|l|l}
\hline
   \multicolumn{1}{|c}{r=0}
 & \multicolumn{1}{|c}{r=1}
 & \multicolumn{1}{|c}{r=2}
 & \multicolumn{1}{|c}{$\ldots$}\\
 \hline
   $|0\rangle$
 &
$d^{\dag}|0\rangle$
 & ${d^{\dag}}^2|0\rangle$
 & $\ldots$\\
   $c_1^{\dagger}   c_{l_1}^{\phd}|0\rangle$
 & $c_1^{\dagger}c_{l_1}^{\phd} d^{\dag}|0\rangle$
 & $c_1^{\dagger}c_{l_1}^{\phd}{d^{\dag}}^2|0\rangle$
 & $\ldots$\\
   $c_{r_1}^{\dagger}c_{l_1}^{\phd}|0\rangle$
 & $c_{r_1}^{\dagger}c_{l_1}^{\phd} d^{\dag}|0\rangle$
 & $c_{r_1}^{\dagger}c_{l_1}^{\phd}{d^{\dag}}^2|0\rangle$
 & $\ldots$\\
   \multicolumn{1}{|c}{$\vdots$} &
   \multicolumn{1}{|c}{$\vdots$} &
   \multicolumn{1}{|c|}{$\vdots$} &
\end{tabular}
\label{eq:SDQSbasis}
\end{equation}

\noindent where $r$ is the number of excitations  of the
oscillator and the empty state $|0\rangle$ represents the empty
dot in its \emph{mechanical ground state} with leads filled up to
their Fermi energies. It is convenient to organize the
coefficients of the expansion of the state vector $|\Psi\rangle$
in the basis (\ref{eq:SDQSbasis}) in vectors: one for each
electronic configuration. The different elements of the vectors
refer to the different excited states for the oscillator. The
Schr\"odinger equation for the state vector $|\Psi\rangle$ is
represented in the basis (\ref{eq:SDQSbasis}) by two recursive
differential equations for the vector coefficients $\vet{b}$'s:

\be
\bs
 i\dot{\vet{b}}_{\{l_j\}_{j=1}^n}&_{\{r_j\}_{j=1}^n} =
         \left[\hat{H}_{\rm osc}
         +\sum_{k=1}^n(\eps_{r_k}-\eps_{l_k})\right]
         \vet{b}_{\{l_j\}_{j=1}^n\{r_j\}_{j=1}^n}\\
        &+\sum_{l_{n+1}}T_{l}(\hat{x})
         \vet{b}_{1\{l_j\}_{j=1}^{n+1}\{r_j\}_{j=1}^n}
         +\sum_{k=1}^n(-1)^{k-n}T_{r}(\hat{x})
         \vet{b}_{1\{l_j\}_{j=1}^n\{r_j\}_{j=1}^n\backslash \{r_k\}}\\
 i\dot{\vet{b}}_{1\{l_j\}_{j=1}^{n+1}}&_{\{r_j\}_{j=1}^n} =
         \left[\hat{H}_{\rm osc} + \eps_1- e\mathcal{E}\hat{x}
         +\sum_{k=1}^n (\eps_{r_k}-\eps_{l_k})
         -\eps_{l_{n+1}}\right]
         \vet{b}_{1\{l_j\}_{j=1}^{n+1}\{r_j\}_{j=1}^n}\\
        &+\sum_{r_{n+1}}T_{r}(\hat{x})
         \vet{b}_{\{l_j\}_{j=1}^{n+1}\{r_j\}_{j=1}^{n+1}}
         +\sum_{k=1}^{n+1}(-1)^{k-n-1}T_{l}(\hat{x})
         \vet{b}_{\{l_j\}_{j=1}^{n+1}\backslash \{l_k\}\{r_j\}_{j=1}^n}\\
\end{split}
\label{eq:SDQSdiffsystem}
\end{equation}
where $\hat{x}$ is given in its  matrix representation in terms of
the occupation number basis ($\hbar = 1$):

\be
\hat{x}_{rs} = \sqrt{\frac{r}{2 m \w}}(\d_{r,s+1}+\d_{r,s-1})
\end{equation}
and the Hamiltonian for the harmonic oscillator in the same basis
reads\footnote{The representation given in equation
(\ref{eq:SDQSdiffsystem}) is actually independent of the basis
for the oscillator Hilbert space. The $\vet{b}$ vectors are
projections of the state vector $|\Psi\rangle$ on the particular
subspace given by the electronic configuration specified by the
subscript.}:

\be
 \hat{H}_{\rm osc} = \w\left(\frac{1}{2} + r\d_{rs}\right)
\end{equation}

All the constants in equation (\ref{eq:SDQSdiffsystem}) are
identity operators in the mechanical Hilbert space.

One of the key assumptions in the derivation of the GME in the
Gurvitz approach is the position of the energy levels of the
system: they must lie well inside the transport window open
between the chemical potentials of the leads. Since the oscillator
spectrum is \emph{not} bounded from above we assume that only a
finite number of mechanical excitations are involved in the
dynamics of the system. We will see that, at least in the presence
of a mechanical bath, this assumption is numerically fulfilled. In
any case a violation of this condition in the final result would
be unacceptable since it would violate the validity condition of
the GME. From the point of view of the experimental realization of
the device this limit is imposed at least by the leads that set an
upper bound to the amplitude of the dot oscillations.

The Laplace transform of (\ref{eq:SDQSdiffsystem}) with the
initial condition $|\Psi(t=0)\rangle = |0\rangle$ reads:

\be \bs &\left[E + \hat{H}_{\rm osc} + \sum_{k=1}^n
(\eps_{r_k}-\eps_{l_k})\right]
         \tilde{\vet{b}}_{\{l_j\}_{j=1}^n\{r_j\}_{j=1}^n}(E)
         -\sum_{l_{n+1}}T_{l}(\hat{x})
         \tilde{\vet{b}}_{1\{l_j\}_{j=1}^{n+1}\{r_j\}_{j=1}^n}(E)\\
         &\phantom{a}-\sum_{k=1}^n(-1)^{k-n}T_{r}(\hat{x})
         \tilde{\vet{b}}_{1\{l_j\}_{j=1}^n\{r_j\}_{j=1}^n\backslash \{r_k\}}(E) =
         i\d_{n0}\vet{v}_0\\
&\left[E + \hat{H}_{\rm osc} +\eps_1- e\mathcal{E}\hat{x}
         + \sum_{k=1}^n (\eps_{r_k}-\eps_{l_k})
         -\eps_{l_{n+1}}\right]
         \tilde{\vet{b}}_{1\{l_j\}_{j=1}^{n+1}\{r_j\}_{j=1}^n}(E)\\
         &\phantom{a}-\!\sum_{r_{n+1}}T_{r}(\hat{x})
         \tilde{\vet{b}}_{\{l_j\}_{j=1}^{n+1}\{r_j\}_{j=1}^{n+1}}(E)
         -\!\sum_{k=1}^{n+1}(-1)^{k-n-1}T_{l}(\hat{x})
         \tilde{\vet{b}}_{\{l_j\}_{j=1}^{n+1}\backslash \{l_k\}\{r_j\}_{j=1}^n}(E)=0
\end{split}
\label{eq:SDQSalgebsystem}
\end{equation}
where $\vet{v}_0$ is the initial condition of the oscillator.

The continuous sums in the  system (\ref{eq:SDQSalgebsystem}) can
be performed with an argument similar to the one used for the
static QD. We have just to be careful with the matrix notation and
change the basis to diagonalize the matrix:

\be
 M = \left[E + \hat{H}_{\rm osc} +\eps_1- e\mathcal{E}\hat{x}
         + \sum_{k=1}^n (\eps_{r_k}-\eps_{l_k})
         -\eps_{l_{n+1}}\right]
\end{equation}
before taking the integral. The sum $\sum_{r_{n+1}}$ in the second
Eq.\ of (\ref{eq:SDQSalgebsystem}) can be treated analogously. As
in the static case the continuous sums are condensed into
``rates''\footnote{These ``rates'' are position dependent and then
in our quantum treatment they are operators. Actual rates can be
recovered by averaging these operators on a given the quantum
state.}:

\be
\bs
&\left[E + \hat{H}_{\rm osc} + \sum_{k=1}^n (\eps_{r_k}-\eps_{l_k})
         -i\frac{\G_L(\hat{x})}{2} \right]
         \tilde{\vet{b}}_{\{l_j\}_{j=1}^n\{r_j\}_{j=1}^n}(E)\\
         &\phantom{aaaaaa}-\sum_{k=1}^n(-1)^{k-n}T_{r}(\hat{x})
         \tilde{\vet{b}}_{1\{l_j\}_{j=1}^n\{r_j\}_{j=1}^n\backslash \{r_k\}}(E) =
         i\d_{n0}\vet{v}_0\\
&\left[E + \hat{H}_{\rm osc} +\eps_1- e\mathcal{E}\hat{x}
         + \sum_{k=1}^n (\eps_{r_k}-\eps_{l_k})
         -\eps_{l_{n+1}} -i\frac{\G_R(\hat{x})}{2}\right]
         \tilde{\vet{b}}_{1\{l_j\}_{j=1}^{n+1}\{r_j\}_{j=1}^n}(E)\\
         &\phantom{aaaaaa}-\sum_{k=1}^{n+1}(-1)^{k-n-1}T_{l}(\hat{x})
         \tilde{\vet{b}}_{\{l_j\}_{j=1}^{n+1}\backslash \{l_k\}\{r_j\}_{j=1}^n}(E)=0
\end{split}
\label{eq:SDQSrates}
\end{equation}
where

\be
 \G_{L,R}(\hat{x}) = 2\pi D_{L,R}T^2_{l,r}(\hat{x})
                   = \G_{L,R}\, e^{\mp 2\hat{x}/\l}
\end{equation}
and we have introduced  the tunneling length $\lambda$ and the
bare injection (ejection) rate $\G_L$ ($\G_R$).

The reduced density matrix for the system contains information
about the electrical occupation of the QD \emph{and} its
mechanical state. Coherencies between occupied and empty state
vanish because they imply coherencies between states with
different particle number in the baths. The equation
(\ref{eq:genRDM}) for the non-vanishing elements written in the
static QD case still holds:

\be
 \s_{ii}(t) = \sum_{n=0}^{\infty}
 \sum_{\{B_n\}}\langle i_S,B_n|\Psi(t)\rangle\!\langle\Psi(t)|B_n,i_S\rangle
 = \sum_{n=0}^{\infty}\s_{ii}^{(n)}(t)
\end{equation}
with the difference that the  ``elements'' $\s_{ii}$ are now full
operators in the mechanical Hilbert space. It is useful to express
them in terms of the vectors $\vet{b}$:

\be
 \bs
  \s_{00}^{(n)}(t) &=\sum_{\{l_k\}\{r_k\}}
   \vet{b}_{\{l_k\}_{k=1}^n\{r_k\}_{k=1}^n}(t)
   \vet{b}_{\{l_k\}_{k=1}^n\{r_k\}_{k=1}^n}^{\dag}(t)\\
  \s_{11}^{(n)}(t) &=\sum_{\{l_k\}\{r_k\}}
   \vet{b}_{1\{l_k\}_{k=1}^{n+1}\{r_k\}_{k=1}^n}(t)
   \vet{b}_{1\{l_k\}_{k=1}^{n+1}\{r_k\}_{k=1}^n}^{\dag}(t)
 \end{split}
 \label{eq:SDQSnRDM}
\end{equation}
The notation of  equation (\ref{eq:SDQSnRDM}) can be understood in
terms of the Dirac notation: $\vet{b}^{\dag}$ is the \emph{bra} of
the corresponding vector $\vet{b}$ (the \emph{ket})\footnote{In
other terms the linear operator $\s_{ii}$ is the tensor product of
the vector $\vet{b}$ and the linear form $\vet{b}^{\dag}$.}. The
inverse Laplace transform brings us back to the vectors
$\tilde{\vet{b}}$:

\be
 \bs
  \s_{00}^{(n)}(t) &=\int \frac{dE dE'}{4 \pi^2}\sum_{\{l_k\}\{r_k\}}
   \tilde{\vet{b}}_{\{l_k\}_{k=1}^n\{r_k\}_{k=1}^n}(E)
   \tilde{\vet{b}}_{\{l_k\}_{k=1}^n\{r_k\}_{k=1}^n}^{\dag}(E')
   e^{-i(E-E')t}\\
  \s_{11}^{(n)}(t) &=\int \frac{dE dE'}{4 \pi^2}\sum_{\{l_k\}\{r_k\}}
   \tilde{\vet{b}}_{1\{l_k\}_{k=1}^{n+1}\{r_k\}_{k=1}^n}(E)
   \tilde{\vet{b}}_{1\{l_k\}_{k=1}^{n+1}\{r_k\}_{k=1}^n}^{\dag}(E')
   e^{-i(E-E')t}
 \end{split}
\end{equation}

The case with $n=0$ must be treated separately. The inverse
Laplace transform of the first equation in the system
(\ref{eq:SDQSrates}) specialized for $n=0$ reads:

\be
 i\dot{\vet{b}}_0 = \hat{H}_{\rm osc}\vet{b}_0
                    -i\frac{\G_L(\hat{x})}{2}\vet{b}_0
\label{eq:dotb0}
\end{equation}
and its Hermitian conjugate:
\be
 -i\dot{\vet{b}}_0^{\dag} = \vet{b}_0^{\dag}\hat{H}_{\rm osc}
                    +i\vet{b}_0^{\dag}\frac{\G_L(\hat{x})}{2}
\label{eq:dotb0dag}
\end{equation}
where we have used the property of adjoint of vectors and
operators $(AB)^{\dag} = B^{\dag}A^{\dag}$ and the fact that the
oscillator Hamiltonian and position operator are Hermitian on
the mechanical Hilbert space. The combination of equations
(\ref{eq:SDQSnRDM}), (\ref{eq:dotb0}), (\ref{eq:dotb0dag}) and the
Leibnitz rule for derivatives extended to the tensor product
between vectors and linear forms lead to the first component of
the GME for the SDQS:

\be
 \dot{\s}_{00}^{(0)}(t) = -i\Big[\hat{H}_{\rm osc},\s_{00}^{(0)}\Big]
                          -\frac{\G_L}{2}
                          \Big\{e^{-2\hat{x}/\lambda},\s_{00}^{(0)}\Big\}
\label{eq:SDQSGME0}
\end{equation}
where $[A,B]\equiv AB - BA$ is the commutator and $\{A,B\} \equiv
AB + BA$ the anticommutator between the operators $A$ and $B$.
Equation (\ref{eq:SDQSGME0}) already contains the essence of the
driving part of the GME: a coherent evolution represented by the
commutator with the oscillator Hamiltonian and a non-coherent term
due to the interaction with the bath. In this second contribution
the quantum features are given by the particular ordering of the
operators.

For the general case with $n \ne 0$ the procedure is to take the
first equation of (\ref{eq:SDQSrates}) evaluated in $E$ and make
the tensor product with
$\tilde{\vet{b}}_{\{l_j\}_{j=1}^n\{r_j\}_{j=1}^n}^{\dag}(E')$
$e^{-i(E - E')t}$, subtract side by side the adjoint of the first
equation in (\ref{eq:SDQSrates}) evaluated in $E'$ multiplied
(from the right) with the vector
$\tilde{\vet{b}}_{\{l_j\}_{j=1}^n\{r_j\}_{j=1}^n}(E)$
$e^{-i(E-E')t}$, integrate in $dE$ and $dE'$ and sum over the
possible bath configuration with $n$ extra electrons in the right
lead. Using the properties of the Laplace transform and the
representation of the reduced density matrix in terms of the
vectors $\vet{b}$ (\ref{eq:SDQSnRDM}) we get:

\be
\bs
 i\dot{\s}&_{00}^{(n)}- \Big[\hat{H}_{\rm osc},\s_{00}^{(n)}\Big]
                        -i\frac{\G_L}{2} \Big\{e^{-2\hat{x}/\l},\s_{00}^{(n)}\Big\}\\
  &-\!\!\sum_{\{l_j\}\{r_j\}}\!\!\int\frac{dE dE'}{4 \pi^2}\!\sum_{k=1}^{n}
  (-1)^{k-n}\Big[
  T_r(\hat{x})
  \tilde{\vet{b}}_{1\{l_j\}_{j=1}^n\{r_j\}_{j=1}^n\backslash \{r_k\}}(E)
  \tilde{\vet{b}}_{\{l_j\}_{j=1}^n\{r_j\}_{j=1}^n}^{\dag}(E')\\
  &-\tilde{\vet{b}}_{\{l_j\}_{j=1}^n\{r_j\}_{j=1}^n}(E)
  \tilde{\vet{b}}_{1\{l_j\}_{j=1}^n\{r_j\}_{j=1}^n\backslash \{r_k\}}^{\dag}(E')
  T_r(\hat{x})
  \Big]e^{-i(E-E')t} = 0
  \end{split}
\label{eq:SDQSrate2}
\end{equation}

We solve the first equation in (\ref{eq:SDQSrates}) with respect
to $\tilde{\vet{b}}_{\{l_j\}_{j=1}^n\{r_j\}_{j=1}^n}$. Then we
insert the result and its adjoint in (\ref{eq:SDQSrate2}) and, as
in the static QD, we are left with the only non-vanishing
continuous sum in the ``missing'' variable $r_n$. The result is
the matrix equation:

\be
 \dot{\s}_{00}^{(n)}= -i\Big[\hat{H}_{\rm osc},\s_{00}^{(n)}\Big]
 -\frac{\G_L}{2}\Big\{e^{-2\hat{x}/\l},\s_{00}^{(n)}\Big\}
 +\G_R e^{\hat{x}/\l}\s_{11}^{(n-1)}e^{\hat{x}/\l}
\end{equation}

The treatment of the second equation in (\ref{eq:SDQSrates}) is
totally analogous and brings us to  an equation of motion for the
charged component of the density matrix ($\s_{11}^{(n)}$).
Collecting all the results we can write the $n$-resolved GME:

\be
\bs
 \dot{\s}_{00}^{(n)}&= -i\Big[\hat{H}_{\rm osc},\s_{00}^{(n)}\Big]
        -\frac{\G_L}{2}\Big\{e^{-2\hat{x}/\l},\s_{00}^{(n)}\Big\}
        +\G_R e^{+\hat{x}/\l}\s_{11}^{(n-1)}e^{+\hat{x}/\l}\\
 \dot{\s}_{11}^{(n)}&= -i\Big[\hat{H}_{\rm osc} -e\mathcal{E}\hat{x},\s_{11}^{(n)}\Big]
        -\frac{\G_R}{2}\Big\{e^{+2\hat{x}/\l},\s_{11}^{(n)}\Big\}
        +\G_L e^{-\hat{x}/\l}\s_{00}^{(n)}e^{-\hat{x}/\l}
\end{split}
\label{eq:nGMESDQS}
\end{equation}

In order to complete the description of the dynamics  of the SDQS
we have to take into account the mechanical bath and its
interaction with the system. We derive the mechanical component of
the GME starting from the general formulation for the equation of
motion of the reduced density matrix (\ref{eq:QOpGME}). We
consider the problem described by the Hamiltonian (at the moment
independent of the electronic dynamics):

\be
 H = H_{\rm sys} + H_{\rm bath} + H_{\rm int}
\end{equation}
where
\be
\bs
 H_{\rm sys} &= \hbar \w \left(\frac{1}{2} + d^{\dag}d\right)\\
 H_{\rm bath} &= \sum_{\vet{q}} \hbar \w_{\vet{q}}
 {d_{\vet{q}}}^{\dagger} d_{\vet{q}}\\
\end{split}
\end{equation}
and the generic interaction contribution:

\be
 H_{\rm int} = \hbar\sum_a G_a F_a
 \label{eq:GenInt}
\end{equation}
$G_a$ being a system operator and $F_a$ a bath  operator and $a$ a
generic quantum number. We will specialize later the interaction in the
form we have introduced in the chapter dedicated to the model:

\be
 H_{\rm int} = \hbar \sum_{\vet{q}}
              g(d_{\vet{q}} + d^{\dagger}_{\vet{q}})
              (d + d^{\dagger})
\label{eq:nRWAint}
\end{equation}
and with the rotating wave approximation

\be
 H_{\rm int}^{\rm (RWA)} = \hbar \sum_{\vet{q}}
              g(d^{\dagger}_{\vet{q}}d + d^{\dagger}d_{\vet{q}})
\label{eq:RWAint}
\end{equation}

We start recalling the GME \eqref{eq:QOpGME} derived  in the weak
coupling using the quantum optical formalism:

\be
 \dot{\tilde{\s}} = -\frac{1}{\hbar^2} \int_{0}^{\infty} d\t
 {\rm Tr_B}\{[\tilde{H}_{\rm int}(t),[\tilde{H}_{\rm int}(t-\t),
 \tilde{\s}(t)\otimes \rho_{\rm B}]]\}
\end{equation}
With the generic form for the interaction Hamiltonian (\ref{eq:GenInt}) we get:

\be
\bs
 \dot{\tilde{\s}}(t) = - \int_0^{\infty}\!\!d\t \sum_{ab}
 \{
  &[\tilde{G}_a(t)\tilde{G}_b(t-\t) \tilde{\s}(t)
 -\tilde{G}_b(t-\t) \tilde{\s}(t) \tilde{G}_a(t)]
  \langle \tilde{F}_a(\t)\tilde{F}_b(0) \rangle\\
 +&[\tilde{\s}(t)\tilde{G}_a(t-\t)  \tilde{G}_b(t)
  -\tilde{G}_b(t) \tilde{\s}(t) \tilde{G}_a(t-\t)]
  \langle \tilde{F}_a(0)\tilde{F}_b(\t) \rangle
 \}
\end{split}
\end{equation}
where the tilde indicates the  interaction picture and $\langle
\bullet \rangle \equiv {\rm Tr_B}\{\rho_B \bullet\}$. We can
easily go to the Schr\"odinger picture:

\be
\bs
 \dot{\s} = -\frac{i}{\hbar}[H_{\rm sys},\s] +
  \sum_{ab}\int_0^{\infty} d \t
  \{
  &[\tilde{G}_b(-\t)\langle \tilde{F}_a(\t)\tilde{F}_b(0)\rangle\s,G_a]\\
   +
  &[G_b,\s\tilde{G}_a(-\t)\langle \tilde{F}_a(\t)\tilde{F}_b(0)\rangle]
  \}
\end{split}
\end{equation}
Following \cite{koh-jcp-97} we introduce the compact notation:

\be
 \dot{\s} = -\frac{i}{\hbar}[H_{\rm sys},\s] +
  \sum_{a}\{[G_a^+\s,G_a] + [G_a,\s G_a^-]\}
\label{eq:KohGME}
\end{equation}
where

\be \bs G_a^+ &= \sum_b\int_0^{\infty} d\tau
         \tilde{G}_b(-\t)\langle \tilde{F}_a(\t)\tilde{F}_b(0)\rangle\\
G_a^- &= \sum_b\int_0^{\infty} d\tau
         \tilde{G}_b(-\t)\langle \tilde{F}_b(0)\tilde{F}_a(\t)\rangle\\
\end{split}
\end{equation}
This formalism is very efficient since,  for a given interaction,
it requires for the GME simply the calculation of the two
operators $G^+$ and $G^-$. For the interaction Hamiltonian
(\ref{eq:nRWAint}) we identify

\be
\bs
 G_{\vet{q}} &= d + d^{\dag}\\
 F_{\vet{q}} &=g(d_{\vet{q}}^{\phd} + d_{\vet{q}}^{\dag})
\end{split}
\end{equation}
and calculate

\be
\bs
 G^+_{\vet{q}} =& \sum_{\vet{q}'}\int_0^{\infty} d\t
 \tilde{G}_{\vet{q}'}(-\t)
 \langle \tilde{F}_{\vet{q}}(\t)\tilde{F}_{\vet{q}'}(0)\rangle\\
 =& d^{\phd}
 g^2 \int_0^{\infty} d\t e^{i\w\t}
 \{e^{ i\w_{\vet{q}}\t }n_B(\w_{\vet{q}}) +
   e^{-i\w_{\vet{q}}\t }[1+ n_B(\w_{\vet{q}})] \}\\
 &+ d^{\dag}
 g^2  \int_0^{\infty} d\t e^{-i\w\t}
 \{e^{ i\w_{\vet{q}}\t }n_B(\w_{\vet{q}}) +
   e^{-i\w_{\vet{q}}\t }[1+ n_B(\w_{\vet{q}})] \}\\
   =& \pi g^2 \{d [1+n_B(\w_{\vet{q}})] + d^{\dag}n_B(\w_{\vet{q}}) \}
   \d(\w - \w_{\vet{q}})
\end{split}
\end{equation}
where we assumed the bath in thermal equilibrium and calculated the average:

\be
 \langle d_{\vet{q}}^{\dag}d_{\vet{q}}^{\phd}\rangle
 = n_B(\w_{\vet{q}}) \equiv \frac{1}{e^{\b\hbar\w_{\vet{q}}}-1}
\end{equation}
and we integrated the exponentials
\be
 \int_0^{\infty} d\t e^{i(\w \pm \w_{\vet{q}})\t}
 = \pi\d(\w \pm \w_{\vet{q}})
 + \mathcal{P} \left(\frac{i}{\w \pm \w_{\vet{q}}}\right)
 \approx \pi\d(\w \pm \w_{\vet{q}})
\end{equation}
where $\mathcal{P}$ denotes the principal  value. We neglected the
contribution due the principal value that only slightly shifts the
oscillator frequency $\w$. Terms proportional to $\d(\w +
\w_{\vet{q}})$ vanish since both $\w$ and $\w_{\vet{q}}$ are
positive. The operator $G^-$ can be calculated in a similar way
and reads:

\be
 G^-_{\vet{q}}= \pi g^2 \{ d n_B(\w_{\vet{q}}) +
   d^{\dag} [1+n_B(\w_{\vet{q}})] \}
   \d(\w - \w_{\vet{q}}) = (G^+_{\vet{q}})^{\dag}
\end{equation}
Substituting all the $G$ operators, the GME (\ref{eq:KohGME}) takes the form:

\be
 \dot{\s} = -\frac{i}{\hbar}[H_{\rm sys},\s] +
 \frac{\g}{2}[1 + n_B(\w)] [d + d^{\dag}, \s d^{\dag} -d \s ]
 +
 \frac{\g}{2}n_B(\w) [d + d^{\dag}, \s d -d^{\dag} \s ]
\label{eq:GMEnRWA}
\end{equation}
where $\gamma = 2 \pi D(\w) g^2$ is the damping  rate and $D(\w)$
is the density of states of the phonon bath at the frequency of
the oscillator. We rewrite the previous equation in the form:

\be
 \dot{\s} = \mathcal{L}[\s] = \mathcal{L}_{\rm coh}[\s]
 + \mathcal{L}_{\rm damp} [\s]
\end{equation}
The linear (super) operator $\mathcal{L}$ is  also known as
the Liouvillean and is a linear operator defined on the space of
linear operators on the Hilbert space of the system. The first
term $\mathcal{L}_{\rm coh} = -\frac{i}{\hbar}[H_{\rm sys},\s]$
describes the coherent dynamics of the isolated system. The terms
proportional to $\gamma$ and grouped in $\mathcal{L}_{\rm damp}
\s$ represent the interaction with the bath which is damping the
oscillator. This interaction introduces decoherence in the system
in the sense that no matter how we prepare the initial quantum
state of the oscillator ($\s(t=0)$), in absence of other driving
forces, the stationary state reached at long times ($\s(t =
\infty)$) is a thermal distribution that corresponds to a diagonal
density matrix with no coherencies left. The thermal stationary
solution is a typical property required from a generalized master
equation. The GME (\ref{eq:GMEnRWA}) is also translationally
invariant as can be more directly checked introducing the position
and momentum operators for the oscillator $x,p$\footnote{From now
on we drop for simplicity the hat for the operators. It will
be clear from the context if we are dealing with operators or
with classical variables.}:

\be
 \dot{\s} = -\frac{i}{\hbar}[H_{\rm sys},\s]
            -\frac{i\g}{2\hbar} [x,\{p,\s\}]
            -\frac{\g m \w}{\hbar}\left[n_B(\w)+\frac{1}{2}\right]
             [x,[x,\s]]
\label{eq:TrasInvLiou}
\end{equation}

Unfortunately a translationally invariant  GME with a thermal
stationary solution is generating density matrices which are not
\emph{a priori} always positive definite. This is a general
problem of the GME \cite{koh-jcp-97}. In our specific case though
we checked numerically that in the relevant cases the positivity
was not broken within numerical errors.

The interaction Hamiltonian in the rotating wave approximation
(\ref{eq:RWAint}) can be treated in a similar way. One has to
extend the space of quantum numbers and define:

\be
\begin{array}{ll}
 G_{\vet{q}_1} = d & G_{\vet{q}_2} = d^{\dag}\\
 F_{\vet{q}_1} = g d_{\vet{q}}^{\dag} & F_{\vet{q}_2} = g d_{\vet{q}}
\end{array}
\end{equation}
The corresponding $G^+$ and $G^-$ operators read:

\be
\bs
G_{\vet{q}_1}^+ &= d^{\dag}
\pi g^2 n_B(\w_{\vet{q}}) \d(\w - \w_{\vet{q}})\\
G_{\vet{q}_2}^+ &= d
\pi g^2 [1 + n_B(\w_{\vet{q}})] \d(\w - \w_{\vet{q}})\\
G_{\vet{q}_1}^- &=(G_{\vet{q}_2}^+)^{\dag}\\
G_{\vet{q}_2}^- &=(G_{\vet{q}_1}^+)^{\dag}
\end{split}
\end{equation}
We insert the operators in the general GME (\ref{eq:KohGME}) and
obtain:

\be
\bs
 \dot{\s} = -\frac{i}{\hbar}[H_{\rm sys},\s]
 &+ \frac{\g}{2} n_B(\w)([d^{\dag}\s,d] + [d^{\dag},\s d])\\
 &+ \frac{\g}{2} [1+ n_B(\w)]([d,\s d^{\dag}] + [d\s, d^{\dag}])
\end{split}
\label{eq:RWAdampkernel}
\end{equation}
where we have defined as usual the damping rate $\g \equiv 2 \pi D g^2$.
Collecting all the terms we can write the GME for the SDQS in the form:

\be
\begin{tabular}{|rl|}
\hline
 \phantom{a} & \phantom{a}\\
 $\dot{\s}_{00}=$& $\!\!\!-i\Big[H_{\rm osc},\s_{00}\Big]
        -\frac{\G_L}{2}\Big\{e^{-\frac{2x}{\l}},\s_{00}\Big\}
        +\G_R e^{\frac{x}{\l}}\s_{11}e^{\frac{x}{\l}}
        + \mathcal{L}_{\rm damp}[\s_{00}]$\\
 \phantom{a} & \phantom{a}\\
 $\dot{\s}_{11}=$& $\!\!\!-i\Big[H_{\rm osc} -e\mathcal{E}x,\s_{11}\Big]
        -\frac{\G_R}{2}\Big\{e^{\frac{2x}{\l}},\s_{11}\Big\}
        +\G_L e^{-\frac{x}{\l}}\s_{00}e^{-\frac{x}{\l}}
        + \mathcal{L}_{\rm damp}[\s_{11}]$\\
 \phantom{a} & \phantom{a}\\
\hline
\end{tabular}
\label{eq:SDQSGME}
\end{equation}
where we have taken the sum over the number of electrons  collected
in the resevoir and we have introduced the generic damping
Liouvillean $\mathcal{L}_{\rm damp}$. One can use for example one
of the two damping  Liouvillean we have just derived. In the rest
of the thesis we will always adopt for the SDQS the
translationally invariant damping Liouvillean (\ref{eq:GMEnRWA}).
We will also refer to the electronic part of the equation
(\ref{eq:SDQSGME}) as to the driving term. In a compact form:

\be
 \dot{\s} = \mathcal{L}_{\rm coh}[\s] + \mathcal{L}_{\rm driv}[\s] +
 \mathcal{L}_{\rm damp}[\s]
\end{equation}
where we have introduced a block matrix structure to  take care of
the mechanical and electrical degrees of freedom simultaneously.

\subsection{Triple Dot Quantum Shuttle}
The driving term of the Liouvillean operator for the Triple  Dot
Quantum Shuttle can be derived in strict analogy with the fixed double
dot. The major simplification with respect to the SDQS is
the drop of the position dependence in the coupling to the leads
as one can see from the Hamiltonian for the model:

\be
    H =H_{\rm sys}+ H_{\rm leads} + H_{\rm bath}
      +H_{\rm tun}+ H_{\rm int}
\end{equation}
where
\be
 \bs
  H_{\rm sys} =&
   \e_0 |0\rangle\!\langle 0|
   + \frac{\D V}{2} |L\rangle \! \langle L|
   - \frac{\D V}{2 x_0}x  |C\rangle\!\langle C|
   - \frac{\D V}{2} |R\rangle \! \langle R|
   + \hbar \w \left(d^{\dag}d + \frac{1}{2}\right)\\
   &+t_R(x)(|C \rangle\!\langle R| + |R \rangle\!\langle C|)
    +t_L(x)(|C \rangle\!\langle L| + |L \rangle\!\langle C|)\\
  H_{\rm leads} =&
  \sum_{k}(\eps_{l_k}
c^{\dagger}_{l_k}c^{\phd}_{l_k}
          +\eps_{r_k}
c^{\dagger}_{r_k}c^{\phd}_{r_k})\\
H_{\rm bath} =&
 \sum_{\vet{q}} \hbar \w_{\vet{q}}{d_{\vet{q}}}^{\dagger} d_{\vet{q}}\\
  H_{\rm tun} =&
  \sum_{k}T_{l}( c^{\dagger}_{l_k}|0\rangle\!\langle L| +
                 c^{\phd}_{l_k}   |L\rangle\!\langle 0|)
         +T_{r}( c^{\dagger}_{r_k}|0\rangle\!\langle R| +
                 c^{\phd}_{r_k}   |R\rangle\!\langle 0|) \\
 H_{\rm int } =&
 \hbar \sum_{\vet{q}} g(d^{\dagger}_{\vet{q}}d + d^{\dagger}d_{\vet{q}})
 \end{split}
\end{equation}
The reduced density matrix for the  triple dot system takes into
account mechanical and electrical degrees of freedom. As in the
case of the fixed double dot we can organize the density matrix
in a block representation:

\be
\s =
\left[
\begin{array}{cccc}
\s_{00} & \s_{0L} & \s_{0C} & \s_{0R}\\
\s_{L0} & \s_{LL} & \s_{LC} & \s_{LR}\\
\s_{C0} & \s_{CL} & \s_{CC} & \s_{CR}\\
\s_{R0} & \s_{RL} & \s_{RC} & \s_{RR}
\end{array}
\right]
\end{equation}
Due to the incoherent leads the  elements $\s_{0\a}$ and
$\s_{\a0}$ with $\a= L,C,R$ identically vanish\footnote{The
incoherent leads do not have coherent superposition of states with
different particle number. Only this kind of ``forbidden'' bath
states would allow coherent superposition of states with 0 and 1
particle in the device and thus non-vanishing coherencies
$\s_{\a0}$ or $\s_{0\a}$.}. In the same matrix representation we
write\footnote{This equation was first derived in the Gurvitz
scheme by Armour and MacKinnon in \cite{arm-prb-02}.} the driving
Liouvillean:

\be
 \mathcal{L}_{\rm driv}[\s] =
 \left[
 \begin{array}{cccc}
  \G_R \s_{RR}-\G_L \s_{00} & 0            & 0 & 0 \\
  0 & \G_L \s_{00} & 0 & -\textstyle{\frac{1}{2}} \G_R \s_{LR} \\
  0 & 0 & 0 & -\textstyle{\frac{1}{2}}\G_R \s_{CR} \\
  0 & -\textstyle{\frac{1}{2}}\G_R \s_{RL}
    & -\textstyle{\frac{1}{2}}\G_R \s_{RC}
    & -\G_R \s_{RR}
 \end{array}
 \right]
\end{equation}
where the injection and ejection rates have the form:

\be
 \G_{L,R} = 2 \pi D_{L,R} T_{l,r}^2
\end{equation}

The overall structure of the driving  component of the Liouvillean
can be understood in terms of ``decaying channels'', since the
interaction with the continuum of states in the leads gives rise
to incoherent tunneling processes. This concept is underlying the
following formulation of the incoherent dynamics
\cite{gur-prb-96}:

\be
\bs
 \Big(\mathcal{L}_{\rm driv}[\s]\Big)_{\a\b} =&
  -\frac{1}{2}\s_{\a\b}\left(
  \sum_{\d \ne \a} \G_{\a \to \d}+
  \sum_{\d \ne \b} \G_{\b \to \d}
  \right)\\
  &+\frac{1}{2}\sum_{\a'\b' \ne \a\b} \s_{\a'\b'}
  (\G_{\a' \to \a} + \G_{\b' \to \b})
 \end{split}
 \label{eq:genrates}
\end{equation}
where $\a,\a',\b,\b',\d = 0,L,C,R$ and $\G_{\a \to \b}$  is the
probability per unit time for the system to make a transition from
state $|\a\rangle$ to state $|\b\rangle$. The generic state $|\a
\rangle \! \langle \b|$ is emptied (first line in equation
\ref{eq:genrates}) and pumped (second line) with different $\G$
rates. The central dot does not contribute to this incoherent
dynamics since it is coupled only to the left and right dot
\emph{discrete} states. Due to the left-right asymmetry only the
rates $\G_{0 \to L} \equiv \G_L$ and $\G_{R \to 0} \equiv \G_R$
are non-zero. The mechanical state of the system is not involved
in the equation (\ref{eq:genrates}) but plays an active role in
the coherent dynamics. In block matrix notation the system
Hamiltonian takes the form:
\be
 H_{\rm sys} = \left[
 \begin{array}{cccc}
 \eps_0 + H_{\rm osc} & 0 & 0 & 0\\
 0 & \frac{\D V}{2}+ H_{\rm osc} & t_L(x) & 0\\
 0 & t_L(x) & -\frac{\D V}{2 x_0}x + H_{\rm osc}& t_R(x)\\
 0 & 0 & t_R(x) & -\frac{\D V}{2}+ H_{\rm osc}
 \end{array}
 \right]
\end{equation}
The corresponding coherent Liouvillean reads:
\be
 \Big(\mathcal{L}_{\rm coh}[\s]\Big)_{\a\b} =
 -i\sum_{\d}\left[
 (H_{\rm sys})_{\a\d} \s_{\d\b} -  \s_{\a\d}(H_{\rm sys})_{\d\b}
 \right]
\end{equation}
We assume for the damping term the same used by Armour and
MacKinnon.  It is the standard quantum optical damping in RWA that
we derived in the previous section (see eq.
\eqref{eq:RWAdampkernel}):
\be
\bs
 \mathcal{L}_{\rm damp}[\s] =&
  -\frac{\g}{2}n_B(\w)(dd^{\dag}\s - 2d^{\dag}\s d + \s dd^{\dag})\\
 &-\frac{\g}{2}[n_B(\w)+ 1](d^{\dag}d\s - 2d\s d^{\dag} + \s d^{\dag}d)
\end{split}
\end{equation}
We write for completeness the system of equations for the 10
non-vanishing sub-matrices of the reduced density matrix:
\be
\begin{tabular}{|rl|}
\hline
\phantom{.} & \\
 $\dot{\s}_{00} =$ & $-i[H_{\rm osc},\s_{00}]
                      +\G_R \s_{RR} - \G_L \s_{00}
                      +\mathcal{L}_{\rm damp}[\s_{00}]$\\
\phantom{.} & \\
 $\dot{\s}_{LL} =$ & $-i[H_{\rm osc},\s_{LL}]
                      -it_L(x)\s_{CL} + i\s_{LC}t_L(x)
                      +\G_L \s_{00}
                      +\mathcal{L}_{\rm damp}[\s_{LL}]$\\
\phantom{.} & \\
 $\dot{\s}_{CC} =$ & $-i\left[H_{\rm osc}
                      - \frac{\D V}{2 x_0}x,\s_{CC}\right]
                      -it_L(x)\s_{LC} + i\s_{CL}t_L(x)$\\
%\phantom{.} &  \\
                &$-it_R(x)\s_{RC} + i\s_{CR}t_R(x)
                 +\mathcal{L}_{\rm damp}[\s_{CC}]$\\
\phantom{.} & \\
 $\dot{\s}_{RR} =$ & $-i[H_{\rm osc},\s_{RR}]
                      -it_R(x)\s_{CR} + i\s_{RC}t_R(x)
                      -\G_R \s_{RR}
                      +\mathcal{L}_{\rm damp}[\s_{RR}]$\\
\phantom{.} & \\
 $\dot{\s}_{LC} =$ & $-i[H_{\rm osc},\s_{LC}]
                      -i\frac{\D V}{2}\s_{LC}
                      \left(1 + \frac{x}{x_0}\right)
                      -it_L(x)\s_{CC} + i\s_{LL}t_L(x)$\\
%\phantom{.} & \\
                &$+i\s_{LR}t_R(x) +\mathcal{L}_{\rm damp}[\s_{LC}]$\\
\phantom{.} & \\
 $\dot{\s}_{CL} =$ & $-i[H_{\rm osc},\s_{CL}]
                      +i\frac{\D V}{2}\left(1 + \frac{x}{x_0}\right)\s_{CL}
                      +i\s_{CC}t_L(x) - it_L(x)\s_{LL}$\\
%\phantom{.} & \\
                &    $-it_R(x)\s_{RL} +\mathcal{L}_{\rm damp}[\s_{CL}]$\\
\phantom{.} & \\
 $\dot{\s}_{LR} =$ & $-i[H_{\rm osc},\s_{LR}]
                      -i\D V\s_{LR} -i t_L(x)\s_{CR}
                      +i\s_{LC}t_R(x)$\\
%\phantom{.} & \\
                &$-\frac{\G_R}{2}\s_LR
                 +\mathcal{L}_{\rm damp}[\s_{LR}]$\\
\phantom{.} & \\
 $\dot{\s}_{RL} =$ & $-i[H_{\rm osc},\s_{RL}]
                      +i\D V\s_{RL} +i\s_{RC} t_L(x)
                      -it_R(x)\s_{CL}$\\
%\phantom{.} & \\
                & $-\frac{\G_R}{2}\s_RL
                   +\mathcal{L}_{\rm damp}[\s_{RL}]$\\
\phantom{.} & \\
 $\dot{\s}_{CR} =$ & $-i[H_{\rm osc},\s_{CR}]
                      -it_L(x)\s_{LR}
                     -i\frac{\D V}{2}\left(1-\frac{x}{x_0}\right)\s_{CR}$\\
%\phantom{.} & \\
                &$-i t_R(x)\s_{RR} + i \s_{CC}t_R(x)
                 +\mathcal{L}_{\rm damp}[\s_{CR}]$\\
\phantom{.} & \\
 $\dot{\s}_{RC} =$& $-i[H_{\rm osc},\s_{RC}]
                     +i\s_{RL}t_L(x)
                     +i\frac{\D V}{2}\s_{RC}\left(1-\frac{x}{x_0}\right)$\\
%\phantom{.} & \\
                &$+i \s_{RR}t_R(x) - i t_R(x)\s_{CC}
                 +\mathcal{L}_{\rm damp}[\s_{RC}]$\\
\phantom{.} & \\
\hline
\end{tabular}
\label{eq:TDQSGME}
\end{equation}

\section{The stationary solution: a numerical challenge}\label{sec:numericalGME}

The master equation generally describes the irreversible dynamics
due to the coupling between the system and the infinite number of
degrees of freedom of the environment. It is reasonable to require
that in absence of a driving mechanism the system tends
asymptotically to thermalize with the bath. This property is
reflected in the evolution of the reduced density matrix that in
this case has a thermal stationary solution. In the case of
shuttle devices the oscillator is driven by the electrical
dynamics: every time an electron jumps onto the moving island it
feels an electrostatic force that excites the oscillator. For this
reason, at least for small enough damping we do \emph{not} expect
the oscillator to relax to the stationary thermal distribution.
Nevertheless since both the electronic and the mechanical degrees
of freedom of the system are coupled to baths we \emph{do} expect
a stationary solution for the GMEs (\ref{eq:SDQSGME}) and
(\ref{eq:TDQSGME}), i.e. a matrix $\s^{stat}$ that fulfills the
condition:

\be
 \mathcal{L} \s^{stat} = 0
\label{eq:stationary}
\end{equation}

\subsection{A matter of matrix sizes}
We calculate the stationary matrix numerically: we have to find
the null vector of the matrix representation for the Liouvillean
super-operator $\mathcal{L}$. The challenge arises from the matrix
sizes. In principle the reduced density matrix has infinite size
due to the mechanical degree of freedom. In order to treat the
problem numerically we truncate the corresponding Hilbert space
retaining only the first $N$ states of the harmonic oscillator
basis\footnote{The case of a different mechanical potential is not
more difficult in principle, once we know the eigenvectors and
eigenvalues of the corresponding one-dimensional Schr\"odinger
equation.}. The size of the reduced density matrix $\s$ is in the
SDQS and TDQS respectively $2N \times 2N$ and $4N \times 4N$:
prefactors 2 and 4 come from the size of the electrical Hilbert
space which is spanned by the vectors $|0\rangle, |1\rangle$ for the
single-dot device and $|0\rangle, |L\rangle,|C\rangle,|R\rangle$
for the triple dot.

The Liouvillean is a linear operator on the space of
Hilbert-Schmidt operators\footnote{Given some Hilbert space
$\mathcal{H}$ an operator  $A$ is of Hilbert-Schmidt if it is
linear and  ${\rm Tr}\{A^{\dag}A\}$ is finite.} on the Hilbert
space for the system (Liouville space). Equipped with the scalar
product:

\be
 (A,B) \equiv {\rm Tr}\{A^{\dag}B\}
\end{equation}
the Liouville space  becomes a Hilbert space. One can therefore
introduce an orthonormal basis and represent linear operators as
matrices. The truncated Hilbert space for the system gives rise to
finite size Liouvillean matrices: for the SDQS we reach the size
of $4N^2 \times 4N^2$ while the richer electronic structure of
the TDQS is reflected in a $16N^2 \times 16N^2$ elements
Liouvillean. Even with the condition of incoherent baths that
prevents coherencies within states with different electron number
in the system and therefore sets to zero all sub-matrices in the
form $\s_{01}$ or $\s_{10}$ in the SDQS and $\s_{0\a}$ or $\s_{\a
0}$ with $\a = L,C,R$ in the TDQS we can not reduce the size of
the Liouvillean matrix to more than $2N^2 \times 2N^2$ and $10N^2
\times 10N^2$ respectively.

The description of the shuttle device dynamics requires
(especially in the shuttling regime) amplitude oscillations of the
vibrating dot between 5 and 10 times larger than the zero point
fluctuations. For this reason, in both devices, we are left to
study the null space of matrices of typical size of $2 \cdot 10^4 \times
2\cdot 10^4$. To indicate that we are treating the truncated matrix
representation of the original stationary state problem we change
slightly notation and formulate the numerical problem:

\be
  \vet{L}\vet{p}^{stat} = 0
\end{equation}
with ${\rm Tr}\{\vet{p}^{stat}\} = 1$. The solution to this
numerical problem came from prof.~Timo Eirola, Helsinki University
of Technology, in the form of a suggestion and implementation for
the SDQS of the iterative Arnoldi scheme. We successfully extended
the method to the TDQS.

The Arnoldi scheme applied to the calculation of the null space
has advantages with respect to the singular value decomposition
both in terms of computational speed and memory
consumption\footnote{For the theory and implementation of the
Arnoldi scheme we refer to the lecture notes ``Numerical Linear
Algebra; Iterative Methods" by Eirola and Nevanlinna
\cite{eirola}.}. First we do not need to store the Liouvillean
matrix and we can always work with operators on the system Hilbert
space only; second we iteratively look for the best approximation
to the null vector $\vet{p}^{stat}$ in spaces which are typically
much smaller that the Liouville space. Good introductions to the
Arnoldi scheme can be found in different places in the literature
(for example \cite{golub} or \cite{eirola}). We dedicate the next
section to a detailed analysis of the Arnoldi iteration scheme
referring for examples to the SDQS Liouvillean. Some of the
material can be found also in the appendix of
\cite{fli-preprint-04}.

\subsection{The Arnoldi scheme}

The central r\^ole in the Arnoldi scheme is played by Krylov spaces.
For a given Liouvillean $\vet{L}$ and a vector $\vet{x}_0$ of the
Liouville space we define the Krylov space as:
\be
 \mathcal{K}_j(\vet{L},\vet{x}_0)
 \equiv {\rm span}(\vet{x}_0,\vet{Lx}_0,\ldots,\vet{L}^{j-1}\vet{x}_0)
\end{equation}
where $j$ is a small\footnote{We mean small compared to the
dimension of the Liouville space. We used  $j = 20$ for a
Liouville space dimension of roughly $2\cdot 10^4$.} natural
number. It is important to note that for the construction of the
Krylov space all what we need are the vectors $\vet{x}_0$,
$\vet{Lx}_0$, $\vet{L}^2\vet{x}_0$, $\ldots$ and not explicitly
the matrix $\vet{L}$. In the SDQS device we would take an
arbitrary state represented by the two matrices $\s_{00}$ and
$\s_{11}$ and choose a reshaping procedure to map them into a
single vector $\vet{x}_0 \in \mathbb{C}^{2N^2 \times 1}$. The
vector $\vet{Lx}_0$ is obtained applying the operator defined in
equation (\ref{eq:SDQSGME}) to the density matrices $\s_{00}$ and
$\s_{11}$ and then reshaping with the same procedure used for
$\vet{x}_0$.

The Arnoldi iteration starts with the construction of an
orthonormal basis $\{\vet{q}_k\}_{k=1}^j$ for the Krylov space
using standard Gram-Schmidt orthonormalization:

\be \bs
 \vet{q}_1 &= \frac{\vet{x}_0}{\|\vet{x}_0\|}\\
 \vet{q}_{k+1} &= \frac
   {\vet{Lq}_k-\sum_{i=1}^k[\vet{q}_i^{\dag}\cdot \vet{Lq}_k]\vet{q}_i}
 {\|\vet{Lq}_k-\sum_{i=1}^k[\vet{q}_i^{\dag}\cdot \vet{Lq}_k]\vet{q}_i\|}
 ,k=1,\ldots,j-1
\end{split}
\end{equation}
where the norm in the Liouville space is defined from the
canonical Hermitian product  $\|\vet{a}\|_2 \equiv
\sqrt{\vet{a}^{\dag}\vet{a}}$ and has nothing to do with the
scalar product we introduced to demonstrate that the Liouville
space is a Hilbert space. Each new tentative basis vector
$\vet{Lq}_k$ is first orthogonalized (by subtracting all
components in the preceding vectors of the basis) and then
normalized. This requires the calculation of the quantities:

\be
 h_{i,k} = \vet{q}_i^{\dag}\cdot \vet{Lq}_k,
 i=1,\ldots,k;k=1,\ldots,j
\end{equation}
and
\be
 h_{k+1,k} = \|\vet{Lq}_k-\sum_{i=1}^k[\vet{q}_i^{\dag}\cdot
 \vet{Lq}_k]\vet{q}_i\|, k=1,\ldots,j
\end{equation}
that are stored into the upper Hessenberg matrix

\be
 \vet{H}_j =
 \left[
 \begin{array}{ccccc}
 h_{1,1} & h_{1,2} & h_{1,3} & \cdots & h_{1,j}\\
 h_{2,1} & h_{2,2} & h_{2,3} & \cdots & h_{2,j}\\
 0 & h_{3,2} & h_{3,3} & \cdots & h_{3,j}\\
 0 & 0 & h_{4,3} & \cdots & h_{4,j}\\
\vdots & \vdots & \vdots & \ddots & \vdots\\
0 & 0 & 0 & \cdots & h_{j+1,j}
 \end{array}
 \right] \in \mathbb{C}^{j+1 \times j}
\end{equation}
while the basis vectors are stored as columns in the matrix

\be
 \vet{Q}_j = [\vet{q}_1,\vet{q}_2, \ldots, \vet{q}_j] \in
 \mathbb{C}^{2N^2 \times j}
\end{equation}
which fulfills $\vet{Q}_j^{\dag}\vet{Q}_j = \vet{I}$, $\vet{I} \in
\mathbb{C}^{j \times j}$ being the identity matrix, since the
basis is orthonormal. The method proceeds by looking for the best
approximation of the null vector for the Liouvillian within the
Krylov space $\mathcal{K}_j(\vet{L},\vet{x}_0)$. We call this
vector $\vet{x}_j$. In terms of its $j$ components in the
orthonormal basis $\vet{x}_j = \vet{Q}_j\vet{v}_j$. The $j$
coordinates in the Krylov space $\vet{v}_j$ solve the minimum
problem:

\be
 \min_{\scriptsize \begin{array}{c}
   \vet{x}\in \mathcal{K}_j \\
   \|x\|_2 = 1 \\
 \end{array} }  \!\!\! \|\vet{Lx}\|_2 = \|\vet{Lx}_j\|_2 =
 \|\vet{LQ}_j\vet{v}_j\|_2
\end{equation}
In the process of finding these coordinates a key r\^ole is played
by the Hessenberg matrix since the following relation holds:

\be
 \vet{LQ}_j = \vet{Q}_{j+1}\vet{H}_j
\label{eq:Arnoldikernel}
\end{equation}
We refer to the notes by Eirola \cite{eirola} for a rigorous
mathematical proof of the relation (\ref{eq:Arnoldikernel}) and we
only limit ourselves to exploit here some of its consequences:

\be \bs
  \min_{\scriptsize \begin{array}{c}
   \vet{x}\in \mathcal{K}_j \\
   \|\vet{x}\|_2 = 1 \\
 \end{array} }  \!\!\!  \|\vet{Lx}\|_2
 &= \|\vet{LQ}_j\vet{v}_j\|_2
  = \|\vet{Q}_{j+1}\vet{H}_j\vet{v}_j\|_2\\
 &=\sqrt{(\vet{Q}_{j+1}\vet{H}_j\vet{v}_j)^{\dag}\cdot
         (\vet{Q}_{j+1}\vet{H}_j\vet{v}_j)}\\
 &=\sqrt{(\vet{H}_j\vet{v}_j)^{\dag}
         (\vet{Q}_{j+1}^{\dag}\vet{Q}_{j+1})
         (\vet{H}_j\vet{v}_j)}\\
 &=\sqrt{(\vet{H}_j\vet{v}_j)^{\dag}\vet{I}
         (\vet{H}_j\vet{v}_j)} = \|\vet{H}_j\vet{v}_j\|_2
         =  \min_{\scriptsize \begin{array}{c}
   \vet{v}\in \mathbb{C}^j \\
   \|\vet{v}\|_2 = 1 \\
 \end{array} }  \!\!\! \|\vet{H}_j\vet{v}\|_2
\end{split}
\end{equation}

We started with a minimum problem involving the vector $\vet{x}$
of length $2N^2$ and a matrix $\vet{L}$ of size $2N^2 \times 2N^2$
and we have reduced it to the minimum problem in the last line
that only involves a vector $\vet{v}$ of length $j$ and a matrix
$\vet{H}_j$ of size $(j+1) \times j$. Since $j=20$ the latter is
absolutely not demanding neither from a computational or a memory
point of view on a modern computer. We proceed to the minimization
using singular value decomposition (SVD) \cite{eirola}. Given a
complex matrix $\vet{A} \in \mathbb{C}^{m \times n}$ with $m \geq
n$, there exist two unitary matrices $\vet{U} \in \mathbb{C}^{m
\times m}$ and $\vet{V}^{n \times n}$ such that $\vet{A} =
\vet{U\Sigma V}^{\dag}$, and $\vet{\Sigma} \in \mathbb{R}^{m
\times n}$ is diagonal with non-negative diagonal elements
(conventionally in decreasing order starting with the highest
singular value in the upper corner \cite{golub}). Thus for the
Hessenberg matrix $\vet{H}_j$
\be
 \vet{H}_j = \vet{U\Sigma V^{\dagger}}
\end{equation}
where $\vet{U} \in \mathbb{C}^{(j+1) \times (j+1)}$, $\vet{V} \in
\mathbb{C}^{j \times j}$ and

\be
 \vet{\Sigma} =
\left[
\begin{array}{ccc}
\l_1 & \cdots & 0 \\
\vdots & \ddots & \vdots \\
0 & \cdots & \l_j \\
0 & \cdots & 0
\end{array}
\right] \in \mathbb{R}^{(j+1) \times j}
\end{equation}
contains the singular values $\l_j \geq 0$. The minimum problem is
solved as follows:

\be
  \min_{\scriptsize \begin{array}{c}
   \vet{v}\in \mathbb{C}^j \\
   \|\vet{v}\|_2 = 1 \\
 \end{array} }  \!\!\!  \|\vet{H}_j\vet{v}\|_2 =
  \min_{\scriptsize \begin{array}{c}
   \vet{v}\in \mathbb{C}^j \\
   \|\vet{v}\|_2 = 1 \\
 \end{array} }  \!\!\!  \|\vet{U \Sigma V}^{\dag}\vet{v}\|_2 =
  \min_{\scriptsize \begin{array}{c}
   \vet{v}\in \mathbb{C}^j \\
   \|\vet{v}\|_2 = 1 \\
 \end{array} }  \!\!\!  \|\vet{\Sigma V}^{\dag}
 \vet{v}\|_2= \l_j
\end{equation}
where we have used the unitarity  of $\vet{U}$ and $\vet{V}$ and
we have chosen the minimizing vector $\vet{v}_j$ to be the column
of $\vet{V}$ corresponding to the smallest singular value. Having
found $\vet{v}_j$ the best approximation of the null vector within
the Krylov space can be calculated $\vet{x}_j = \vet{Q}_j
\vet{v}_j$. Finally we test the result and compare
$\|\vet{L}\vet{x}_j\|_2$ with a given tolerance. If the test is
positive we accept the result and reshape the vector $\vet{x}_j$
into the matrix form as an approximation of the stationary
solution of the GME. Otherwise we use $\vet{x}_j$ as the starting
vector for a new iteration of the Arnoldi scheme. We have chosen
as tolerance parameter $10\eps\|\vet{L}\|_2$ where $\eps$ is the
machine precision and the norm\footnote{We used for the
Liouvillean the norm $\|\vet{L}\|_2 \equiv \ \max_{\|\vet{x}\|_2 =
1}\frac{\|\vet{Lx}\|_2}{\|\vet{x}\|_2}$.} of the Liouvillean was
estimated by T.~Eirola as $\|\vet{L}\|_2 \approx \exp(N/\log N)$.

\subsection{Preconditioning}

The Arnoldi scheme is iterative and can suffer from convergence
problems. It is not a priori clear how many iterations one needs
to converge and fulfill the convergence criterion. A possible
answer to a non-convergent code is to reformulate the problem into
an equivalent and (hopefully) convergent form. This was exactly
the situation we had to face with the Arnoldi scheme applied to
the problem of finding the stationary reduced density matrix for
shuttle devices. The solution to this problem came again from
prof. T.~Eirola in the form of a preconditioner. The basic idea is
to find a regular operator\footnote{To be precise the
preconditioner should be regular only on the image of the
Liouvillian.} $\mathcal{M}$ on the Liouville space, invertible,
easy to implement, such that the original problem
$\mathcal{L}[\s^{stat}] = 0$ can be recast into the form:
\be
 \mathcal{M}[\mathcal{L}[\s^{stat}]] = 0
\end{equation}
and that the finite version of the operator
$\mathcal{M}\mathcal{L}$ gives rise to a (fast) convergent
iteration scheme. The operator $\mathcal{M}$ is also known as the
\emph{preconditioner}.

The Arnoldi scheme is particularly efficient in finding the best
approximation of the eigenvalues and corresponding eigenvectors
for those eigenvalues that are separated from the rest of the
spectrum. Since we want to calculate the null vector it is
important that the preconditioner  moves the non-vanishing part of
the spectrum far from the origin\footnote{In this sense the
philosophy is: ``Invert as much as you can!''}.

In order to understand  the preconditioner that we used for the
shuttling problem, we introduce the generic Sylvester operator
$\phi : \mathbb{C}^{n \times m} \to \mathbb{C}^{n \times m}$:

\be
 \phi(\vet{X})= \vet{AX - XB}
\end{equation}
where $\vet{A} \in \mathbb{C}^{n \times n}$, $\vet{B} \in
\mathbb{C}^{m \times m}$, $\vet{X} \in \mathbb{C}^{n \times m}$.
This operator is invertible if and only if the spectrua of
$\vet{A}$ and $\vet{B}$ have empty intersection and the inversion
routine is computationally light. A part of the Liouvillean
operator is of Sylvester form. In the SDQS for example:

\be
 \mathcal{L}[\s] = \mathcal{L}_{\rm Sylv}[\s] +
 \mathcal{L}_{\rm rest}[\s]
\end{equation}
where
\be
 \mathcal{L}_{\rm Sylv} = \vet{A}\s + \s\vet{A}^{\dag} =
 \left[
 \begin{array}{cc}
A_{00} \s_{00} + \s_{00}A_{00}^{\dag} & 0\\
0 & A_{11} \s_{11} + \s_{11}A_{11}^{\dag}
 \end{array}
 \right]
\end{equation}
and
\be
 \bs
  A_{00} &= -\frac{i}{\hbar}H_{\rm osc} -
  \frac{\G_L}{2}e^{-\frac{2x}{\l}}-  \frac{i \g}{2 \hbar} xp -
  \frac{\g m \w}{\hbar}\left(n_B+ \frac{1}{2}\right)x^2\\
  A_{11} &= -\frac{i}{\hbar}(H_{\rm osc} - e \mathcal{E}x)
  -\frac{\G_R}{2}e^{\frac{2x}{\l}} -  \frac{i \g}{2 \hbar} xp -
  \frac{\g m \w}{\hbar}\left(n_B+ \frac{1}{2}\right)x^2\\
 \end{split}
\end{equation}
where $n_B$ is the average occupation number of the energy states
of the harmonic oscillator in equilibrium with the bath. The rest
of the Liouvillean in the non-Sylvester form reads:

\be
 \mathcal{L}_{\rm rest}
 \left[\begin{array}{c}
 \s_{00}\\
 \s_{11}
 \end{array} \right]
 =
 \left[ \begin{array}{c}
 \G_R e^{\frac{x}{\l}}\s_{11}e^{\frac{x}{\l}}
 -\frac{i\g}{2\hbar}(x\s_{00}p - p\s_{00}x)
 +\frac{\g m \w}{\hbar}(2n_B + 1) x\s_{00}x\\
 \G_L e^{-\frac{x}{\l}}\s_{00}e^{-\frac{x}{\l}}
 -\frac{i\g}{2\hbar}(x\s_{11}p - p\s_{11}x)
 +\frac{\g m \w}{\hbar}(2n_B + 1) x\s_{11}x\\
 \end{array}\right]
\end{equation}

We use as the preconditioner for the Arnoldi iteration scheme
$\mathcal{M} = \mathcal{L}_{\rm Sylv}^{-1}$. The check of empty
intersection between the spectrum of $\vet{A}$ and
$-\vet{A}^{\dag}$ must be done numerically due to the presence of
the exponentials of position operators.

The stationary solution of the GME represents the starting point
for the analysis of the properties of shuttle devices that we will
present in the following chapters. The Arnoldi scheme allows the
calculation of the stationary solution $\s^{stat}$ of the GME for
a given set of parameters and a fixed dimension $N$ of the Hilbert
space of the oscillator. The number $N$ has been chosen from
considerations on the phase space distribution for the stationary
solution\footnote{See next chapter for details on the Wigner
distribution and its meaning.}.
%%%%%%%%%%%%%%%%%%%%%%%%%%%%%%%%%%%%%%%%%%%%%%%%%%%%%%%%%%%%%%%%%%%
% Figure
%%%%%%%%%%%%%%%%%%%%%%%%%%%%%%%%%%%%%%%%%%%%%%%%%%%%%%%%%%%%%%%%%%%
\begin{figure}
 \begin{center}
  \includegraphics[angle=0,width=\textwidth]{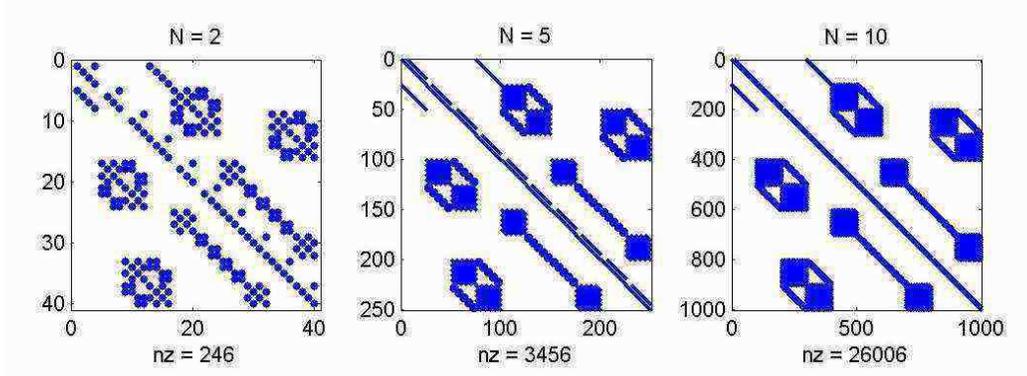}
  \caption{\small  \textit{Example of matrix representation of the
  Liouvillean for the TDQS. The Hilbert space of the harmonic
  oscillator has dimension respectively $N = 2,5,10$ ($nz$ is the number of no-zero elements
  of the mtrix).
   The non-zero elements are represented by a blue
   dot. The exponentials of position operator present
   in the tunneling amplitudes make the Liouvillean
   non-sparse (large squares progressively emerging
   in the figures from left to right). For large $N$ the SVD is not
   viable and we use the Arnoldi iteration scheme to find the null
    space of the Liouvillean.}}
 \end{center}
\end{figure}
%%%%%%%%%%%%%%%%%%%%%%%%%%%%%%%%%%%%%%%%%%%%%%%%%%%%%%%%%%%%%%%%%%%

\clearpage{\pagestyle{empty}\cleardoublepage}

%%%%%%%%%%%%%%%%%%%%%%%%%%%%%%%%%%%%%%%%%%%%%%%%%%%%%%%%%%%
%                      WF
%%%%%%%%%%%%%%%%%%%%%%%%%%%%%%%%%%%%%%%%%%%%%%%%%%%%%%%%%%%
\chapter{Wigner function distribution}

The shuttle dynamics has, especially in the single dot device, an
appealing simple classical interpretation and one can say that the
name itself of ``shuttle'' suggests the idea of sequential and
periodical loading, mechanical transport and unloading of
electrons between a source and a drain lead. Semiclassical models
typically treat the mechanical degree of freedom classically
combined with incoherent sequential tunneling of electrons
between leads and oscillating dot
\cite{gor-prl-98,isa-phb-98,wei-epl-99,nis-prb-01,pis-prb-04}.
Motivated by the small size of the oscillations of a nanoscale
shuttle, and thus by the possibility to observe signatures of
quantum dynamics of the mechanical degree of freedom, we decided,
following the suggestion of Armour and MacKinnon
\cite{arm-prb-02}, to explore the model with a quantized
oscillator. Nevertheless we also wanted to keep as much as
possible the intuitive classical picture and to handle the
quantum-classical correspondence. The Wigner function distribution
seemed to us a good answer to all these requirements. It allows a
clear visualization of the numerical results obtained within the
framework of the GME and it shows in its equation of motion (the
Klein-Kramers equation) an explicit quantum-classical
correspondence (expansion in powers of $\hbar$) \cite{hil-phr-84}.

\section{A quantum phase-space distribution}
\label{sec:QuantumDistribution}

The concept of a probability distribution in phase space is
problematic in quantum mechanics due to the uncertainty principle.
It is not clear for example the meaning of the probability that a
particle has a well defined position and momentum at a certain
time and thus it seems difficult to define a probability
distribution for such ``non-observable events''. Nevertheless one
can demonstrate that it is possible to associate with a state
defined by a density operator $\rho(t)$ a unique function
$P(q,p,t)$ on the phase space $(q,p) \times [t_0, \infty)$ that
satisfies some desirable properties. We list the required
properties as stated by Hillary et al.~in the review
``Distribution functions in Physics: fundamentals''
\cite{hil-phr-84}:

\begin{enumerate}
\item{$P(q,p,t)$ should be real;}

\item{$P(q,p,t)$ should reduce to a probability distribution once
the conjugate variable is integrated:
\be
\bs
 &\int dp\, P(q,p,t) = \langle q|\rho(t) |q\rangle\\
 &\int dq\, P(q,p,t) = \langle p|\rho(t) |p\rangle\\
\end{split}
\end{equation}
and thus $P(q,p)$ is normalized
\be
 \int dp \int dq\, P(q,p,t) = 1
\end{equation}}

\item{$P(q,p,t)$ should be Galilei invariant;}

\item{$P(q,p,t)$ should be invariant under space and time reflection;}

\item{In the force free case $P(q,p,t)$ should obey the classical equation
of motion: \be
 \frac{\partial P}{\partial t} = -\frac{p}{m}\frac{\partial P}{\partial q}
\end{equation}}

\item{If $P_1(q,p,t)$ and $P_2(q,p,t)$ are the distributions
corresponding to the states $\rho_1(t)$ and $\rho_2(t)$ the
following relation should hold:

\be
 \int dq \langle q |\rho_1 \rho_2 | q \rangle =
 2 \pi \hbar \int dq \int dp\, P_1(q,p,t)P_2(q,p,t)
\end{equation}}
\end{enumerate}
Condition 6 is a requirement that comes from quantum mechanical
considerations as can be seen from the following consequences.
Assume $\rho_1=\rho_2= \r$ being statistical mixtures of pure
states $\rho_{\a}$ with weights $w_{\a}$

\be \r = \sum_{\a}w_{\a}\rho_{\a}
\end{equation}
and $P(q,p,t)$ the associated distribution, then from condition 6
we get

\be
 \int dq \int dp [P(q,p,t)]^2 =
 \frac{\sum_{\a}w_{\a}^2}{2\pi\hbar}
\end{equation}
This relation is discarding any too peacked distribution. In
particular the classical distribution $P(q,p,t) =
\d(q-q_0(t))\d(p-p_0(t))$ does not fulfill condition 6. If we
assume now that $\rho_1$ and $\rho_2$ are pure states associated
with orthogonal vectors, then:

\be
 \int dq \int dp P_1(q,p,t)P_2(q,p,t) = 0
\end{equation}
which implies that the two distributions are \emph{not} positive
everywhere. For this reason the distribution we are axiomatically
constructing will take the name of quasi-probability distribution.

The distribution function  that satisfy properties (1-5) or (1-4
and 6) exists unique \cite{oco-pla-81}, is defined as

\be
 W(q,p,t) =
 \frac{1}{2 \pi \hbar}
 \int_{-\infty}^{+\infty}\!\!\!d \xi
 \left\langle q-\frac{\xi}{2}\right| \r(t) \left|q+\frac{\xi}{2}\right\rangle
 \exp\left(\frac{ip\xi}{\hbar}\right)
 \label{eq:WF}
\end{equation}
and is commonly called Wigner distribution. The parallel with
classical mechanics can be pushed even further associating to
every observable on the Hilbert space the corresponding Wigner
function:

\be
 A(q,p) =
 \int_{-\infty}^{+\infty}\!\!\!d \xi
 \left\langle q-\frac{\xi}{2}\right| A \left|q+\frac{\xi}{2}\right\rangle
 \exp\left(\frac{ip\xi}{\hbar}\right)
 \label{eq:WFA}
\end{equation}
Two important relations can be derived from the definitions
(\ref{eq:WF}) and (\ref{eq:WFA}).
\begin{enumerate}
\item The trace of an operator in proportional to the integral on
the phase space of its corresponding Wigner function:

\be
 {\rm Tr}A = \frac{1}{2 \pi \hbar} \int  dq \int dp \, A(q,p)
\end{equation}
\item Given two operators $A$ and $B$ and the associate Wigner
functions $A(q,p)$ and $B(q,p)$, the Wigner function associated to
the product of the two operators $F = AB$ reads
\cite{oco-pla-81}\footnote{This relation was first derived by
Groenewold in 1946 in the context of an rigorous mathematical
correspondence between commutator and poisson bracket in the limit
$\hbar \to 0$.}:
\end{enumerate}

\be
 F(q,p) = A(q,p)\exp \left(\frac{\hbar \Lambda}{2 i}\right) B(q,p)
 \label{eq:Groe}
\end{equation}
where $\Lambda$ is the differential operator:
\be
 \Lambda = \frac{\overleftarrow{\partial}}{\partial p}
           \frac{\overrightarrow{\partial}}{\partial q} -
           \frac{\overleftarrow{\partial}}{\partial q}
           \frac{\overrightarrow{\partial}}{\partial p}
\end{equation}
The arrows indicate in which direction the derivative should be
taken. A better insight in the expression (\ref{eq:Groe}) is given
by the expansion of the exponential. Up to the first two terms it
reads:

\be \bs
 F(q,p) =&
 A(q,p)B(q,p)\\
 +& \frac{\hbar}{2i}
  \left(\frac{\partial A}{\partial p}
        \frac{\partial B}{\partial q}
       -\frac{\partial A}{\partial q}
        \frac{\partial B}{\partial p}
  \right)\\
 +& \left(\frac{\hbar}{2i}\right)^2
  \left(\frac{\partial^2 A}{\partial p^2}
  \frac{\partial^2 B}{\partial q^2}
  -2\frac{\partial^2 A}{\partial p \partial q}
    \frac{\partial^2 B}{\partial q \partial p}
  +\frac{\partial^2 A}{\partial q^2}
  \frac{\partial^2 B}{\partial p^2}\right)\\
 +&\left(\frac{\hbar}{2i}\right)^3\ldots
 \end{split}
\end{equation}
The expansion shows the power of the formalism to treat the
quantum-classical correspondence: given a length, a mass, and a
time scale for the system we can rescale the coordinates and the
expansion will appear as an expansion in $\hbar/S_{\rm sys}$ where
$S_{\rm sys}$ is the typical action of the system. Classical
systems have a large action $S_{sys} \gg \hbar$ and only the first
term in the expansion is relevant. In the opposite limit $S_{sys}
\approx \hbar$ the full expansion should be considered. We have
now the tools to translate into the Wigner formalism the
expression for the expectation value of an operator: We start from
the definitions:
\be
 \langle A(t) \rangle \equiv {\rm Tr}\{\rho(t) A\} =
 \int  dq \int dp \, W(q,p,t)
 \exp\left(\frac{\hbar \Lambda}{2 i}\right)
 A(q,p)
\end{equation}
Inserting the expansion in $\hbar$ of the exponential and using
the regularity of the Wigner distribution at infinity we obtain
the ``classical'' result:

\be
 \langle A(t) \rangle = \int dq \int dp\, W(q,p,t)A(q,p)
\end{equation}

A second important consequence of the Groenewold equation
(\ref{eq:Groe}) is the quantum-classical correspondence for the
dynamics of the distribution function $W(q,p,t)$. In the Wigner
formalism the equation of Liouville-von Neumann reads:
\be
 \frac{\partial W}{\partial t} =
  -\frac{i}{\hbar}
  \left[H\exp \left(\frac{\hbar \Lambda}{2 i}\right) W -
   W\exp \left(\frac{\hbar \Lambda}{2 i}\right) H\right]
   \label{eq:anticommutator}
\end{equation}
where $H\equiv H(q,p)$ is the Wigner function associated with the
Hamiltonian for the system. Due to the fact that the differential
operator $\Lambda$ is odd with respect of the exchange of the two
functions on which it is acting only odd powers of the expansion in
$\hbar/S_{\rm sys}$ survive in (\ref{eq:anticommutator}). In the
limit $\hbar/S_{\rm sys} \to 0$ we get the classical Liouville
equation:

\be
  \frac{\partial W^c}{\partial t} = \{H,W^c\}_P
\end{equation}
where $\{,\}_P \equiv \frac{\overleftarrow{\partial}}{\partial q}
\frac{\overrightarrow{\partial}}{\partial p}
-\frac{\overleftarrow{\partial}}{\partial p}
           \frac{\overrightarrow{\partial}}{\partial q} = -\Lambda$
are the Poisson brackets and the superscript $c$ indicates the
classical limit for the Wigner distribution\footnote{As can be
seen from this derivation, in the framework of the Wigner
distribution the quantum-classical limit attributed to Dirac
$-\frac{i}{\hbar}[,] \to \{,\}_P $ for $\hbar \to 0$ acquires a
precise mathematical meaning.}.

\section{Klein-Kramers equations for the SDQS}

With the knowledge acquired in the previous section we will now
concentrate on the translation in the Wigner formalism of the GME for
shuttle devices that we derived in the previous chapter. We
concentrate only on the SDQS first because the technique is
completely illustrated in this simpler case and also because
contrary to the single dot device for the TDQS we will use the
Wigner distribution only as a visualization tool.
In the derivation of the GME for the single dot device (see~\S
\ref{sec:GMESDQS}) we identified three components of the
Liouvillean super-operator:

\be
 \mathcal{L} = \mathcal{L}_{\rm coh} + \mathcal{L}_{\rm driv} + \mathcal{L}_{\rm damp}
\end{equation}
distinguishing the coherent contribution that describes the
evolution of the system without the electrical and mechanical
baths from a driving contribution where the coupling to the leads
is inserted and a damping term that takes into account the
interaction with the dissipative environment of the mechanical
degree of freedom. We separately treat their translation into the
Wigner formalism in the following sections.

\subsection{Coherent Liouvillean}

We start recalling explicitly the action of the coherent part of
the Liouvillean  on the two non-vanishing\footnote{For an
explanation on the vanishing electrical coherencies $\s_{01}$ and
$\s_{10}$ see \S \ref{sec:ReducedDensityMatrix}.} electrical
components of the reduced density matrix ($\s_{00}$ and $\s_{11}$):

\be
 \mathcal{L}_{\rm coh}
 \left(
 \begin{array}{c}
 \s_{00}\\
 \s_{11}
 \end{array}
 \right) = -\frac{i}{\hbar}\left(
 \begin{array}{c}
 {[H_{\rm osc},\s_{00}]}\\
 {[H_{\rm osc} - e\mathcal{E}x,\s_{11}]}
 \end{array}
 \right)
\end{equation}

It is easy to verify from the definition (\ref{eq:WFA}) that for
operators which are sum of functions of momentum or position
operator only, the associated Winger functions coincide with the
corresponding classical dynamical variables. For example the
Wigner function of the oscillator Hamiltonian reads:

\be
 H_{\rm osc}(q,p) = \frac{p^2}{2m} + \frac{1}{2}m\w^2q^2
\end{equation}
and analogously the electrostatic component becomes
$e\mathcal{E}q$.

As already mentioned the commutator structure of the Liouville-von
Neumann equation implies that only the odd components of the
$\hbar$ expansion of the Groenewold operator $\exp(\hbar \Lambda
/2i)$ should be retained. Since the third order in the expansion
contains already derivatives in $q$ and $p$ up to the third order,
and the oscillator Hamiltonian is quadratic in $q$ and $p$ only
the first order survives and the coherent part of the Liouvilliean
acquires the classical form\footnote{This result is more general
and resembles the Ehrenfest theorem for Hamiltonian up to second
order in x and p. In that case one gets classical equations of
motion for the expectation values $\langle x\rangle$ and $\langle
p\rangle$.}:

 \be
 \mathcal{L}_{\rm coh}
 \left(
 \begin{array}{c}
 W_{00}\\
 W_{11}
 \end{array}
 \right) =
 \left(
 \begin{array}{c}
 {m \w^2 q\frac{\partial W_{00}}{\partial p}
 -\frac{p}{m}\frac{\partial W_{00}}{\partial q}}\\
 {m \w^2(q-d)\frac{\partial W_{11}}{\partial p}
 -\frac{p}{m}\frac{\partial W_{11}}{\partial q}}
 \end{array}
 \right)
\end{equation}
where the displacement $d = \frac{e \mathcal{E}}{m\w^2}$ is the
equilibrium position of the oscillator subject to a constant force
$e\mathcal{E}$ and we have introduced a Wigner distribution  for
each of the electrical states of the system. In this charge
resolved Wigner functions one can read the probability density for
the quantum dot to have a specific position and (average) momentum
when in a specific electrical state. In this way we can monitor
the correlation between charge and position (momentum) of the
oscillating dot, important to discriminate the shuttling regime
from other transport regimes.

\subsection{Driving Liouvillean}

The driving term of the Liouvillean for the SDQS is a rate
equation somehow modified to take into account the operator form
of the position-dependent rates:

\be
 \mathcal{L}_{\rm driv}
 \left(
 \begin{array}{c}
 \s_{00}\\
 \s_{11}\\
 \end{array}
 \right) =
 \left(
 \begin{array}{c}
 -\frac{\G_L}{2}\{e^{-2x/\l},\s_{00}\} +
        \G_R e^{ x/\l}\s_{11}e^{ x/\l}\\
 -\frac{\G_R}{2}\{e^{ 2x/\l},\s_{11}\} +
        \G_L e^{-x/\l}\s_{00}e^{-x/\l}\\
 \end{array}
 \right)
\end{equation}
where $\{A,B\} \equiv AB + BA$ is the anticommutator.

We calculate first the generic form of the symmetric component of
the driving Liouvillian in the Wigner representation. It is easier
in this case to start directly from the definition:
\be
\bs
 &\frac{1}{2\pi\hbar}
 \int_{-\infty}^{+\infty}\!\!\!d \xi
 \left\langle q-\frac{\xi}{2}\right|
 \G_{L,R}e^{\mp x/\l} \s_{ii}(t) e^{\mp x/\l}
 \left|q+\frac{\xi}{2}\right\rangle
 \exp\left(\frac{ip\xi}{\hbar}\right)\\
 =&\, \G_{L,R}e^{\mp 2q/\l}\int_{-\infty}^{+\infty}\!\!\!d \xi
 \left\langle q-\frac{\xi}{2}\right|
 \s_{ii}(t)
 \left|q+\frac{\xi}{2}\right\rangle
 \exp\left(\frac{ip\xi}{\hbar}\right)\\
 =& \G_{L,R}e^{\mp 2q/\l} W_{ii}(q,p,t)
\end{split}
\end{equation}
where $i=0,1$ in accordance with the sign of  the exponential
respectively $-,+$ and the subscript $L,R$ in the tunneling rate
$\G$. Only the classical component contributes to this term
since the Wigner representation is just the product of the three
Wigner functions.

Quantum correction are present instead in the  anticommutator
component of the driving Liouvillean. Again we calculate the
generic form :

\be
\bs
 &-\frac{1}{2\pi\hbar}
  \int_{-\infty}^{+\infty}\!\!\!d \xi
  \left\langle q-\frac{\xi}{2}\right|
  \frac{\G_{L,R}}{2}\{e^{\mp 2x/\l}, \s_{ii}(t) \}
  \left|q+\frac{\xi}{2}\right\rangle
  \exp\left(\frac{ip\xi}{\hbar}\right)\\
%%%%%%%%%%%%%%%%%%
 =&\, -\frac{\G_{L,R}}{2}\left[
  e^{\mp 2q/\l}
 \exp \left(\frac{\hbar \Lambda}{2 i}\right)
  W_{ii}(q,p,t)
 +W_{ii}(q,p,t)
  \exp \left(\frac{\hbar \Lambda}{2 i}\right)
  e^{\mp 2q/\l}\right]\\
%%%%%%%%%%%%%%%%%%
 =& -\G_{L,R}e^{\mp 2q/\l}\sum_{n=0}^{\infty}
 \frac{1}{(2n)!}
 \left(\frac{\hbar \Lambda}{2i}\right)^{2n}
 W_{ii}(q,p,t)\\
%%%%%%%%%%%%%%%%%%
 =& -\G_{L,R}e^{\mp 2q/\l}\sum_{n=0}^{\infty}
 \frac{(-1)^n}{(2n)!}
 \left(\frac{\hbar}{\l}\right)^{2n}
 \frac{\partial^{2n}}{\partial p^{2n}}
 W_{ii}(q,p,t)
\end{split}
\label{eq:Kleinder}
\end{equation}
where because of the symmetry of the anticommutator only  even
powers of the Groenewold operator give non vanishing contributions
(line 2 to 3 in \eqref{eq:Kleinder} ) and they are equal for both terms of the
anticommutator. In the last line we appreciate the control
parameter for the quantum-classical correspondence. Especially in
the shuttling regime the typical length, mass and time scales of
the system are the tunneling length $\l$ and the harmonic
oscillator mass $m$ and inverse frequency $1/\w$. Rescaling the
phase space coordinates to the adimensional $X = q/\l$ and $P =
p/(m\w\l) $ we obtain the expansion parameter:

\be
 \frac{\hbar}{m\w\l^2} = 2\left(\frac{\D x_z}{\l}\right)^2
\end{equation}
that is the ratio between the  zero point fluctuation of the
oscillator position ($\D x_z$) and the tunneling length ($\l$).
The bigger $\l$ the more classical is the behavior of the harmonic
oscillator.

All together the driving Liouvillean in the Wigner representation takes the form:

\be
 \mathcal{L}_{\rm driv}
\left(
 \begin{array}{c}
 W_{00}\\
 W_{11}
 \end{array}
\right) =
\left(
 \begin{array}{c}
 {\G_{R}e^{ 2q/\l} W_{11}
 -\G_{L}e^{- 2q/\l}\sum_{n=0}^{\infty}
 \frac{(-1)^n}{(2n)!}
 \left(\frac{\hbar}{\l}\right)^{2n}
 \frac{\partial^{2n}}{\partial p^{2n}}
 W_{00}}\\
 {\G_{L}e^{- 2q/\l} W_{00}
 -\G_{R}e^{2q/\l}\sum_{n=0}^{\infty}
 \frac{(-1)^n}{(2n)!}
 \left(\frac{\hbar}{\l}\right)^{2n}
 \frac{\partial^{2n}}{\partial p^{2n}}
 W_{11}}
 \end{array}
\right)
 \end{equation}

\subsection{Damping Liouvillean}

The damping Liouvillean is identical for the two electrical
components of the reduced density matrix and we give a unified
derivation of the corresponding Wigner formulation. In terms of
density matrices it reads:
\be
\mathcal{L}_{\rm damp}(\s) =
  -\frac{i\g}{2\hbar}
  [x,\{p,\s\}]
  -\frac{\g m \w}{2\hbar}(2n_B+1)
  [x,[x,\s]]
\end{equation}
Once again we apply the Groenewold modified  product\footnote{The
product of three operator can be handled easily since one can
split the product in two parts and the property $(A*B)*C =
A*(B*C)$ holds, where we have indicated with $*$ the Groenewold
modified product.} (\ref{eq:Groe}) to obtain the Wigner
formulation. Since the damping Liouvillean is quadratic in the
operators $x$ and $p$ we expect only the first two terms of the
expansion in the operator $\Lambda$ to contribute.
\be
 \mathcal{L}_{\rm damp}W =
 \g \frac{\partial}{\partial p}p W
 +\g m \hbar \w \left(n_B + \frac{1}{2} \right)
 \frac{\partial^2}{\partial p^2} W
 \label{eq:dampFP}
\end{equation}

The term of first order in the derivative is  called damping term
while the second diffusion term. One way to understand the meaning
of this names is to derive this equation for the distribution
function starting from the corresponding Langevin equations
describing the brownian motion of a particle in an harmonic
potential:
\be
\bs
 \dot{x} &= \frac{p}{m}\\
 \dot{p} &= -m\w^2x-\g p + \xi(t)
\end{split}
\end{equation}
The first term in the RHS of equation ({\ref{eq:dampFP}) is
related to the velocity dependent friction ($-\g p$) while the
second comes from the stochastic force $\xi(t)$. The analogy with
classical brownian motion can be pushed even further considering
the rescaling:
\be
X=\frac{x}{\ell}, \qquad
P=\frac{p}{m\w\ell}
\end{equation}
where $\ell = \sqrt{\frac{\hbar}{2m\w}(2n_B +1)}$. The damping
Liouvillian in terms of the rescaled variables reads:
\be
\mathcal{L}_{\rm damp}(W)=
 \g \frac{\partial}{\partial P}
 \left(P + \frac{\partial}{\partial P}\right)\,W
\end{equation}
and only the scaling\footnote{Quantum contribution (e.g regions of
negative distribution $W(q,p,t)<0$) could be contained in the
coherence of the initial condition. These will anyway be decohered
by the damping Liouvillean itself.} distinguishes the quantum from
the classical description While in the low temperature limit $2n_B
\ll 1$ and the length scale $\ell$ tends to the oscillator zero
point fluctuation $\D x_z$, in the high temperature limit $\hbar$
drops out from the equations and $\ell$ is given by the the thermal
length $\l_{th} = \sqrt{\frac{k_BT}{m\w^2}}$.

%\vspace{1cm}

Finally we collect the results of the previous sections and write
the equation of motion for the SDQS in the Wigner function
formulation:
\be
\begin{tabular}{|rl|}
\hline
\phantom{.} & \\
 $\displaystyle \frac{\partial W_{00}}{\partial t} =$&
%%%%%%%%%%
 $ \displaystyle
 \!\!\!\left[m \w^2 q\frac{\partial}{\partial p}
 -\frac{p}{m}\frac{\partial}{\partial q}
%%%%%%%%%%
 +\g \frac{\partial}{\partial p}p
 +\g m \hbar \w \left(n_B + \frac{1}{2} \right)
 \frac{\partial^2}{\partial p^2}\right]W_{00} $\\
 \phantom{.} & \\
%%%%%%%%%%
 &$\displaystyle
 \!\!\!+\G_{R}e^{ 2q/\l} W_{11}
 -\G_{L}e^{- 2q/\l}\sum_{n=0}^{\infty}
 \frac{(-1)^n}{(2n)!}
 \left(\frac{\hbar}{\l}\right)^{2n}
 \frac{\partial^{2n}W_{00}}{\partial p^{2n}}
 $\\
\phantom{.} & \\
%%%%%%%%%%%%%%%%%%%%%%%%%%%%%%%%%%%%%%%%%%%%%%%
 $\displaystyle
 \frac{\partial W_{11}}{\partial t} =$&
%%%%%%%%%%
 $\displaystyle
 \!\!\!\left[m \w^2(q-d)\frac{\partial }{\partial p}
 -\frac{p}{m}\frac{\partial}{\partial q}
%%%%%%%%%%%
 +\g \frac{\partial}{\partial p}p
 +\g m \hbar \w \left(n_B + \frac{1}{2} \right)
 \frac{\partial^2}{\partial p^2}\right]W_{11} $\\
 \phantom{.} & \\
%%%%%%%%%%
 &$\displaystyle
 \!\!\!+\G_{L}e^{- 2q/\l} W_{00}
 -\G_{R}e^{2q/\l}\sum_{n=0}^{\infty}
 \frac{(-1)^n}{(2n)!}
 \left(\frac{\hbar}{\l}\right)^{2n}
 \frac{\partial^{2n}W_{11}}{\partial p^{2n}}
$\\
\phantom{.} & \\
 \hline
\end{tabular}
\label{eq:KleinKramers}
\end{equation}

%%%%%%%%%%%%%%%%%%%%%%%%%%%%%%%%%%%%%%%%%%%%%%%%%
% Figure
%%%%%%%%%%%%%%%%%%%%%%%%%%%%%%%%%%%%%%%%%%%%%%%%%
\begin{figure}[ht]
 \begin{center}
  \includegraphics[angle=0,width=\textwidth]{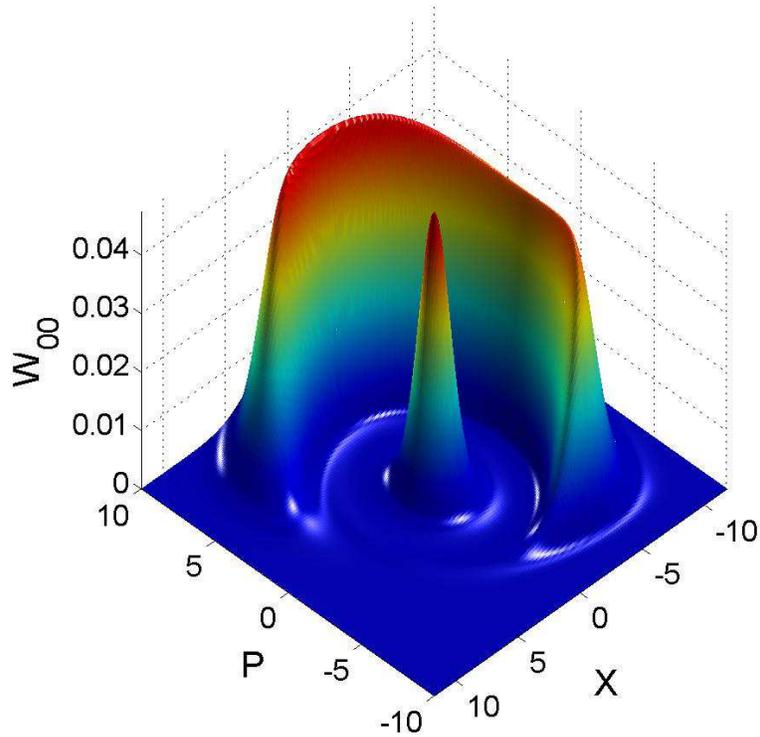}
  \caption{\small  \textit{Example of 3-D visualization of the stationary solution of the
  Klein-Kramers equations for the SDQS. The surface represents the quasi-probability
  distribution of the empty dot $W^{stat}_{00}(q,p)$.
  Coordinate and momentum are expressed in adimensional units: $X = q/x_0$ and
  $P = px_0/\hbar$, where $x_0 = \sqrt{\hbar/(m\w)}$. The distribution shows two distinct
  features. The probability is high near the origin of the phase space and represents
  the oscillator in the ground state. The other structure
  corresponds to the part of the limit cycle drawn by the empty dot
  in the shuttling regime.}}
 \end{center}
\end{figure}
%%%%%%%%%%%%%%%%%%%%%%%%%%%%%%%%%%%%%%%%%%%%%%%%%%%

\clearpage{\pagestyle{empty}\cleardoublepage}

%%%%%%%%%%%%%%%%%%%%%%%%%%%%%%%%%%%%%%%%%%%%%%%%%%%%%%%%%%%
%                   INVESTIGATION TOOLS
%%%%%%%%%%%%%%%%%%%%%%%%%%%%%%%%%%%%%%%%%%%%%%%%%%%%%%%%%%%

\chapter{Investigation tools}\label{sec:investigation_tools}

The dynamics of the shuttling devices or at least of the
simplified models for them that we consider, is all contained in
the GMEs for the reduced density matrices (\ref{eq:SDQSGME}) and
(\ref{eq:TDQSGME}) or in the corresponding Klein-Kramers equations
for the Wigner distributions (e.g.\ equation \eqref{eq:KleinKramers}
for the SDQS). The coupling to the continuum of states in the
electrical and mechanical baths introduces irreversibility in the
system and its statistical properties (represented by the reduced
density matrix or Wigner distribution) tend to a stationary limit.
This does \emph{not} mean that the system itself is in a static
condition. In particular an electric  current sustained by the
bias always flows through the device driving the movable dot into
a variety of different dynamical regimes. In order to analyze the
properties of these different operating conditions of the
electronic shuttles we use three investigation tools. We study the
phase space distribution functions associated with the stationary
solution of the GME, the corresponding current and the
current-noise, taking as control parameters the damping rate of
the mechanical degree of freedom, or the gate voltage of the
different dots, or the injection and ejection rates. Through the
phase-space analysis or the current and noise calculation we gain
somehow complementary information about the system. These
approaches are described in detail in the next three sections.

\section{Phase space distributions: dots, rings and bananas}

The phase space analysis of the shuttle dynamics is based on the
Wigner representation of the stationary solution of the
GME:
\be
 W_{ij}^{\rm stat}(q,p) =
 \frac{1}{2 \pi \hbar}
 \int_{-\infty}^{+\infty}\!\!\!d \xi
 \left\langle q-\frac{\xi}{2}\right| \s_{ij}^{\rm stat} \left|q+\frac{\xi}{2}\right\rangle
 \exp\left(\frac{ip\xi}{\hbar}\right)
\end{equation}
where the subscript $ij$ indicates the electrical state that we
are considering. Since typically we are not interested in
coherencies\footnote{The probabilistic interpretation of the Wigner distribution would not be acceptable for electronic coherencies.}  between electronic states we set $i=j$ and the
resulting Wigner distribution represents the joint probability for
the system to be in the specific mechanical state $(q,p)$ and
electrical state $i$. The total Wigner distribution
\be
 W^{\rm stat}_{\rm tot}(q,p) = \sum_{i} W^{\rm stat}_{ii}(q,p)
\end{equation}
where  the sum is extended to all the electrical states of the
system, gives information about the mechanical state independently
from the electronic state of the device.

\subsection{Shuttling instability}

For the SDQS we  analyze first the behavior of the Wigner
distribution as  a function of the mechanical damping. The
mechanical dynamics shows a typical feature qualitatively
independent from the other parameters of the model. At high
damping rates the total Wigner distribution is concentrated around
the origin of the phase space and represents the harmonic
oscillator in the ground state. While reducing the mechanical
damping a ring progressively develops and the central ``dot''
eventually disappears (Fig. \ref{fig:WFall}). We interpret the
ring as a smeared limit cycle trajectory: by reducing the damping
the equilibrium position becomes unstable and a stable limit cycle
develops. The threshold for this transition is given by the
effective tunneling rates of the electrons\footnote{A more precise
definition of this electronic time can be found the next
chapter.}.

This interpretation is represented in the following  picture:
every time an electron jumps on the movable grain the latter is subject
to the electrostatic force $e\mathcal{E}$ that accelerates it
towards the right. Energy is pumped into the mechanical
system and the dot starts to oscillate. If the damping is high
compared to the tunneling rates the oscillator dissipates this
energy into the environment before the next tunneling event: on
average the dot remains in its ground state. On the contrary for
very small damping the relaxation time of the oscillator is long
and multiple ``forcing events'' happen before the relaxation takes
place. This continuously drives the oscillator far from
equilibrium and a stationary state is reached only when the energy
pumped per cycle into the system is dissipated during the same
cycle in the environment.

It is not difficult to realize that, like for a macroscopic swing,
in order to sustain the motion one needs coordination between
forcing (here related to the electrical dynamics) and
oscillations. This coordination is revealed by the charge resolved
Wigner distributions $W_{00}$ and $W_{11}$. The ring that appears
in the total distribution is asymmetrically shared by the empty
and charged-dot distributions (Fig. \ref{fig:WFall}).
%%%%%%%%%%%%%%%%%%%%%%%%%%%%%%%%%%%%%%%%%%%%%%%%%%%%%%%%%%%%%%%%%
% Figure
%%%%%%%%%%%%%%%%%%%%%%%%%%%%%%%%%%%%%%%%%%%%%%%%%%%%%%%%%%%%%%%%%
\begin{figure}[ht]
 \begin{center}
  \includegraphics[angle=0,width=.7\textwidth]{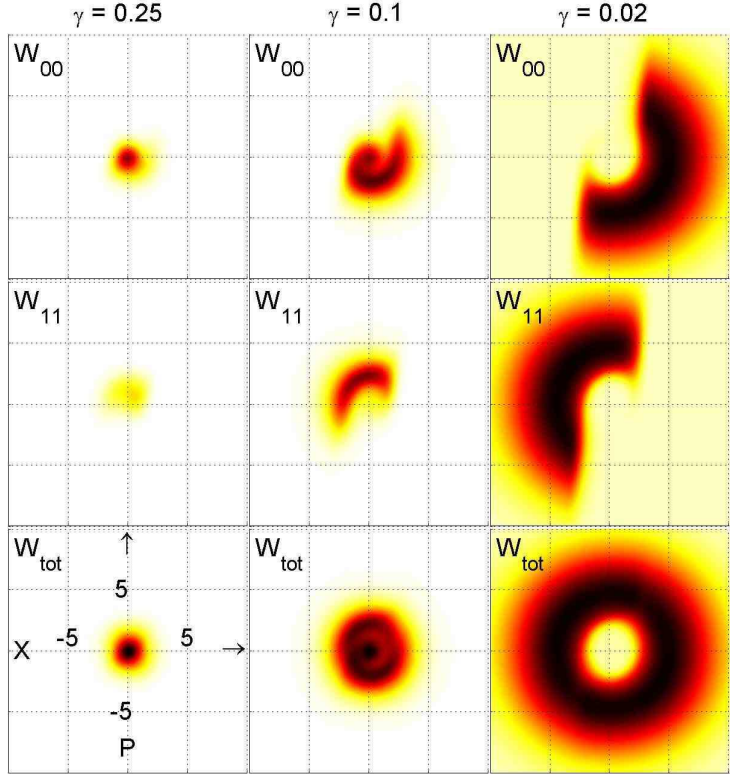}
  \caption{\small  \textit{Phase space picture of the tunneling-to-shuttling
  crossover. The respective rows show the Wigner
  distribution functions for the discharged ($W_{00}$),  charged
  ($W_{11}$) and sum  ($W_{\rm tot}$) states of the oscillator in
   the phase space (horizontal axis--coordinate in units of
   $x_0 = \sqrt{\hbar/(\hbar \w)}$, vertical axis -- momentum in
   $\hbar/x_0$). The values of the parameters are:
   $\l = x_0$, $T = 0$,  $d = 0.5x_0$, $\G = 0.05\w$. The values
   of $\g$ are in units of $\w$. The Wigner functions are normalized
    within each column.}}
  \label{fig:WFall}
 \end{center}
\end{figure}
%%%%%%%%%%%%%%%%%%%%%%%%%%%%%%%%%%%%%%%%%%%%%%%%%%%%%%%%%%%%%%%%%

The Wigner distributions reveal the charge-position (momentum)
correlation typical of the shuttling regime: for negative
displacements and positive momentum (i.e. leaving the source lead)
the dot is prevalently charged while it is empty for positive
displacements and negative momentum (coming from the drain lead).
Self-sustained mechanical oscillations and shuttling electron
transport are part of the same phenomenon.

\subsection{Classical vs. Quantum Shuttle}

The instability we showed in the previous section was predicted
already  by Gorelik {\it et al.}~\cite{gor-prl-98} (even if as a
function of the external bias) in the form of a sharp transition
between an equilibrium and a limit cycle dynamics of the
oscillator and is called \emph{shuttle instability}. The quantum
mechanical description of the harmonic oscillator that we give
introduces two effects in the model: it smears the transition into
a crossover and reveals the possibility to trigger the shuttling
regime even at zero electric field \cite{nov-prl-03} when
classically  the electrical and mechanical dynamics would be
decoupled and the device would behave as a damped harmonic
oscillator that with no feedback influences a single electron
transistor (Fig. \ref{fig:zerofield}).
%
%%%%%%%%%%%%%%%%%%%%%%%%%%%%%%%%%%%%%%%%%%%%%%%%%%%%%%%%%%%%%%%%%
% Figure
%%%%%%%%%%%%%%%%%%%%%%%%%%%%%%%%%%%%%%%%%%%%%%%%%%%%%%%%%%%%%%%%%
\begin{figure}[h]
 \begin{center}
  \includegraphics[angle=-90,width=.6\textwidth]{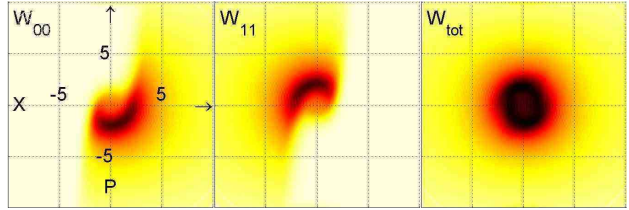}
  \caption{\small  \textit{Shuttling at zero electric field. The three
  Wigner distributions for the empty, charged, and sum states are
  represented in the usual adimensional phase space $(X,P)$ (see Fig.
  \ref{fig:WFall}). The parameter values are $\l = x_0$, $T = 0$,
  $d = 0$, $\G = 0.05\w$, $\g=0$.}}
  \label{fig:zerofield}
 \end{center}
\end{figure}
%%%%%%%%%%%%%%%%%%%%%%%%%%%%%%%%%%%%%%%%%%%%%%%%%%%%%%%%%%%%%%%%%

The amplitude of the shuttling oscillations of the quantum dot is
also dependent on the tunneling length and bare injection rate. In
particular it grows with the tunneling length $\l$ and with the
inverse of the bare injection rate.
%
%%%%%%%%%%%%%%%%%%%%%%%%%%%%%%%%%%%%%%%%%%%%%%%%%%%%%%%%%%%%%%%%%
% Figure
%%%%%%%%%%%%%%%%%%%%%%%%%%%%%%%%%%%%%%%%%%%%%%%%%%%%%%%%%%%%%%%%%
\begin{figure}[hb]
 \begin{center}
  \includegraphics[angle=-90,width=.6\textwidth]{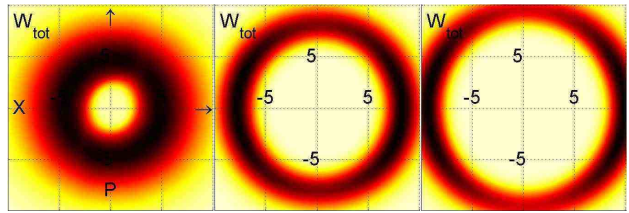}
  \caption{\small  \textit{Transition from ``quantum'' to ``classical'' shuttling.
  The figure shows the total Wigner function for three different
  sets of parameters. We keep constant damping ($\g = 0.02\w$)
  temperature ($T = 0$) and electric field ($d = 0.5x_0$) and change
  the injection rate and tunneling length. From left to right
  respectively: $(\G =0.05\w,\, \l = x_0)$, $(\G =0.05\w ,\, \l = 2x_0)$,
  $(\G =0.01\w,\, \l = 2x_0)$.}}
  \label{fig:WFquantclass}
 \end{center}
\end{figure}
%%%%%%%%%%%%%%%%%%%%%%%%%%%%%%%%%%%%%%%%%%%%%%%%%%%%%%%%%%%%%%%%%
%
This behavior can be understood as a self-consistent compensation
to accomplish the non-adiabatic condition that the effective
injection rate $\bar{\G}$ must be comparable in the shuttling
regime with the oscillator frequency $\w/2 \pi$. The larger the
amplitude of the oscillation the more classical is the  behavior
of the harmonic oscillator. This concept can be understood also
from a geometric point of view: the ring gets closer and closer to
a circle since the width, given by the amount of thermal and shot
noise remains constant (or decreases as we will see in the
coexistence regime semianalytics) while the diameter increases.
Another natural parameter that reveals this quantum to classical
transition is the ratio between $\hbar$ and the area enclosed by
the ring. This tends to zero in the classical limit (Fig.
\ref{fig:WFquantclass}).

%\begin{figure}
% \begin{center}
%  \includegraphics[angle=-90,width=\textwidth]{Figures/WFtriple.eps}
%  \caption{\small  \textit{.}}
%  \label{fig:WFtriple}
% \end{center}
%\end{figure}

\subsection{Different electronic processes in the TDQS}

The Wigner distribution analysis of the TDQS is performed in
analogy with the SDQS on the  states corresponding to the empty
and charged central dot.  Namely we consider:

\be
\bs
 W_{UU}^{\rm stat}(q,p) &=
 \frac{1}{2 \pi \hbar}
 \int_{-\infty}^{+\infty}\!\!\!d \xi
 \left\langle q-\frac{\xi}{2}\right|
 \s_{00}^{\rm stat} +\s_{LL}^{\rm stat} +\s_{RR}^{\rm stat} \left|q+\frac{\xi}{2}\right\rangle
 \exp\left(\frac{ip\xi}{\hbar}\right)\\
  W_{CC}^{\rm stat}(q,p) &=
 \frac{1}{2 \pi \hbar}
 \int_{-\infty}^{+\infty}\!\!\!d \xi
 \left\langle q-\frac{\xi}{2}\right| \s_{CC}^{\rm stat} \left|q+\frac{\xi}{2}\right\rangle
 \exp\left(\frac{ip\xi}{\hbar}\right)
\end{split}
\end{equation}

As suggested by the work of Armour and  MacKinnon
\cite{arm-prb-02} we consider as control parameter the difference
between the gate voltages on the left and right dot (the so-called
device bias $\D V$). It was found in  \cite{arm-prb-02} that the
triple dot system exhibits different regimes of transport at
different device biases. The current peaks at $\D V \approx
n\hbar\w$ were identified as effects of electromechanical
resonances within the device. Yet, the different peaks may
correspond to different physical mechanisms -- while the peak
around $\D V \approx \hbar\w$ is mainly due to the incoherent
oscillator-assisted tunneling the peak at $\D V \approx 2\hbar\w$
reveals a clear shuttling component. This finding by Armour and
MacKinnon based on indirect evidence of parametric dependencies of
the current curves (e.g. the dependence of the current curve on
the tunneling length $\l$) is confirmed by the direct inspection
of the phase space distribution (Fig. \ref{fig:modena}). The
half-moon-like shape characteristic for the shuttling transport is
only present around $\D V \approx 2\w$ while all other plots show
just the fuzzy spot indicative of incoherent tunneling. However,
our direct criterion for detecting the shuttling regime reveals a
close similarity between the resonances. For increasing injection
rate we can see that the shuttling regime gradually sets in also
in the vicinity of the first resonance peak. This reveals the
cross over character of the onset of the shuttling instability
found also for the SDQS.

%%%%%%%%%%%%%%%%%%%%%%%%%%%%%%%%%%%%%%%%%%%%%%%%%%%%%%%%%%%%%%%%%
% Figure
%%%%%%%%%%%%%%%%%%%%%%%%%%%%%%%%%%%%%%%%%%%%%%%%%%%%%%%%%%%%%%%%%
\begin{figure}[h]
 \begin{center}
  \includegraphics[angle=0,width=.7\textwidth]{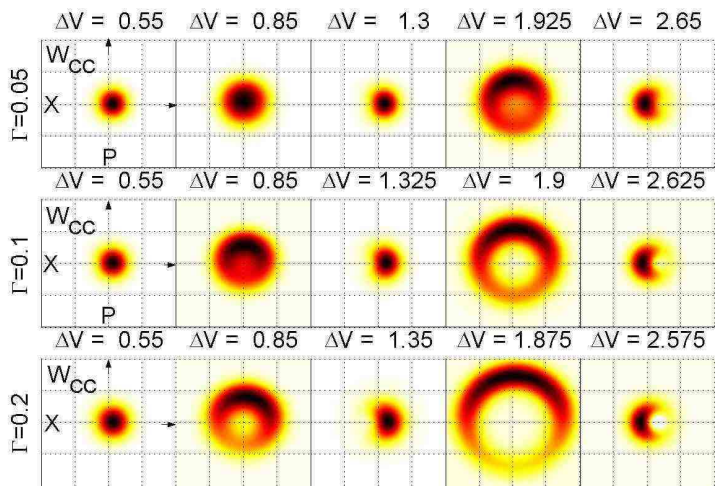}
  \caption{\small  \textit{Wigner distributions for the central dot
   in the charged state. The half-moon shuttling signature, always present
  at the second resonance ($\D V \approx 2\hbar\w$) progressively emerges
  also in the first resonance ($\D V \approx \hbar\w$). The other parameters are
  $V_0 = 0.757\hbar\w,\,\l = 5 \D x_z,\,x_0 = 7.071\D x_z,\,\g = 0.0125\w$.}}
  \label{fig:modena}
 \end{center}
\end{figure}
%%%%%%%%%%%%%%%%%%%%%%%%%%%%%%%%%%%%%%%%%%%%%%%%%%%%%%%%%%%%%%%%%

As can be seen in more detail in the work  by C.~Flindt et al.
\cite{fli-preprint-04} and in the Master thesis by C.~Flindt
\cite{christian} the TDQS exhibits at higher device biases another
transport mechanism that competes with the shuttling transport.
For $\D V \approx 3\w,4\w$ \emph{inelastic co-tunneling} takes
place and a clear signature of this phenomenon can be seen in the
vanishing of the charged central dot Wigner function (see Fig.\ 5
in \cite{fli-preprint-04}).

\section{Current}

The Wigner distribution functions are a very clean theoretical
tool to investigate the dynamical properties of shuttle devices,
but unfortunately it seems very difficult (if not impossible) to
access experimentally these functions. For this reason, with in
mind the distribution function description, we investigate the
electronic transport properties in the shuttle devices (more
realistically measurable) starting with the stationary
current.

\subsection{Calculation of the stationary current}

We already encountered in section \ref{subsec:GME} the definition of
right and left lead currents:
\be
\bs
 I_R(t) &= \dot{N}_R(t)\\
 I_L(t) &= \dot{N}_L(t)
 \end{split}
\end{equation}
where $N_R(t)$ ($N_L(t)$) are the total number of electrons
(holes) collected in the right (left) lead at time $t$. We have
also related these functions to the elements of the
$n$($h$)-resolved density matrix and corresponding GME (Eq.\
\eqref{eq:e-current} and \eqref{eq:h-current}). Since we are
interested exclusively in the stationary current\footnote{The
stationary current is equal on both leads due to charge
conservation.} we analyze now in detail only the electron current
in the right lead in presence of a shuttle device (SDQS or TDQS).
The total number of electrons collected in the right lead $N_R(t)$
can be written as:
\be
 N_R(t) = \sum_{n=0}^{\infty} nP_n(t)
\end{equation}
where $P_n(t)$ is the probability for $n$ electrons to be
collected into the right lead at time $t$. We express the probability
$P_n(t)$ as the trace of the $n$-resolved reduced
density matrix over the system degrees of freedom:

\be
 P_n(t) = {\rm Tr}_{\rm sys}\left\{\s^{(n)}(t)\right\}
 = {\rm Tr}_{\rm osc} \left\{\textstyle{\sum_i}\s_{ii}^{(n)}(t)\right\}
\end{equation}
where $i$ is the electrical state of the device: and in the SDQS
$i = 0,1$ while in the TDQS $i = 0,L,C,R$. The $n$-resolved GME
($n$-GME) comes now into play and the current is expressed as a
function of $\s_{ii}^{(n)}(t)$:

\be
 \bs
 I_R(t)&= \sum_{n=0}^{\infty} \sum_i n{\rm Tr}_{\rm osc}
 \left\{ \dot{\s}_{ii}^{(n)}(t)\right\}\\
 &= \sum_{n=0}^{\infty} n \sum_i {\rm Tr}_{\rm osc}
 \left\{
 \mathcal{L}_{\rm coh}\big[\s^{(n)}(t)\big]_{ii} +
 \mathcal{L}_{\rm driv}\big[\s(t)\big]^{(n)}_{ii}+
 \mathcal{L}_{\rm damp}\big[\s^{(n)}_{ii}(t)\big]
 \right\}
 \end{split}
\end{equation}
where we have implicitly specified with the position of the square
parentheses the different components of the density matrix
addressed by the different  Liouvilleans\footnote{The damping
Liouvillean acts only in the Hilbert space of the harmonic
oscillator and does not mix different electronic states or states
with different number of particle in the collector. The coherent
Liouvillean is sensitive to the charge state of the system and does
not care about the particular bath state, while the driving term is
mixing electronic states \emph{and} bath states with different
number of particles since it describes the electron tunneling
between system and baths. All of them are local in time (Markov
approximation).}.

We calculate first the SDQS right (particle) current. The coherent
and damping part of the Liouvillean have a commutator structure
(see Eq. (\ref{eq:TrasInvLiou})) and vanish under trace. We are
left with the contribution given by the driving Liouvillean:
\be
 I_R(t)
 = {\rm Tr}_{\rm osc}\left\{ \G_R e^{2x/\l}
  \sum_{n = 0}^{\infty}n \left[\s_{11}^{(n-1)}(t)-\s_{11}^{(n)}(t)\right]\right\}
\end{equation}
where we have used the cyclic property of the trace to evaluate
the sum over the two electrical components. The particular
ordering of the position dependent injection and ejection
operators $\G_{L,R}(x)$ is irrelevant under trace and the sum over
the number of electron in the resevoir is formally identical to
the one obtained for the static single dot (see Eq.\
\eqref{eq:e-current}). We evaluate it and obtain:

\be
 I_R(t) = {\rm Tr}_{\rm osc}\left\{\G_R
 e^{2x/\l}\s_{11}(t)\right\}
 \label{eq:IRSDQS}
\end{equation}
With completely analogous arguments one derives the expression for
the left current:

\be
 I_L(t) = {\rm Tr}_{\rm osc}\left\{\G_L
 e^{-2x/\l}\s_{00}(t)\right\}
 \label{eq:ILSDQS}
\end{equation}

The stationary current is obtained by substituting in
(\ref{eq:IRSDQS}) or (\ref{eq:ILSDQS}) the stationary solution
$\s^{\rm stat}$ of the GME. It is
interesting that the stationary particle current is written as the
average injection (ejection) rate:
\be
 I^{\rm stat}=
\begin{array}{c}
 \nearrow\\
 \searrow
 \end{array}
\begin{array}{c}
 {\rm Tr}_{\rm osc}
 \left\{\G_R
 e^{2x/\l}\s^{\rm stat}_{11}\right\}
=
 {\rm Tr}_{\rm sys}\left\{\G_R(x) |1\rangle\!\langle 1|
 \s^{\rm stat}\right\} = \bar{\G}_R\\
 \\
 \\
 {\rm Tr}_{\rm osc}
 \left\{\G_L
 e^{-2x/\l}\s^{\rm stat}_{00}\right\}
=
 {\rm Tr}_{\rm sys}\left\{\G_L(x) |0\rangle\!\langle 0|
 \s^{\rm stat}
 \right\}
  = \bar{\G}_L
 \end{array}
\begin{array}{c}
 \searrow\\
 \nearrow
 \end{array}
= \bar{\G}
\end{equation}
where we have introduced the projectors $|1\rangle\!\langle 1|
\equiv c_1^{\dag}c_1^{\phd}$ and $|0\rangle\!\langle 0| \equiv 1 -
c_1^{\dag}c_1^{\phd}$ on the electronic Hilbert  space and we
calculate  the average injection and ejection rates taking into
account  Coulomb blockade and the single level structure. Namely
the injection (ejection) rate average is calculated only on system
states with empty (charged) dot. A special remark concerns
correlations: the shuttling regime is characterized by strong
electromechanical correlations and they should be recorded in the
current since in general:
\be
 I^{\rm stat} =
 {\rm Tr}_{\rm sys}\{\G_L(x) |0\rangle\!\langle 0|
 \s^{\rm stat}\}
 \ne
 {\rm Tr}_{\rm sys}\{\G_L(x)\s^{\rm stat}\}
 {\rm Tr}_{\rm sys}\{|0\rangle\!\langle 0|\s^{\rm stat}\}
\end{equation}

\subsection{Current characteristics of the SDQS}

We study the current as a function of the mechanical damping. The
results for different values of the bare injection (ejection) rate
and of the tunneling length are presented in Fig.
\ref{fig:SDQSCurrent}. At low damping the current saturates,
independently from the value of the other parameters, at  the
``magic'' value $I\approx 0.16\w$ (numerical approximation of the
frequency of the harmonic oscillator $\w/2 \pi$). Increasing the
damping the stationary current drops, more or less rapidly, to a
plateau, dependent this time on both the bare tunneling rate and
length. Further increase of the damping does not change the
scenario. The transition between the plateau and the saturation
point is faster the more ``classical'' the other parameters
are
(i.e. small injection rate and large tunneling length\footnote{See
previous section for the meaning of the term ``classical''
associated to a parameter configuration.}).

%%%%%%%%%%%%%%%%%%%%%%%%%%%%%%%%%%%%%%%%%%%%%%%%%%%%%%%%%%%%%%%%%
% Figure
%%%%%%%%%%%%%%%%%%%%%%%%%%%%%%%%%%%%%%%%%%%%%%%%%%%%%%%%%%%%%%%%%
\begin{figure}[h]
 \begin{center}
  \includegraphics[angle=0,width=.55\textwidth]{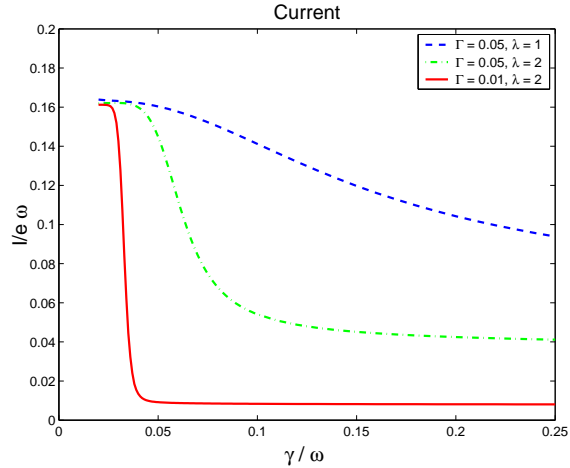}
  \caption{\small  \textit{Particle stationary current of the SDQS
  plotted as a function of the damping rate. The bare electronic rate $(\Gamma)$ is
  given in units of the oscillator frequency. The tunneling
  length is given in units of $x_0 = \sqrt{\hbar/m\w}$.
  The current and the damping rate are expressed in adimensional
  units also by rescaling with the oscillator frequency $\w$. The saturation
  of the current to the limit of one electron per cycle characteristic of
  the shuttling regime at low damping is visible. The transition to the
  the high damping limit (oscillator in ground state) gets sharper the more classical
  is the dynamics as is indicated also by the parameters (see text for explanation).}}
  \label{fig:SDQSCurrent}
 \end{center}
\end{figure}
%%%%%%%%%%%%%%%%%%%%%%%%%%%%%%%%%%%%%%%%%%%%%%%%%%%%%%%%%%%%%%%%%

If we compare the current results with the Wigner function
distribution we can recognize a correspondence between the
shuttling charge-position (momentum) correlation and the
saturation point, as well as a progressive disappearing of the
ring structure in correspondence with the current transition. The
high damping plateau in the current sets in when the mechanical
oscillator lays into its ground state and the Wigner distribution
function is reduced to a fuzzy spot close to the origin of the
phase space (Fig. \ref{fig:SDQSCurrentWigner}).

%%%%%%%%%%%%%%%%%%%%%%%%%%%%%%%%%%%%%%%%%%%%%%%%%%%%%%%%%%%%%%%%%
% Figure
%%%%%%%%%%%%%%%%%%%%%%%%%%%%%%%%%%%%%%%%%%%%%%%%%%%%%%%%%%%%%%%%%
\begin{figure}[h]
 \begin{center}
  \includegraphics[angle=0,width=.65\textwidth]{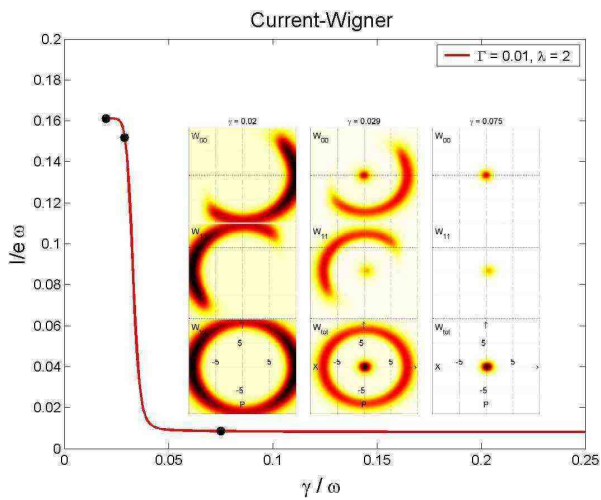}
  \caption{\small  \textit{Most ``classical'' current-damping curve. The Wigner
  function distributions in the insets correspond to the three dots indicated
  on the curve and show how the transition looks like in the phase space.}}
  \label{fig:SDQSCurrentWigner}
 \end{center}
\end{figure}
%%%%%%%%%%%%%%%%%%%%%%%%%%%%%%%%%%%%%%%%%%%%%%%%%%%%%%%%%%%%%%%%%

\subsection{Electromechanical resonances of the TDQS}

The calculation of the TDQS stationary current follows the same
argument line we adopted for the SDQS. Also the analytic expression
shares essentially the same structure, the only difference being
the projector that appears in the right lead current and the
tunneling rate that in this case is \emph{not}  position
dependent. The electrons tunnel to the resevoir only from
the right dot and the current reads:
\be
 I_R^{\rm stat} =  \G_R {\rm Tr}_{\rm osc}
 \left\{\s^{\rm stat}_{RR}\right\}
=
 \G_R {\rm Tr}_{\rm sys}\left\{ |R\rangle\!\langle R|
 \s^{\rm stat}\right\} = \bar{\G}_R
\end{equation}
Since there are not position operators in the left or right
lead current one could argue that the stationary current of
the TDQS do not record signs of charge-position correlation.
The solution to this problem comes from the structure of
the device. Instead of looking at the system-lead tunneling
current one can concentrate on the tunneling current within the
device: in the stationary limit charge conservation ensures they
are equal. For example the left-center current reads in the
stationary limit:
\be
 I_{LC}^{\rm stat} = {\rm Tr}_{\rm osc}\{it_L(x)
 (\s^{\rm stat}_{CL} - \s^{\rm stat}_{LC})\}
\end{equation}
as can be derived tracing the second equation of (\ref{eq:TDQSGME}).

A detailed analysis of the current in the TDQS with  the device
bias\footnote{We recall that the device bias is defined as the
difference between the gate voltages of the left and right dot and
is represented in the model (Eq.~\eqref{eq:TdS-Ham}) as the mismatch
between the left and right dot energy levels ($\D V$).} as control
parameter can be found in the work by Armour and MacKinnon
\cite{arm-prb-02}. Extension and analysis of the same model in
other parameter regimes are given in \cite{modena} and extensively
in \cite{christian} and \cite{fli-preprint-04}. Qualitatively we
can describe the current-device bias characteristics as  series of
peaks (electromechanical resonancies) at $\D V \approx n \hbar
\w$. It is difficult to establish only from the analysis of the
current the nature of these peaks. Armour and MacKinnon have an
interpretation of the different mechanisms underlying these
resonances based on the analysis of the spectrum of the decoupled
three dot device + harmonic oscillator. This method can visualize
very well the basic mechanisms of elastic or inelastic phonon
assisted tunneling. Other information can be gained from the
behaviour of the current at different resonances as a function of
the tunneling length or mechanical damping. We acknowledge the
validity of this analysis but we also think that an independent
source of information (such as the phase space or the
current-noise) is needed to discriminate directly between
different operating regimes. In Fig.\ \ref{fig:TDQSCurrent} we
report a set of current curves in which one can recognize the
first three electromechanical resonances.

%%%%%%%%%%%%%%%%%%%%%%%%%%%%%%%%%%%%%%%%%%%%%%%%%%%%%%%%%%%%%%%%%
% Figure
%%%%%%%%%%%%%%%%%%%%%%%%%%%%%%%%%%%%%%%%%%%%%%%%%%%%%%%%%%%%%%%%%
\begin{figure}[h]
 \begin{center}
  \includegraphics[angle=0,width=.65\textwidth]{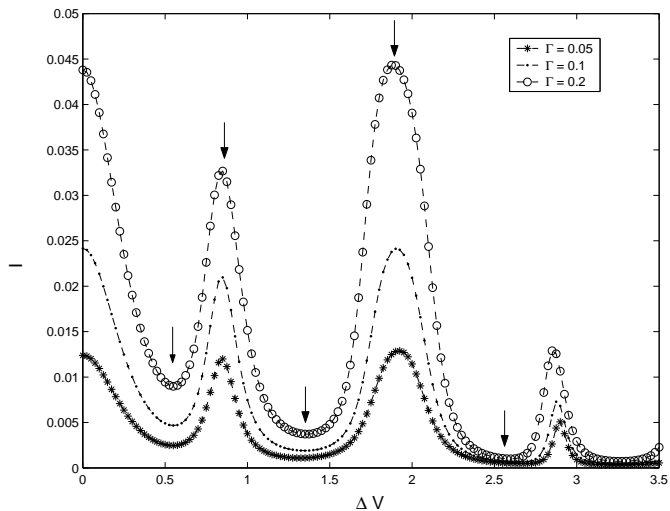}
  \caption{\small  \textit{Current as function of device bias in TDQS. Different
  curves correspond to different injection rates $\Gamma$. The particle current
  and the injection rate $\Gamma$ are scaled with the mechanical frequency $\w$.
  The first three electromechanical resonances are visible. The
  arrows indicate points of minimum or maximum current  and the corresponding Wigner distributions
  are reported in Fig.~\ref{fig:modena}. The other parameters are $V_0 = 0.757\hbar\w,
  \, \l = 5 \D x_z,\, x_0 = 7.071\D x_z ,\, \g = 0.0125\w$.}}
  \label{fig:TDQSCurrent}
 \end{center}
\end{figure}
%%%%%%%%%%%%%%%%%%%%%%%%%%%%%%%%%%%%%%%%%%%%%%%%%%%%%%%%%%%%%%%%%

The temperature influences the device characteristics  in
different ways since different competing transport mechanisms
(phonon assisted-tunneling, co-tunneling, shuttling) depend on
temperature in very different ways. A generic feature can though
be recognized: at finite temperature electromechanical resonances
are found also at negative device biases (see for example Fig.\ 2
in \cite{fli-preprint-04}) when electrons can tunnel through the
device only by gaining energy from the oscillator. This
observation suggests the idea of an active phonon cooling
device\footnote{This idea was suggested by prof.\ K.\ M\o lmer
during private communications.} useful to prepare for example a
macroscopic harmonic oscillator close to its quantum ground state
more efficiently than by coupling it to a ``very cold'' thermal
resevoir.

\section{Current noise}\label{sec:current-noise}

An unequivocal experimental  observation of the shuttling
transition has not yet been achieved. The IV-curve measured in
recent experiments on a vibrating ${\rm C_{60}}$ can be
interpreted in terms of shuttling \cite{fed-epl-02}, but also
alternative explanations have been promoted \cite{boe-epl-01,
mcc-prb-03, braig-prb-03}. It is therefore natural to look for
more refined experimental tools than just the average current
through the device. An obvious candidate is the current noise
spectrum \cite{bla-phr-00, bee-pht-03}. The measurement of the
noise spectrum or even higher moments (full counting statistics)
reveals more information about the transport through the device
than just the mean current. The theoretical studies of the noise
have attracted much attention recently in NEMS in general
\cite{mit-prb-04, arm-prb-04, cht-preprint-04} as well as for the
shuttling set up \cite{pis-prb-04, isa-epl-04, rom-preprint-04} in
the (semi)classical limit. The quantum mechanical theory for shot
noise in NEMS that we treat here is also exposed in
\cite{nov-prl-04} and in a more general form in
\cite{fli-preprint-04} and is mainly due to T.~Novotn\'y in
collaboration with the author and C.~Flindt.

\subsection{The MacDonald formula}\label{sec:MacDonald}

Let us consider the (particle) current flowing from  the dot to
the right lead of the SDQS. This is defined (in the Heisenberg
picture) as the time derivative of the number operator of the
right lead:

\be
\bs
 \hat{I}_R(t) &= \dot{\hat{N}}_R(t) =
  -\frac{i}{\hbar}[\hat{H}(t),\hat{N}_R(t)]\\
  &= -\frac{i}{\hbar} \hat{T}_r(\hat{x})
  \sum_k \left( \hat{c}^{\dag}_{r_k}\hat{c}^{\phd}_1
              + \hat{c}^{\dag}_1\hat{c}^{\phd}_{r_k}\right)
\end{split}
\label{eq:IRheis}
\end{equation}
where
\be
 N_R = \sum_{k}c^{\dag}_{r_k}c^{\phd}_{r_k},
\end{equation}
the symbol $\hat{\bullet}$ indicates the Heisenberg  picture and
$H$ is the Hamiltonian presented in equation (\ref{eq:SdS-Ham0},\ref{eq:SdS-Ham}).
The current-current correlation function of right currents reads:
\be
 C_{I_R,I_R}(t,t') =
 \frac{1}{2}\langle \{\hat{I}_R(t),\hat{I}_R(t')\} \rangle
 -\langle \hat{I}_R(t)\rangle\langle \hat{I}_R(t')\rangle
 \label{eq:IIcorrelation}
\end{equation}
and the brackets $\langle \phantom{a}\rangle$  indicate the
quantum mechanical average\footnote{Since we work in Heisenberg
picture the quantum average is given by the expectation value on
the initial state -- i.e. at the time when the Schr\"odinger and
Heisenberg pictures coincide.}. The noise spectrum is defined as
the Fourier transform of the correlation function\footnote{Some
authors define this Fourier with an extra factor of 2 at the
numerator. A balancing factor of 2 at the denominator is then
inserted in the definition of the Fano factor.}:

\be
 S_{RR}(\w) = \lim_{t \to \infty}\int_{-\infty}^{+\infty} d\t \,
 C_{I_R,I_R}(t+\t,t) \,e^{i\w\t}
 \label{eq:noisedef}
\end{equation}
It is possible to demonstrate (see for example
\cite{fli-preprint-04}) that the zero frequency component of the
current noise is independent from the particular junction
considered. We will for this reason speak in general of current
noise $S(\w = 0)$ even if the specific realization of the
calculation refers to the right junction current. Multitime
averages (e.g. (\ref{eq:IIcorrelation})) can be evaluated in
general using the Quantum Regression Theorem \cite{gardiner}.
Necessary condition is though the fact that the operators averaged
must belong to the \emph{system}. This condition is obviously
violated by the operator $\hat{I}_R(t)$ defined in equation
(\ref{eq:IRheis}) that contains \emph{system} and \emph{bath}
operators. An alternative way of calculating the current noise
relates it to the probability $P_n(t)$ that $n$ electrons have
tunnelled through the device by the time $t$.

We now present this alternative formulation\footnote{One of the
main reasons for the extensive derivation of the $n$-GME that we
gave in chapter \ref{sec:GME} is this formulation of the SDQS
current noise.}. Let us take the definition of noise spectrum
(\ref{eq:noisedef}) in the zero frequency limit and consider the
symmetry relation

\be \lim_{t \to \infty}C_{I_R,I_R}(t+\t,t) = \lim_{t \to
\infty}C_{I_R,I_R}(t-\t,t)
\end{equation}
and the definition of the current operator (\ref{eq:IRheis}). We can
rewrite the noise in the form:
\be
\bs
 S(0)=& \lim_{t \to \infty}
 \Big[\langle \big\{\lim_{\t \to \infty}[\hat{N}_R(\t) - \hat{N}_R(0)],
 \frac{d}{dt}[\hat{N}_R(t) - \hat{N}_R(0)]\big\}\rangle\\
 &-2\langle \lim_{\t \to \infty}[\hat{N}_R(\t) - \hat{N}_R(0)]\rangle
 \langle \frac{d}{dt}[\hat{N}_R(t) - \hat{N}_R(0)]\rangle \Big]
\end{split}
\label{eq:Macdonald1}
\end{equation}
where we have also used the property:
\be
 \lim_{\t \to \infty} \lim_{t \to \infty} \hat{N}_R(t+\t) - \hat{N}_R(t)
 \sim \lim_{\t \to \infty} \hat{N}_R(\t) - \hat{N}_R(0)
\end{equation}
where with the symbol $\sim$ we mean that they have the same
asymptotic behaviour. This property is a  direct consequence of a
large times stationary current.

Finally, by inserting the definition of transferred charge
operator $\hat{Q}_R(t) \equiv \hat{N}_R(t) - \hat{N}_R(0)$ into
(\ref{eq:Macdonald1}), we obtain:

\be
 S(0) = \lim_{t \to \infty}\frac{d}{dt}
\Big[\langle\hat{Q}_R^2(t)\rangle
-\langle\hat{Q}_R(t)\rangle^2\Big]
\end{equation}
In the basis in which the transferred charge operator is diagonal
we can express the quantum expectation values\footnote{In what
follows we assume the equivalence between the Heisenberg and
Schr\"odinger picture and evaluate the averages in the latter
representation.} of $Q$ and $Q^2$ using the corresponding diagonal
density matrix\footnote{Note that in this basis the density matrix
is diagonal since the state of the right bath is a statistical
mixture of states with different particle numbers.} $P_n(t)$ and
we obtain:

\be
 S(0) = \lim_{t \to +\infty}\frac{d}{dt}
 \Big[\sum_{n = 0}^{\infty} n^2P_n(t) -
 \big(\sum_{n= 0}^{\infty}nP_n(t)\big)^2\Big]
 \label{eq:MacDonald}
\end{equation}
known as the MacDonald  formula \cite{ela-pla-02}. This formula
represents the starting point for the calculation of the noise in
the shuttle devices. We analyze in detail the SDQS case. The TDQS
case is the main subject of a detailed article
\cite{fli-preprint-04} and of the Master thesis of C.~Flindt
\cite{christian} and we direct the interested reader to that
literature.

\subsection{Current noise in the SDQS}\label{sec:NoiseSDQS}

The shot noise is completely determined by the evaluation of large
times asymptotics of the functions:

\be
  A(t) = \sum_{n = 0}^{\infty}nP_n(t),\quad
  B(t) = \sum_{n = 0}^{\infty}n\dot{P}_n(t), \quad
  C(t) = \sum_{n = 0}^{\infty}n^2\dot{P}_n(t)
\end{equation}
that appear in the MacDonald formula if one explicitly calculates
the time derivative. We recognize in $A(t)$ the average number of
electrons $N_R(t)$ tunnelled to the right lead by the time $t$ and
in $B(t)$ the right lead current $I_R(t)$. From the previous
section we know the asymptotic behaviour of the the current at
large times:

\be
 \lim_{t \to \infty}I_R(t) =
 {\rm Tr}_{\rm osc} \left\{\G_R
 e^{2x/\l}\s^{\rm stat}_{11}\right\}
\end{equation}
Using the $n$-GME we can express also the $C$ function in terms of
the reduced density matrices $\s_{11}^{(n)}$

\be
 \bs
 C(t) = \sum_{n = 0}^{\infty}n^2\dot{P}_n(t)
 &= \sum_{n = 0}^{\infty} n^2 \G_R {\rm Tr}_{\rm osc}
 \big[e^{2x/\l}(\s_{11}^{(n-1)} - \s_{11}^{(n)}) \big]\\
 &= \sum_{n=0}^{\infty} 2n  {\rm Tr}_{\rm osc}
 \big[\G_R e^{2x/\l}\s_{11}^{(n)}\big]
 + \G_R {\rm Tr}_{\rm osc}\big[e^{2x/\l}\s_{11}\big]
\end{split}
\label{eq:Cfunc}
\end{equation}
where we have used the boundary condition $\s_{11}^{(-1)}\equiv 0$
due to the high bias limit that does not allow electrons to
\emph{enter} from the right lead. It is convenient for the
calculation of the $t \to \infty$ limit in (\ref{eq:MacDonald}) to
introduce the generating functions\footnote{These are in fact an
operator-valued generalization of the generating function
concept.}:

\be
 F_{ii}(t;z) = \sum_{n=0}^{\infty}\s_{ii}^{(n)}(t)z^n
\end{equation}
with the properties:
\be
 \bs
 &F_{ii}(t;1) = \s_{ii}(t)\\
 &\frac{\partial}{\partial
 z}F_{ii}(t;z)\Big|_{z=1} = \sum_{n=0}^{\infty} n
 \s_{ii}^{(n)}(t)
 \end{split}
\end{equation}
The  equations of motion for $F_{ii}(t;z)$ are easily derived from
the GME for the $n$ resolved density matrix (\ref{eq:nGMESDQS}) by
multiplying both sides with $z^n$ and summing over $n$. The
resulting Liouvillean for the $F$'s is very similar to the one we
calculated for the density matrix\footnote{The only small (but
important) modification is the $z$ in the driving term of the
first equation. It corresponds to the change of the electron
counting exponent to $n-1$ in the original $n$-GME (Eq. \
\eqref{eq:nGMESDQS}).}:
\begin{equation}
\begin{split}
    \frac{\partial}{\partial t}F_{00}(t;z) &= -\frac{i}{\hbar} [H_{\rm osc},F_{00}(t;z)]
    + \mathcal{L}_{\rm damp}\,F_{00}(t;z)\\
    &- \frac{\Gamma_L}{2}\{e^{-\frac{2x}{\lambda}},F_{00}(t;z)\}
    + z \Gamma_R e^{\frac{x}{\lambda}}F_{11}(t;z)e^{\frac{x}{\lambda}}\\
    &=\mathcal{L}_{00} F_{00}(t;z) + z \mathcal{L}_{01}F_{11}(t;z)
    \ , \\
    \frac{\partial}{\partial t} F_{11}(t;z) &= -\frac{i}{\hbar}
    [H_{\rm osc}-e\mathcal{E}x,F_{11}(t;z)]+ \mathcal{L}_{\rm damp}\,F_{11}(t;z)\\
    &- \frac{\Gamma_R}{2}\{e^{\frac{2x}{\lambda}},F_{11}(t;z)\}
    + \Gamma_L e^{-\frac{x}{\lambda}}F_{00}(t;z)e^{-\frac{x}{\lambda}}\\
    &= \mathcal{L}_{10} F_{00}(t;z) + \mathcal{L}_{11}F_{11}(t;z)
    \ ,
\end{split}
\label{eq:GMEgen}
\end{equation}
where we have introduced the block structure of the Liouvillean
(super)operator $\mathcal{L}=
\left(\begin{smallmatrix}\mathcal{L}_{00}&\mathcal{L}_{01}
\\ \mathcal{L}_{10}&\mathcal{L}_{11}\\\end{smallmatrix}\right)$.
Using the $F$'s and (\ref{eq:Cfunc}) the shot noise formula
\eqref{eq:MacDonald} can be rewritten as

\begin{equation}
 \begin{split}
 S(0)=&\bigg(\mathrm{Tr_{osc}}\Big[\G_R e^{\frac{2x}{\lambda}}\Big(2\frac{\partial}
 {\partial z}F_{11}(t;z)\Big|_{z=1}+F_{11}(t;1)\Big)\Big]\\
 &\left.-2\mathrm{Tr_{osc}}\Big[\G_R e^{\frac{2x}{\lambda}}F_{11}(t;1)\Big]
 \times\mathrm{Tr_{osc}}\Big[\frac{\partial}{\partial z}
 \sum_{i=0}^1 F_{ii}(t;z)\Big|_{z=1}\Big]\bigg)\right|_{t\to\infty}.
\end{split}
\label{eq:noisegen}
\end{equation}
The Laplace transform of \eqref{eq:GMEgen} yields
\begin{equation}
    \begin{pmatrix}
    \tilde{F}_{00}(s;z)\\
    \tilde{F}_{11}(s;z)
    \end{pmatrix} =
    \begin{pmatrix}
    s - \mathcal{L}_{00} & -z\mathcal{L}_{01}\\
    -\mathcal{L}_{10} & s - \mathcal{L}_{11}
    \end{pmatrix}^{-1}
    \begin{pmatrix}
    f_{00}^{\rm init}(z)\\
    f_{11}^{\rm init}(z)
    \end{pmatrix}
\end{equation}
where $f_{ii}^{\rm init}(z) = \sum_n \s_{ii}^{(n)}(0)z^n$ depends
on the initial conditions. Defining the resolvent
$\mathcal{G}(s)=(s-\mathcal{L})^{-1}$ of the full Liouvillean we
arrive at the following expressions for the operator-valued
generating functions $F_{ii}$'s
\be
    \begin{pmatrix}
    \tilde{F}_{00}(s;1)\\
    \tilde{F}_{11}(s;1)
    \end{pmatrix} = \mathcal{G}(s)
    \begin{pmatrix}
    \s_{00}^{\rm init}\\
    \s_{11}^{\rm init}
    \end{pmatrix}
    \label{eq:Ffunction}
\end{equation}
and their derivatives\footnote{It can be useful in the calculation
to remember the matrix identity : $ \frac{\partial}{\partial z}[
A^{-1}(z)B(z)] = A^{-1}(z)\frac{\partial}{\partial z}B(z) -
A^{-1}\frac{\partial}{\partial z}A(z)[A^{-1}B(z)]$.}
$\partial_zF_{ii}(z)$
\be
    \left.
    \frac{\partial}{\partial z}
    \begin{pmatrix}
    \tilde{F}_{00}(s;z)\\
    \tilde{F}_{11}(s;z)
    \end{pmatrix}\right|_{z=1} =
    \mathcal{G}(s)
    \begin{pmatrix}
      0 & \mathcal{L}_{01} \\
      0 & 0
    \end{pmatrix}
    \mathcal{G}(s)
    \begin{pmatrix}
    \s_{00}^{\rm init}\\
    \s_{11}^{\rm init}
    \end{pmatrix}
    +\mathcal{G}(s)
    \begin{pmatrix}
    f_{00}^{'\rm init}(1)\\
    f_{11}^{'\rm init}(1)
    \end{pmatrix}.
 \label{eq:Ffunctder}
\end{equation}
In order to extract the large-$t$ behavior we study the
asymptotics of the above expressions as $s\to 0+$\footnote{Given a
function $f(t)$ and its Laplace transform $\tilde{f}(s$) the
following identity holds $\lim_{t \to +\infty}f(t) = \lim_{s \to
0+} s\tilde{f}(s)$.}. This is entirely determined by the resolvent
$\mathcal{G}(s)$ in the small-$s$ limit. Since $\mathcal{L}$ is
singular (recall $\mathcal{L}\s^{\rm stat}=0$) the resolvent is
singular at $s=0$. To extract the singular behavior we introduce
the projector $\mathcal{P}$ on the null space of the Liouvillean:\
$\mathcal{P}\bullet=\left(\begin{smallmatrix}\s_{00}^{\rm stat}&
\\ \s_{11}^{\rm stat}\\\end{smallmatrix}\right)\mathrm{Tr_{sys}}(\bullet)$. We
also need the complement $\mathcal{Q}=1-\mathcal{P}$. The
projectors $\mathcal{P},\,\mathcal{Q}$ and  the Liouvillean
$\mathcal{L}$ fulfill the relations:
\be
 \bs
\mathcal{P}\mathcal{L} &=\mathcal{L}\mathcal{P}=0\\
\mathcal{L}&=\mathcal{Q}\mathcal{L}\mathcal{Q}
 \end{split}
 \label{eq:LPQproperties}
\end{equation}
These relations have a simple geometric interpretation. The
Liouvillean (super)operator annihilates the null vector component
of the operator on which it is applied. This component is instead
the only one preserved by $\mathcal{P}$. Every vector is thus sent
to zero by the combined action of $\mathcal{L}$ and $\mathcal{P}$
in whatever order. The second relation in (\ref{eq:LPQproperties})
reflects the complementary behaviour of the projector
$\mathcal{Q}$ that extracts the operator components which are also
preserved by the Liouvillean and does not affects the action of
the latter. The resolvent can be expressed as:
\be
 \bs
 \mathcal{G}(s)
 &=(s\mathcal{P}+s\mathcal{Q}-\mathcal{Q}\mathcal{L}\mathcal{Q})^{-1}\\
 &=\frac{1}{s}\mathcal{P}+ \mathcal{Q}\,\frac{1}{s-\mathcal{L}}\,\mathcal{Q}
 \approx \frac{1}{s}\mathcal{P}-\mathcal{Q}\mathcal{L}^{-1}\mathcal{Q}
 \end{split}
\end{equation}
in leading order for small $s$. The object
$\mathcal{Q}\mathcal{L}^{-1}\mathcal{Q}$ (the pseudoinverse of
$\mathcal{L}$) is regular as the ``inverse" is performed on the
Liouville subspace spanned by $\mathcal{Q}$ where $\mathcal{L}$ is
regular (no null vectors).

We can now calculate the $s \to 0+$ limit of (\ref{eq:Ffunction})
and (\ref{eq:Ffunctder}) and via inverse Laplace transform find
their large $t$ asymptotics:
\be
    \left.
    \begin{pmatrix}
    F_{00}(t;1)\\
    F_{11}(t;1)
    \end{pmatrix}\right|_{t\to\infty}  \!\!\!\to \mathcal{P}
    \begin{pmatrix}
    \s_{00}^{\rm init}\\
    \s_{11}^{\rm init}
    \end{pmatrix}=\begin{pmatrix}
    \s_{00}^{\rm stat}\\
    \s_{11}^{\rm stat}
    \end{pmatrix}
\end{equation}
and
\be
    \frac{\partial}{\partial z}\left.\begin{pmatrix}
    F_{00}(t;z)\\
    F_{11}(t;z)
    \end{pmatrix}\right|_{z=1,t\to\infty} \!\!\!\to
    \begin{pmatrix}
    \s_{00}^{\rm stat}\\
    \s_{11}^{\rm stat}
    \end{pmatrix} \big(tI^{\rm stat}+C^{\rm init}\big)
    -\mathcal{Q}\mathcal{L}^{-1}\mathcal{Q}
    \begin{pmatrix}
    \Gamma_R e^{\frac{x}{\lambda}}\s_{11}^{\rm stat}e^{\frac{x}{\lambda}}\\
    0
    \end{pmatrix}\
\end{equation}
where $C^{\rm init}$ depends on initial conditions. We insert
these in \eqref{eq:noisegen} and obtain:
\be
 \bs
 S(0) =
 & \left. 2{\rm Tr}_{\rm osc}\Big[\G_R e^{2x/\l}\s_{11}^{\rm stat}
  (tI^{\rm stat} + C^{\rm init})\Big]\right|_{t \to \infty}\\
 & -2{\rm Tr}_{\rm osc} \left\{\G_R
  e^{2x/\l}\left[\mathcal{QL}^{-1}\mathcal{Q}
  \begin{pmatrix}\G_R e^{x/\l}\s_{11}^{\rm stat}\\0\end{pmatrix}\right]_{11} \right\}\\
 & + {\rm Tr}_{\rm osc}\Big[\G_R e^{2x/\l}\s_{11}^{\rm
 stat}\Big]\\
 & \left. - 2{\rm Tr}_{\rm osc}\Big[\G_R e^{2x/\l}\s_{11}^{\rm stat}\Big]
  \times {\rm Tr}_{\rm osc}\sum_{i=0}^1\Big[\s_{ii}^{\rm stat}
  (tI^{\rm stat} + C^{\rm init})\Big]\right|_{t \to \infty}\\
 & + 2{\rm Tr}_{\rm osc}\Big[\G_R e^{2x/\l}\s_{11}^{\rm stat}\Big]
  \times {\rm Tr}_{\rm
  osc}\sum_{i=0}^1\left\{\left[\mathcal{QL}^{-1}\mathcal{Q}
  \begin{pmatrix}\G_R e^{x/\l}\s_{11}^{\rm stat}\\0\end{pmatrix}\right]_{ii}\right\}
\end{split}
\label{eq:Sfinal}
\end{equation}
The first term cancels identically the fourth and the noise (as it
should) is independent from the initial conditions. Also the last
term vanishes since the linear form ${\rm Tr}_{\rm osc}\sum_i
\equiv {\rm Tr}_{\rm sys}$ tries in vain to extract the component in
the direction of null vector for the Liouvillian which is absent after the projector
$\mathcal{Q}$ has been applied.

\subsection{The Fano factor}

We choose to represent the noise of the SDQS
 by the adimensional quantity $F = S(0)/I^{\rm stat}$
called Fano factor. In terms of the stationary solution of the GME
the Fano factor for the SDQS reads
\begin{equation}
    F = 1 - \frac{2\Gamma_R}{I^{\rm stat}}
    \mathrm{Tr_{osc}}\left\{e^{\frac{2x}{\lambda}}
    \left[\mathcal{Q}\mathcal{L}^{-1}\mathcal{Q}\begin{pmatrix}
    \Gamma_R e^{\frac{x}{\lambda}}\s_{11}^{\rm
    stat}e^{\frac{x}{\lambda}}\\
    0 \end{pmatrix}\right]_{11}\right\}\ .
    \label{Fano}
\end{equation}
as can be proved dividing (\ref{eq:Sfinal}) by the stationary
current $I^{\rm stat}$. For the numerical evaluation of the Fano
factor we introduce the auxiliary quantity $\Sigma$:
\be
 \Sigma = \mathcal{Q}\mathcal{L}^{-1}\mathcal{Q}\begin{pmatrix}
    \Gamma_R e^{\frac{x}{\lambda}}\s_{11}^{\rm
    stat}e^{\frac{x}{\lambda}}\\
    0 \end{pmatrix}
\end{equation}
which satisfies the equation.
\be
 \mathcal{L}\Sigma =
 \begin{pmatrix}
 \G_R e^{x/\l} \s_{11}^{\rm stat} e^{x/\l} \s_{11}^{\rm stat} -I^{\rm stat}\s_{00}^{\rm stat}\\
 -I^{\rm stat} \s_{11}^{\rm stat}
 \end{pmatrix}
 \label{eq:auxSigmaEq}
\end{equation}
Once the stationary solution of the GME $\s^{\rm stat}$ is known,
equation (\ref{eq:auxSigmaEq}) is just an inhomogeneous (linear) equation
for $\Sigma$. Once again the dimension of the matrix
representation of the Liouvillean (super)operator $\mathcal{L}$
makes the numerical problem non-trivial. The solution comes from a
reformulation of the ideas and concepts of the Arnoldi iteration
scheme applied to inhomogeneous equations called Generalized
Residual Method (GMRes) \cite{eirola}. The problem is to find, for
a given vector $\vet{b}$ in the Liouville space, the solution
$\vet{x}$ of the equation $\vet{Lx} = \vet{b}$. We start from a
Krylov space $\mathcal{K}_j(\vet{L}, \vet{r}_0)$, where $\vet{r}_0
= \vet{Lx}_0 - \vet{b}$ and $\vet{x}_0$ represents the initial
guess for the solution. GMRes finds the vector $\vet{x}_j \in
\mathcal{K}_j(\vet{L}, \vet{r}_0)$ that minimizes the norm of the
residual $\vet{r}_j = \vet{b} - \vet{Lx}_j$. We represent the
vector $\vet{x}_j$ in terms of $j$ coordinates of the Krylov space
and the matrix $\vet{Q}_j$ of the basis vectors: $\vet{x}_j =
\vet{x}_0 + \vet{Q}_j\vet{v}_j$. We shift the minimum problem from
the Krylov space $\mathcal{K}_j$ to $\mathbb{C}^{j}$:
\be
 \bs
 \min_{\vet{x} \in \mathcal{K}_j}\|\vet{b} - \vet{Lx}\|_2
 &=\|\vet{b} - \vet{Lx}_j\|_2\\
 &= \|\vet{b} - \vet{L}(\vet{x}_0 + \vet{Q}_j\vet{v}_j)\|_2\\
 &=\|\vet{r}_0 - \vet{LQ}_j\vet{v}_j\|_2\\
 &=\|\vet{r}_0 - \vet{Q}_{j+1}\vet{H}_j\vet{v}_j\|_2
 =\min_{\vet{v} \in \mathbb{C}^{j}}\|\vet{r}_0 - \vet{Q}_{j+1}\vet{H}_j\vet{v}\|_2
 \end{split}
\end{equation}
where we have used the fundamental relation
(\ref{eq:Arnoldikernel}) to get rid of the Liouvillian operator.
Since the Krylov space is constructed from $\vet{r}_0$ the first
column of $\vet{Q}_{j+1}$ is $\frac{\vet{r}_0}{\|\vet{r}_0\|_2}$
and we thus have:
\be
 \|\vet{r}_0 - \vet{Q}_{j+1}\vet{H}_j\vet{v}_j\|_2 =
 \|\vet{Q}_{j+1}(\vet{e}_1r_0 - \vet{H}_j\vet{v}_j)\|_2 =
 \|\vet{e}_1r_0 - \vet{H}_j\vet{v}_j\|_2
\end{equation}
with $r_0 = \|\vet{r}_0\|_2$ and $\vet{e}_1 = [1,0,\ldots,0]^T \in
\mathbb{C}^{(j+1)\times 1}$.  The problem of minimizing
$\|\vet{e}_1r_0 - \vet{H}_j\vet{v}_j\|_2$ can be solved using
$QR$-decomposition of $\vet{H}_j$. This method is fast since the
dimension of $\vet{H}_j$ is of the order of the Krylov space
dimension. The method is, like Arnoldi scheme  iterative and the
convergence is strongly dependent from the problem at hand. The
preconditioning is again a good method to cure non-convergent
problems. We used in this case the same Sylvester-like
preconditioner that we introduced for the calculation of the
stationary solution of the GME in Sec.\ \ref{sec:numericalGME}.

Before presenting the results of the numerical evaluation of the
Fano  factor in the SDQS we want to discuss its typical behaviour
for simpler systems. This will provide us with some expectations
and ideas to interpret the numerical results.

For a system that can be treated in the Landauer-B\"uttiker
formalism (see for example \cite{bruus}) the scattering states
and the transmission probabilities between in and out states
define the transport properties. In particular the average
(particle) current through the device is given as:

\be
 I = \frac{2e}{h}V\sum_{n=1}^NT_n
\end{equation}
where $T_n$ is the transmission probability for the $n$th
conducting channel and V is the bias across the system. The zero
frequency current-current correlation function reads:

\be
 S(\w = 0)= \frac{2e}{h}V\sum_{n=1}^NT_n(1-T_n).
\end{equation}
For these systems the Fano factor takes the form:

\be
 F = \frac{S(0)}{I} = \frac{\sum_{n=1}^NT_n(1-T_n)}{\sum_{n=1}^NT_n}
 \label{eq:FLandButt}
\end{equation}
which is a number between 0 and 1. For small  transition
probabilities ($T_n \ll 1$) the Fano factor tends to 1. The
tunneling event is so rare that is reasonable to think that there
is no correlation between two subsequent tunneling events.
The Fano factor 1
is also called Poissonian since it is possible to demonstrate that
in that regime the number of electrons $N(\D t)$ transmitted in a
time interval $\D t$ is distributed according to a Poisson
distribution:

\be
 P\{N(\D t) = k\} = \frac{(\l \D t)^k}{k!}e^{-\l \D t}
\end{equation}
where $\l$ is the intensity of the process. Davies {\it et al.}\
\cite{dav-prb-92} described  the transport process for a single
channel as a classical stochastic process of particles with a
definite probability $T$ of going through a barrier. It is a
classical point of view that condenses the dynamics into a
probability distribution of success for a tunneling event. In
their analysis the Fano factor naturally appears as a measure of
the randomness of the process. Totally uncorrelated tunneling
events give a Fano factor 1. Resonant tunneling through a
symmetric double barrier give Fano factor $F=1/2$ whereas a Fano
factor $F=0$ corresponds to transmission with no randomness.

The strong position-charge correlation of  the shuttling regime
suggests that the tunneling events are almost deterministically
determined by the mechanical dynamics. Consequently we expect that
the low degree of randomness in the electron transport is
reflected in low Fano factors. This conjecture was recently
confirmed by Pistolesi \cite{pis-prb-04} that has predicted a
vanishing Fano factor for a driven classical charge shuttle at
large amplitudes.

Landauer-B\"uttiker formalism can treat  only sub-poissonian noise
since $F \leq 1$ in \eqref{eq:FLandButt}. However, in interacting
systems, general theoretical prediction and numerical calculation
have demonstrated the existence of super-poissonian noise. The
Fano factor may even diverge as discussed by Blanter and
B\"uttiker \cite{bla-phr-00}.

In figure \ref{fig:NoiseSDQS} we present  the Fano factor as a
function of the mechanical damping $\g$ for different values of
the bare injection rate  $\G$ and tunneling length $\l$. We
recognize common features in the three curves. At high damping the
Fano factor is of order 1 and (at least for the ``most classical''
set of parameters $\l = 2x_0$ and $\G = 0.05\w, 0.01\w$) close to
what we expect from a resonant tunneling system. We will see that
the discrepancies from the symmetric double barrier are due to the
quantum fuzziness in the position of the dot that influences the
injection and ejection rates and to the charge dependent
equilibrium position of the dot. Diminishing the damping the Fano
factors encounter a more or less pronounced maximum and drop
finally to very low values for small damping. The maximum at
intermediate damping rates is more pronounced and sharper the more
classical are the parameters and can reach values of $F \approx
600$ for the most classical case.
%%%%%%%%%%%%%%%%%%%%%%%%%%%%%%%%%%%%%%%%%%%%%%%%%%%%%%%%%%%%%%%%%
% Figure
%%%%%%%%%%%%%%%%%%%%%%%%%%%%%%%%%%%%%%%%%%%%%%%%%%%%%%%%%%%%%%%%%
\begin{figure}[h]
 \begin{center}
  \includegraphics[angle=0,width=.7\textwidth]{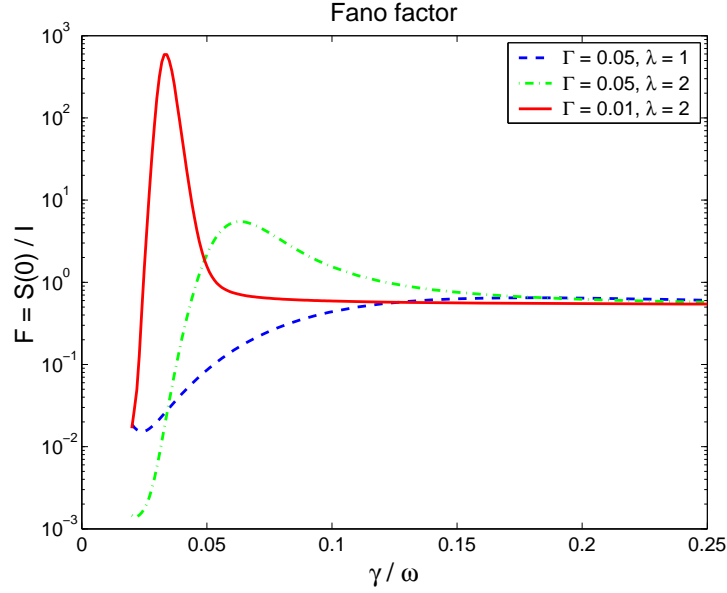}
  \caption{\small  \textit{Fano factor for the SDQS vs. damping $\gamma$.
  The mechanical dissipation rate $\gamma$ and the electrical rate $\G = \G_L = \G_R$ are
  given in units of the mechanical frequency $\w$.
  The tunneling length $\l$ in terms of $x_0 = \sqrt{\hbar/(m\w)}$. The
  other parameters are $d = 0.5x_0$ and $T=0$. The very low noise in the shuttling
  (low damping) regime is a sign of ordered transport. The huge super-poissonian
  Fano factors corresponds to the onset of the coexistence regime.}}
  \label{fig:NoiseSDQS}
 \end{center}
\end{figure}
%%%%%%%%%%%%%%%%%%%%%%%%%%%%%%%%%%%%%%%%%%%%%%%%%%%%%%%%%%%%%%%%
A comparison with the current curves (fig. \ref{fig:SDQSCurrent})
shows that the peak in the Fano factor corresponds to the
transition region from the tunneling to the shuttling current.
Similarly the correspondence can be established also with the
Wigner function distribution: the region of damping in which
tunneling (dot in $W_{\rm tot}$) and shuttling (ring in $W_{\rm
tot}$) features coexist is associated with a super-poissonian Fano
factor. The very low ($F \approx 0.01$) Fano factors for low
damping are a signature of the deterministic transport that takes
place in the shuttling regime. It is interesting to note that this
regularity persists also deep in the quantum regime as can be seen
for $\G = 0.05\w$ and $\l = x_0$. The relative uncertainty in
the amplitude of the oscillation (see Fig. \ref{fig:WFall}) does
not seem to influence the current noise\footnote{It must be
noticed that this behaviour is fragile and strongly dependent on
the degree of unharmonicity of the oscillator. The current is
given by the mechanical oscillation period that, for a harmonic
oscillation is \emph{independent} of the amplitude. Thus amplitude
fluctuations do not influence the electrical dynamics. Increase of
the Fano factor in the shuttling regime has been reported for
unharmonic potentials \cite{isa-epl-04}.}.

\clearpage{\pagestyle{empty}\cleardoublepage}

%%%%%%%%%%%%%%%%%%%%%%%%%%%%%%%%%%%%%%%%%%%%%%%%%%%%%%%%%%%%%%
%                           REGIMES
%%%%%%%%%%%%%%%%%%%%%%%%%%%%%%%%%%%%%%%%%%%%%%%%%%%%%%%%%%%%%%

\chapter{The three regimes}\label{sec:the_three_regimes}

The shuttle devices are characterized by a  strong interplay
between electrical and mechanical degrees of freedom, the
electrostatic force modifying the mechanical dynamics of the
oscillating dot and the position dependent tunneling amplitudes
representing the back-action of the mechanical degree of freedom
on the electrical dynamics. Despite the complexity expected
because of the non-linear couplings, the dynamics of shuttle
devices (at least in the SDQS realization) can be classified into
three operating regimes defined by different relations between the
typical time scales in the device. We make in the next sections a
qualitative analysis of these regimes leaving to the next chapter
the definition of the simplified models and the quantitative
comparison between full and approximate description in terms of
the investigation tools that we introduced in the previous
chapter.

\section{Tunneling}\label{sec:tunneling-reg}

The tunneling regime is the high damping regime  in which the
relaxation time of the mechanical oscillation (inverse of the
damping rate $\gamma$) is much shorter than the typical electronic
time (inverse of the average tunneling rate $\bar{\G}$). We also
assume that the oscillator dynamics is under-damped:

\be
 \w \gg \gamma \gg \bar{\G} =
 {\rm Tr}_{\rm sys}\left\{\G_R(x) |1\rangle\!\langle 1|
 \s^{\rm stat}\right\}
\end{equation}

\noindent The dynamics of the device can be described by a four
steps cycle (see Fig. \ref{fig:tunneling}):

%%%%%%%%%%%%%%%%%%%%%%%%%%%%%%%%%%%%%%%%%%%%%%%%%%%%%%%%%%%%%%%%%
% Figure
%%%%%%%%%%%%%%%%%%%%%%%%%%%%%%%%%%%%%%%%%%%%%%%%%%%%%%%%%%%%%%%%%
\begin{figure}[h]
 \begin{center}
  \includegraphics[angle=0,width=.7\textwidth]{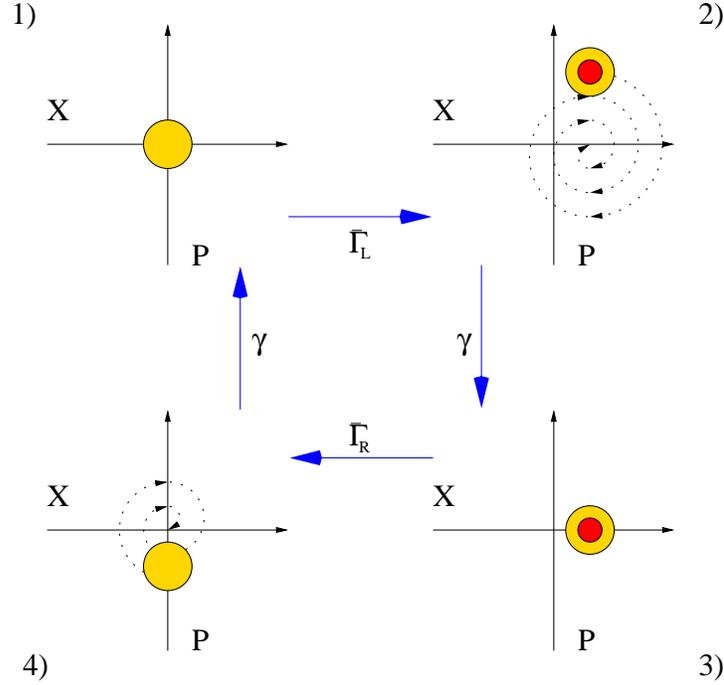}
  \caption{\small  \textit{Schematic picture of the tunneling regime.
  In the scheme we represent the motion of quantum dot in the phase
  space of the oscillator (position $X$ and momentum $P$) and at the
  same time we keep track of the charge degree of freedom. The four
  steps on the cycle are described in detail in the text.}}
  \label{fig:tunneling}
 \end{center}
\end{figure}
%%%%%%%%%%%%%%%%%%%%%%%%%%%%%%%%%%%%%%%%%%%%%%%%%%%%%%%%%%%%%%%%%

\begin{enumerate}
 \item The dot is empty and in its mechanical ground state. We represent
 it in the phase space at rest in the origin and neglect in this simplified
 description the quantum indetermination and/or thermal noise that would blur
 the mechanical phase space distribution.

 \item A tunneling event (from the source lead) and consequent electrostatic
 force $e\mathcal{E}$ perturbs the mechanical state of the dot which gets a
 positive position and momentum and starts to oscillate due to the harmonic
 restoring forces. The amplitude of the dot oscillations diminishes due to the
 dissipative environment. The $1) \to 2)$ transition takes place randomly but
 with a defined rate $\bar{\G}_L$: the average injection rate\footnote{In the
 choice of $\bar{\G}_L$ as injection rate implies that we are taking into account
 the average occupation of the dot. In other words we are treating the stationary
 state. See the next chapter for more details.}.

 \item The dot is at rest in the charged state equilibrium position
 ($d = \frac{e\mathcal{E}}{m\w^2}$) given by the combination of the harmonic
 potential and the electrostatic force.  The transition $2) \to 3)$ takes place
 in the short relaxation time $1/\gamma$.

 \item A second tunnelling event (now towards the drain lead) brings back the dot
 into the original harmonic potential. The empty dot  (distribution function)
 spirals towards the origin of the phase space and the cycle is closed. The time
 scale of the $3) \to 4)$ transition is given by the inverse ejection rate $1/\bar{\G}_R$
 while the mechanical relaxation $4) \to 1)$ takes place within a time $1/\gamma$.
\end{enumerate}

\noindent Since the mechanical damping rate is much larger than
the average injection (ejection) rate the system will stay most of
the time in the configuration 1 and 3. A description with a coarse
grained time do not see the mechanical transients 2 and 3 and is
given in terms of a two state model: empty dot in the origin and
charged dot in the shifted equilibrium position.

\section{Shuttling}
\label{sec:shuttling-reg}

The shuttling regime is characterized by a non-adiabatic interplay
between electrical and mechanical degrees of freedom of the
device. In particular the average injection and ejection rate
exactly matches the mechanical frequency of the oscillator. The
mechanical relaxation rate is much smaller than the mechanical
frequency:

\be
 \gamma \ll \frac{\w}{2 \pi} = \bar{\G}
\end{equation}
In the shuttling regime the quantum dot oscillates in a
self-sustained stable limit cycle: all the energy pumped per cycle
into the system by the electrostatic force is dissipated in the
environment during the oscillation itself. Also the shuttling
regime can be described by a four step cycle (see Fig.
\ref{fig:shuttling}):

\begin{enumerate}

\item The quantum dot, close to the source lead, is charged with an
extra electron. The Coulomb blockade prevents any other charging
event. The high chemical potential on the left lead makes the
charging incoherent and irreversible.

\item  Under the combined effect of the mechanical restoring  force
and the electrostatic force the dot is accelerated towards the
right lead. The tunneling parameters have been chosen such that
the dot is effectively decoupled from the left and right lead
while close to the center of the oscillation. Thus the electrical
state of the device remains unchanged along most of the
oscillation trajectory.

\item The dot is close to the right lead and the very high ejection
rate makes the electron unloading practically deterministic.

\item The empty dot returns towards the left lead under the action
of the restoring force of the harmonic oscillator and completes
the cycle.
\end{enumerate}

\noindent This sequential ``classical'' description of the
shuttling regime is motivated by the phase space analysis of the
previous chapter. The current saturation of one electron per cycle
and the low noise seem to confirm this interpretation. At least
for the most classical parameters range we will see that the
dynamics is well described by a ``trajectory'' in the phase space
of the \emph{system} where a mean field charge coordinate has been
added to the mechanical position and momentum\footnote{See section
\ref{sec:shuttling} }. \vspace{2cm}

%%%%%%%%%%%%%%%%%%%%%%%%%%%%%%%%%%%%%%%%%%%%%%%%%%%%%%%%%%%%%%%%%
% Figure
%%%%%%%%%%%%%%%%%%%%%%%%%%%%%%%%%%%%%%%%%%%%%%%%%%%%%%%%%%%%%%%%%
\begin{figure}[h]
 \begin{center}
  \includegraphics[angle=0,width=.7\textwidth]{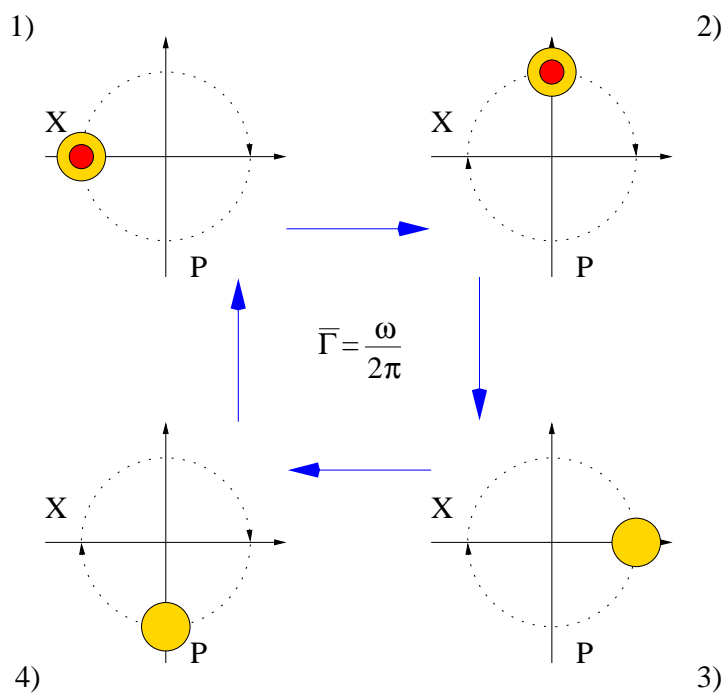}
  \caption{\small  \textit{Schematic picture of the shuttling regime.}}
  \label{fig:shuttling}
 \end{center}
\end{figure}
%%%%%%%%%%%%%%%%%%%%%%%%%%%%%%%%%%%%%%%%%%%%%%%%%%%%%%%%%%%%%%%%%
\vspace{1cm}

\section{Coexistence} \label{sec:coexistence-reg}

At intermediate damping rates the shuttle device shows
\emph{bistability}. Under these conditions neither the tunneling or
the shuttling are globally stable regimes and the noise present in
the system (shot noise of the charge transfer, thermal noise,
quantum mechanical indetermination) causes the random switching
between the two processes. The bistability is represented in the
stationary distribution  function by the coexisting tunneling and
shuttling features. We showed that tunneling and shuttling regimes
are associated with different average currents. It is not
difficult to imagine that the current of the coexistence regime is
an average of the shuttling and tunneling currents weighted on the
probability for the device to be in one or in the other regime.
The slow switching between these two currents generates the
peculiar huge super-poissonian Fano factor that also characterized
the coexistence regime. It is possible to demonstrate that the
zero frequency current noise associated to a slow dichotomous
switching between two states with average currents $I_+$ and $I_-$
reads:

\be
 S(0) = \frac{\a\b}{(\a+\b)^3}(I_+ -I_-)^2
 \label{eq:noisedich}
\end{equation}
where $\a$ ($\b$) is the switching rate from state 1 to 2 (2 to
1). It is important that the average transition time between
shuttling and tunneling is the longest time scale in the system
dynamics. This fact ensures:

\begin{enumerate}
\item Many tunneling or shuttling events before the regime transition take
place and thus there exists a well defined average current for the two ``states'';

\item A huge Fano factor $F \approx I_{sh}/\a$.
\end{enumerate}

The time resolved dynamics for the coexistence regime is depicted
in figure \ref{fig:coexistence} where the transition rates between
the two regime are represented respectively by $\G_{out}$
(tunneling $\to$ shuttling) and $\G_{in}$ (shuttling $\to$
tunneling). The names ($in$ and $out$) associated to the
transition rates are  motivated by the  dynamics of the mechanical
amplitude during the transition. The average oscillation amplitude
are very different in the shuttling and tunneling regime and can
be used to identify the two states. In fact we will describe the
switching dynamics in terms of a bistable potential with the
amplitude as the effective coordinate\footnote{See section
\ref{sec:coexistence} for details.}.
%%%%%%%%%%%%%%%%%%%%%%%%%%%%%%%%%%%%%%%%%%%%%%%%%%%%%%%%%%%%%%%%%
% Figure
%%%%%%%%%%%%%%%%%%%%%%%%%%%%%%%%%%%%%%%%%%%%%%%%%%%%%%%%%%%%%%%%%
\begin{figure}[h]
 \begin{center}
  \includegraphics[angle=0,width=.6\textwidth]{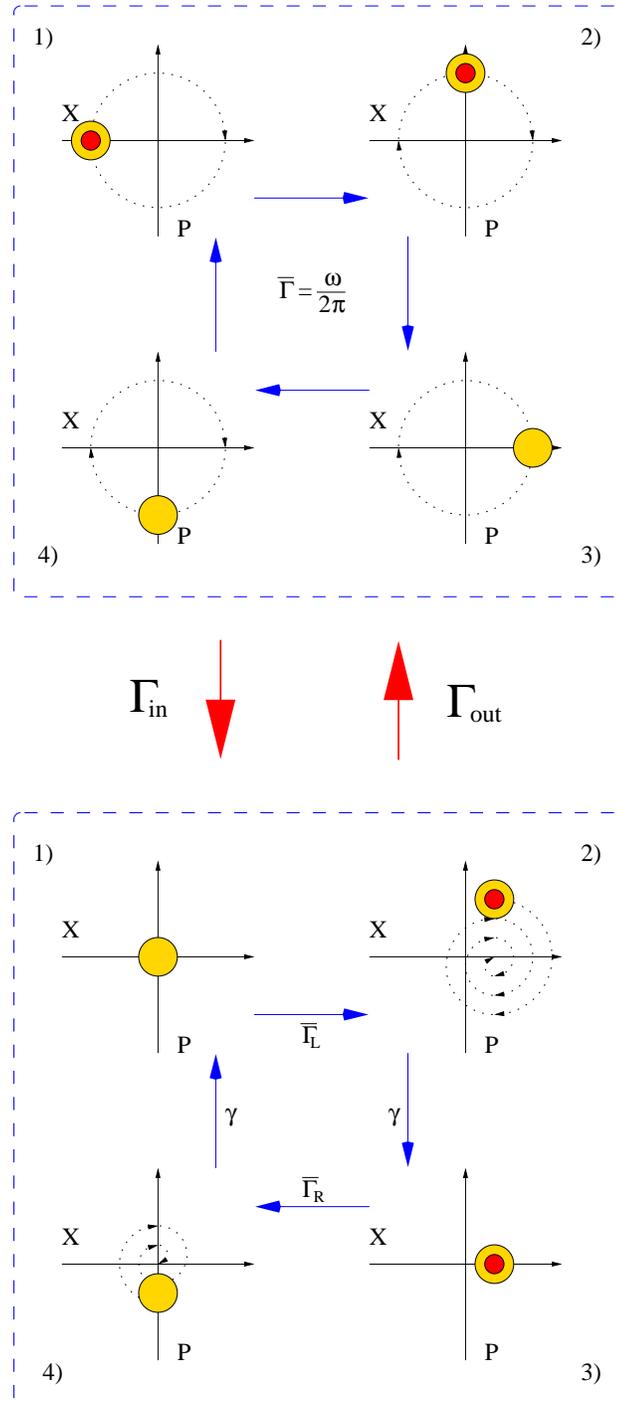}
  \caption{\small  \textit{Schematic picture of the coexistence regime.}}
  \label{fig:coexistence}
 \end{center}
\end{figure}
%%%%%%%%%%%%%%%%%%%%%%%%%%%%%%%%%%%%%%%%%%%%%%%%%%%%%%%%%%%%%%%%%

\clearpage{\pagestyle{empty}\cleardoublepage}

%%%%%%%%%%%%%%%%%%%%%%%%%%%%%%%%%%%%%%%%%%%%%%%%%%%%%%%%
%       SIMPLIFIED MODELS
%%%%%%%%%%%%%%%%%%%%%%%%%%%%%%%%%%%%%%%%%%%%%%%%%%%%%%%%

\chapter{Simplified models} \label{sec:simplified_models}

We qualitatively described in the previous  chapter three possible
operating regimes for shuttle-devices. They represent for the SDQS
the whole scenario of possible dynamics. Part of the complexity of
the TDQS is also captured in this scheme\footnote{For a further
insight into this particular model  see \cite{arm-prb-02,
christian, fli-preprint-04}.}. The specific separation of time
scales allows us to identify the relevant variables and describe
each regime by a specific simplified model. Models for the
tunneling, shuttling and coexistence regime are analyzed
separately in the three following sections. We also give a
comparison with the full description in terms of Wigner
distributions, current and current-noise to prove that the models,
at least in the limits set by the chosen investigation tools,
capture the relevant dynamics.

\section{Renormalized resonant tunneling} \label{sec:tunneling}

The electrical dynamics is definitely the slowest in the tunneling
regime since the mechanical relaxation time (much longer in itself
than the oscillation period) is much shorter than the average
injection or ejection time. We already noticed (sec.
\ref{sec:tunneling-reg}) that, because of this time-scale
separation, the observation of the device dynamics would most
of the time show two mechanically frozen states:

\begin{description}
 \item[0.] Empty dot at rest in the oscillator equilibrium position
 \item[1.] Charged dot in the shifted equilibrium position of the
 harmonic oscillator perturbed by the constant electrostatic force
 $e\mathcal{E}$.
\end{description}

We combine this observation with a quantum description  of the
mechanical oscillator and possible thermal noise under the
assumption that the $n$-resolved reduced density matrix of the
device can be written in the form:

\be
 \bs
  \s_{00}^{(n)}(t) &= p^{(n)}_{00}(t)\s_{\rm th}(0)\\
  \s_{11}^{(n)}(t) &= p^{(n)}_{11}(t)\s_{\rm th}(e\mathcal{E})
 \end{split}
 \label{eq:tun-ansatz}
\end{equation}
where
\be
 \s_{\rm th}(\mathcal{F}) =
 \frac{e^{-\b(H_{\rm osc} -\mathcal{F} x)}}
 {{\rm Tr}_{\rm osc}\left[e^{-\b(H_{\rm osc} -\mathcal{F} x)}\right]}
\end{equation}
is the thermal density matrix of a harmonic oscillator  subject
to an external force $\mathcal{F}$. The functions
$p^{(n)}_{00}(t)$ and $p^{(n)}_{11}(t)$ represent the probability
to find the system respectively in the state 0 or 1 that we
described above when $n$ electrons have tunnelled through the dot.
The sum over all possible device and bath configurations is
$\sum_n p^{(n)}_{00}(t)+ p^{(n)}_{11}(t) =1$ at all times. The
equations of motion for the probabilities $p^{(n)}_{ii}(t)$ can be
derived by inserting the assumption  (\ref{eq:tun-ansatz}) in the
$n$-GME (\ref{eq:nGMESDQS}) and taking the trace over the
mechanical degrees of freedom\footnote{The damping Liouvillean
$\mathcal{L}_{\rm damp}$ is missing in \ref{eq:nGMESDQS}. Anyway,
together with the coherent Liouvillean $\mathcal{L}_{\rm coh}$, it
does not contribute to the equation of motion for the
$p^{(n)}_{ii}$. They vanish under the trace ${\rm Tr}_{\rm osc}$
because of their commutator structure.}:

\be
\bs
 \dot{p}^{(n)}_{00} &
 = \tilde{\G}_R\, p^{(n-1)}_{11} -\tilde{\G}_L\, p^{(n)}_{00} \\
 \dot{p}^{(n)}_{11} &
 = \tilde{\G}_L\, p^{(n)}_{00} - \tilde{\G}_R\, p^{(n)}_{11}
\end{split}
\label{eq:nME-tun}
\end{equation}
where
\be
\bs
 \tilde{\G}_L &= {\rm Tr}_{\rm osc}\left[
  \G_L e^{-\frac{2x}{\l}}\s_{\rm th}(0)\right]\\
 \tilde{\G}_R &= {\rm Tr}_{\rm osc}\left[
  \G_R e^{\frac{2x}{\l}}\s_{\rm th}(e\mathcal{E})\right]
\end{split}
\end{equation}
are the injection and ejection rates renormalized to take into
account the quantum/thermal distribution of the harmonic
oscillator position. The master equation (\ref{eq:nME-tun}) summed
over the number $n$ of electrons collected in the right lead
reads:

\be
 \bs
 \dot{p}_{00} &
   = \tilde{\G}_R\, p_{11} - \tilde{\G}_L\, p_{00} \\
 \dot{p}_{11} &
   = \tilde{\G}_L\, p_{00} - \tilde{\G}_R\, p_{11}
 \end{split}
 \label{eq:ME-tun}
\end{equation}
where we have introduced the occupation  probabilities of the
state $i$ $p_{ii} \equiv \sum_{n=0}^{\infty} p_{ii}^{(n)}$.
Equations (\ref{eq:ME-tun}) describe the dynamics of a resonant
tunneling device. The effects of the movable grain are
contained in the effective rates $\tilde{\G}_L,\tilde{\G}_R$.

\subsection{Electrical rates}

The renormalized electrical  rates can be calculated analytically
as functions of the model parameters. The calculation gives the
exact contribution of the quantum mechanical and thermal noise to
the electrical dynamics. We first consider the limit of very low
temperatures $k_BT \ll \hbar\w$. Only the ground state of the
harmonic oscillator is occupied and the renormalized
\emph{injection} rate reads:

\be
\bs
 \tilde{\G}_L
 &= {\rm Tr}_{\rm osc} \left[ |0\rangle\!\langle 0|
 \G_L e^{-2x/\l}\right]
 = \G_L \langle 0|e^{-\frac{2x}{\l}}|0\rangle\\
 &= \frac{\G_L}{\sqrt{\pi}}\int_{-\infty}^{+\infty}d\xi\,
  \exp\left[-\left(\xi^2 + \frac{2\xi x_0}{\l}\right)\right]
 = \G_L e^{(\frac{x_0}{\l})^2}
\end{split}
\label{eq:GLT0}
\end{equation}
where $|0\rangle$ is the  vector representing the ground state of
the harmonic oscillator and $x_0 = \sqrt{\hbar/(m\w)}$. For the
calculation of the renormalized \emph{ejection} rate it is useful
to introduce the displacement unitary operator:
\be
 D(l) = e^{-\frac{i}{\hbar}lp}
\end{equation}
with the properties\footnote{These properties can be easily
derived from the definition of the operator $D(l)$ and the
Baker-Hausdorff-Campbell formula \cite{destri}.}:
\be
\bs
 D(l) d D^{\dag}(l) &= d - l\sqrt{{\textstyle\frac{m\w}{2\hbar}}}\\
 D(l) d^{\dag} D^{\dag}(l) &= d^{\dag} - l\sqrt{{\textstyle\frac{m\w}{2\hbar}}}\\
 D(l) x D^{\dag}(l) &= x - l
\end{split}
\end{equation}
where $x,p$ are the position and  momentum operators and
$d^{\dag},d$ are the creation and annihilation operators for the
harmonic oscillator. In terms of the displacement operator $D$ the
density matrix $\s_{\rm th}(\mathcal{F})$ can be written:

\be
 \s_{\rm th}(\mathcal{F}) =
 D\left({\textstyle\frac{\mathcal{F}}{m\w^2}}\right)
 \s_{\rm th}(0)
 D^{\dag}\left({\textstyle\frac{\mathcal{F}}{m\w^2}}\right)
\end{equation}
Only the displaced ground state  contributes to the renormalized
ejection rate:

\be
 \bs
 \tilde{\G}_R
 &= {\rm Tr}_{\rm osc} \left[
  D(d)|0\rangle\!\langle 0| D^{\dag}(d)\,
  \G_R e^{-2x/\l}\right]\\
 &= \G_R \langle 0|D^{\dag}(d)\, e^{\frac{2x}{\l}}\,D(d)|0\rangle\\
 &=\G_R e^{\frac{2d}{\l}}\langle 0|e^{\frac{2x}{\l}}|0\rangle
 = \G_R e^{\frac{2d}{\l}+\left(\frac{x_0}{\l}\right)^2}
\end{split}
\label{eq:GRT0}
\end{equation}
where $d= \frac{e\mathcal{E}}{m\w^2}$.

As expected the ejection rate  is modified by the ``classical''
shift $d$ of the equilibrium position due to the electrostatic
force on the charged dot. Somehow unexpectedly both rates are also
\emph{enhanced} by the quantum uncertainty in the position present
in the oscillator ground state. The relevance of the quantum
correction is given by the ratio between the zero point position
dispersion $x_0$ and the tunneling length $\l$: the smaller the
ratio the smaller the correction.

The calculation of the finite temperature rates is problematic due
to the presence in the trace of the product of the exponentials of two non commutative
operators ($H_{\rm osc},x$). We solve the problem using the low
temperature calculation just performed and some symmetry
arguments. First we notice that the density matrix $\s_{\rm th}$
is the stationary solution of the GME\footnote{We have assumed
this property from the very beginning. For a detailed proof of
this statement see for example ``\emph{The Theory of Open Quantum
Systems}'' by H.-P.~Breuer and F.~Petruccione \cite{breuer}.}:

\be
 \dot{\s} = -\frac{i}{\hbar}[H_{\rm osc},\s]
            -\frac{i\g}{2\hbar} [x,\{p,\s\}]
            -\frac{\g m \w}{\hbar}\left[n_B(\w)+\frac{1}{2}\right]
             [x,[x,\s]]
\label{eq:GMEosc}
\end{equation}

We express the trace in the position (generalized) basis
$|q\rangle$ and the renormalized rates take the form:
\be
 \bs
  \tilde{\G}_L &= \G_L \int dq\, P_{\rm th}(q) e^{-\frac{2q}{\l}}\\
  \tilde{\G}_R &= \G_R e^{\frac{2d}{\l}}\int dq\, P_{\rm th}(q) e^{\frac{2q}{\l}}
 \end{split}
 \label{eq:Ratesgeneral}
\end{equation}
where we have introduced the coordinate probability distribution
$P_{\rm th}(q)\equiv \langle q| \s_{\rm th} | q \rangle$. We
encountered this distribution already in section
(\ref{sec:QuantumDistribution}) in the axiomatical definition
of the Wigner function. We required specifically:
\be
 \int dp W(q,p,t) = \langle q| \s | q \rangle
\end{equation}
The problem is thus shifted to the calculation of the Wigner
function associated with the thermal density matrix $\s_{\rm th}$.
The dynamics of the Wigner distribution is given by the
Fokker-Planck equation\footnote{The Klein-Kramers equations of
second order are also called Fokker-Plank equations.} related
to the GME (\ref{eq:GMEosc}):

\be
 \frac{\partial W}{\partial \t} =
 \left[X\frac{\partial}{\partial P} - P\frac{\partial}{\partial X} +
  \frac{\g}{\w} \frac{\partial}{\partial P}
 \left(P + \frac{\partial}{\partial P}\right)\right]\,W
 \label{eq:FPinvariant}
\end{equation}
where we have defined the adimensional variables:
\be
 \t = \w t, \quad X = \frac{q}{\ell}, \quad P = \frac{p}{m \w \ell}
\end{equation}
and $\ell = \sqrt{\frac{\hbar}{2m\w}(2n_B +1)}$. The Fokker-Planck
equation (\ref{eq:FPinvariant}) does \emph{not} depend on the
temperature. The stationary solution is the Gaussian\footnote{This
statement can be easily verified by substitution. In order to
derive it one considers that the Gaussian in $P$ is the solution
for the dissipative part of the Liouvillean (proportional to $\g$)
while the coherent part is solved by spherically symmetric
distributions since in polar coordinates $(A,\phi)$ we have
$X\partial_P - P\partial_X \to -\partial_{\phi}$.} in $X$ and $P$
\be
  W^{\rm stat}(X,P) = \frac{1}{2\pi} \exp\left(-\frac{X^2+P^2}{2}\right)
 \label{eq:WFthermal}
\end{equation}
with \emph{no parameters}. The position distribution function
$P_{\rm th}(q)$ thus depends on the temperature only in the
scaling factor and has the same functional form (Gaussian) of the
zero temperature case:
\be
 P_{\rm th}(q)= \frac{1}{\sqrt{2\pi  \ell^2}}\exp\left(-\frac{q^2}{2\ell^2}\right)
 \label{eq:Pq}
\end{equation}
Finally we insert (\ref{eq:Pq}) into (\ref{eq:Ratesgeneral}) and
perform the gaussian integrals to obtain the finite temperature
rates:

\be
 \bs
 \tilde{\G}_L &= \G_L e^{2\left(\frac{\ell}{\l}\right)^2}\\
 \tilde{\G}_R &= \G_R e^{\frac{2d}{\l} + 2\left(\frac{\ell}{\l}\right)^2}
 \end{split}
 \label{eq:Ratesfinal}
\end{equation}

In the low temperature limit equations (\ref{eq:Ratesfinal})
obviously  reduce to (\ref{eq:GLT0}) and (\ref{eq:GRT0}). The high
temperature limit $k_BT \gg \hbar\w$ is also interesting:

\be
 \bs
 \tilde{\G}_L &= \G_L e^{2\left(\frac{\l_{\rm th}}{\l}\right)^2}\\
 \tilde{\G}_R &= \G_R e^{\frac{2d}{\l} + 2\left(\frac{\l_{\rm th}}{\l}\right)^2}
 \end{split}
 \label{eq:Ratesthermal}
\end{equation}
In this case the renormalization is the more effective the larger is
the thermal length $\l_{\rm th} = \sqrt{k_BT/(m\w^2)}$ compared to
the tunneling length $\l$.

\subsection{Phase-space distribution}

The phase space distribution for the stationary state of the
simplified model for the tunneling regime is built on the Wigner
representation of the thermal density matrix $\s_{\rm th}$ and the
stationary solution of the system (\ref{eq:ME-tun}) for the
occupation $p_{ii}$ of the electromechanical states $i$. The
latter is easily expressed in terms of the renormalized rates
$\tilde{\G}_{L,R}$:
\be
 \bs
  p_{00}^{\rm stat} &= \frac{\tilde{\G}_R}{\tilde{\G}_L + \tilde{\G}_R}\\
  p_{11}^{\rm stat} &= \frac{\tilde{\G}_L}{\tilde{\G}_L + \tilde{\G}_R}
 \end{split}
\end{equation}
We have already met the Wigner distribution corresponding to the
thermal density matrix $\s_{\rm th}$ in Eq.~(\ref{eq:WFthermal}).
We now rewrite it in terms of the canonical variables $(q,p)$:
\be
\s_{\rm th}(0) \to W_{\rm th}(q,p) = \frac{1}{ 2 \pi m\w\ell^2}
\exp\left\{-\frac{1}{2}\left[\left(\frac{q}{\ell}\right)^2 +
                     \left(\frac{p}{\ell m\w}\right)^2 \right]\right\}
\end{equation}
The calculation of the Wigner function for the displaced thermal
distribution $\s_{\rm th}(e\mathcal{E})$ can be performed using
the properties of the displacement operators $D$:
\be
\bs
 \s_{\rm th}(e\mathcal{E}) &\to \frac{1}{2\pi \hbar} \int d\xi
 \left\langle q -\frac{\xi}{2} \right |
  D(d) \s_{\rm th}(0) D^{\dag}(d)
 \left| q +\frac{\xi}{2} \right\rangle \exp\left(\frac{i}{\hbar}p\xi\right)\\
&=\frac{1}{2\pi \hbar} \int d\xi
 \left\langle q - d -\frac{\xi}{2} \right |
  \s_{\rm th}(0)
 \left| q - d +\frac{\xi}{2} \right\rangle \exp\left(\frac{i}{\hbar}p\xi\right)\\
&= W_{\rm th}(q-d,p)
\end{split}
\end{equation}
where we have used the property $D(l)|a\rangle = |a-l\rangle$ that
we can be proven as follows: given $|a\rangle$ eigenvector for the
position operator $x$ with eigenvalue $a$
\be
\bs
 x D(l)|a\rangle &= D(l)\,D^{\dag}(l)\,x\,D(l)|a\rangle\\
 &=D(l)(x-l)|a\rangle = (a-l)D(l)|a\rangle
\end{split}
\end{equation}

The stationary solution of the Klein-Kramers equation
(\ref{eq:KleinKramers}) in the tunneling regime limit thus reads:
\be
\bs
 W_{00}^{\rm stat}(q,p) &=
  \frac{\tilde{\G}_R}{\tilde{\G}_L + \tilde{\G}_R} W_{\rm th}(q,p)\\
 W_{11}^{\rm stat}(q,p) &=
  \frac{\tilde{\G}_L}{\tilde{\G}_L + \tilde{\G}_R} W_{\rm th}(q-d,p)
\end{split}
\end{equation}

%%%%%%%%%%%%%%%%%%%%%%%%%%%%%%%%%%%%%%%%%%%%%%%%%
% Figure
%%%%%%%%%%%%%%%%%%%%%%%%%%%%%%%%%%%%%%%%%%%%%%%%%
\begin{figure}[ht]
 \begin{center}
  \includegraphics[angle=-90,width=.6\textwidth]{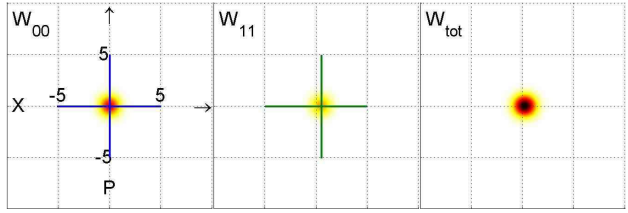}
  \caption{\small  \textit{Wigner distribution functions for the SDQS in the tunneling regime.
   calculated numerically using the full description. Coordinates (horizontal axis) are measured
   in terms of $x_0$, momenta (vertical axis) in terms of $m\w x_0$. The crosses indicates
   the cuts in the phase space along which we perform the comparison between numerical
   and analytical solution shown in figure \ref{fig:Cutstun}. The parameters are:
   $\G_L = \G_R = 0.01\w$, $\g=0.25\w$, $d = 0.5x_0$, $\l = 2x_0$, $T=0$. }
   \label{fig:WFtun}}
 \end{center}
\end{figure}
%%%%%%%%%%%%%%%%%%%%%%%%%%%%%%%%%%%%%%%%%%%%%%%%%
%
\vspace{5cm}
%%%%%%%%%%%%%%%%%%%%%%%%%%%%%%%%%%%%%%%%%%%%%%%%%
% Figure
%%%%%%%%%%%%%%%%%%%%%%%%%%%%%%%%%%%%%%%%%%%%%%%%%
\begin{figure}[ht]
 \begin{center}
  \includegraphics[angle=-90,width=.8\textwidth]{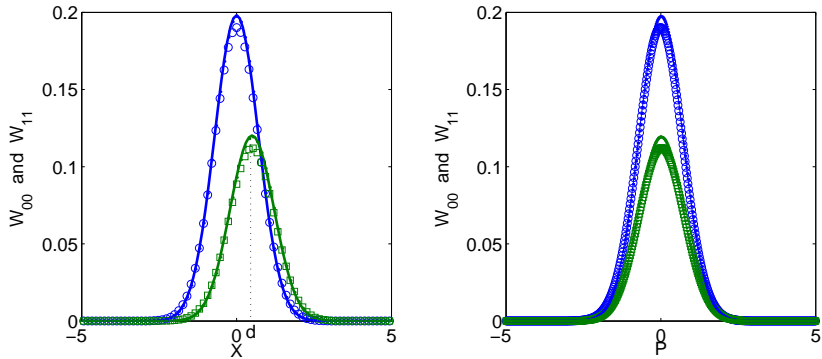}
  \caption{\small  \textit{Comparison between the numerical and the analytical results
  for the Wigner distribution functions. The squares (circles) are numerical results for the parameters
  mentioned in figure \ref{fig:WFtun} in the charged (empty) dot configuration, and the full lines
  represent the analytical calculations.
  We also plotted with dots the numerical results for $\G = 0.001\omega$.}
  \label{fig:Cutstun}}
 \end{center}
\end{figure}
%%%%%%%%%%%%%%%%%%%%%%%%%%%%%%%%%%%%%%%%%%%%%%%%%%%

The stationary distribution of the tunneling model is completely
determined by the length $\ell$ and associated momentum $m\w\ell$,
the equilibrium position shift $d$ and the tunneling length $\l$.
The electronic and mechanical relaxation rates $\gamma$ and
$\G_{L,R}$ drop out from the solution and only set the range of
applicability of the simplified model.

In figures \ref{fig:WFtun}, \ref{fig:Cutstun} and
\ref{fig:CutstunTemp} we compare the Wigner functions calculated
both analytically and numerically in the tunneling regime. They
show in general a very good agreement. In particular we notice in
Fig. \ref{fig:Cutstun} that the convergence to the tunneling regime is not yet
complete for $\gamma = 0.25\omega$ and $\G_{L,R} = 0.01\omega$. To check the
theory we diminished the bare electrical rate further (the quantum
corrections are negligible since $\l = 2x_0$, the classical
shift can be taken into account and gives a factor $e\approx 2.7$ which does
not change the analysis) and kept the mechanical damping fixed to
``maintain'' the condition $\gamma \ll \w$. For $\G = 0.001\omega$ we
got perfect agreement. We also analyze the temperature dependence
of the stationary Wigner function distribution (Fig.\
\ref{fig:CutstunTemp}) and verify the scaling given by the
temperature dependent length $\ell$.
%
%%%%%%%%%%%%%%%%%%%%%%%%%%%%%%%%%%%%%%%%%%%%%%%%%
% Figure
%%%%%%%%%%%%%%%%%%%%%%%%%%%%%%%%%%%%%%%%%%%%%%%%%
\begin{figure}[ht]
 \begin{center}
  \includegraphics[angle=-90,width=.8\textwidth]{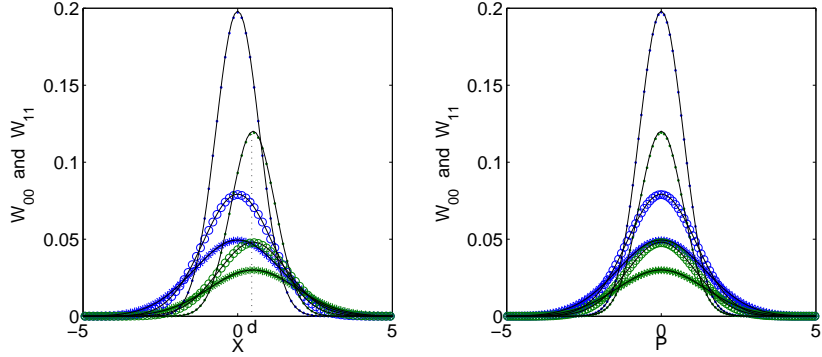}
  \caption{\small  \textit{Tunneling Wigner distributions as a function of the temperature.
  The  relevant parameters are: $\gamma = 0.25\w$, $\G = 0.001\omega$, $n_B = 0,0.75,1.5$ respectively
  represented by dots, circles and asterisks. Full lines are the analytical results.
  See Fig: \ref{fig:WFtun} for the other parameters and the scales. }
  \label{fig:CutstunTemp}}
 \end{center}
\end{figure}
%%%%%%%%%%%%%%%%%%%%%%%%%%%%%%%%%%%%%%%%%%%%%%%%%%%

\subsection{Current}

The current through the SDQS towards the right lead is given by the
general formula:
\be
 I_R(t) = \sum_{n=0}^{\infty} n\dot{P}_n(t)
\end{equation}
In the tunneling limit the probabilities $P_n(t)$ are be expressed in terms of $p^{(n)}_{ii}(t)$ since:
\be
P_n(t) =  {\rm Tr}_{\rm osc}[p^{(n)}_{00}(t)\s_{\rm th}(0) +
p^{(n)}_{11}(t)\s_{\rm th}(e\mathcal{E}) ] = p_{00}^{(n)}(t) + p_{11}^{(n)}(t)
\end{equation}
Apart from the renormalized electronic rates the system has no
sign of the oscillator degree of freedom and can be treated
formally as a static quantum dot. We write in complete analogy
with (\ref{eq:e-current}) the right (left) lead time dependent
current:
\be
\bs
 I_R(t) &= \tilde{\G}_R p_{11}(t)\\
 I_L(t) &= \tilde{\G}_L p_{00}(t)
\end{split}
\end{equation}
In the stationary limit they coincide:
\be
 I^{\rm stat}=
\begin{array}{c}
 \nearrow\\
 \searrow
 \end{array}
\begin{array}{c}
  \lim_{t \to \infty} I_R(t)
= \tilde{\G}_R p^{\rm stat}_{11}
= \bar{\G}_R\\
 \\
 \\
  \lim_{t \to \infty} I_L(t)
= \tilde{\G}_L p^{\rm stat}_{00}
= \bar{\G}_L
 \end{array}
\begin{array}{c}
 \searrow\\
 \nearrow
 \end{array}
= \frac{\tilde{\G}_R \tilde{\G}_L}{\tilde{\G}_L + \tilde{\G}_R}
=\bar{\G} \label{eq:Currentun}
\end{equation}
We stress the difference between $\bar{\G}_{L(R)}$ and
$\tilde{\G}_{L(R)}$: while the first is the average injection
(ejection rate) and thus represents the current through the
device, the second is an average over the \emph{mechanical}
distribution function of the position dependent rate operator.
Despite being an average rate it does not take into account
the charge state of the dot (and the Coulomb blockade) and thus cannot
represent the current through the device. We show in figure
\ref{fig:Currentun} the current calculated numerically and the
asymptotic value of the tunneling regime given by Eq.
\eqref{eq:Currentun}.
%%%%%%%%%%%%%%%%%%%%%%%%%%%%%%%%%%%%%%%%%%%%%%%%%
% Figure
%%%%%%%%%%%%%%%%%%%%%%%%%%%%%%%%%%%%%%%%%%%%%%%%%
\begin{figure}
 \begin{center}
  \includegraphics[angle=0,width=.6\textwidth]{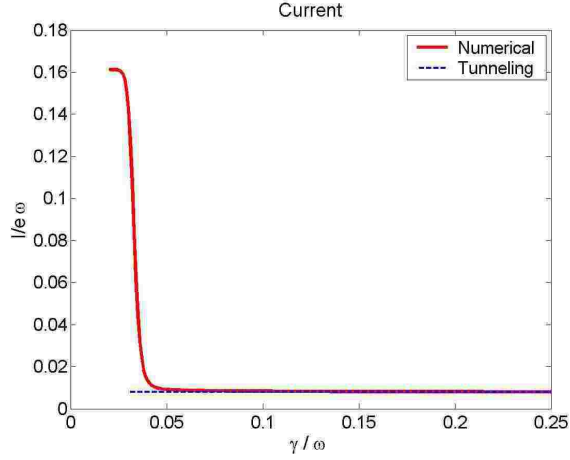}
  \caption{\small  \textit{Current as a function of the damping for the SDQS.
  The asymptotic tunneling limit is indicated. The parameters are:
  $\G_L = \G_R = 0.01\w$, $\g=0.25\w$, $d = 0.5x_0$, $\l = 2x_0$, $T=0$.}
   \label{fig:Currentun}}
 \end{center}
\end{figure}
%%%%%%%%%%%%%%%%%%%%%%%%%%%%%%%%%%%%%%%%%%%%%%%%%%%

\subsection{Current-noise}

The calculation of the current noise and Fano factor in the
tunneling regime is interesting since it gives us the opportunity to
explain the concepts and methods we introduced in section
\ref{sec:current-noise} still keeping a computational level analytically
affordable due to the much lesser number of degrees of freedom of the
effective model.

We start with the MacDonald formula for the zero frequency current
noise:

\be
 S(0) = \lim_{t \to +\infty}\frac{d}{dt}
 \Big[\sum_{n = 0}^{\infty} n^2P_n(t) -
 \big(\sum_{n= 0}^{\infty}nP_n(t)\big)^2\Big]
\end{equation}
where $P_n(t)=p_{00}^{(n)}+p_{11}^{(n)}$ and the equation of
motion for $p_{ii}^{(n)}$ are:

\be
 \bs
   \dot{p}^{(n)}_{00} &
 = \tilde{\G}_R\, p^{(n-1)}_{11} -\tilde{\G}_L\, p^{(n)}_{00} \\
 \dot{p}^{(n)}_{11} &
 = \tilde{\G}_L\, p^{(n)}_{00} - \tilde{\G}_R\, p^{(n)}_{11}
 \end{split}
\end{equation}
We identify the effective Liouvillean $\mathcal{L}$ for the
relevant variables $p_{ii}(t)$ in the model:

\be
 \mathcal{L} =
 \begin{pmatrix}
 -\tilde{\G}_L & \tilde{\G}_R\\
 \tilde{\G}_L & -\tilde{\G}_R
 \end{pmatrix}
\end{equation}
The evaluation of the different terms of the current-noise can be
carried out by introducing the generating functions $F_{ii}(t;z) =
\sum_n p_{ii}^{(n)}(t)z^n$ and following the same reasoning we
indicated in section \ref{sec:NoiseSDQS}. The result is formally
the same\footnote{The perfect matching can be better appreciated
in a more general super-operator formulation in which the current
noise reads in both cases $\langle\!\langle\tilde{0}| I_R -2I_R
\mathcal{QL}^{-1}\mathcal{Q} I_R|0 \rangle\!\rangle$. See for
example \cite{fli-preprint-04} for details.}: we express the Fano
factor in terms of the stationary probabilities $p_{ii}^{\rm
stat}$ and the pseudoinverse of the the Liouvillean
$\mathcal{QL}^{-1}\mathcal{Q}$.
\be
 F = 1 - \frac{2}{I^{\rm stat}}
 \left\|\begin{pmatrix}
 0 & \tilde{\G}_R\\
 0 & 0
 \end{pmatrix}
 \mathcal{QL}^{-1}\mathcal{Q}
 \begin{pmatrix}
 0 & \tilde{\G}_R\\
 0 & 0
 \end{pmatrix}
 \begin{pmatrix}
 p_{00}^{\rm stat}\\
 p_{11}^{\rm stat}
 \end{pmatrix}\right\|_1
 \label{eq:Fanotun}
\end{equation}
where $\|\bullet\|_1$ is the sum of the elements of the vector.
The projectors $\mathcal{P}$ and $\mathcal{Q}$, built from their
definition (see sec. \ref{sec:NoiseSDQS}), have the matrix form:

\be
 \mathcal{P} =
 \begin{pmatrix}
 p_{00}^{\rm stat} & p_{00}^{\rm stat}\\
 p_{11}^{\rm stat} & p_{11}^{\rm stat}
 \end{pmatrix}, \quad
 \mathcal{Q} =
 \begin{pmatrix}
 \phantom{-}p_{11}^{\rm stat} & -p_{00}^{\rm stat}\\
 -p_{11}^{\rm stat} & \phantom{-}p_{00}^{\rm stat}
 \end{pmatrix}.
\end{equation}

For the explicit calculation of Eq.~(\ref{eq:Fanotun}) we define
the auxiliary vector $\Sigma$:

\be
 \Sigma =
 \begin{pmatrix}
 \Sigma_{00}\\
 \Sigma_{11}
 \end{pmatrix}
 =
 \mathcal{QL}^{-1}\mathcal{Q}
 \begin{pmatrix}
 0 & \tilde{\G}_R\\
 0 & 0
 \end{pmatrix}
 \begin{pmatrix}
 p_{00}^{\rm stat}\\
 p_{11}^{\rm stat}
 \end{pmatrix}
\end{equation}
that satisfies the equation
\be
 \mathcal{L}\Sigma = \mathcal{Q}
 \begin{pmatrix}
 0 & \tilde{\G}_R\\ 0&0
 \end{pmatrix}
 \begin{pmatrix}
 p_{00}^{\rm stat}\\
 p_{11}^{\rm stat}
 \end{pmatrix} =
 \tilde{\G}_R (p_{11}^{\rm stat})^2
 \begin{pmatrix}
 1\\
 -1
 \end{pmatrix}
\end{equation}
Together with the condition $\Sigma_{00}+\Sigma_{11} =
0$\footnote{This condition is readily proven since the last
operator to the left in the definition of $\Sigma$ is
$\mathcal{Q}$ and the sum on the elements of $\Sigma$ would
extract the component in the direction of the null vector of the
Liouvillean. This component has been already projected out by
$\mathcal{Q}$.} we can calculate $\Sigma$ and by substitution into
(\ref{eq:Fanotun}) find the Fano factor $F$:

\be
 F = 1 - \frac{2\tilde{\G}_R\Sigma_{11}}{I^{\rm stat}}
   = 1 - \frac{2\tilde{\G}_R(\tilde{\G}_L + \tilde{\G}_R)}
              {\tilde{\G}_L\tilde{\G}_R}
         \frac{\tilde{\G}_R\tilde{\G}_L^2}
              {(\tilde{\G}_L + \tilde{\G}_R)^3}
   = \frac{\tilde{\G}_L^2 + \tilde{\G}_R^2}
          {(\tilde{\G}_L + \tilde{\G}_R)^2}
\end{equation}

%%%%%%%%%%%%%%%%%%%%%%%%%%%%%%%%%%%%%%%%%%%%%%%%%
% Figure
%%%%%%%%%%%%%%%%%%%%%%%%%%%%%%%%%%%%%%%%%%%%%%%%%
\begin{figure}
 \begin{center}
 \includegraphics[angle=0,width=.6\textwidth]{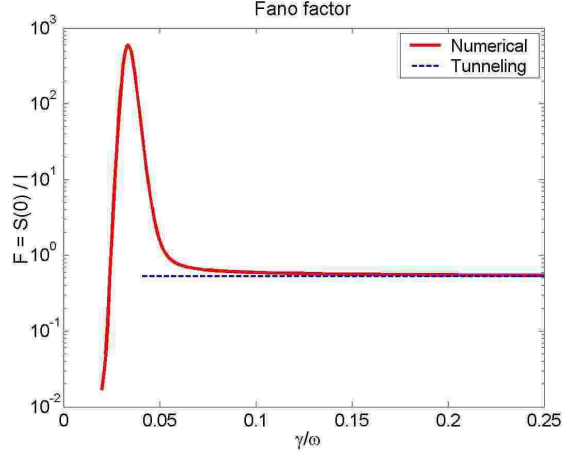}
  \caption{\small  \textit{Current-noise as a function of the damping for the SDQS.
  The asymptotic tunneling limit is indicated. The parameters are:
  $\G_L = \G_R = 0.01\w$, $\g=0.25\w$, $d = 0.5x_0$, $\l = 2x_0$, $T=0$.}
  \label{fig:Fanotun}}
 \end{center}
\end{figure}
%%%%%%%%%%%%%%%%%%%%%%%%%%%%%%%%%%%%%%%%%%%%%%%%%%%

\section{Shuttling: a classical transport regime} \label{sec:shuttling}

The simplified model for the shuttling dynamics is based on the
observation --extracted from the full description-- that the
system exhibits in this operating regime extremely low Fano
factors ($F \approx 10^{-2}$). In the simplified model we assume
that there is \emph{no noise at all} in the system. Its state is
represented by a point  that moves on a trajectory in the
3-dimensional device phase-space of position, momentum and charge
of the oscillating dot. It is hard not to imagine the charge on
the oscillating dot as a stochastic variable governed by
tunnelling processes. Nevertheless in the shuttling regime the
tunneling events are made effectively deterministic since they are
very highly probable at specific times settled by the mechanical
dynamics. On the other hand, due to the electrostatic force
$e\mathcal{E}$ acting on the charged dot that mirrors the
electrical dynamics into a mechanical acceleration only
``deterministic'' tunneling events can produce a regular
mechanical dynamics. We discuss in the next session the derivation
of the equations of motion for this semiclassical model\footnote{
The derivation is not rigorous and the equations of motion
(\ref{eq:shuttlingfin}) should be probably viewed as an ``educated
guess'' to describe the shuttling device. Nevertheless the effort
for a mathematical derivation from the Klein-Kramers equations
(\ref{eq:KleinKramers}) has been motivated by the very good
agreement between the extremely simple model and the full
description.} and then compare the results with the full quantum
description.

\subsection{Equation of motion for the relevant variables}

The dynamics of the SDQS is described by the set of coupled
Klein-Kramers equations (\ref{eq:KleinKramers}). In order  to
implement the zero noise assumption we first  set $T=0$ and
simplify the equations further by neglecting all the terms of the
$\hbar$ expansion since we assume the classical action of the
oscillator to be much larger than the Planck
constant\footnote{This condition is fulfilled in the shuttling
regime since the area enclosed by the ringlike structure of the
total Wigner distribution in the oscillator phase space is much
larger than$\hbar$.}. We obtain:

\be
 \bs
 \frac{\partial W_{00}^{\rm cl}}{\partial \t} =&
 \left[X\frac{\partial}{\partial P} - P\frac{\partial}{\partial X} +
  \frac{\g}{\w} \frac{\partial}{\partial P}P\right]\,W_{00}^{\rm cl}\\
 &-\frac{\G_L}{\w} e^{-2X} W_{00}^{\rm cl} +\frac{\G_R}{\w} e^{2X} W_{11}^{\rm cl}\\
\frac{\partial W_{11}^{\rm cl}}{\partial \t} =&
 \left[\left(X - \frac{d}{\l}\right)\frac{\partial}{\partial P} - P\frac{\partial}{\partial X} +
  \frac{\g}{\w} \frac{\partial}{\partial P}
 P\right]\,W_{11}^{\rm cl}\\
 &-\frac{\G_R}{\w} e^{ 2X} W_{11}^{\rm cl}
  +\frac{\G_L}{\w} e^{-2X} W_{00}^{\rm cl}
\end{split}
\label{eq:FPshuttling}
\end{equation}
where we have introduced the adimensional variables:
\be
  \t = \w t, \quad X = \frac{q}{\l}, \quad P = \frac{p}{m \w \l}
\end{equation}
The superscript ``cl'' indicates that we are dealing with the
classical limit of the Wigner function justified by the complete
elimination of the quantum ``diffusive'' terms from the
Fokker-Planck equations and thus indefinitely sharp distribution
functions are now allowed. We can at this point introduce a
separation Ansatz:

\be
 \bs
 W_{00}^{\rm cl}(X,P,\t) = p_{00}(\t)\d(X-X^{\rm cl}(\t))\d(P-P^{\rm cl}(\t))\\
 W_{11}^{\rm cl}(X,P,\t) = p_{11}(\t)\d(X-X^{\rm cl}(\t))\d(P-P^{\rm cl}(\t))
 \end{split}
 \label{eq:separation}
\end{equation}
where the trace over the system phase-space sets $p_{00} + p_{11}
= 1$. The variables $X^{\rm cl}$ and $P^{\rm cl}$ represent the
position and momentum of the (center of mass) of the oscillating
dot; $p_{11(00)}$ is the probability for the quantum dot to be charged (empty).

By inserting the Ansatz (\ref{eq:separation}) into equation
(\ref{eq:FPshuttling}) summing over the charge index $i$ and
matching the coefficients of  the different distribution functions
on the left and right hand side we obtain:

\be
 \bs
 \dot{X}^{\rm cl} &= P^{\rm cl}\\
 \dot{P}^{\rm cl} &= -X^{\rm cl} + \frac{d}{\l}p_{11} - \frac{\g}{\w} P^{\rm cl}\\
 \end{split}
 \label{eq:mechanical}
\end{equation}
In order to close the system of  differential equations we need
the equations of motion for $p_{ii}$ \footnote{The dynamics of
$p_{00}$ and $p_{11}$ are obviously coupled.} provided by the
integration over $X$ and $P$ of (\ref{eq:FPshuttling}) with the
separation Ansatz:
\be
 \bs
 \dot{p}_{00} &=-\frac{\G_L}{\w}e^{-2X^{\rm cl}}p_{00}
                +\frac{\G_R}{\w}e^{2X^{\rm cl}}p_{11}\\
 \dot{p}_{11} &= \frac{\G_L}{\w}e^{-2X^{\rm cl}}p_{00}
                -\frac{\G_R}{\w}e^{2X^{\rm cl}}p_{11}
 \end{split}
 \label{eq:electrical}
\end{equation}

Unfortunately the system of equations (\ref{eq:mechanical}) and
(\ref{eq:electrical}) is \emph{not} equivalent to
(\ref{eq:FPshuttling}): integration and  summation have introduced
spurious solutions. If we substitute the Ansatz
(\ref{eq:separation}) into the equations (\ref{eq:FPshuttling})
and use the equations of motion (\ref{eq:mechanical}) and
(\ref{eq:electrical}) we obtain the condition:

\be
  p_{00}p_{11} = 0
  \label{eq:condition}
\end{equation}

The only differentiable solution for this equation is $p_{00} = 0$
or $p_{11} = 0$ for all times and is not satisfying the equation
of motion (\ref{eq:electrical}). Strictly speaking there are no
solutions of (\ref{eq:FPshuttling}) in the form suggested in the
Ansatz\footnote{First order partial differential  equations can be
solved by the characteristics method that calculate the flow along
trajectories: this is not true for system of partial differential
equations. More physically: the switching process in itself is
introducing noise.}.

Nevertheless we can imagine that the electrical switching time
between the only allowed charged ($p_{11}=1;\, p_{00}=0$) and
empty ($p_{00}=1;\, p_{11}=0$) states is much shorter than the
shortest mechanical time (the oscillator period $T = 2\pi /\w$ ). A
solution of the system of equations (\ref{eq:mechanical}) and
(\ref{eq:electrical}) with this time scale separation could
``almost everywhere'' satisfy the condition (\ref{eq:condition})
and, inserted into (\ref{eq:separation}) represent a solution for
(\ref{eq:FPshuttling}).

We rewrite the set of equations (\ref{eq:mechanical}) and
(\ref{eq:electrical}) as:

\be
 \bs
  \dot{X} &= P\\
  \dot{P} &= -X + d^* Q - \gamma^* P\\
  \dot{Q} &= \G_L^* e^{-2X}(1-Q) -\G_R^* e^{2X}Q
 \end{split}
 \label{eq:shuttlingfin}
\end{equation}
where we have dropped for simplicity the ``cl'' superscript from
the mechanical variables, we have renamed $p_{11} \equiv Q$ and
used the trace condition $p_{00} = 1-p_{11}$. We have also defined
the rescaled parameters:

\be
 d^* = d/\l, \quad \g^* = \g/\w, \quad \G_{L,R}^* = \G_{L,R}/\w
\end{equation}

\subsection{Stable limit cycles}

The set of coupled non-linear differential equations
(\ref{eq:shuttlingfin}) represents the starting point of the
analysis of the simplified model for the shuttling regime. We
calculate the numerical solution for different values of the
parameters and different  initial condition. For the parameter
values that correspond in the full description to the shuttling
regime, the system has a limit cycle solution with the desirable
time scale separation we discussed in the previous section. We
report in figure \ref{fig:Limitcycles} the typical appearance of
the limit cycle. In the first graph (a) we show the charge $Q(\t)$
as a function of time. The charge value is jumping periodically
from 0 to 1 and back with a period equal to the mechanical period.
The transition  itself is almost instantaneous. The limit cycle is
a trajectory in the 3D phase-space of the device $(X,P,Q)$. In the
graphs (b), (c) and (d) of figure \ref{fig:Limitcycles} three
different projections of the trajectory are reported. The $X,P$
projection shows the characteristic circular trajectory of
harmonic oscillations. In the $X,Q$ ($P,Q$) projection the
position(momentum)-charge correlation is visible. From the
combination of the two projection we can argue that the trajectory
in the $X,Q$ graph is drawn clockwise during the cycle. The
oscillating dot gets charged on the left, it is then carrying the
charge towards the right and finally it unloads the electron to
the right lead before returning empty towards the left. The
amplitude of the oscillation is several times the tunneling
length. All the qualitative features of the shuttling regime are
present.

The full description of the SDQS in the shuttling  regime has a
phase space visualization in terms of a ring shaped
\emph{stationary} total Wigner distribution function. We can
interpret this fuzzy ring as the probability distribution obtained
from many different noisy realizations of (quasi) limit cycles.
The stationary solution for the Wigner distribution is the result
of a diffusive dynamics on an effective ``Mexican hat''
potential\footnote{See the next section for details on the
derivation of this effective potential.}  that involves both
amplitude and phase of the oscillations. In the noise-free
semiclassical approximation we turn off the diffusive processes
and the point-like state describes in the shuttling regime a
single trajectory with a definite constant amplitude and
\emph{periodic} phase. We expect this trajectory to be the average
of the noisy trajectories represented by the Wigner distribution.
In the third column of figure \ref{fig:rings} the total Wigner
function corresponding to different parameter realization of the
shuttling regime is presented. The white circle is the
semiclassical trajectory. In the first two columns the asymmetric
sharing of the ring between the charged and empty states is also
compared with the corresponding $Q=1$ and $Q=0$ portions of the
semiclassical trajectory.

%%%%%%%%%%%%%%%%%%%%%%%%%%%%%%%%%%%%%%%%%%%%%%%%%
% Figure
%%%%%%%%%%%%%%%%%%%%%%%%%%%%%%%%%%%%%%%%%%%%%%%%%
\begin{figure}
 \begin{center}
 \includegraphics[angle=0,width=.8\textwidth]{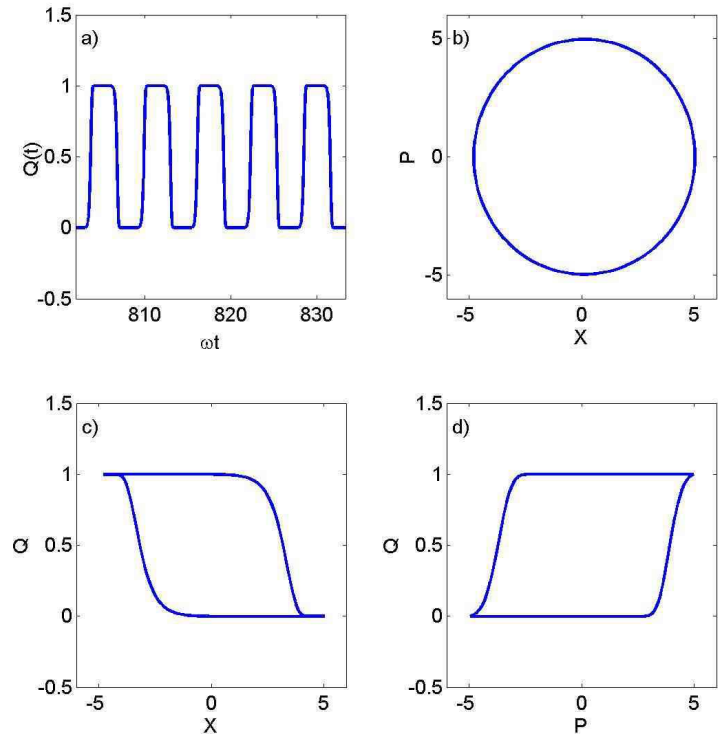}
  \caption{\small  \textit{Different representations of the limit cycle solution
  of the system of differential equations (\ref{eq:shuttlingfin}) that describes the
  shuttling regime. a) Charge on the dot as a function of time.
  The charging and discharging times are the shortest time scales. b) Circular trajectory
  in the mechanical phase space: the motion of the dot is harmonic. c) Projection of the
  limit cycle in the charge position plane: the trajectory shows charge-position correlation.
  d) Projection of the limit cycle in the charge momentum plane: the trajectory shows
  charge-momentum correlation. $X$ is the coordinate in units of the tunneling length $\l$,
  while $P$ is the momentum in units of $m\w\l$.}
  \label{fig:Limitcycles}}
 \end{center}
\end{figure}
%%%%%%%%%%%%%%%%%%%%%%%%%%%%%%%%%%%%%%%%%%%%%%%%%

%%%%%%%%%%%%%%%%%%%%%%%%%%%%%%%%%%%%%%%%%%%%%%%%%
% Figure
%%%%%%%%%%%%%%%%%%%%%%%%%%%%%%%%%%%%%%%%%%%%%%%%%
\begin{figure}
 \begin{center}
 \includegraphics[angle=0,width=.7\textwidth]{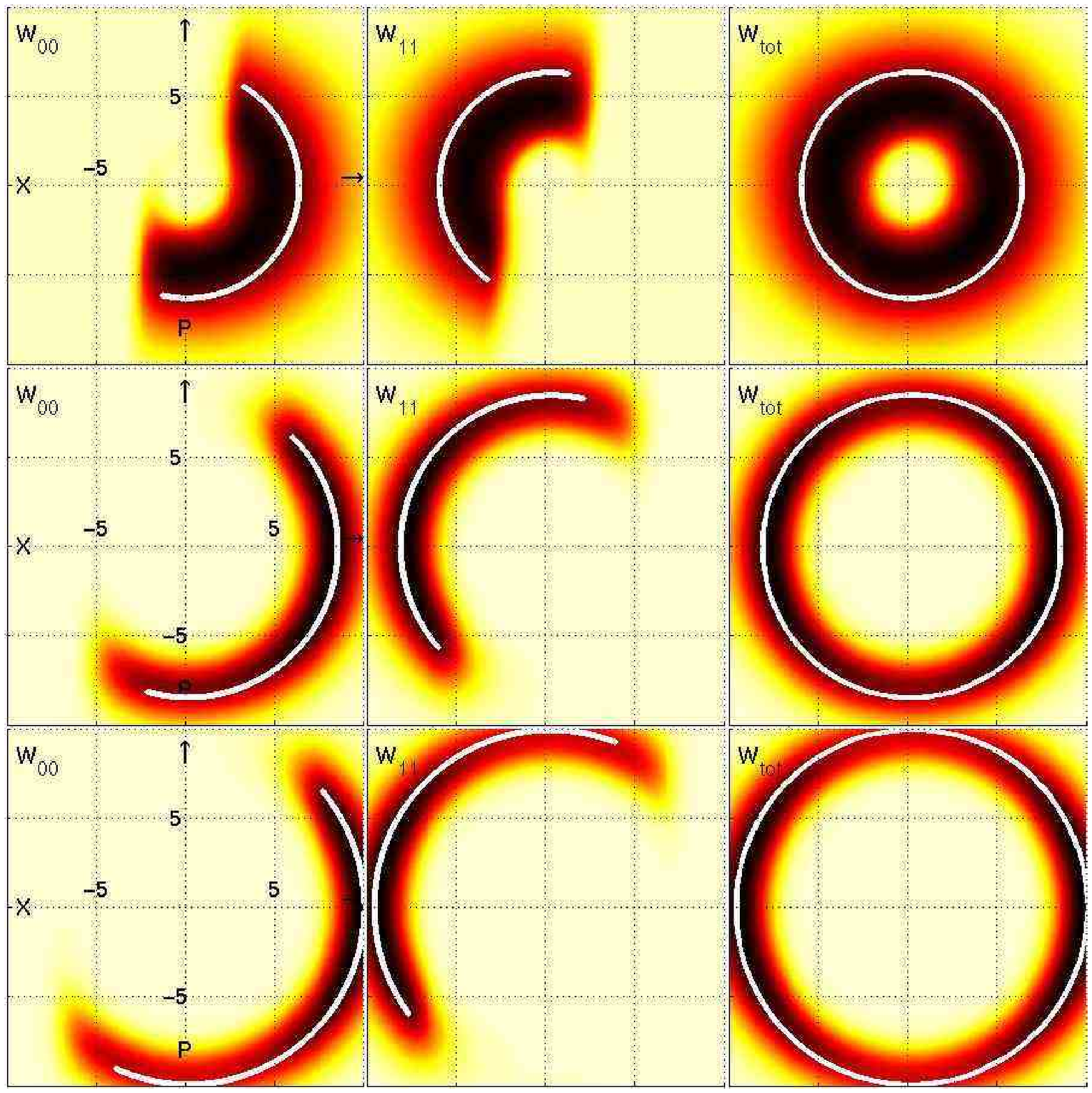}
  \caption{\small  \textit{Correspondence between the Wigner function representation
  and the simplified trajectory limit for the shuttling regime. The white ring is the ($X,P$)
  projection of the limit cycle. The $Q=1$ and $Q=0$ portions of the trajectory are visible in the
  charged and empty dot graphs respectively. The parameters are $\gamma = 0.02\w$, $d = 0.5x_0$,
  $\G=0.05\w$ in the first two rows and $\G=0.01$ in the last one. The tunneling length is $\l =
  x_0$ in the first row and $\l=2x_0$ in the second and the third.}
  \label{fig:rings}}
 \end{center}
\end{figure}
%%%%%%%%%%%%%%%%%%%%%%%%%%%%%%%%%%%%%%%%%%%%%%%%%%%

In the semiclassical description we also have  direct access to
the current as a function of the time. For example the right lead
current reads:

\be
 I_R(\t) = Q(\t)\G_R e^{2X(\t)}
\end{equation}
In figure \ref{fig:shutcurr} the right current is presented as a
function of time for a few mechanical oscillation periods. The
current is given by a series of spikes that well represent the
single electron released to the right lead after being shuttled by
the oscillating dot.

%%%%%%%%%%%%%%%%%%%%%%%%%%%%%%%%%%%%%%%%%%%%%%%%%
% Figure
%%%%%%%%%%%%%%%%%%%%%%%%%%%%%%%%%%%%%%%%%%%%%%%%%
\begin{figure}
 \begin{center}
 \includegraphics[angle=0,width=.6\textwidth]{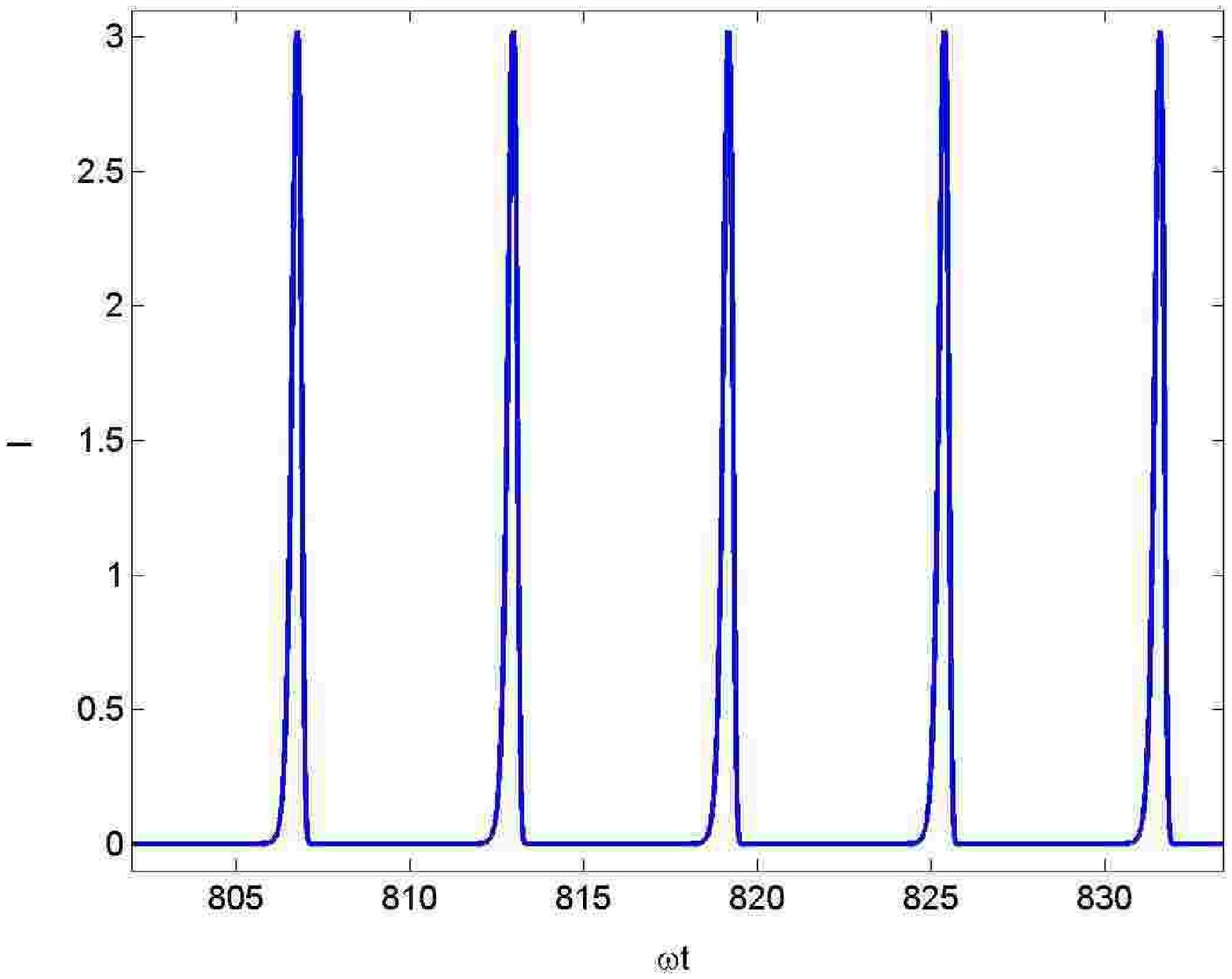}
  \caption{\small  \textit{Time resolved shuttling current calculated on the limit
  cycle solution of the system of differential equations (\ref{eq:shuttlingfin}).
  The regular spikes well represent the single electron shuttled per cycle in this regime.}
  \label{fig:shutcurr}}
 \end{center}
\end{figure}
%%%%%%%%%%%%%%%%%%%%%%%%%%%%%%%%%%%%%%%%%%%%%%%%%%%

The average current is a coarse grained measurement of the time
dependent current. Since the current is periodic, if the
measurement time is long enough we obtain a stationary current:

\be
 I^{\rm stat} = \frac{1}{2\pi}\int_{\t}^{\t + 2\pi}\!\!d\t'\,I_R(\t')
\end{equation}
Since the current is a periodic function of time the average over
one period is equal to the average over an infinite time. The
numerical integration of the function plotted in figure
\ref{fig:shutcurr} gives

\be
 I^{\rm stat} = 0.15916
\end{equation}
which is with impressive accuracy the ``magic value'' of $1/2\pi$
(i.e. one electron transferred per oscillation period).

Also for the TDQS we tried a semiclassical  analysis of the
shuttling regime. For the derivation of the  semiclassical
equations we start again from the Hamiltonian (\ref{eq:TdS-Ham})
but we consider $x$ and $p$ as classical variables. The GME
equation for the reduced density matrix $\s$ is derived by keeping
$x$ as a parameter ($p$ does not enter explicitly the formulation
of the GME).  Since the oscillator is treated classically also the
elements of the density matrix $\s_{ij}$ are now real numbers and
the trace sum rule can be used to reduce the effective number of
equation for the electrical dynamics\footnote{Also in this case
the elements $\s_{0i}$ and $\s_{i0}$ with $i=L,C,R$ vanish
identically reducing the effective number of equations from 16 to
10.} from 10 to 9. The GME in matrix notation reads:
\be
 \dot{\s} = -\frac{i}{\hbar}[H_{sys},\s] + \Xi[\s]
 \label{eq:TDQSelectrical}
\end{equation}
where
\be
 H_{\rm sys} = \frac{1}{2}m\w^2x^2 + \frac{p^2}{2m} + \begin{pmatrix}
 \frac{\D V}{2} & t_L(x) & 0\\
 t_L(x) & -\frac{\D V}{2x_0}x & t_R(x)\\
 0 & t_R(x) & -\frac{\D V}{2}
 \end{pmatrix}
\end{equation}
and the driving Liouvillean $\Xi$ has the form
\be
 \Xi[\s]= \G
\begin{pmatrix}
1-\s_{LL} -\s_{CC} -\s_{RR} & 0 & -\s_{LR}/2\\
0                           & 0 & -\s_{CR}/2\\
-\s_{RL}/2 & -\s_{RC}/2 & -\s_{RR}
\end{pmatrix}
\end{equation}
where we have taken $\G_L = \G_R = \G$. As equation of motion for
the mechanical variables $x$ and $p$ we take the Hamilton
equations derived from the effective Hamiltonian

\be
\langle H_{\rm sys} \rangle= {\rm Tr}_{\rm el} [\s H_{\rm sys}]
\end{equation}
where ${\rm Tr}_{\rm el}$ is the trace on the electronic states,
combined with a phenomenological friction:
\be
 \bs
 \dot{x} =& \frac{\partial \langle H_{\rm sys}\rangle}{\partial p}
 = \frac{p}{m}\\
 \dot{p} =&-\frac{\partial \langle H_{\rm sys} \rangle}{\partial x} -\g p =
 -m\w^2x + \frac{\D V}{2x_0}\s_{CC} -\g p\\
 &+\frac{t_L}{\l}(\s_{CL} +\s_{LC}) -\frac{t_R}{\l}(\s_{CR} + \s_{RC})
 \end{split}
 \label{eq:TDQSmechanical}
\end{equation}

The phase space has now 11 dimensions  (9 electrical and 2
mechanical) but the principle is the same as in the SDQS. Also in
this case the solution of the system of differential equations
(\ref{eq:TDQSelectrical}) and (\ref{eq:TDQSmechanical}) exhibits
for parameters that correspond to the shuttling regime a stable
limit cycle solution. In figure \ref{fig:TDQSrings} we compare the
Wigner distribution function for the central dot with the projection
on the $(x,p)$  plane of the limit cycle trajectory of the
semiclassical approximation. The probability to find the electron
on the central dot $\s_{CC}$ is not in general oscillating between
0 and 1. The red dots in the figure indicate the position of the
maximum occupation. In the TDQS the half moon characteristic of
the shuttling regime rotates in the phase space \cite{nara}, close
to the first two electromechanical resonances, as a function of
the device bias. The correspondence between the darker areas of
the Wigner function and the red dots in figure
\ref{fig:TDQSrings} shows that also this property is captured by
the semiclassical approximation.

%%%%%%%%%%%%%%%%%%%%%%%%%%%%%%%%%%%%%%%%%%%%%%%%%
% Figure
%%%%%%%%%%%%%%%%%%%%%%%%%%%%%%%%%%%%%%%%%%%%%%%%%
\begin{figure}
 \begin{center}
 \includegraphics*[angle=-180,width=.8\textwidth]{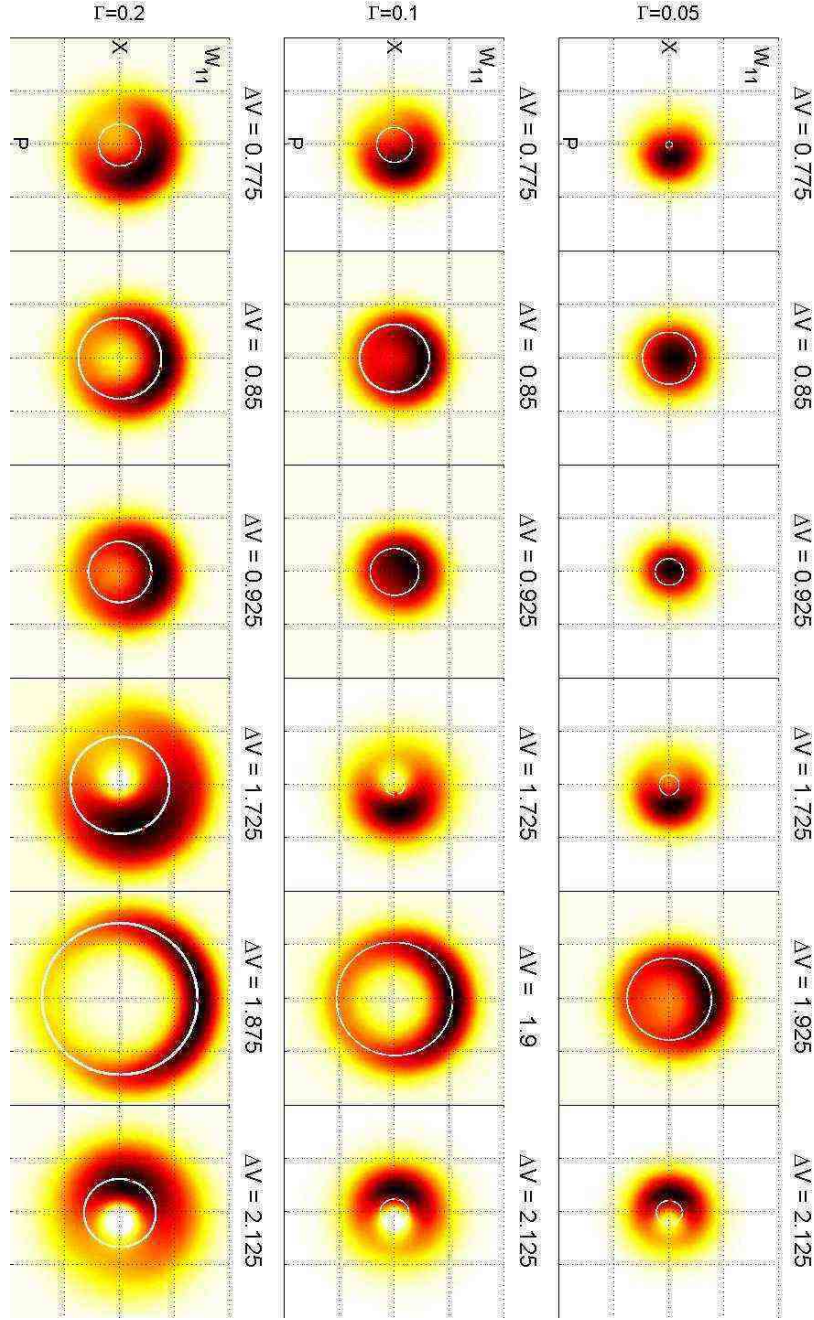}
  \caption{\small  \textit{Comparison between the semiclassical approximation
  and the full description for TDQS. The figure shows the central dot Wigner
  function and the respective limit cycle. The dark dot on the trajectory is in correspondence with the maximum
  of the semiclassical occupation $\s_{CC}$. Also the phase space rotation is captured
   by the simplified model. The matching is best in the shuttling
  regime. See figure \ref{fig:modena} for the details on the parameters.}
  \label{fig:TDQSrings}}
 \end{center}
\end{figure}
%%%%%%%%%%%%%%%%%%%%%%%%%%%%%%%%%%%%%%%%%%%%%%%%%%%

\newpage
\section{Coexistence: a dichotomous process} \label{sec:coexistence}

The  slowest dynamics in the coexistence regime is represented by
the switching process between the shuttling and the tunneling
regime. The amplitude of the dot oscillations is the relevant
variable that is recording this slow dynamics. We analyze this
particular operating regime of the SDQS in three steps. We first
assume the slow switching mode between two different current
channels and explore the consequences of this hypothesis in terms
of current and current noise. We then derive the effective
bistable potential for the amplitude associated with the slowest
dynamics of the shuttle device in the coexistence regime. Finally
we apply Kramers' theory for escape rates to this effective
potential and calculate the switching rates between the two
amplitude equilibrium states corresponding to the local minima of
the potential. We conclude the section comparing the
(semi)analytical results of the simplified model with the
numerical calculations  corresponding to the full description.

%%%%%%%%%%%%%%%%%%%%%%%%%%%%%%%%%%%%%%%%%%%%%%%%%
% Figure
%%%%%%%%%%%%%%%%%%%%%%%%%%%%%%%%%%%%%%%%%%%%%%%%%
\begin{figure}
 \begin{center}
 \includegraphics[angle=0,width=.7\textwidth]{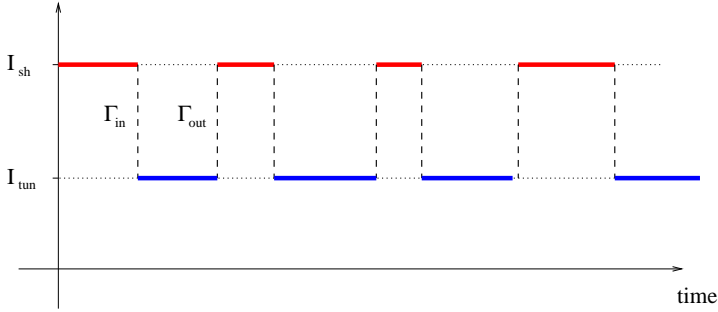}
  \caption{\small  \textit{Schematic representation of the time evolution
  of the current in the dichotomous process between current modes. The scheme
  refers in particular to the SDQS coexistence regime where the currents involved are the
  shuttling ($I_{\rm sh}$) and the tunneling ($I_{\rm tun}$) currents and the names of switching
  rates $\G_{\rm in}$ and $\G_{\rm out}$ indicate the behaviour of
  the dot oscillation amplitude (the relevant variable in the model)
  during the transition.}}
 \end{center}
\end{figure}
%%%%%%%%%%%%%%%%%%%%%%%%%%%%%%%%%%%%%%%%%%%%%%%%%%%

\subsection{Dichotomous process between current modes}\label{sec:dichotomous}

Let us consider a bistable system with two different modes $+$ and $-$ and
two different currents $I_+$ and $I_-$, respectively, associated
with the modes. The system can switch between mode $+$ and $-$
randomly, but with definite rates: namely $\a$ for the process $+
\to -$ and $\b$ for the opposite $- \to +$. We collect this
information in the master equation:

\be
 \dot{\vet{P}} = \frac{d}{dt}\begin{pmatrix} P_+ \\ P_-\\ \end{pmatrix} =
 \begin{pmatrix} -\a & \phantom{-}\b\\ \phantom{-}\a & -\b\\ \end{pmatrix}
 \begin{pmatrix} P_+ \\ P_- \\ \end{pmatrix} = \vet{LP}
 \label{eq:MasterEquation}
\end{equation}
where $P_+$ and $P_-$ are the probabilities to find the system in
the mode ``$+$'' and ``$-$'' respectively.

The matrix notation suggests to read the equation
(\ref{eq:MasterEquation}) at a more general level and embed its
classical rate dynamics into a quantum formalism.

Instead of the modes $+$ and $-$ we consider the state vectors
$|+\rangle$ and $|-\rangle$: they define the Hilbert space of the
quantum system. Density operators defined on this Hilbert space
represent the state of the system. Since we are only interested in
classical states we assume that the density matrix $\s$ is always
diagonal in the $|+\rangle$, $|-\rangle$ basis. In terms of the
probabilities  $P$ the density matrix reads obviously:

\be
 \s = \begin{pmatrix} \s_{++} & \s_{+-} \\ \s_{-+} & \s_{--}\end{pmatrix}
    = \begin{pmatrix} P_+ & 0 \\ 0 & P_-\end{pmatrix}
\end{equation}

Since each of the two states is associated to a well defined
current, also the current operator is diagonal in the $+,-$ basis:

\be
 I = \begin{pmatrix} I_+ & 0\\ 0 & I_- \end{pmatrix}
\end{equation}

The average current is the trace of the current operator
multiplied by the density matrix, namely:

\be
 \langle I(t) \rangle = {\rm Tr}[\s(t) I] = P_+(t) I_+ + P_-(t) I_-
\end{equation}
and it is in general time dependent since the occupation
probabilities evolve in time according to the equation
(\ref{eq:MasterEquation}).

Equation (\ref{eq:MasterEquation}) is the equation of motion
for the density matrix $\s$. In general the Liouville
space\footnote{It is the space of the linear operators on the
original Hilbert space.} for a system with a Hilbert space of
dimension $N$ has dimension $N^2$. In this particular case the
system is simplified: the space of the diagonal linear operators
has again dimension 2. The equation (\ref{eq:MasterEquation}) is
the representation of the equation:

\be
 \dot{\s} = \mathcal{L}[\s]
\label{eq:GMEgeneral}
\end{equation}
in the basis:
\be
 \vet{e}_1 = \begin{pmatrix} 1 & 0\\ 0 & 0 \end{pmatrix}
 , \quad
 \vet{e}_2 = \begin{pmatrix} 0 & 0\\ 0 & 1 \end{pmatrix}
\label{eq:matrixbasis}
\end{equation}
with the identification:

\be
\s \to \vet{P} = \begin{pmatrix} P_+ \\ P_- \end{pmatrix}
, \quad
\mathcal{L} \to \vet{L} =
\begin{pmatrix} -\a & \phantom{-}\b \\ \phantom{-}\a & -\b \end{pmatrix}
\end{equation}

Until now we have used the Schr\"odinger picture in which the
states evolve. For the definition of the current noise we have to
introduce the analogue for open irreversible systems of the
Heisenberg picture defined for closed reversible
system\footnote{For a rigorous treatment of the argument see for
example H.-P.~Breuer {\it et al.}\ \cite{breuer}.}. The equation
(\ref{eq:GMEgeneral}) has the formal solution for a time independent
Liouvillean:

\be
 \s(t) = e^{t\mathcal{L}}\s(0)
\end{equation}

The initial condition $\s(0)$ is evolved to the time $t$ by the
super-operator $e^{t\mathcal{L}}$. The Heisenberg picture for the
operators is designed to preserve the average:

\be
\langle I(t) \rangle
 = {\rm Tr} [I_S \s(t)]
 = {\rm Tr} [I_S e^{t \mathcal{L}}\s(0)]
 = {\rm Tr} [e^{t \mathcal{L}^{\dagger}}I_S \s(0)]
 = {\rm Tr} [I_H(t) \s(0)]
\end{equation}
where the (current) operator appears first in the Schr\"odinger
$I_S$, and then in the Heisenberg $I_H(t)$ picture.
$\mathcal{L}^{\dagger}$ is the adjoint of the superoperator
$\mathcal{L}$. It is useful to consider the representation of the
above equations in the basis (\ref{eq:matrixbasis}):

\be
\langle I(t) \rangle
= (\vet{I}_S,\vet{P}(t))
= (\vet{I}, e^{t\vet{L}}\vet{P}(0))
= (e^{t\vet{L}^{\dagger}}\vet{I},\vet{P}(0))
= (\vet{I}_H(t),\vet{P}(0))
\end{equation}
where $\vet{I}_S = [I_+,I_-]^T$ is the vector representation of
the current operator in Schr\"odinger picture. It is also not
difficult to realize that the trace of the product of two
operators is equivalent to the scalar product in the
vector representation. The current-noise is the spectral density
of the average current fluctuations:

\be
 S(\w) = \lim_{t \to \infty} \int_{-\infty}^{+\infty}
 \!\!\!d\t [\langle I(t+\t)I(t)\rangle -
 \langle I(t+\t)\rangle \langle I(t) \rangle]\,e^{i\w\t}
\end{equation}
In terms of the current operators in the Heisenberg picture, the
noise reads:

\be
\bs
 S(\w) = \lim_{t \to \infty} \int_{-\infty}^{+\infty}
 \!\!\!d\t \{&{\rm Tr}[I_H(t+\t)I_H(t)\s(0)] \\
 -&{\rm Tr}[I_H(t+\t)\s(0)]{\rm Tr}[I_H(t)\s(0)]\}e^{i\w\t}
\end{split}
\end{equation}

The system tends for large times to the stationary state:

\be
\bs
 P_+^{\rm stat} &= \frac{\b}{\a + \b}\\
 P_-^{\rm stat} &= \frac{\a}{\a + \b}
\end{split}
\end{equation}
as can be verified by substitution in (\ref{eq:MasterEquation}).
The corresponding stationary current reads:
\be
 I^{\rm stat} = \lim_{t \to \infty} \langle I(t) \rangle =
 \frac{I_+ \b + I_- \a}{\a + \b}
\end{equation}

We are interested in the zero frequency noise spectrum $\w \to 0$.
We split the integral into two parts: over negative and positive
$\t$ respectively. Using the symmetry under time reversal $\t \to
-\t $ of the integrand we obtain:
\be
 S(0) =  \lim_{\w \to 0}\lim_{t \to \infty}
 2\Re \left[\int_{0}^{\infty} \!\!\! d\t
 \left({\rm Tr}[I_H(t +\t)I_H(t)\s(0)] - {I^{\rm stat}}^2\right)e^{i(\w + i\eps)\t}
 \right]
 \end{equation}
where we have also added to the frequency the small imaginary convergence
factor $i\eps$. We now use the quantum
regression theorem to rewrite the current-current correlator and
perform the limit in $t$:

\be
 S(0) =  \lim_{\w \to 0}
 2\Re \left[\int_{0}^{\infty} \!\!\! d\t
 \left({\rm Tr}[Ie^{\t\mathcal{L}}I\s^{\rm stat}] - {I^{\rm stat}}^2\right)e^{i(\w + i\eps)\t}
 \right]
\end{equation}
We then integrate in $\t$ and obtain:
\be
 S(0) =  \lim_{\w \to 0}
 2\Re \left[
 {\rm Tr}[I(\mathcal{L}+i\w)^{-1}I\s^{\rm stat}]
 - \frac{{I^{\rm stat}}^2}{i\w} \right]
\end{equation}
In the limit $\w \to 0$  the resolvent $(\mathcal{L}+i\w)^{-1}$
is singular since $\mathcal{L}$ has a non-trivial null space. To
handle this singularity we introduce the super-operator
projectors:
\be
 \mathcal{P}[\bullet] = \s^{\rm stat}{\rm Tr}[\bullet]
 ,\quad
 \mathcal{Q} = 1 - \mathcal{P}
 \label{eq:PQdef}
\end{equation}
We already encountered these projectors in section
\ref{sec:NoiseSDQS} where we discussed their properties
(\ref{eq:LPQproperties}). In terms of these projectors the
resolvent can be rewritten:
\be
 (\mathcal{L}+i\w)^{-1} = \mathcal{Q}(\mathcal{L}+i\w)^{-1}\mathcal{Q} + \frac{\mathcal{P}}{i\w}
\end{equation}
The term $\mathcal{Q}(\mathcal{L}+i\w)^{-1}\mathcal{Q}$ is
well-behaved in the limit $\w \to 0$ since the $\mathcal{Q}$
projectors are confining the inverse of the Liouvillean on the
subspace orthogonal to the null space. The term
$\frac{\mathcal{P}}{i\w}$ is still singular.  However, using the
definition (\ref{eq:PQdef}) we can calculate:
\be
 {\rm Tr}\left[I\frac{\mathcal{P}}{i\w}I\s^{\rm stat}\right] =
 {\rm Tr}\left[I\frac{I^{\rm stat}}{i\w}\s^{\rm stat}\right] =
 \frac{{I^{\rm stat}}^2}{i\w}
\end{equation}
and notice that the two diverging contribution in the current
noise exactly cancel. The zero frequency current noise reads:
\be
 S(0) = -2{\rm Tr}(I\mathcal{QL}^{-1}\mathcal{Q}I\s^{\rm stat})
\end{equation}

For the evaluation of the current noise $S(0)$ we introduce the
auxiliary quantity $\Sigma$

\be
 \Sigma = \mathcal{QL}^{-1}\mathcal{Q}I\s^{\rm stat}
\end{equation}
so that
\be
 S(0)= -2{\rm Tr}(I \Sigma) = -2(\vet{I},\vet{\Sigma})
\end{equation}
where we have used in the last equality the vector notation
characteristic of the example at hand. The quantity $\Sigma$
satisfies the equation:
\be
 \mathcal{L}\Sigma
 = \mathcal{Q}I\s^{\rm stat} = (I-I^{\rm stat})\s^{\rm stat}
\end{equation}
that in vector notation reads
\be
 \begin{pmatrix} -\a & \b \\ \a & -\b \end{pmatrix}
 \begin{pmatrix} \Sigma_+ \\ \Sigma_- \end{pmatrix}=
 \frac{(I_+ - I_-)}{\a+\b} \begin{pmatrix}  \a \\-\b \end{pmatrix}
\end{equation}
The components of the vector $\vet{\Sigma}$ also satisfy the
independent relation $\Sigma_+ + \Sigma_- = 0$. We solve then for
$\vet{\Sigma}$ and we are able to give an analytical expression
for the noise of the dichotomous process:

\be
 S(0) = 2\frac{\a\b}{(\a + \b)^3}\D I^2
\end{equation}
where $\D I = I_+ - I_-$.
Summarizing, we give the expression for the stationary current and
Fano factor for a dichotomous process between two current modes
with currents $I_+$ and $I_-$ and switching rates $\a$ and $\b$:

\be
\bs
 &I^{\rm stat} = \frac{I_+\b +I_-\a}{\a + \b}\\
 &F            = \frac{S(0)}{I^{\rm stat}} = 2\frac{\a\b}{(\a + \b)^2}
                 \frac{(I_+ - I_-)^2}{I_+\b + I_-\a}
\end{split}
\end{equation}

The framework of the coexistence simplified model is given by
these results. The task is now to recognize in the dynamics of the
shuttle device the two modes and above all calculate the switching
rates. The answer are in the Kramers' escape rates for a bistable
effective potential.

\subsection{Effective potential}

The stationary total Wigner function for the SDQS evolves as a
function of the mechanical damping from a fuzzy dot close to the
origin of the phase space (tunneling regime) to a growing
ring-shape in the small damping (shuttling) regime. At
intermediate damping rates  the shuttling ring and the tunneling
fuzzy dot coexist. Those properties of the Wigner function can be
understood in terms of an effective stationary potential in the
phase space generated by the non-linear dynamics of the shuttle
device. We show in figure \ref{fig:hats} the three qualitatively
different shapes of the potential guessed from the observation of
the stationary Wigner functions associated with the three
operating regimes. Fedorets {\it et al.}~\cite{fed-prl-04} started
the analytical work for the understanding of the
tunneling-shuttling transition\footnote{Very recently they also
introduced the spin degree of freedom in the oscillating dot
\cite{fed-preprint-04}.} in terms of an effective radial
potential. Taking inspiration from their work we extend the
analysis to the slowest \emph{dynamics} in the device and use
quantitatively the idea of the effective potential for the
description of the coexistence regime.

%%%%%%%%%%%%%%%%%%%%%%%%%%%%%%%%%%%%%%%%%%%%%%%%%
% Figure
%%%%%%%%%%%%%%%%%%%%%%%%%%%%%%%%%%%%%%%%%%%%%%%%%
\begin{figure}
 \begin{center}
 \includegraphics[angle=0,width=.7\textwidth]{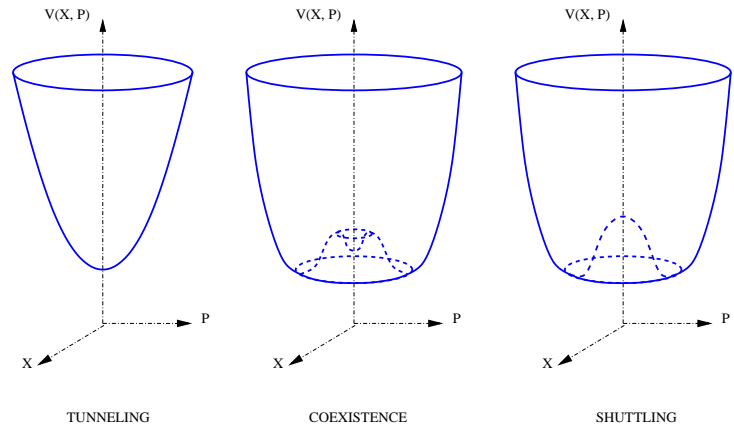}
  \caption{\small  \textit{Schematic representation of the
  effective potentials for the three operating regimes.}
  \label{fig:hats}}
 \end{center}
\end{figure}
%%%%%%%%%%%%%%%%%%%%%%%%%%%%%%%%%%%%%%%%%%%%%%%%%%%

\subsubsection{Elimination of the fast dynamics}\label{sec:nofast}

The starting point are the Klein-Kramers equations for the SDQS
that we rewrite symmetrized by shifting the coordinates origin to
$d/2$:

\be
 \bs
\frac{\partial W_{00}}{\partial t} =&
\left[m \w^2 \left(q + \frac{d}{2}\right)\frac{\partial}{\partial p}
 -\frac{p}{m}\frac{\partial}{\partial q}
 +\g \frac{\partial}{\partial p}p
 +\g m \hbar \w \left(n_B + \frac{1}{2} \right)
 \frac{\partial^2}{\partial p^2}\right]W_{00}\\
 & +\G_{R}e^{ 2q/\l} W_{11}
 -\G_{L}e^{- 2q/\l}\sum_{n=0}^{\infty}
 \frac{(-1)^n}{(2n)!}
 \left(\frac{\hbar}{\l}\right)^{2n}
 \frac{\partial^{2n}W_{00}}{\partial p^{2n}}\\
\frac{\partial W_{11}}{\partial t} =&
\left[m \w^2 \left(q - \frac{d}{2}\right)\frac{\partial}{\partial p}
 -\frac{p}{m}\frac{\partial}{\partial q}
 +\g \frac{\partial}{\partial p}p
 +\g m \hbar \w \left(n_B + \frac{1}{2} \right)
 \frac{\partial^2}{\partial p^2}\right]W_{11}\\
 & +\G_{L}e^{- 2q/\l} W_{00}
 -\G_{R}e^{2q/\l}\sum_{n=0}^{\infty}
 \frac{(-1)^n}{(2n)!}
 \left(\frac{\hbar}{\l}\right)^{2n}
 \frac{\partial^{2n}W_{11}}{\partial p^{2n}}
 \end{split}
 \label{eq:KlKr}
\end{equation}
One by one we will now get rid of the variables that due to
their fast dynamics are not relevant for the description of the
coexistence regime. In equations (\ref{eq:KlKr}) we describe the
electrical state of the dot as empty or charged. We shift to the
description in terms of state $+$ and state $-$ with the
definition:

\be
 W_+ = W_{00} + W_{11}, \quad W_- = W_{00} - W_{11}
\end{equation}
In terms of these new Wigner functions the Klein-Kramers equations
read:

\be
 \bs
  \partial_t W_+ &= \{H_{\rm osc}, W_+\}_P
  + \mathcal{L}_1 W_+ + \mathcal{L}_2W_-\\
  \partial_t W_- &= \{H_{\rm osc}, W_-\}_P
  + (\mathcal{L}_1 +\G_+)W_- + (\mathcal{L}_2+\G_-)W_+\\
 \end{split}
 \label{eq:KlKrplusminus}
\end{equation}
where
\be
 \bs
  \G_+ &= \G_R e^{2q/\l} + \G_L e^{-2q/\l}\\
  \G_- &= \G_R e^{2q/\l} - \G_L e^{-2q/\l}\\
  \mathcal{L}_1 &=
  \g \partial_p p
  +\g m \hbar \w \left(n_B + \frac{1}{2} \right)\partial^2_p
  - \frac{\G_+}{2}\sum_{n=1}^{\infty}\frac{1}{2n!}
  \left(\frac{\hbar}{i\l}\right)^{2n}
  \partial^{2n}_p\\
  \mathcal{L}_2 &= \frac{e\mathcal{E}}{2}\frac{\partial}{\partial p}
  + \frac{\G_-}{2}\sum_{n=1}^{\infty}\frac{1}{2n!}
  \left(\frac{\hbar}{i\l}\right)^{2n}
  \partial_p^{2n}
 \end{split}
\end{equation}
The symbol $\{\bullet,\bullet\}_P$ represents the Poisson brackets
and we have introduced for the partial derivatives the short
notation $\partial_x \equiv \frac{\partial}{\partial x}$. It is
useful for the calculation of the effective potential to treat
position and momentum on equal footing. For this reason we
introduce the adimensional variables:

\be
 X = \frac{q}{\l}, \quad P = \frac{p}{m\l\w}, \quad \t = t\w
\end{equation}
The general structure of the equations (\ref{eq:KlKrplusminus})
remains unchanged but the Liouvilleans and the $\G$ functions take
the form:

\be
 \bs
  \G_+ &= \frac{\G_R}{\w} e^{2X} + \frac{\G_L}{\w} e^{-2X}\\
  \G_- &= \frac{\G_R}{\w} e^{2X} - \frac{\G_L}{\w} e^{-2X}\\
  \mathcal{L}_1 &=
  \frac{\g}{\w} \partial_P P
  +\frac{\g}{\w} \left(\frac{x_0}{\l}\right)^2 \left(n_B + \frac{1}{2}
  \right)\partial^2_P
  - \frac{\G_+}{2}\sum_{n=1}^{\infty}\frac{(-1)^n}{2n!}
  \left(\frac{x_0}{\l}\right)^{4n}
  \partial^{2n}_P\\
  \mathcal{L}_2 &= \frac{d}{2\l}\partial_P
  + \frac{\G_-}{2}\sum_{n=1}^{\infty}\frac{(-1)^n}{2n!}
  \left(\frac{x_0}{\l}\right)^{4n}
  \partial_P^{2n}
 \end{split}
 \label{eq:GammaLiouvill}
\end{equation}

Since in the following we will assume the ``classical'' limit of a
tunneling length much larger than the zero point uncertainty in
the position, only the first term of the sums that appears in the
definition of $\mathcal{L}_1$ and $\mathcal{L}_2$ will be kept.
Following Fedorets et al. \cite{fed-prl-04} we also introduce
polar coordinates:

\be
 X = A\sin \phi, \quad P = A \cos \phi
\end{equation}
The equations (\ref{eq:KlKrplusminus}) are transformed into:

\be
 \bs
  \partial_{\t} W_+ &= (-\partial_{\phi}  + \mathcal{L}_1) W_+ + \mathcal{L}_2W_-\\
  \partial_{\t} W_- &= (-\partial_{\phi}  + \mathcal{L}_1 -\G_+)W_- + (\mathcal{L}_2+\G_-)W_+\\
 \end{split}
 \label{eq:KlKrpolar}
\end{equation}
It is interesting to note that the Poisson brackets for the harmonic
oscillator reduce in the polar coordinates to $-\partial_{\phi}$
since there is no amplitude dynamics in the phase space. The $\G$
functions now read:

 \be
 \bs
  \G_+ &= \frac{\G_R}{\w} e^{2A\sin \phi} + \frac{\G_L}{\w} e^{-2A\sin \phi}\\
  \G_- &= \frac{\G_R}{\w} e^{2A\sin \phi} - \frac{\G_L}{\w} e^{-2A\sin \phi}
 \end{split}
 \end{equation}
For the transformation of the Liouvilleans $\mathcal{L}_{1,2}$ we
used the polar coordinate representation of the following
differential operators:

\be
 \bs
  \partial_P
     =& -\frac{\sin \phi}{A} \partial_{\phi} + \cos \phi \partial_A\\
     =& \frac{1}{A}\cos \phi \partial_A A - \frac{1}{A}\partial_{\phi}\sin \phi\\
  \partial^2_P
     =& \frac{2}{A^2} \cos \phi \sin \phi \partial_{\phi} +
     \frac{\sin^2\phi}{A} \partial_A + \frac{\sin^2\phi}{A^2}
     \partial^2_{\phi}\\
     &- \frac{2}{A}\sin\phi \cos\phi
     \partial^2_{A\phi} + \cos^2\phi \partial^2_A\\
     =& \frac{1}{A}\cos^2\phi\partial^2_A A - \frac{2}{A}
     \partial^2_{A\phi} \sin\phi \cos\phi + \frac{1}{A^2}
     \partial^2_{\phi} \sin^2\phi \\
     &-\frac{1}{A} \partial_A
     \sin^2\phi- \frac{2}{A^2}\partial_{\phi} \cos\phi \sin\phi\\
  \partial_P P
     =& \frac{1}{A}\cos^2\phi \partial_A A^2
     -\partial_{\phi}\sin\phi \cos\phi \\
     =& 1 - \sin\phi \cos\phi \partial_{\phi} + \cos^2\phi
     A\partial_A
    \end{split}
    \label{eq:xp2Aphi}
\end{equation}
where $\partial^2_{A\phi} \equiv \frac{\partial^2}{\partial A \partial \phi}$.
The two different forms (partial differential operator
$\partial_A$ and/or $\partial_{\phi}$ to the extreme left or
right) that for the moment seem redundant will be very useful in
the derivation of the effective potential.

We now start to make approximations on the system of equations
(\ref{eq:KlKrpolar}) with the main idea of eliminating the fast
irrelevant variables. We know that eventually the density
operators $\s_{ii}$ that describe the system will reach a
stationary state. In absence of the harmonic oscillator the state
$+$ would be fixed by the trace sum rule and the state $-$ would
relax to zero on a time scale fixed by the tunneling rates. We
assume that also in the presence of the mechanical degree of
freedom the relaxation dynamics of the $|-\rangle$ state is much
faster than the one of the $|+\rangle$ state. We set
$\partial_{\t} W_-$ to zero in the second equation of (\ref{eq:KlKrpolar})
and formally solve the equation for $W_-$. We insert the result in
the first equation of (\ref{eq:KlKrpolar}) and obtain a closed equation for
the (total) Wigner function $W_+$:

\be
 \bs
 \partial_{\t} W_+ &= [-\partial_{\phi} + \underbrace{\mathcal{L}_1 +
 \mathcal{L}_2
 (1- \hat{G}_0\mathcal{L}_1)^{-1}\hat{G}_0(\mathcal{L}_2 +
 \G_-)}_{\displaystyle\mathcal{L}_+}]W_+\\
 \end{split}
  \label{eq:Kramers1}
\end{equation}
where $\hat{G}_0 \equiv (\partial_{\phi} + \G_+)^{-1}$ and we also
define the (super)operator $\mathcal{L}_+$. At this point the
detailed information about the charge state of the oscillating dot
is integrated out, but the equation (\ref{eq:Kramers1}) still takes
the fast phase dynamics into account. Since we are interested only
in the amplitude dynamics (the slowest in the coexistence regime)
we introduce the projector $\mathcal{P}_{\phi}$ that averages over
the phase and its orthogonal complement $\mathcal{Q}_{\phi} = 1 -
\mathcal{P}_{\phi}$:

\be
 \mathcal{P}_{\phi}[\bullet] = \frac{1}{2\pi}\int_0^{2 \pi} d\phi \bullet
\end{equation}

Using these two operators we decompose the Wigner distribution
function into:

\be
 W_+ = \mathcal{P} W_+ + \mathcal{Q} W_+ = \bar{W}_+ + \tilde{W}_+
\end{equation}
We apply the same projectors also to the equation
(\ref{eq:Kramers1}) and we obtain the set of coupled partial
(integro-\footnote{The operator $\mathcal{L}$ contains the inverse
of differential operators.})differential equations:

\be
 \bs
 \partial_{\t}\bar{W}_+
 &= \mathcal{P}_{\phi} \mathcal{L}_+ \bar{W}_+ +
   \mathcal{P}_{\phi}\mathcal{L}_+ \tilde{W}_+\\
\partial_{\t}\tilde{W}_+
 &= [-\partial_{\phi} + \mathcal{Q}_{\phi}\mathcal{L}_+]\tilde{W}_+
 + \mathcal{Q}_{\phi}\mathcal{L}_+ \bar{W}_+
\end{split}
\label{eq:Kramers2}
\end{equation}
where we have used the properties:
\be
 \bs
 &\mathcal{P}_{\phi}\partial_{\phi} = 0\\
 &\mathcal{Q}_{\phi}\partial_{\phi}
 =
\partial_{\phi}\mathcal{Q}_{\phi} = \partial_{\phi}
\end{split}
\end{equation}
which is readily demonstrated since all the distributions involved
are periodic functions of $\phi$.  We now assume that the phase
relaxation time that governs the dynamics of the phase-dependent
distribution $\tilde{W}_+$ is much shorter than the amplitude
relaxation time of the distribution $W_+$ \footnote{This
assumption is based on the observation that the diffusion dynamics
that governs the relaxation times is largely facilitated along the
transverse direction where no potential barriers (see fig.
\ref{fig:hats}) must be overcome. The potential barriers occur
instead in the radial direction.}. We then set $\partial_{\t}
\tilde{W}_+ = 0$ in the second equation of (\ref{eq:Kramers2}),
solve it for $\tilde{W}_+$ and insert the result in the first
equation. The result is a closed equation for the
(quasi)probability distribution function $\bar{W}_+$:

\be
 \partial_{\t} \bar{W}_+ = \mathcal{P}_{\phi}\mathcal{L}_+
 [1+ (1-\hat{g}_0\mathcal{Q}_{\phi}\mathcal{L}_+)^{-1}
  \hat{g}_0\mathcal{Q}_{\phi}\mathcal{L}_+ ]\bar{W}_+
  \label{eq:Kramers3}
\end{equation}
where $\hat{g}_0 \equiv (\partial_{\phi})^{-1}$. We notice that
the operator $\hat{g}_0$ is in the equations always following the
projector $\mathcal{Q}_{\phi}$. This combination of operators
ensures us as an output a periodic function in the variable
$\phi$.

\subsubsection{Small parameters expansion}\label{sec:expansion}

We believe that equation (\ref{eq:Kramers3}) contains the relevant
dynamics of the SDQS in the coexistence regime, but, despite all
the approximations already introduced the problem still looks
untractable. Again following the already mentioned work by
Fedorets {\it et al.} \cite{fed-prl-04} we consider the perturbation expansion of
(\ref{eq:Kramers3}) in the ``small parameters'':

\be
 \frac{d}{\l},\quad \left(\frac{x_0}{\l}\right)^2, \quad \frac{\g}{\w} \ll 1
 \label{eq:smallparameters}
\end{equation}
These three inequalities describe a limit often encountered in
this thesis and they correspond respectively  to the three
physical assumptions:

\begin{enumerate}

\item The external electrostatic force is a small perturbation of
the harmonic oscillator restoring force in terms of the
sensitivity to displacement of the tunneling rates. This ensures a
quasi oscillator-independent treatment of the tunneling regime.

\item The tunneling length is big compared to the zero point
fluctuations. Since the oscillator dynamics for the shuttling (and
then partially also for the coexistence) regime happens on the
scale of the tunneling length, this condition ensures a
quasi-classical behaviour of the harmonic oscillator.

\item The coupling of the oscillator to the thermal bath is weak
and the oscillator dynamics is under-damped.

\end{enumerate}

We want to expand the equation (\ref{eq:Kramers3}) up to second
order in the three small parameters (\ref{eq:smallparameters}). To
proceed systematically we first notice that they appear at least
to first order in the Liouvilleans $\mathcal{L}_1$ and
$\mathcal{L}_2$ and thus also in the operator $\mathcal{L}_+$. We
can then safely expand (\ref{eq:Kramers3}) up to second order in
$\mathcal{L}_+$ without missing any desired term:

\be
 \partial_{\t}\bar{W}_+
 \approx \mathcal{P}_{\phi} \mathcal{L}_+
 [1 + \hat{g}_0\mathcal{Q}_{\phi}\mathcal{L}_+]\bar{W}_+
\end{equation}
Using the definition of $\mathcal{L}_+$ contained in
(\ref{eq:Kramers1}) we now expand this last equation up to second
order in $\mathcal{L}_1$ and $\mathcal{L}_2$:
\be
 \bs
 \partial_{\t}\bar{W}_+
 \approx
 [&\mathcal{P}_{\phi} \mathcal{L}_1
 +\mathcal{P}_{\phi}\mathcal{L}_2\hat{G}_0 \G_-\\
 +& \mathcal{P}_{\phi}\mathcal{L}_2\hat{G}_0 \mathcal{L}_1\hat{G}_0 \G_-
 + \mathcal{P}_{\phi}\mathcal{L}_2\hat{G}_0\mathcal{L}_2\\
 +& \mathcal{P}_{\phi}\mathcal{L}_1\hat{g}_0\mathcal{Q}_{\phi}\mathcal{L}_1
 + \mathcal{P}_{\phi}\mathcal{L}_2\hat{G}_0
   \G_-\hat{g}_0\mathcal{Q}_{\phi}\mathcal{L}_2\hat{G}_0\G_-\\
 +& \mathcal{P}_{\phi}\mathcal{L}_1\hat{g}_0
   \mathcal{Q}_{\phi}\mathcal{L}_2\hat{G}_0\G_-
 + \mathcal{P}_{\phi}\mathcal{L}_2\hat{G}_0\G_-
   \hat{g}_0\mathcal{Q}_{\phi}\mathcal{L}_1]\bar{W}_+
 \end{split}
 \label{eq:Kramers4}
\end{equation}
The last step is to consider within the equation
(\ref{eq:Kramers4}) only the contributions up to second order in
the expansion parameters (\ref{eq:smallparameters}). In the
definition of the Liouvilleans in the adimensional form
(\ref{eq:GammaLiouvill}) we note that each of the small parameters
(\ref{eq:smallparameters}) is associated with a differential
operator $\partial_P$. We then average out the phase variable from
the equations. We thus expect the second order expansion in the
small parameters (\ref{eq:smallparameters}) to be of second order
in the differential operator $\partial_A$. More precisely the
expansion takes the form:

\be
 \partial_{\t} \bar{W}_+(A,\t)=
 \frac{1}{A}\partial_A A[V'(A) + D(A)\partial_A]\bar{W}_+(A,\t)
 \label{eq:KramersquasiA}
\end{equation}
where $V'(A) = \frac{d}{dA}V(A)$ and $D(A)$ are given functions of
A. Before calculating explicitly all the different contributions
that compose the functions $V'$ and $D$ we want to explore the
consequences of the formulation of the Klein-Kramers equations
(\ref{eq:KlKr}) in the form (\ref{eq:KramersquasiA}). The
stationary solution of the equation (\ref{eq:KramersquasiA})
reads:
\be
 \bar{W}_+^{\rm stat}(A) = \frac{1}{\mathcal{Z}}
 \exp\left(-\int_0^A
dA' \frac{V'(A')}{D(A')}\right)
\end{equation}
where $\mathcal{Z}$ is the normalization that ensures the integral
of the phase-space distribution to be unity: $\int_0^{\infty}
dA'2\pi A' \bar{W}_+^{\rm stat}(A') = 1$. The equation
(\ref{eq:KramersquasiA}) is identical to the Fokker-Planck
equation for a particle in the bidimensional rotationally
invariant potential $V$ (see figure \ref{fig:hats}) with
stochastic forces described by the (position dependent) diffusion
coefficient $D$.

All the differential operators acting on $\bar{W}_+$ in
(\ref{eq:Kramers4}) are at maximum of second order in
$\partial_A$. In order to obtain Eq.~(\ref{eq:KramersquasiA}) we
substitute into (\ref{eq:Kramers4}) the definition of
$\mathcal{L}_1$ and $\mathcal{L}_2$ and we ``push'' the
differential operator $\partial_A$ all the way towards the left or
the right in the sequence of operators acting on $\bar{W}_+$. All
the components in which we are left with only one $\partial_A$
operator on the right define the effective potential $V$. All the
others have the second $\partial_A$ on the extreme right and
define $D(A)$. All contributions to the effective potential $V$
and diffusion coefficient $D$ can be grouped according to the
power of the small parameters that they contain.

\be
 \bs
 V'(A) &=
  \frac{\g}{\w} \frac{A}{2}
 +\frac{d}{2\l} \a_0(A)
 +\left(\frac{x_0}{\l}\right)^4\a_1(A)
 +\left(\frac{d}{2\l}\right)^2\a_2(A)
 +\frac{\g}{\w}\frac{d}{2\l}\a_3(A)\\
 D(A) &=
  \frac{\g}{\w}\left(\frac{x_0}{\l}\right)^2
  \frac{1}{2}\left(n_B + \frac{1}{2}\right)
 +\left(\frac{x_0}{\l}\right)^4\b_1(A)
 +\left(\frac{d}{2\l}\right)^2\b_2(A)
 +\frac{\g}{\w}\frac{d}{2\l}\b_3(A)
 \end{split}
\end{equation}
where the $\a$ functions read:
\be
 \bs
  \a_0 =&
   \mathcal{P}_{\phi}\cos \phi \hat{G}_0 \G_-\\
  \a_1 =&
   -\frac{1}{4}\mathcal{P}_{\phi} \cos \phi \G_-
   \partial_P (\hat{G}_0 \G_-)\\
  \a_2 =&
   \mathcal{P}_{\phi} \cos \phi \hat{G}_0 \G_-
   \hat{g}_0 \mathcal{Q}_{\phi}
   \partial_P (\hat{G}_0 \G_-)\\
  \a_3 =&
   \mathcal{P}_{\phi}\cos \phi \Big[\hat{G}_0^2 \G_- +
   A \hat{G}_0 \partial_P (\hat{G}_0 \G_-)
   - \frac{A}{2} \sin \phi \partial_P (\hat{G}_0 \G_-)\Big]
 \end{split}
\end{equation}
and the $\b$'s can be written as:
\be
 \bs
  \b_1 =& \frac{1}{4}
   \mathcal{P}_{\phi}\cos^2\phi \Big[\G_+ - \G_- \hat{G}_0 \G_-\Big]\\
  \b_2 =&
   \mathcal{P}_{\phi} \cos\phi \Big[\hat{G}_0 \cos\phi
   + \hat{G}_0\G_- \hat{g}_0 \mathcal{Q}_{\phi} \cos\phi \hat{G}_0\G_-\Big]\\
  \b_3 =&
   A \mathcal{P}_{\phi}\cos \phi \Big[ \hat{G}_0 \cos^2\phi
   \hat{G}_0 \G_- + \frac{1}{4} \hat{G}_0 \G_- \sin 2\phi
    - \frac{1}{4} \sin 2\phi \hat{G}_0\G_-\Big]\\
   \end{split}
\end{equation}
In the expression above the $\phi$ variable disappears on the RHS
when we apply the projector $\mathcal{P}_{\phi}$. We have kept the
somehow mixed notation with the differential operator $\partial_P$
to keep the notation lighter. In general the operators are acting
on all what is on their right. Otherwise parentheses are limiting
their range of action (see for example $\partial_P$ and in
$\a_1$).
As examples of the arguments that appear in the calculation of the
$\a$ and $\b$ functions we give first the derivation of a
``missing'' contribution, then we explicitly derive $\alpha_3$
 and $\beta_3$.

Some second order contributions in the parameters
(\ref{eq:smallparameters}) simply do not appear in the final
expansions\footnote{It is the case of the contributions in
$\left(\frac{x_0}{\l}\right)^2$ and
$\frac{d}{2\l}\left(\frac{x_0}{\l}\right)^2$.} because the small
parameter combinations which they represent is not present. The
contribution in $\left(\frac{\g}{\w}\right)^2$ instead is formally
there but vanishes identically.
From the expansion in terms of the operators $\mathcal{L}_1$ and
$\mathcal{L}_2$ this contribution reads:
\be
 \mathcal{L}_{\g\g} \equiv \mathcal{P}_{\phi} \partial_P P \hat{g}_0 \mathcal{Q}_{\phi}
 \partial_P P
\end{equation}
We insert the polar coordinate expressions for the differential
operators $\partial_P P$ (see Eq.~\ref{eq:xp2Aphi}) choosing the
left operator form for the first and the right form for the second
and we obtain:
\be
 \mathcal{L}_{\g\g} = \frac{1}{A}\partial_A A^2 \mathcal{P}_{\phi}
 \cos^2\phi\hat{g}_0 \mathcal{Q} ( 1 + \cos^2\phi A \partial_A)
\end{equation}
where we have used also the fact that the operator is applied to
$\bar{W}_+$ which does not depend on $\phi$. Remembering that
$\mathcal{Q}_{\phi} = 1- \mathcal{P}_{\phi}$ and $\hat{g}_0$ is
the indefinite integral in the variable $\phi$ we conclude the
calculation:
\be
 \bs
 \mathcal{L}_{\g\g} &=
 \frac{1}{A}\partial_A A^2 \mathcal{P}_{\phi} \cos^2\phi \hat{g}_0
 \left(\cos^2\phi - \frac{1}{2}\right) A \partial_A\\
  &=
 \frac{1}{2A}\partial_A A^2 \mathcal{P}_{\phi} \cos^3\phi \sin\phi
 A \partial_A = 0
 \end{split}
\end{equation}

The functions $\alpha_3$ and $\beta_3$ are both derived from the
contribution $\frac{\g}{\w}\frac{d}{2\l}$ of the small parameter
expansion \eqref{eq:Kramers4}. The corresponding Liouvillean
reads:

\be \bs
 \mathcal{L}_{\g d} =
   &\mathcal{P}_{\phi}\partial_P \hat{G}_0 \partial_P P \hat{G}_0 \G_- +\\
   &\mathcal{P}_{\phi}\partial_P P  \hat{g}_0 \mathcal{Q}_{\phi} \partial_P \hat{G}_0 \G_- +\\
   &\mathcal{P}_{\phi}\partial_P \hat{G}_0 \G_- \hat{g}_0 \mathcal{Q}_{\phi} \partial_P P
 \end{split}
\end{equation}
We now use polar coordinates and take into account that the
Liouvillean is applied to a function $\bar{W}_+$ independent of
the variable $\phi$. We obtain:

\be \bs
 \mathcal{L}_{\g d} =
 \frac{1}{A} \partial_A A & \mathcal{P}_{\phi} \cos \phi \\
 \Big[ &
   \hat{G}_0^2 \G_-
 + \hat{G}_0 A \cos \phi \partial_P (\hat{G}_0\G_-)
 + \hat{G}_0 A \cos^2 \phi \hat{G}_0 \G_- \partial_A +\\
 &
   A \cos \phi  \hat{g}_0 \mathcal{Q}_{\phi} \partial_P(\hat{G}_0\G_-)
 + A \cos \phi  \hat{g}_0 \mathcal{Q}_{\phi} \cos \phi \hat{G}_0\G_- \partial_A +\\
 &
 \hat{G}_0\G_- \hat{g}_0 \mathcal{Q}_{\phi} A \cos^2 \phi
 \partial_A
 \Big]
 \end{split}
\end{equation}
The $\alpha_3$ and $\beta_3$ contributions can be split to
obtain\footnote{In this passage we have also used the projector
$\mathcal{P}_{\phi}$ to define a scalar product
$\mathcal{P}_{\phi} f(\phi) g(\phi) \equiv (f,g)$ and the adjoint
relation: $(f, \hat{O}g ) = (\hat{O}^{\dagger}f,g)$.}:

\be
 \bs
 \mathcal{L}_{\g d} = &
  \frac{1}{A} \partial_A A
   \left\{
   \mathcal{P}_{\phi} \cos \phi
   \Big[
     \hat{G}_0^2 \G_- +
   A \hat{G}_0 \partial_P (\hat{G}_0 \G_-)
   - \frac{A}{2} \sin \phi \partial_P (\hat{G}_0 \G_-)
   \Big]
   \right\} +\\
  &\frac{1}{A} \partial_A A
   \left\{
   \mathcal{P}_{\phi} \cos \phi
   \Big[
   \hat{G}_0 \cos^2\phi
   \hat{G}_0 \G_- + \frac{1}{4} \hat{G}_0 \G_- \sin 2\phi
    - \frac{1}{4} \sin 2\phi \hat{G}_0\G_-
   \Big]
   \right\}\partial_A
 \end{split}
\end{equation}
Since we have projected out the phase $\phi$ we are effectively
working in a one-dimensional phase space given by the amplitude
$A$. Equation (\ref{eq:KramersquasiA}) though is \emph{not} as it
is a Kramers equation for a single variable. This is related to
the fact that also the distribution $\bar{W}_+$ is \emph{not} the
amplitude distribution function, but, so to speak, a cut at fixed
phase of a two dimensional rotationally invariant distribution.
The difference is a geometrical factor $A$ that it turns out is
also simplifying the equation. We define the amplitude probability
distribution $\mathcal{W}(A,\t) = A \bar{W}_+(A,\t)$ and insert
this definition in equation (\ref{eq:KramersquasiA}). We obtain:
\be
 \bs
 \partial_{\t} \mathcal{W}(A,\t)
 &= \partial_A A[V'(A) +
 D(A)\partial_A]\frac{1}{A}\mathcal{W}(A,\t)\\
 &= \partial_A [\mathcal{V}'(A) + D(A)\partial_A]\mathcal{W}(A,\t)
 \end{split}
 \label{eq:KramersA}
\end{equation}
where we have defined the geometrically corrected potential
\be
 \mathcal{V}(A) = V(A) - \int_{A_0}^{A} \!\! \frac{D(A')}{A'} dA'
\end{equation}
which for an amplitude independent diffusion coefficient gives a
corrected potential logarithmically divergent in the origin. The
integration extremal is arbitrary and reflects the arbitrary
constant in the definition of the potential. The equation
(\ref{eq:KramersA}) is the one-dimensional Kramers equation that
constitutes the starting point for the calculation of the
switching rates that characterize the coexistence regime.

%%%%%%%%%%%%%%%%%%%%%%%%%%%%%%%%%%%%%%%%%%%%%%%%%
% Figure
%%%%%%%%%%%%%%%%%%%%%%%%%%%%%%%%%%%%%%%%%%%%%%%%%
\begin{figure}
 \begin{center}
 \includegraphics[angle=0,width=.5\textwidth]{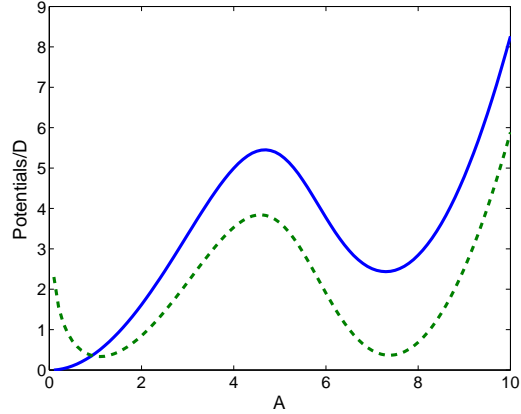}
  \caption{\small  \textit{Example of the potential $V$ (full) and the
  geometrically corrected potential $\mathcal{V}$ (dashed)  for the SDQS in the coexistence regime.
  Notice for the corrected potential the logarithmic divergence in the origin.}
  \label{fig:Copotexample}}
 \end{center}
\end{figure}
%%%%%%%%%%%%%%%%%%%%%%%%%%%%%%%%%%%%%%%%%%%%%%%%%%%

\subsubsection{Evaluation of the $\a_i$ and $\b_i$ functions}

We evaluate the $\a_i$ and $\b_i$ functions numerically.
Fortunately there is a rather natural discretization of the
operators that compose the $\a$ and $\b$. This discretization is
given by the operator $\mathcal{P}_{\phi}\cos\phi$ that opens the
operator string of all the components and that can be considered
as a sort of Fourier transform of the function that follows. We
specify better the concept and the method by reducing explicitly
to the ``numerical form'' the contributions $\a_0$ and $\a_2$:

\be
\bs
 \a_0 =&
   \mathcal{P}_{\phi}\cos \phi \hat{G}_0 \G_-\\
 \a_2 =&
   \mathcal{P}_{\phi} \cos \phi \hat{G}_0 \G_-
   \hat{g}_0 \mathcal{Q}_{\phi}
   \partial_P (\hat{G}_0 \G_-)\\
\end{split}
\label{eq:a0a2}
\end{equation}
where we remind $\hat{G}_0 = (\partial_{\phi} + \G_+)^{-1}$.  We
define the function $G(A,\phi) \equiv \hat{G}_0\G_-(A,\phi)$. If
$\tilde{G}_n(A)$ is the discrete Fourier transform  of $G$
(periodic function of $\phi$)

\be
 \tilde{G}_n(A) = \frac{1}{2\pi}\int_0^{2\pi} G(A,\phi)e^{in\phi}
\end{equation}
with $n \in \mathbb{Z}$, then we can easily recognize the Fourier
transform in the definition of $\a_0$ and write:
\be
 \a_0(A) = \frac{1}{2}(\tilde{G}_{1} + \tilde{G}_{-1})
\end{equation}
We still need to identify the structure of the functions
$\tilde{G}_n(A)$. It is useful for this purpose to calculate the
Fourier transform of the functions $\G_+$ and $\G_-$ and define
the vectors $\tilde{\bar{\G}}_+$ and  $\tilde{\bar{\G}}_-$ whose
components are:

\be
 \bs
  (\tilde{\bar{\G}}_+)_n = \frac{\G}{\w}[J_n(-2iA) + J_n(2iA)]\\
  (\tilde{\bar{\G}}_-)_n = \frac{\G}{\w}[J_n(-2iA) - J_n(2iA)]
 \end{split}
\end{equation}
where we have assumed for simplicity that $\G_L=\G_R=\G$ and the
Bessel  function $J_n(z)$ is defined by:
\be
 J_n(z) = \frac{1}{2\pi}\int_0^{2\pi}e^{in\phi + iz\sin\phi}d\phi
\end{equation}
At the operator level we can write:

\be
 \G_-(A,\phi) = [\partial_{\phi} + \G_+(A,\phi)]G(A,\phi)
\end{equation}
We take the Fourier transform in $\phi$ on both sides and we
obtain:
\be
 (\tilde{\bar{\G}}_-)_n = \sum_{m = -\infty}^{\infty}
 [\underbrace{in\d_{mn} + (\tilde{\bar{\bar{\G}}}_+)_{nm}}_{\displaystyle M_{nm}}] \tilde{G}_m
\end{equation}
where we have defined the  matrix $(\tilde{\bar{\bar{\G}}}_+)_{nm}
\equiv (\tilde{\bar{\G}}_+)_{n-m}$ and the matrix $M$. Finally we
also define the Fourier transform of the cosine $\bar{C} =
[\ldots,\frac{1}{2}, 0, \frac{1}{2}, \ldots]^T$ with the $1/2$ in
position $\pm 1$ and we write the $\a_0$ function in the compact
numerical form:

\be
 \a_0= (\bar{C},M^{-1}\tilde{\bar{\G}}_-)
\end{equation}

where the symbol $(\bullet,\bullet)$ indicates the scalar product
between the two vectors. All the vectors must be truncated in the
numerical evaluation, but fortunately the Bessel functions $J_n$
decay fast with $n$ and typically we verified numerical
convergence taking only 40 terms around the zero position of the
discrete Fourier transform ``momentum space''.

If the derivation of the numerical form for $\a_0$ is revealing
the spirit of the calculation, the $\a_2$ case contains instead
some of the typical ``tricks''. We start from the definition
(\ref{eq:a0a2}) and express the differential operator $\partial_P$
in polar coordinates. We obtain:

\be
 \a_2 =
 \mathcal{P}_{\phi}\cos\phi
   \hat{G}_0\G_-\hat{g}_0\mathcal{Q}_{\phi}
   \left(-\frac{\sin\phi}{A}\partial_{\phi}
   +\cos\phi\partial_A\right) \hat{G}_0\G_-
\end{equation}
We absorb the partial derivative $\partial_{\phi}$ with the
identification:

\be
 \partial_{\phi} = \hat{G}_0^{-1} - \G_+
\end{equation}
Another important element is the combination
$\hat{g}_0\mathcal{Q}_{\phi}$. It is important to study the
Fourier transform of such an operator. Since $\hat{g}_0 =
(\partial_{\phi})^{-1}$ we obtain:
\be
(\hat{g}_0)_{mn} = -\frac{i}{n}\d_{nm}
\end{equation}
The singularity for $n=0$ is cured by the projector
$\mathcal{Q}_{\phi}$ that removes the $0^{\rm th}$ component in
the Fourier transform of the vector on which it is acting. We call
for this reason the combination $\hat{g}_0\mathcal{Q}_{\phi}$ the
pseudoinverse\footnote{From the arguments above is clear that $
\hat{g}_0 \mathcal{Q}_{\phi} \equiv \mathcal{Q}_{\phi} \hat{g}_0
\mathcal{Q}_{\phi}$.} of the differential operator
$\partial_{\phi}$ and we use the symbol
$(\partial_{\phi}^{-1})_{PS}$.

The last problem that must be faced is the differential operator
$\partial_A$ that  we always encounter in the expression
$\partial_A G$. Using the definition of $G$ and the results
obtained in the evaluation of $\a_0$ we can write:

\be
 \partial_A \tilde{G} = M^{-1}\partial_A \tilde{\bar{\G}}_- - M^{-1}\partial_A (M)M^{-1}\tilde{\bar{\G}}_-
\end{equation}
For the matrix $M$ and the vector $\tilde{\bar{\G}}_-$ we have the
explicit formulation in terms of the Bessel functions. Using the
recursive relation between the functions $J$ and their derivatives
\be
 \partial_zJ_n(z)=\frac{1}{2}[J_{n-1}(z) - J_{n+1}(z)]
\end{equation}
we obtain:
\be
 \bs
(\partial_A M)_{mn} = \frac{i\G}{\w}
[&J_{n-m+1}(-2iA) - J_{n-m-1}(-2iA)\\
-&J_{n-m+1}(+2iA) + J_{n-m-1}(+2iA)]\\
(\partial_A \tilde{\bar{\G}}_-)_n = \frac{i\G}{\w}
[&J_{n+1}(-2iA) - J_{n-1}(-2iA)\\
+&J_{n+1}(+2iA) - J_{n-1}(+2iA)]=
(\partial_A \tilde{\bar{\bar{\G}}}_-)_{n0}
\end{split}
\end{equation}
The ``numerical form'' of the function $\a_2$ reads:

\be
\bs
\a_2 =
 -\frac{1}{A}\Big(\bar{C},\,&M^{-1}\tilde{\bar{\bar{\G}}}_-
 (\partial_{\phi})^{-1}_{PS}
 \big(\bar{\bar{S}}\tilde{\bar{\G}}_-
 -\bar{\bar{S}}\tilde{\bar{\bar{\G}}}_+\tilde{G}\\
 &-A\bar{\bar{C}}M^{-1}(\partial_A\tilde{\bar{\G}}_--\partial_A (M)\tilde{G})\big)\Big)
\end{split}
\end{equation}
where we have also introduced the $\sin$ and $\cos$ Fourier matrices:

\be
 \bar{\bar{S}} = \frac{1}{2i}
 \begin{pmatrix} \phantom{-}0 & \phantom{-}1 & 0 & \cdots\\
                           -1 & \phantom{-}0 & 1 & \cdots\\
                 \phantom{-}0 &           -1 & 0 & \ddots\\
                 \phantom{-}\vdots & \phantom{-}\vdots & \ddots &\ddots\\        \end{pmatrix}, \quad
 \bar{\bar{C}} = \frac{1}{2}
 \begin{pmatrix} 0 & 1 & 0 & \cdots\\
                 1 & 0 & 1 & \cdots\\
                 0 & 1 & 0 & \ddots\\
                 \vdots & \vdots & \ddots & \ddots \\        \end{pmatrix}
\end{equation}
The other $\a$ and $\b$ functions are evaluated in a similar way
and the calculations involved do not present any  further
complication.

\subsection{Switching Rates}

We have proven in section \ref{sec:nofast} that the SDQS dynamics in the coexistence
regime can be described by a Kramers equation for the amplitude
probability distribution $\mathcal{W}$ with a given potential
$\mathcal{V(A)}$ and diffusion ``constant'' $D(A)$. We then
dedicated the last section to describe an affordable and reliable
numerical evaluation of the $\a$ and $\b$ functions that appear in
the definition of the potential and diffusion constant. We are now
able to identify completely the SDQS operating in the coexistence
regime as a particular realization of the model for a dichotomous
process between current modes presented in section
\ref{sec:dichotomous}. The shuttling and tunneling regimes with
their characteristic currents are the two modes. Within the
framework of the Kramers escape time theory we can now calculate
the switching rates between these two modes.

The effective potential $\mathcal{V}$ that we obtained has for
parameters that correspond to the coexistence regime, a typical
double well bistable shape (Fig.~\ref{fig:Copotexample}). We
assume for a while the diffusion constant to be independent of
the amplitude $A$. In this approximation the stationary solution
of the equation (\ref{eq:KramersA}) reads:

\be
 \mathcal{W}^{\rm stat}(A)
 = \frac{1}{\mathcal{Z}} \exp\left(-\frac{\mathcal{V}(A)}{D}\right)
\end{equation}
where $\mathcal{Z}$ is the normalization $\mathcal{Z} =
\int_0^{\infty}\mathcal{W}^{\rm stat}(A) dA $. The probability
distribution is concentrated around the minima of the potential
and has a minimum in correspondence of the potential barrier. If
this potential barrier is high enough (i.e.\ $\mathcal{V}_{\rm max}
- \mathcal{V}_{\rm min} \gg D$) we clearly identify two distinct
states with definite average amplitude: the lower amplitude state
corresponding to the tunneling regime and the higher to the
shuttling (see fig. \ref{fig:Codisexample}).

%%%%%%%%%%%%%%%%%%%%%%%%%%%%%%%%%%%%%%%%%%%%%%%%%
% Figure
%%%%%%%%%%%%%%%%%%%%%%%%%%%%%%%%%%%%%%%%%%%%%%%%%
\begin{figure}
 \begin{center}
 \includegraphics[angle=0,width=.6\textwidth]{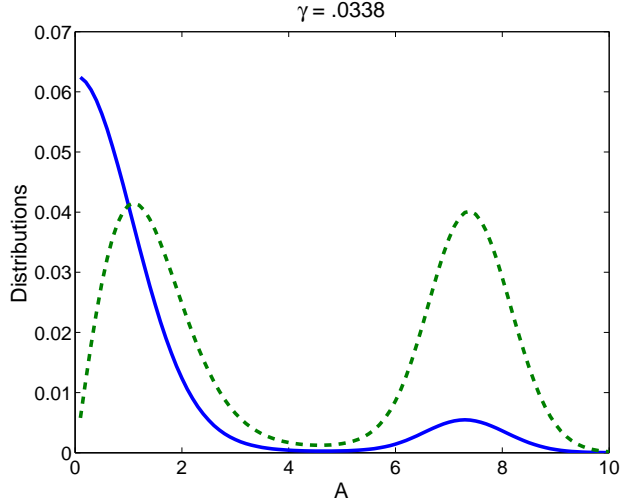}
  \caption{\small  \textit{Example of the stationary distribution  $\bar{W}_+^{\rm stat}$ (full) and the
  amplitude distribution $\mathcal{W}^{\rm stat}$ (dashed)  for the SDQS in the coexistence regime.
  The tunneling and shuttling states are in both cases well separated.}
  \label{fig:Codisexample}}
 \end{center}
\end{figure}
%%%%%%%%%%%%%%%%%%%%%%%%%%%%%%%%%%%%%%%%%%%%%%%%%%%

The Kramers equation (\ref{eq:KramersA}) describes the coexistence
regime of a SDQS by mapping it into a classical model for a
particle moving in a bistable potential $\mathcal{V}$ with random
forces described by the diffusion constant $D$. It is this random
force that allows the particle to overcome the barrier and jump
between the two minima with definite rates. These rates are in the
model for the SDQS the switching rates between the tunneling and
shuttling mode $\G_{\rm in}$ and $\G_{\rm out}$ that we introduced
in section \ref{sec:coexistence-reg}. The bistable potential model
and the problem of calculating the average escape time (i.e. the
average time necessary to leave one potential well) has been
object of intense study because, despite its simplicity, it finds
application, in different local variations, in many branches of
science\footnote{For a review on this problem-- also called the
exit problem or the escape time problem-- see for example
\cite{nae-sia-90} or \cite{lud-sia-75}.}. The first result on this
problem, published in 1899 is the Arrhenius law for the chemical
reaction rate $\kappa$ (which Arrhenius attributed to van't Hoff)
\be
 \kappa = \W e^{-E_C/k_BT}
\end{equation}
where $E_C$ is the activation energy, $k_B$ the Boltzmann constant,
$T$ is the absolute temperature and $\W$ the attempt frequency or
stearic factor. We base our derivation of the escape rates
$\G_{\rm in}$ and $\G_{\rm out}$ on the general treatment given in
the book ``The Fokker-Planck Equation Methods of Solution and
Applications'' by H.~Risken \cite{risken}.

\subsubsection{Mean First Passage Time (MFPT) for a random variable}

Given a stochastic variable $\xi(t)$ (in our case the amplitude
$A(t)$) we can ask the question when this variable first leaves a
certain domain (e.g.\ one of the potential wells). This time is
called first-passage time. For definiteness we choose the initial
condition for the amplitude $A$ to be in the tunneling (lower
amplitude) well and we ask when the amplitude is leaving for the first
time the tunneling well to pass into the shuttling well. Since $A$
is a random variable, also the first-passage time is a random
variable since it depends on the specific walk of the amplitude.
We want to calculate the distribution function for the
first-passage times and in particular its first moment, the
mean-first passage time. This time would constitute in our
specific case the inverse of the tunneling to shuttling switching
rate $\G_{\rm out}$. We set the upper border $A_{\rm out}$ between the
saddle point amplitude $A_S$ and the shuttling amplitude $A_{\rm
shut}$. We also set an arbitrary lower reflecting border in $A_{\rm min}$. \footnote{The exact position of the borders is not relevant
as far as they are far from the minimum or the maximum of the
potential.}(see Fig. \ref{fig:ratesscheme}).

%%%%%%%%%%%%%%%%%%%%%%%%%%%%%%%%%%%%%%%%%%%%%%%%%
% Figure
%%%%%%%%%%%%%%%%%%%%%%%%%%%%%%%%%%%%%%%%%%%%%%%%%
\begin{figure}
 \begin{center}
 \includegraphics[angle=0,width=.6\textwidth]{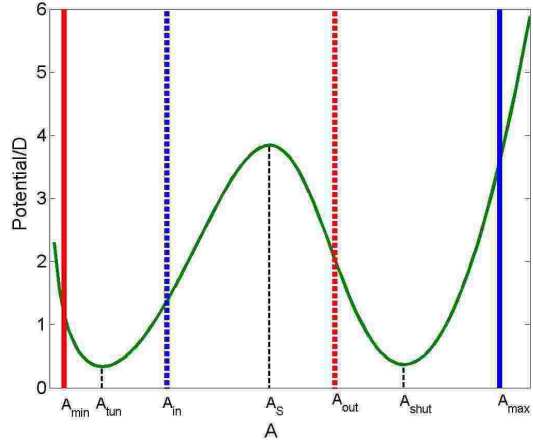}
  \caption{\small  \textit{Bistable effective potential for the SDQS coexistence regime.
  The important amplitudes for the calculation of the rates are indicated.
  The reflecting (full) and absorbing (dashed) borders for the calculation of the
  escape rates.}
  \label{fig:ratesscheme}}
 \end{center}
\end{figure}
%%%%%%%%%%%%%%%%%%%%%%%%%%%%%%%%%%%%%%%%%%%%%%%%%%%

We define the probability density $P(x,t|A_{\rm tun},0)$ for the
stochastic variable $A(t)$ starting at $t=0$ with $A(0)=A_{\rm
tun}$ to be in $x$ at time $t$ without reaching the upper border
$A_{\rm out}$. It is possible to demonstrate
\cite{risken} that the probability $P$ must obey the Kramers
equation (\ref{eq:KramersA}) if $A_{\rm min}<x<A_{\rm out}$, is
identically zero for $x \geq A_{\rm out}$ since the walks that touch the upper border
are not taken into account and satisfy reflecting boundary conditions at the lower border, namely:

\be
 \bs
 &\frac{\partial P}{\partial t} =\mathcal{L}_K P;
 \quad
 P(x,0|A_{\rm tun},0)= \d(x-A_{\rm tun})
 \quad \textrm{for} \quad
 A_{\rm min}<x<A_{\rm out}\\
 & P(x,t|A_{\rm tun},0) = 0
 \quad \textrm{for} \quad
 x=A_{\rm out}\\
 & \partial_x e^{\frac{\mathcal{V}(x)}{D}}
 P(x,t|A_{\rm tun},0)|_{x = A_{\rm min}}
= 0
 \end{split}
\end{equation}
where $\mathcal{L}_K$ is the Kramers' Liouvillean of equation
(\ref{eq:KramersA}). The probability $W(A_{\rm tun},t)$ of
realizations which have started at $A_{\rm tun}$ and which have
not yet reached the upper boundary up to the time $t$ is
given by:

\be
 W(A_{\rm tun},t) = \int_{A_{\rm min}}^{A_{\rm out}} P(x,t|A_{\rm
 tun},0)dx
\end{equation}
The probability $-dW$ of those realizations which reach the
upper boundary in the time interval $(t,t+dt)$ thus reads

\be
 -dW(A_{\rm tun},t) = -\int_{A_{\rm min}}^{A_{\rm out}} \dot{P}(x,t|A_{\rm
 tun},0)dxdt
\end{equation}
The distribution $w(T)$ function for the first passage time T is
therefore given by

\be
 w(A_{\rm tun},T) = -\frac{dW(A_{\rm tun},T) }{dT} =
 -\int_{A_{\rm min}}^{A_{\rm out}} \dot{P}(x,T|A_{\rm
 tun},0)dx
\end{equation}
The moments of the first-passage time distribution are
\be
 T_n(A_{\rm tun})
 = \int_0^{\infty} T^nw(A_{\rm tun},T)dT
 = \int_{A_{\rm min}}^{A_{\rm out}}p_n(x,A_{\rm tun})dx
\end{equation}
where $p_n(x,A_{\rm tun})$ is defined by
\be
 p_n(x,A_{\rm tun})
 = -\int_0^{\infty}T^n \dot{P}(x,T|A_{\rm
tun},0)dT
\end{equation}
We obviously have
\be
 p_0(x,A_{\rm tun})
 = -\int_0^{\infty} \dot{P}(x,T|A_{\rm
tun},0)dT
 = P(x,0|A_{\rm tun},0) = \d(x-A_{\rm tun})
\end{equation}
Performing a partial integration gives the relation
\be
 p_n(x,A_{\rm tun})
 = n\int_0^{\infty}T^{n-1}P(x,T|A_{\rm
tun},0)dT
 \label{eq:recursiveT}
\end{equation}
Applying the Kramers' Liouvillean $\mathcal{L}_K$ to equation
(\ref{eq:recursiveT}) we obtain the recursive relation:
\be
 \mathcal{L}_K[p_n(x,A_{\rm tun})] = -np_{n-1}(x,A_{\rm tun})
\end{equation}
and in particular
\be
 \mathcal{L}_K[p_1(x,A_{\rm tun})] = -\d(x-A_{\rm tun})
 \label{eq:LKp1}
\end{equation}
In other words the function $p_1(x,A_{\rm tun})$ that enters the
definition of the average first-passage time is the Green's
function of the Kramers' Liouvillean $\mathcal{L}_K$. The Kramers'
Liouvillean can be written in the form:

\be
 \mathcal{L}_K = D \partial_A e^{-\frac{\mathcal{V}(A)}{D}}
                   \partial_A e^{+\frac{\mathcal{V}(A)}{D}}
\end{equation}
as can be easily proven by acting with the last partial derivative
on the exponential. It is also possible\footnote{Basically it is a
double integration of equation (\ref{eq:LKp1}).}  to give an
analytic expression for the Green's function $p_1(x,A_{\rm tun})$
of the Kramers' Liouvillean $\mathcal{L}_K$:

\be
 p_1(A,A_{\rm tun}) =
 \frac{1}{D} e^{-\frac{\mathcal{V}(A)}{D}}
 \int_A^{A_{\rm out}}e^{\frac{\mathcal{V}(B)}{D}}
 \left[
 \int_{A}^{A_{\rm out}} \d(C - A_{\rm tun})dC\right]dB
\end{equation}
The mean-escape time from the tunneling well reads:
\be
 T_1(A_{\rm tun}) = \frac{1}{D}
 \int_{A_{\rm min}}^{A_{\rm out}}\left\{e^{-\frac{\mathcal{V}(A)}{D}}
 \int_A^{A_{\rm out}}e^{\frac{\mathcal{V}(B)}{D}} \left[
 \int_{A_{\rm min}}^B \d(C - A_{\rm tun})dC\right]dB\right\}dA
 \label{eq:averagetime}
\end{equation}

The escape time from the shuttling well can be calculated exactly
in the same way using $A_{\rm shut}$ as starting point for the
random process and $A_{\rm in}$ and $A_{\rm max}$ as boundaries
(see Fig.~\ref{fig:ratesscheme}). The calculation of the integrals
in (\ref{eq:averagetime}) can be simplified considering that the
integrand has a sharp maximum in the region $(A,B)\approx (A_{\rm
tun},A_{\rm S})$ due to the behaviour of the effective potential
$\mathcal{V}$ around those two points. The escape time from the
shuttling well can be calculated in a similar way. We obtain in
the end the switching rates:

\be
 \bs
 \G_{\rm out}&= D
\left(
 \int_{A_{\rm tun}}^{A_{\rm out}} dB \,
  e^{ \frac{{\mathcal V}(B)}{D}}
 \int_{A_{\rm min}}^{B} dA \,
  e^{-\frac{{\mathcal V}(A)}{D}}
\right)^{-1}\\
 \G_{\rm in}&= D
\left(
 \int_{A_{\rm in}}^{A_{\rm shut}}dB \,
 e^{\frac{{\mathcal V }(B)}{D}}
 \int_{B}^{A_{\rm max}}dA\,
 e^{-\frac{{\mathcal V}(A)}{D}}
 \right)^{-1}\\
 \end{split}
\end{equation}

We recall here the equation for the current and the Fano factor
for a dichotomous process inserting now the particular currents
and switching rates characteristic of the SDQS coexistence regime:

\be
 \bs
 I^{\rm stat} &= \frac{I_{\rm sh} \G_{\rm out} + I_{\rm tun}\G_{\rm in}}{\G_{\rm in}+ \G_{\rm out}}\\
 F &= \frac{S(0)}{I^{\rm stat}} =
  2\frac{(I_{\rm sh}-I_{\rm tun})^2}
        {I_{\rm sh} \G_{\rm out} + I_{\rm tun}\G_{\rm in}}
   \frac{\G_{\rm in}\G_{\rm out}}{(\G_{\rm in}+\G_{\rm out})^2}
 \end{split}
\end{equation}

They represent, together with the stationary distribution

\be
 \mathcal{W^{\rm stat}}(A) = \frac{1}{\mathcal{Z}} \exp\left(-\frac{\mathcal{V}(A)}{D}\right)
\end{equation}
the starting point for a quantitative comparison between the
simplified model and the full description.

\subsection{Comparison}

\subsubsection{The classical (but ``noisy'') limit}

The comparison between the dichotomous process model and the full
description for the coexistence regime is based, as for the other
regimes, on the three investigation tools: phase-space
distribution, current and current-noise.

The phase space distribution is the most sensitive method to
compare the model and the full description. One of the basic
procedures adopted in the derivation of the Kramers equation
(\ref{eq:KramersA}) is the expansion to second order in the small
parameters (\ref{eq:smallparameters}). In order to test the
reliability of the model we simplify as much as possible the
description reducing the model to a classical description: namely
taking the zero limit for the parameter
$\left(\frac{x_0}{\l}\right)$. We realize physically this
condition assuming a large temperature and a tunneling length $\l$
of the order of the thermal length $\l_{\rm
th}=\sqrt{\frac{k_BT}{m\w^2}}$. We partially discussed this limit
in the previous chapters. Also the full description is slightly
changed, but not qualitatively: the three regimes are still
clearly present with their characteristics. The numerical
calculation is though based on a totally different
approach\footnote{The continued fraction method was applied. See
for example \cite{risken} for the application of this method to
the solution of Fokker-Planck equations. The code was developed by
Dr.~T.~Novotn\'y.}.

%%%%%%%%%%%%%%%%%%%%%%%%%%%%%%%%%%%%%%%%%%%%%%%%%
% Figure
%%%%%%%%%%%%%%%%%%%%%%%%%%%%%%%%%%%%%%%%%%%%%%%%%
\begin{figure}[h]
 \begin{center}
 \includegraphics[angle=0,width=.45\textwidth]{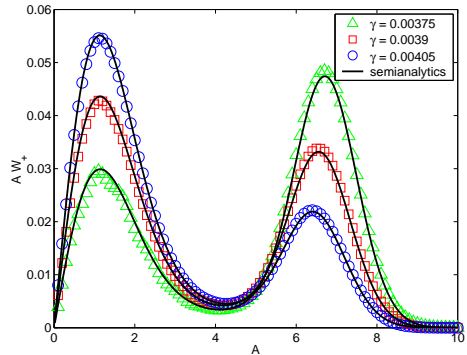}
  \caption{\small  \textit{Stationary amplitude probability distribution $\mathcal{W}$
  for the SDQS in the coexistence regime. We compare the results obtained from the simplified
  model (full line) and from the full description (circles, triangles and squares).
  These results are obtained in the classical high temperature regime $k_BT \gg
  \hbar\w$. The amplitude is measured in units of
  $\l_{\rm th} = \sqrt{\frac{k_BT}{m\w^2}}$. The mechanical damping $\g$ in units of the mechanical
  frequency $\w$. The other parameter values are $d = 0.05\l_{\rm th}$ and $\G =
 0.015\w$, $\lambda = 2 \lambda_{\rm th}$.}
  \label{fig:CompWF}}
 \end{center}
\end{figure}
%%%%%%%%%%%%%%%%%%%%%%%%%%%%%%%%%%%%%%%%%%%%%%%%%%%

In figures \ref{fig:CompWF},
\ref{fig:Compcurr} and \ref{fig:CompNoise} we present respectively
the results for the stationary Wigner function, the current and
the Fano factor in the semiclassical approximation and full
description.
%%%%%%%%%%%%%%%%%%%%%%%%%%%%%%%%%%%%%%%%%%%%%%%%%
% Figure
%%%%%%%%%%%%%%%%%%%%%%%%%%%%%%%%%%%%%%%%%%%%%%%%%
\begin{figure}[h]
 \begin{center}
 \includegraphics[angle=0,width=.51\textwidth]{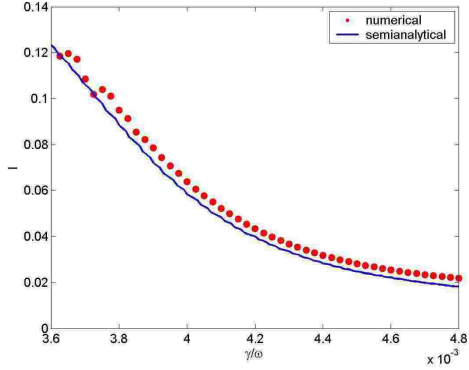}
  \caption{\small  \textit{Current in the coexistence regime of SDQS.
  Comparison between semianalytical and full numerical description.
  For the parameter values see Fig.~\ref{fig:CompWF}.}
  \label{fig:Compcurr}}
 \end{center}
\end{figure}
%%%%%%%%%%%%%%%%%%%%%%%%%%%%%%%%%%%%%%%%%%%%%%%%%%%
%%%%%%%%%%%%%%%%%%%%%%%%%%%%%%%%%%%%%%%%%%%%%%%%%
% Figure
%%%%%%%%%%%%%%%%%%%%%%%%%%%%%%%%%%%%%%%%%%%%%%%%%
\begin{figure}[h]
 \begin{center}
 \includegraphics[angle=0,width=.51\textwidth]{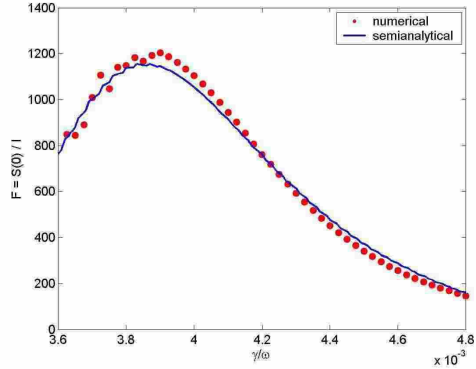}
  \caption{\small  \textit{Fano factor in the coexistence regime of SDQS.
  Comparison between semianalytical and full numerical description.
  For the parameter values see Fig.~\ref{fig:CompWF}.}
  \label{fig:CompNoise}}
 \end{center}
\end{figure}
%%%%%%%%%%%%%%%%%%%%%%%%%%%%%%%%%%%%%%%%%%%%%%%%%%%

\subsubsection{The quantum limit}

The coexistence regime in the parameters corner typically
presented in this thesis (e.g.\ Fig.~\ref{fig:SDQSCurrent}) is not
captured by the model we introduced. The concept of elimination of
the fast dynamics is still correct, but the small parameters
expansion fails. The effective potential calculated from a second
order expansion still gives the position of the ring structure
with reasonable accuracy but the overall stationary Wigner
function is not reproduced due to a non convergent diffusion
function $D(A)$. It is clear that we need to consider higher order
terms in the parameter $(x_0/\l)^2$. This represents nevertheless
a fundamental problem since it would produce terms with higher
order derivatives with respect to the amplitude $A$ in the
Fokker-Planck equation and consequently, to our knowledge, the
break down of the escape time theory.

It has nevertheless been demonstrated  with the help of the higher
cumulants of the current that the dichotomous process description
of the coexistence regime is valid also more in the quantum regime
($\l = 1.5x_0$), the only necessary condition being a separation
of the ring and dot structures in the stationary Wigner function
distribution \cite{fli-preprint-05}.

Motivated by the fact that the effective potential was able to
give the correct position of the shuttling ring we used the
diffusion constant as a fitting parameter. The current and the Fano
factor are very nicely reproduced in terms of a fitted diffusion
constant approximately doubled with respect to the one calculated
at zero temperature with no $\left(\frac{x_0}{\l}\right)^4$
corrections\footnote{Even if at this level it could appear
premature we wonder if this correction can be attributed to an
effective temperature felt by the oscillator in contact with the
electrical system lead-dot-lead far from equilibrium. A.~Armour
{\it et al.} support this hypothesis in similar devices
\cite{arm-prb-04} even if in slightly different regimes.} (Figs.
\ref{fig:fittedcurr} and \ref{fig:fittednoise}).

%%%%%%%%%%%%%%%%%%%%%%%%%%%%%%%%%%%%%%%%%%%%%%%%%
% Figure
%%%%%%%%%%%%%%%%%%%%%%%%%%%%%%%%%%%%%%%%%%%%%%%%%
\begin{figure}[h]
 \begin{center}
 \includegraphics[angle=0,width=.5\textwidth]{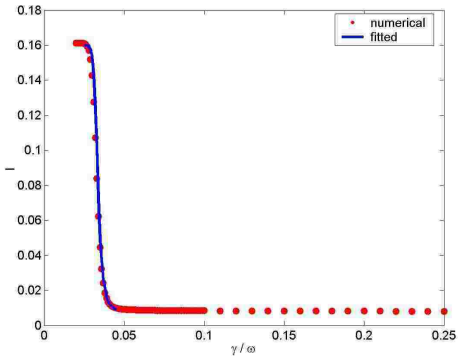}
  \caption{\small  \textit{Current in the coexistence regime of SDQS.
  Comparison between semianalytical description with an effective fitted
  diffusion constant (see the text for explanation) and full numerical description.
  The parameter values are $\G = 0.01\w$, $\l=2x_0$, $d =
  0.5x_0$, $T=0$.}
  \label{fig:fittedcurr}}
 \end{center}
\end{figure}
%%%%%%%%%%%%%%%%%%%%%%%%%%%%%%%%%%%%%%%%%%%%%%%%%%%

%%%%%%%%%%%%%%%%%%%%%%%%%%%%%%%%%%%%%%%%%%%%%%%%%
% Figure
%%%%%%%%%%%%%%%%%%%%%%%%%%%%%%%%%%%%%%%%%%%%%%%%%
\begin{figure}
 \begin{center}
 \includegraphics[angle=0,width=.5\textwidth]{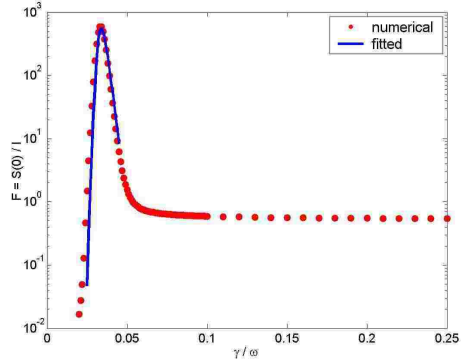}
  \caption{\small  \textit{Fano factor in the coexistence regime of SDQS.
  Comparison between semianalytical description with an effective fitted
  diffusion constant (see the text for explanation) and full numerical description.
  The parameter values are $\G = 0.01\w$, $\l=2x_0$, $d =
  0.5x_0$, $T=0$.}
  \label{fig:fittednoise}}
 \end{center}
\end{figure}
%%%%%%%%%%%%%%%%%%%%%%%%%%%%%%%%%%%%%%%%%%%%%%%%%%%

The full numerical description presented in this thesis is
relatively fragile with respect to the modification of the
tunneling length\footnote{We refer to changes of the tunneling
length of one order of magnitude.} at least for two reasons: the
enlargement of the shuttling amplitude enlarges the number of
oscillator states relevant for the dynamics to untractable
numbers; the preconditioning loses effectiveness and the problem
is not convergent. For this reason we are (still) lacking a
numerical benchmark for the intermediate range of values of the
tunneling length $\l > x_0$ but which is still non-classical where
a second order expansion in the small parameter
$\left(\frac{x_0}{\l}\right)^2$ is probably a good approximation.

\clearpage{\pagestyle{empty}\cleardoublepage}

%%%%%%%%%%%%%%%%%%%%%%%%%%%%%%%%%%%%%%%%%%%%%%%%%%%%%%%
%               CONCLUSIONS
%%%%%%%%%%%%%%%%%%%%%%%%%%%%%%%%%%%%%%%%%%%%%%%%%%%%%%%
\chapter{Conclusions}

We summarize in this chapter the main results presented in the
thesis. The richness of the shuttle model  still leaves many
points open in the description and understanding of the device
dynamics. We close the chapter listing some of the open questions
that could encourage a continuation of the present work.

\section{Summary}

We analyzed in this thesis the dynamics of two different shuttle
devices: the single- and the triple-dot quantum shuttles. Shuttle
devices are nanoelectromechanical systems (NEMS) in which an
oscillating quantum dot can transfer electrons one per cycle from
source to drain lead. The dynamics of such devices is the result
of a strong interplay between their mechanical and electrical
degrees of freedom.

The electrical dynamics is permeated of quantum effects: electrons
can go from the source to the quantum dot and finally to the drain
lead only via tunneling events; the small capacitance of the
oscillating dot and the corresponding large charging energy
reveals the discrete nature of the electrical charge; finally the
presence of three coupled quantum dots is also a source of
coherent electrical dynamics in the TDQS. The small size of the
device makes quantum effects important also for the mechanical
degree of freedom. The resonance frequency in such small (and
relatively stiff) devices can be in the order of GHz. Consequently
the quantum of energy of the harmonic oscillator is comparable
with other energies in the device.

For these reasons for the SDQS we developed \emph{quantum models}:
we extended for the SDQS the classical model proposed by Gorelik {et
al.}~\cite{gor-prl-98} and we adopted for the TDQS the one already
existing invented by A.~Armour and A.~MacKinnon \cite{arm-prb-02}.

Among the different approaches for the description of the dynamics
of a quantum system we chose the \emph{Generalized Master
Equation} (GME). GME's  are particularly well-suited to study
small open quantum systems (i.e. systems with a small number of
degrees of freedom in contact with resevoirs but still conserving
some coherent dynamics) since they neglect the many (irrelevant)
degrees of freedom of the resevoirs still keeping track of the
residual quantum coherencies of the small system.

The shuttle devices are small quantum systems in contact with two
kinds of resevoirs: the leads and the thermal bath. The strength
of the coupling is very different and calls for a different
treatment. The weak coupling between the thermal bath and the
mechanical degree of freedom of the device justifies the
perturbation theory and the quantum-optical derivation of a GME in
Born-Markov approximation. The tunneling coupling of the shuttling
devices to their electrical baths (the leads) is \emph{not} weak.
It sets, on the contrary, the time scale of the electrical
dynamics that in the shuttling regime is comparable with the
period of the mechanical oscillations in the system. This strong
coupling condition is  realized by a tunneling amplitude modulated
by the displacement of the quantum dot from the equilibrium
position in the SDQS while is directly set by a model parameter in
the TDQS. In order to handle the strong coupling we adopted the
\emph{Gurvitz method} for the derivation of the GME and extended
the original formulation to take into account the mechanical
degree of freedom.

The shuttle dynamics has, especially in the single dot device, an
appealing simple classical interpretation and one can say that the
name itself of ``shuttle'' suggests the idea of sequential and
periodical loading, mechanical transport and unloading of
electrons between a source and a drain lead. For this reason,
while  preserving the complete quantum treatment that we achieved
with the GME, we wanted to keep as much as possible the intuitive
classical picture and the possibility to handle the
quantum-classical correspondence. The \emph{Wigner function
distribution} seemed to us a good answer to all these
requirements. It allows a clear visualization of the numerical
results obtained within the framework of the GME and it shows in
its equation of motion (the Klein-Kramers equation) an explicit
quantum-classical correspondence (expansion in powers of $\hbar$)
\cite{hil-phr-84}.

Having the methods for the analysis of the shuttle  device
dynamics we looked for \emph{investigation tools}. We first
realized that a clear finger-print of the shuttling regime that
distinguishes it from other kinds of phonon assisted tunneling is
the charge-position(momentum) correlation. The Wigner distribution
function on the \emph{device} phase space (i.e. also charge
resolved to take into account the two electrical states of the
quantum dot) is a perfect tool to visualize this property. The
only problem with the  Wigner distribution functions is that it
seems very difficult (if not impossible) to access experimentally
these functions. For this reason, with the guidance of the
distribution function description, we investigated the electronic
transport properties in the shuttle devices (more realistically
measurable): first the stationary current and then the
current-noise\footnote{There is already a work that, in the
semiclassical regime, addresses the aspect of full counting
statistics for these devices \cite{pis-prb-04}. Results for the
quantum case will soon appear \cite{fli-preprint-05}.}. A special
r\^{o}le in the calculation of the current-noise for the SDQS was
played by a specific characteristic of the GME obtained in the
Gurvitz approach. In this method the information on the state of
the leads in not completely traced out and the number $n$ of
electrons that passed through the device after a specific time
appears in the equation of motion for the so-called $n$-resolved
reduced density matrix.

With these tools we investigated the properties of the quantum
shuttle devices. The sharp tunneling-shuttling transition found by
Gorelik {\it et al.} for the semiclassical model turned into a
smooth crossover. We also could recognize the onset of the shuttling
regime triggered by quantum mechanical noise with no external
electric field acting on the oscillating quantum dot. The shuttle
regime also revealed the characteristic current quantization (in
our set-up a saturation due to Coulomb blockade to one electron
per cycle in the cleanest shuttling regime) and the extremely low
noise typical of a quasi-deterministic transport regime. With all
the three investigation tools we adopted we could recognize
\emph{three operating regimes} for the shuttle device: tunneling,
shuttling and coexistence regime. They represent for the SDQS the
whole scenario of possible dynamics. Part of the complexity of the
TDQS is also captured in this scheme\footnote{For a further
insight into this particular model see \cite{arm-prb-02, nara,
christian, fli-preprint-04}.}.

The specific separation of time scales in the different regimes
allowed us to identify the relevant variables and describe each
regime by a specific \emph{simplified model}. In the tunneling
(high damping) regime the mechanical degree of freedom is almost
frozen and all the features revealed by the Wigner distribution,
the current and the current noise can be reproduced with a
resonant tunneling model with tunneling rates renormalized due to
the movable quantum dot. Most of the features of the shuttling
regime (self-sustained oscillations, charge-position correlation
and current quantization) are captured by a simple model derived
as the zero-noise limit of the full description. Finally for the
coexistence regime we proposed a dynamical picture in terms of
slow dichotomous switching between the tunneling and shuttling
modes. This interpretation was mostly suggested by the presence in
the stationary  Wigner function distributions of both the
characteristic features of the tunneling and shuttling dynamics
and by a corresponding gigantic peak in the Fano factor. We based
the derivation of the simplified model on the fast variables
elimination from the Klein-Kramers equations for the
Wigner function\footnote{The analytical derivation of the effective
potential is an extension of the work done by Fedorets {\it et
al.} \cite{fed-prl-04}.} and a consequent derivation of an effective
bistable potential for the amplitude of the dot oscillation (the
relevant slow variable in this regime).

\section{Some open questions}

During the developing of the present work many questions arose
about possible phenomena to investigate in different shuttle
devices or other possible regimes that the systems we investigate
could exhibit in some other parameter corner that we did not
reach. This questions could be of interest for example in
understanding or proposing future experiments. We list them here
as a spur for possible continuation of the present work sure that
some of the theoretical tools presented in this thesis are
definitely useful for those tasks.
\begin{itemize}

\item{
The mechanical  bath introduces decoherence in the mechanical
degree of freedom, the leads in the electrical. Correlations are
left (in the SDQS) only in the off-diagonal elements of the charge
resolved density matrices. Can we imagine an electromechanical
system that instead exhibits quantum coherence in the mechanical
degree of freedom? Could we use the electromechanical interaction
to ``pump'' continuously coherence in the device?}

\item{
The stationary solution for the GME is incoherent. Is there any
possibility to reduce the GME to an effective Master equation? We
imagine that especially in the coexistence regime this could help
to avoid the Wigner function small parameter expansion for the
calculation of the effective bistable potential and lead directly
to the calculation of the rates for the slow switching process.
This would make the calculation of the parameters for the
dichotomous process description much more general.}

\item{
Taking into account the spin degrees of freedom and the Coulomb blockade the
incoming bare rate should be considered as double with respect to
the out going rate. This fact does not change qualitatively the
picture since the switching is, in first approximation a
non-dissipative process and the electrostatic force is a small
perturbation of the oscillator restoring force. The mechanical
trajectories are not significantly changed (maybe slightly
decreased in radius) and only the symmetry between the two charged
resolved Wigner functions is modified in favour of $W_{11}$:
the dot is charged for a longer time. But what happens if we
consider a spin dynamics \emph{on} the dot due for example to a
coupling to an external magnetic field\footnote{A partial answer
to this questions is contained in the very recent paper by
Fedorets {\it et al.}~\cite{fed-preprint-04}.}? }

\item{
At high bare injection rates ($\G \gg \w$) the shuttle device is
more stable and hardly leaves the tunneling regime. The mechanical
motion cannot in fact follow the fast electrical forcing and the
effect is generally uncorrelated resonant tunneling. Nevertheless
at extremely low damping the shuttling regime arises. The transition
is very different from the small injection rate shuttling
instability: it is very smooth and the ring structure gradually
emerges from the central fuzzy spot typical for the tunneling
regime. The general question is: can we draw a phase diagram for
shuttling devices as a function of their parameters?}

\item{
A big issue regards the treatment of the bias in the quantum
description. All our equations were derived in the limit of high
external bias. We did not have access to this parameter and we
used for this reason the mechanical damping as a control
parameter. Is it possible to derive a GME at low bias and for
strong couplings to the leads? Probably the dynamics will get
non-Markovian and all kinds of coherent transport effects will
also enter the game. We know from the semiclassical treatment that
the shuttling instability should be present. Can we understand
something new from the quantum treatment?}

\end{itemize}

\clearpage{\pagestyle{empty}\cleardoublepage}

%\include{Papers}

%% Adds the line "Bibliography" to TOC
\addcontentsline{toc}{chapter}{Bibliography}
\bibliography{shuttle}
\end{document}